\documentclass[aps,prl,reprint,superscriptaddress,showpacs]{revtex4-1}
\usepackage{graphicx}  
\usepackage{dcolumn}   
\usepackage{bm}        
\usepackage{verbatim}   

\usepackage[version=3]{mhchem} 
\usepackage{amsmath,amssymb}
\usepackage{textcomp} 
\usepackage{enumitem}
\usepackage{color}
\usepackage{psfrag}
\usepackage[breaklinks,colorlinks,citecolor=blue,linkcolor=blue,urlcolor=blue]{hyperref}
\usepackage{url}

\usepackage{amsfonts}
\usepackage[raggedright]{titlesec}

\usepackage{titlesec}

\titleformat{\paragraph}
{\normalfont\normalsize\bfseries}{\theparagraph}{1em}{}
\titlespacing*{\paragraph}
{0pt}{3.25ex plus 1ex minus .2ex}{1.5ex plus .2ex}

\DeclareUnicodeCharacter{2212}{-}

\begin{document}

\title{Molecular Collisions: from Near-cold to Ultra-cold}

\author{Yang Liu}
\email[]{liuyang59@mail.sysu.edu.cn}
\altaffiliation{School of Physics and Astronomy, Sun Yat-Sen University, Zhuhai, 519082, China}

\author{Le Luo}
\email[]{luole5@mail.sysu.edu.cn}
\affiliation{School of Physics and Astronomy, Sun Yat-Sen University, Zhuhai, 519082, China}

\date{\today}

\begin{abstract}
In the past two decades, the revolutionary technologies of creating cold and ultracold molecules have provided cutting-edge experiments for studying the fundamental phenomena of collision physics. To a large degree, the recent explosion of interest in the molecular collisions has been sparked by dramatic progress of experimental capabilities and theoretical methods, which permit molecular collisions to be explored deep in the quantum mechanical limit. Tremendous experimental advances in the field has already been achieved, and the authors, from an experimental perspective, provide a review of these studies for exploring the nature of molecular collisions occurring at temperatures ranging from the Kelvin to the nanoKelvin regime, as well as for applications of producing ultracold molecules.



\end{abstract}

\maketitle

\tableofcontents

\section{Introduction}

\vspace{-1mm}

Collision processes governed by intra-particle interaction play a pivotal role in nuclear physics, condensed matter physics, atomic and molecular physics, and chemistry. Understanding the quantum nature of these collisions is important for controlling chemical reaction, precision measurement, improving energy efficiency, and exploring novel phases of matter. Quantum collisions related to chemical kinetics and reaction dynamics have been studied using crossed-molecular beam collision \cite{lee1969molecular,herschbach1987molecular}, and then were extended to the cold and ultracold regimes with laser-cooled atoms \cite{weiner1999experiments}, leading to Nobel prizes in both chemistry \cite{chem86} and physics \cite{phys97}.

In recent years, cold and ultracold molecules, due to its complexity of the internal energy level comparing with atoms, have offered exciting opportunities for precision measurements \cite{hudson2006cold,zelevinsky2008precision,chin2009ultracold,baron2014order,kobayashi2019measurement}, many-body physics \cite{baranov2008theoretical,eisert2015quantum}, quantum computation \cite{demille2002quantum,rabl2006hybrid}, quantum simulation \cite{micheli2006toolbox,gorshkov2011tunable,yan2013observation}, and quantum chemistry\cite{balakrishnan2001chemistry,krems2008cold,bell2009ultracold,ni2010dipolar,ospelkaus2010quantum,stuhl2014cold,dulieu2017cold}. To provide a concise introduction of this rapidly developing field,
this article is motivated to review recent progress in near-cold, cold and ultracold collisions associating molecules from experimental perspective. To better categorize the rich collisional physics at low temperature with subtle details, we adopt the technical terminology in \cite{naulin2014experimental, carr2009cold}, $T\geq 1 K$ as near cold, $1\mu K < T < 1 K$ as cold, $T\leq 1\mu K$ as ultracold.

The studies of molecular collision can be dated back to 1970s when molecular reactive collisions have been extensively studied with crossed-molecular beam techniques with great success in obtaining precise information of reaction dynamics for simple atom-molecule system, such as \ce{H + H2} \cite{harich2002forward} and \ce{F + H2} \cite{yang2008dynamical} systems. It is noted that, in most of such studies, the collision energy, i.e. relative kinetic energy is in the regime of room temperature or higher. At these temperature, the de Broglie wavelengths of molecules are on the order of a few thousandths of a nanometer, and the collisions are usually
interpreted as the interaction of hard spheres moving along potential surfaces determined by the electronic charge distribution. Although a few reaction scattering systems could be studied with quantum-state resolution, quantum state-resolved differential cross-sections are extremely difficult to obtain, because the collisional process is usually considered as the combination of collisions of many partial waves which are severely overlapped.
Meanwhile, the averaging over collision energies, initial and final states as well as scattering angles, blurs and smears out any resonance pattern associated with quantum effects. For this high temperature collision, contribution of a single partial wave is hardly to be resolved experimentally.

Thus, near-cold, cold and ultracold molecular collisions became an exciting research field for physicists and chemists with curiosity. On the other hand, the developing of the technology for trapping and controlling colliding molecular species with quantum state resolution gives us the capability to track how molecular species approach each other, interact along their potential energy landscape, and transform to final products through transition state. This capability allows people to understand collision processes with the first principle of quantum mechanics and to control of both coherent and dissipative processes in the collisions.

Much effort has been made to observe the quantum-mechanical effects in molecular collisions. Coralie Berteloite et al. \cite{berteloite2010kinetics} and Astrid Bergeat et al. \cite{bergeat2015quantum} have achieved nearly 4 Kelvin equivalent collision energy by reducing the scattering angle between crossed beams. Joop J. Gilijamse et al. \cite{gilijamse2006near} have demonstrated state- and velocity-controlled crossed-beam collisions for Xe-OH system by using Stark deceleration technique \cite{bethlem1999decelerating}, which has later been extended to many other systems such as OH-NO \cite{kirste2012quantum} and NO-He \cite{vogels2015imaging}. Selective vibrational \cite{liu2001crossed,liu2016vibrational} or rotational excitation \cite{liu2014rotational,shagam2015molecular} have also been investigated, resulting in enhanced or suppressed reaction rates. Despite being quite successful, the relevant energy scale are still limited to few Kelvins. In the past two decades, the creation of cold and ultracold atoms and molecules by laser cooling and trapping techniques also help to create the emergent field of cold and ultracold chemistry, focusing on the developments to the studies of cold and ultracold atom-molecule and molecule-molecule collisions.

There already exist a number of reviews of cold molecular collisions \cite{carr2009cold,wm2008low,krems2009cold,dulieu2017cold, naulin2014experimental,yang2007state,stuhl2014cold}, some of which cover the whole research field, while others focus on a specific temperature regime. This article will not be a full review of the whole field or for physics in a limited temperature range. We would rather map out the main research directions from experimental perspective and focus on important experimental advances which have greatly improve the fundamental understanding of the collision physics for cold molecules and have potentials to be used for exploring the collisions of ultracold molecules.

The article will be organized in following: In the second section, we will review certain relevant theories for experimentally studying near-cold, cold and ultracold molecular collisions. It should be noted that we will not try to give a full list of the theoretical methods, but those that are required to understanding the experiments. Then, in the third, fourth, and fifth section, we review experimental studies of atom-molecule and molecule-molecule collisions in near cold, cold, and ultracold respectively. In the sixth section, we give a brief summary of the recent advances of creating ultracold molecules. In the end, we conclude the progress and present our envision of the future developments in this field.

\section{Theory of molecular collisions}

Since this article reviews the molecular collisions from near-cold to ultracold regime, where the de Broglie wavelengths for translational motion become quite large, and considering collision partners as classical particles becomes meaningless, we would focus our scope on quantum collision theory instead of classical collision theory, detailed treatment of which is given in various textbooks \cite{bernstein1979atom, child1996molecular, levine2009molecular}. In the following, we will give detailed introduction on construction of an accurate potential energy surface, and the treatment of collisions using different scattering theories, leading to the full understanding of energy dependence of the collisonal cross-sections. 

\subsection{Potential energy surface}

The potential energy surface (PES) play a central role in concepts and theories of molecular interactions, and is the key ingredient in the computation and prediction of structure and dynamics. In the Born-Oppenheimer approximation, there exists a PES for each electrnic state of the molecular system. The great obstacle to successful scattering calculations, always the case, is the scarcity of accurate ab initio potentials for systems involving three or more atoms. Additionally, the elastic and inelastic collision cross sections have been found very sensitive to the details of the interaction potential in many collisions by various theoretical models, due to the presence of (quasi)bound state resonance. This subsection will show how the PESs calculation is carried out and how do they influence the collision outcome in some typical molecular systems. 

Usually, in order to obtain a high-precision and accurate PES around region of interest (e.g. van der Waals region of the colliding complex), ab initio electronic structure calculations need to be performed \cite{chalasinski1994origins, chalasinski2000state}. To begin, geometry optimizations on the relevant electronic states are carried out, for example employing a restricted-spin coupled-cluster method with single, double and non-iterative triple excitations \{RCCSD(T)\} \cite{raghavachari1989fifth}, and/or complete-active-space self-consistent-field (CASSCF) \cite{roos1987complete} multireference internally contracted configuration interaction (MRCI) methods \cite{werner1985second}. Basis sets such as augmented correlation-consistent polarized-valence triple-zeta (aug-cc-pVTZ) \cite{dunning1989gaussian} are used in such optimizations together with the frozen core approximation \cite{sachs1975frozen}. Thus, minimum-energy geometry and relative electronic energies of the specific electronic states are computed. In addition, different basis sets can be used to investigate the effects of basis set extension to the complete basis set (CBS) limit and core electron correlation on the computed quantities (namely, equilibrium geometrical parameters and relative electronic energies). Currently, MOLPRO programs \cite{werner2012molpro} has been popular for perfoming these calculations. 

Then, ab initio potential energy functions for the relevant electronic states can be obtained by fitting the polynomial to calculated ab initio electronic energies \cite{schatz1989analytical, hollebeek1999constructing}. Several methods can be used to give reasonably accurate fits to ab initio data, such as rotating Morse spline (RMOS) \cite{kuntz1972ion}, the reproducing kernel Hilbert-space (RKHS) interpolation method \cite{ho2000proper}, the distributed approximating functional method \cite{frishman1997distributed}, many-body expansion approach \cite{garcia1984fit}, and the generalized London-Eyring-Polanyi-Sato double-polynomial (GLDP) method \cite{murrell1984molecular}. These fitting methods should satisfy three criteria: the fitting procedure should be simple, the resulting surface should be smooth and agree with the ab initio data, without oscillations between ab initio points, and the mathematical form should be physically realistic in order to minimize the number of ab initio geometries needed to obtain a surface with the correct features and topology.

In the above calculation, the uncertainties can come from various sources, which are the level of ab initio theory used to approximately solve Schr\"odinger's equation, the size of basis set employed in the calculations, the dimensionality of the PES, the number of grid points used (i.e. the number of ab initio calculations performed), and the accuracy of the analytic fit to the computed energies. Since an accurate PES is the first ingredient for an accurate scattering study, it is therefore very important to control the uncertainties from various sources to obtain optimal accuracy of the PES. 

The calculation of PES for \ce{NH-NH} would be exemplary. The PES can be used to determine the cross sections for elastic and spin-changing \ce{NH-NH} collisions, which in turn determines the feasibility for the evaportive cooling process. Guillaume S. F. Dhont et al. constructed four-dimensional ab initio PESs for \ce{NH-NH} complex, which correlate with two separate NH molecules in their $^{3}\Sigma^{-}$ electronic ground state \cite{dhont2005ab}. The potential surface of the quintet state (S=2) was calculated at the coupled cluster level with RCCSD(T) method. The triplet and singlet states described with multrireference wave functions and the exchange splittings between the potential surfaces were calculated with CASSCF method followed by second or third order perturbation theory. Full potential energy surfaces were computed as a function of the four intermolecular Jacobi coordinates, with an aug-cc-pVTZ basis on the N and H atoms. The analytical representation of these potentials was given by expanding their dependence on the molecular orientations in coupled spherical harmonics, and representing the dependence of the expansion coefficents on the intermolecular distance $R$ by the RKHS method. A following study presented new RCCSD(T) calculations, which corrected the erroneous behavior in the long range due to a lack of size consistency in the open-shell RCCSD(T) method \cite{janssen2009ab}. 

Another example would be \ce{O(^{3}P) + H2} \cite{li1997theoretical, alexander1998theoretical, rogers2000chemically, brandao2004potential, atahan2006ab}, which is the object of theoretical study for many years due to the importance of the \ce{O(^{3}P) + H2 \to OH + H} reaction in combustion and its role as the prototype elementary insertion reaction. Early, PESs for \ce{O(^{3}P) + H2} were calculated for investigation of the structure and reaction dynamics of \ce{O(^{3}P, ^{1}D, ^{1}S)} doped in solid \ce{H2/D2} \cite{li1997theoretical}. The calculations were of the MRCI variety and CASSCF calculations were carried out to optimize the orbitals using the MOLPRO program. Subsequently, Millard H. Alexander determined the PESs for both \ce{H2} and \ce{D2} using CASSCF supplemented by MRCI calculations \cite{alexander1998theoretical}. Stephanie Rogers et al. performed a study for chemically accurate ab initio PESs \cite{rogers2000chemically}. CASSCF internally contracted configuration interaction (ICCI) calculations \cite{knowles1988efficient, werner1988efficient} using correlation-consistent basis sets \cite{dunning1989gaussian} were performed, and the electronic energies of the lowest $^{3}A'$ and $^{3}A"$ electronically adiabatic states were computed for 951 geometries using MOLPRO. The PESs have chemical accuracy about 0.3 kcal/mol, and are devoid of apparent unphysical features. 

Jo$\tilde{a}$o Brand$\tilde{a}$o et al. built a new potential surface for the \ce{O(^{3}P) + H2} system in the lowest $^{3}A"$ state, using the same ab initio data and the double many-body expansion formalism \cite{brandao2004potential}. They found a deeper van der Waals region would translate into a four times higher cross section than that of \cite{rogers2000chemically}, which reflects the importance of a correct description of van der Waals forces on potential energy surfaces, especially for low-energy collision studies. Sule Atahan et al. reported PESs determined by RCCSD(T) calculations with an aug-cc-pVQZ basis \cite{atahan2006ab}. In addition, the relative contribution of the electrostatic, exchange, and dispersion interaction to the overall \ce{O-H2} van der Waals interaction, as well as the effect of the spin-orbit coupling on the position and depth of the van der Waals minima were investigated by an open-shell implementation of symmetry-adapted perturbation theory \cite{jeziorski1994perturbation}. The studies has been extended to similar system of \ce{S(^{1}D) + H2/D2}, which also provides the prototype for hydrogen-atom abstraction and exchange-reaction studies, and play an important role in combustion and atmospheric chemistry \cite{zyubin2001reaction, ho2002globally}. 

\begin{figure}
\begin{center}
\includegraphics[width=1.0\linewidth]{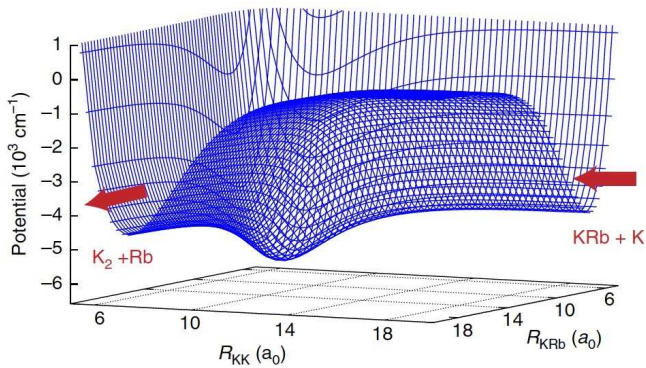}
\caption{ Figure adapted from  \cite{croft2017universality}. K-KRb potential energy surface for collinear geometry. A two-dimensional cut through the energetically lowest $^2A'$ adiabatic potential energy surface of the KRbK trimer as a function of the K-Rb and K-K bond lengths along the collinear geometry. }
\label{figure:K2Rb}
\end{center}
\end{figure}

Using the atom-dimer scattering formalism \cite{pack1987quantum}, which uses adiabatically adjusting, principal axes hyperspherical (APH) coordinates, J. F. E. Croft et al. present exact quantum-mechanical calculations \cite{makrides2015ultracold} within the close-coupling scheme for the \ce{KRb ($^{1}\Sigma^{+}, v=0, j=0$) + K($^{2}S$) \rightarrow K_{2}($^{1}\Sigma_{g}^{+}, v', j'$) + Rb($^{2}S$)} chemical reaction \cite{croft2017universality}. They map out an accurate ab initio groud state potential energy surface of the \ce{K2Rb} complex in full dimensionality and report its quantum-mechanical reaction dynamics, as can be seen in Figure ~\ref{figure:K2Rb}.

As the system become more complex (i.e. involving polyatomic molecule), the calculations become challenging due to the high dimensionality of the PES. The main difficulty lies in the more efficient and accurate fitting technique with much less number of the calculated ab initio electronic energies than 10$^{3N-6}$ required by available fitting methods \cite{schatz1989analytical, hollebeek1999constructing}. There has been great progress in the developing fitting methods, leading to the development of PESs for polyatomic systems with up to 10 atoms \cite{braams2009permutationally, bowman2011high}. Subsequently, quasi-classical trajectory calculations on the \ce{Cl + CHD3 \to HCl + CD3} reaction using the global PES thus attained achieve excellent agreement with experiment \cite{czako2011dynamics}. 

\subsection{Scattering theory}

Quantum scattering calculations of atomic and molecular collisions, in particular atom-diatom chemical reactions, have traditionally been performed using both time-independent and time-dependent quantum-mechanical approaches. Time-dependent wave packet methods is popular for handling molecular collisions at temperatures around room temperature and higher \cite{althorpe2003quantum}. However, it become hard to implement at very low collision energies, because the time evolution of a wavepacket is very slow and propagating it until it reaches the asymptotic region takes very long time. Furthermore, it is difficult to converge wavepacket calculations at very low scattering energies and it tend to cause unphysical reflections. Thus, time-independent methods are preferred to handle the low-energy collisions. In the following, we will give a detailed review on a few of the well-developed theoretical approaches, including quantum capture theory, Wigner threshold laws, and quantum defect theory. 

\subsubsection{Quantum capture theory}

Quantum capture theory is a modification of calssical capture theory \cite{child1996molecular}. In the classical capture theory, trajectories are assumed reactive when the energy is above the centrifugal barrier, and nonreactive otherwise. In quantum mechanics, tunneling may lead to reaction at energies below the barrier, while reflection may occur even if the energy is above the barrier. For a system consisting of two molecules in a certain rovibrational state $|n\rangle = |v_{i}j_{i}m_{i}\rangle$ (i = A, B), the asymptotical scattering wavefunction at large r can be written as \cite{brouard2015tutorials}
\begin{equation}
\begin{split}
 \Psi_{\mathbf{n}}^{SC} &\thickapprox \frac{2\pi}{ik_{\mathbf{n}}r}\sum_{lm_l}\sum_{\mathbf{n}^{\prime}}\sum_{l^{\prime}m_l^{\prime}}|\mathbf{n}^{\prime}\rangle v_{\mathbf{n}^{\prime}}^{-1/2}Y_{l^{\prime}m_{l}^{\prime}}(\hat{\mathbf{r}}) \\
&\times [-e^{-i(k_{\mathbf{n}}r-l\pi/2)}\delta_{\mathbf{n}^{\prime}\mathbf{n}}\delta_{l^{\prime}l}\delta_{m_{l}^{\prime}m_l} \\
&+ e^{i(k_{\mathbf{n}}r-l\pi/2)}S_{\mathbf{n}^{\prime}l_{\prime}m_{l}^{\prime}; \mathbf{n}lm_l}]i^{l}Y_{lm_l}(\hat {\mathbf{k}})^{\ast},
\end{split}
\end{equation}
Solving the time-independent scattering problem, is to find solutions of the Schr\"odinger equation that satisfy the S-matrix boundary conditions for large r,
\begin{equation}
\begin{split}
 \Psi_{\mathbf{n},l,m_l} &= \frac{1}{r}\sum_{\mathbf{n}^{\prime}}\sum_{l^{\prime}m_l^{\prime}}|\mathbf{n}^{\prime}\rangle v_{\mathbf{n}^{\prime}}^{-1/2}Y_{l^{\prime}m_{l}^{\prime}}(\hat{\mathbf{r}}) \\
&\times [-e^{-i(k_{\mathbf{n}}r-l\pi/2)}\delta_{\mathbf{n}^{\prime}\mathbf{n}}\delta_{l^{\prime}l}\delta_{m_{l}^{\prime}m_l} \\
&+ e^{i(k_{\mathbf{n}}r-l\pi/2)}S_{\mathbf{n}^{\prime}l_{\prime}m_{l}^{\prime}; \mathbf{n}lm_l}].
\end{split}
\end{equation}
These solutions are called partial waves, and the metrix with elements $S_{\mathbf{n}^{\prime}l_{\prime}m_{l}^{\prime}; \mathbf{n}lm_l}$ is called the S-matrix. The scattering wave function can be reorganized into an incoming plane wave plus an outgoing spherical wave
\begin{equation}
\begin{split}
 \Psi_{\mathbf{n}}^{SC} &\thickapprox |\mathbf{n}\rangle v_{\mathbf{n}}^{-1/2}e^{ik_{\mathbf{n}}r} \\
&+ \sum_{\mathbf{n}^{\prime}}|\mathbf{n}^{\prime}\rangle v_{\mathbf{n}^{\prime}}^{-1/2}\frac{e^{ik_{\mathbf{n}^{\prime}}r}}{r}f_{\mathbf{n}^{\prime}\gets \mathbf{n}}(\hat{r};\hat{k}) ,
\end{split}
\end{equation}
where the scattering amplitude is given by
\begin{align}
f_{\mathbf{n}^{\prime}\gets \mathbf{n}}(\hat{r};\hat{k}) &= \frac{2\pi}{ik_{\mathbf{n}}}\sum_{lm_{l}l^{\prime}m_l^{\prime}}i^{l-l^{\prime}}Y_{l^{\prime}m_l^{\prime}}(\hat{r})T_{\mathbf{n}^{\prime}l^{\prime}m_l^{\prime}; \mathbf{n}lm_l}Y_{lm_l}(\hat{k})^{\ast}.
\end{align}
where T-matrix is related to S-matrix by $\mathbf{T} = \mathbf{1} - \mathbf{S}$. The state-to-state differential cross section for a particular incident direction $\hat{k}$ is given by
\begin{align}
\sigma_{\mathbf{n}^{\prime}\gets \mathbf{n}}(\hat{r};\hat{k}) = |f_{\mathbf{n}^{\prime}\gets \mathbf{n}}(\hat{r};\hat{k})|^{2}.
\end{align}

So far, quantum capture theory has been widely used in atomic and molecular collision studies, and a number of variations of the method has been developed in scattering treatment for different collision systems at various energy scales, such as coupled channels capture theory and quantum adiabatic capture theory. We will briefly describe these methods in the following paragraphs. 

\paragraph{Coupled channels capture theory}

In quantum capture theory it is assumed that the capture cross section can be found by solving the Schr\"odinger equation in a restricted region that is located entirely in the reactant arrangement, and the scattering calculations can be performed using the coupled channels approach \cite{brouard2015tutorials}. The Hamiltonian can be written as the sum of the radial kinetic energy operator and the remainder,
\begin{align}
\hat{H} = -\frac{\hbar^2}{2\mu}r^{-1}\frac{d^2}{dr^2}r + \Delta \hat{H},
\end{align}
and the Schr\"odinger equation in the reactant arrangement is written as
\begin{align}
\frac{\hbar^2}{2\mu}r^{-1}\frac{d^2}{dr^2}r\Psi = (\Delta \hat{H} - E) \Psi.
\end{align}
The wave function is expanded in channel function $|n^{\prime}\rangle$,
\begin{align}
\Psi_{n} = r^{-1}\sum_{n^{\prime}}|n^{\prime}\rangle U_{n^{\prime}n}(r).
\end{align}
By substituting the expansion into the Schr\"odinger equation and projecting onto the channel eigenfunctions, a set of coupled equations is obtained
\begin{align}
\mathbf{U}^{"}(r) = \mathbf{W}(r)\mathbf{U}(r),
\end{align}
where the coupling matrix is given by
\begin{align}
\mathbf{W_{n^{\prime}n}}(r) = \frac{2\mu}{\hbar^2}\langle \mathbf{n^{\prime}}|\Delta \hat{H} - E|\mathbf{n}\rangle.
\end{align}
The boundary condition $\mathbf{U}(r = 0) = 0$, together with the coupled channels equation, defines a linear relation
\begin{align}
\mathbf{U^{\prime}}(r) = \mathbf{Y}(r)\mathbf{U}(r),
\end{align}
where $\mathbf{Y}(r)$ is the log-derivative matrix. By properly propagating the log-derivative matrix, we can obtain the relation between the $\mathbf{S}$ matrix to the log-derivative matrix. Then, the capture probability for a given incoming channel can be found by
\begin{align}
P_{\mathbf{n}lm_l}(E) = 1- \sum_{\mathbf{n}^{\prime}l^{\prime}m_{l}^{\prime}}|S_{\mathbf{n}^{\prime}l^{\prime}m_{l}^{\prime}; \mathbf{n}lm_l}(E)|^2.
\end{align}
The capture cross section for incoming channel $\mathbf{n}$ is
\begin{align}
\sigma_{\mathbf{n}}(E) = \frac{\pi}{k_{\mathbf{n}}^2}\sum_{lm_l}P_{\mathbf{n}lm_l}(E).
\end{align}

\begin{figure}
\begin{center}
\includegraphics[width=1.0\linewidth]{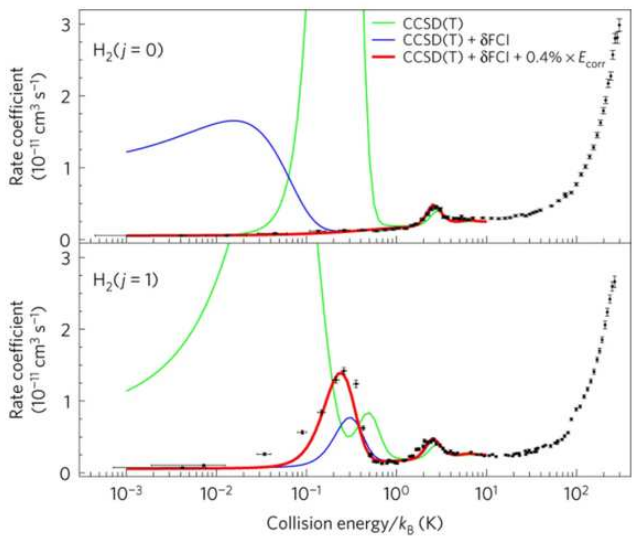}
\caption{Figure adapted from  \cite{klein2017directly}.  Experimental and theoretical Penning ionization rate coefficients of \ce{H2} molecules in ground and excited rotational states by \ce{He^*}. The black dots represent the experimental rate coefficient, the green line represents the calculated rate coefficient using the interaction potential obtained with [CCSD(T)/aug-cc-pV6Z + mid-bond], the blue line represents correction of the interaction potential by including the FCI/aug-cc-pVQZ, and the red line represents further correction with a scaling factor.  }
\label{figure:PotentialCorrection}
\end{center}
\end{figure}

In the study of effect of rotational excitation on the reaction rate \cite{klein2017directly}, Ayelet Klein et al. performed close-coupling quantum scattering calculations \cite{arthurs1960theory} for collisions of He$^{\ast}$ with both para-\ce{H2}(j=0) and ortho-\ce{H2}(j=1), using a two-dimensional potential energy surface calculated by the CCSD(T). As shown in Figure ~\ref{figure:PotentialCorrection}, the calculation results demonstrate that collisions with para-\ce{H2} are not sensitive to the anisotropic part of the interaction potential, while the rate coefficient with ortho-\ce{H2} is very sensitive to the anisotropy, and will shift the position of the orbiting resonance, in stark contrast with para-\ce{H2}. This sensitivity of quantum resonance to small changes in the interaction again emerge when comparing calculation results to experimental data. The interaction potential obtained with CCSD(T) in an aug-cc-pV6Z basis set with additional set of mid-bond functions accounting for the dispersion interaction, erroneously predicts a low-energy resonance for para-hydrogen and two low-energy resonances for ortho-hydrogen. Including the full configuration interaction (FCI) in an aug-cc-pVQZ basis set allows for agreement down to collision energies of a few hundred mK. Further improvement of the interaction potential is achieved by uniformly scaling the correlation energy by 0.4\%, resulting in better agreement with the measured resonance positions and the overall behaviour of the rate coefficient.

\paragraph{Quantum adiabatic capture theory}

At low temperatures the collision time is long compared to characteristic vibrational and rotational timescales in the colliding molecules. This allows us to introduce an approximation \cite{quack1974specific, clary1985calculations, clary1987rate, ramillon1994adiabatic} analogous to the Born-Oppenheimer approximation, which explots the difference in time scales of electronic and nuclear motion. Solving the Schr\"odinger equation for the fast motion amounts to diagonalising the coupling $\mathbf{W}(r)$ matrix on a grid of r points. For each molecular state asymptotically allowed at an energy $E$, the capture probability is computed by solving the one-dimensional quantum capture problem. This is done exactly as the coupled channels equation. 

In an early review on fast chemical reactions, D. C. Clary gives a full description of quantum capture theory \cite{clary1990fast}, especially rotationally adiabatic capture theory which requires less computing power. The prediction of rate constant for reaction such as \ce{He^* + HCl \rightarrow He + HCl+} has been compared to experimental measurements, and the agreement is excellent in the measurement temperature range of 27 - 300 K. The \ce{N + NH \rightarrow N_{2} +H} reaction has been investigated by adiabatic capture theory over the temperature range 2-300 K using an existent potential energy surface, but the resultant thermal rate constants are significantly larger than the experimentally derived rate \cite{frankcombe2007adiabatic}.

Timur V Tscherbul et al. have used both quantum and classical adiabatic capture theories to study reaction of \ce{Li + CaH \rightarrow LiH + Ca} \cite{tscherbul2015adiabatic}. The calculated reaction rate agrees well with the measured value at 1 K \cite{singh2012chemical}. Mariusz Pawlak et al. develop an adiabatic theory for cold anisotropic collisions between slow atoms and cold molecules, using two alternative theoretical approaches, i.e. first-order adiabatic perturbation theory (APT) and adiabatic variational theory (AVT) \cite{pawlak2015adiabatic}. The calculation results with maximum angular momentum $l_{max}=30$ from both theories are compared to reaction rate of \ce{He^* + HD \rightarrow He + HD^+ + e^-}, and good agreements have been found. It has also been found that the couplings between the projection states of the rotational motion of the atom about the molecular axis of the diatom are enough to introduce the effect of anisotropy and play a crucial role in reproducing the shape and positions of scattering resonances. The authors further apply AVT to the Penning ionization involving both rotational ground and excited states \cite{pawlak2017adiabatic}. It shows that the anisotropic long-range interactions cause dramatic differences in the measured reaction rates. In collisions of metastable helium with para-\ce{H2}, the isotropic potential term dominates the dynamics, while the anisotropic interaction play an important role in the dynamics when the collision is with ortho-\ce{H2}.  Debarati Bhattacharya et al. incorporate complex potential energy surfaces into AVT, thereby reducing the multidimensional scattering process to a series of uncoupled 1D problems \cite{bhattacharya2017polyatomic, bhattacharya2019quantum}. By applying this approach to the case of Penning ionization of \ce{He(2^3P)} with para/ortho-\ce{H2} and \ce{HD}, remarkable agreement between calculation reaction rate coefficients and experimental measurements has been achieved. It is shown the presence of the single electron in the excited \emph{p}-orbital of the He atom changes the entire system and resulting spatial polarization brings in a pronounced anisotropic effect in the interaction potential.

Mariusz Pawlak et al. have studied the effect of nonrigidity of the \ce{H2} molecule on the profile of the scattering resonance using AVT \cite{pawlak2019nonrigidity}, and show the calculated reaction rate coefficients are in excellent agreement with the experimental data after proper inclusion of the flexibility of the molecule into the interaction potential, eliminating the need of empirical adjustment of the interaction potential to match the calculation with the experiment. This finding demonstrates the importance of the flexibility of the interacting molecule, which may be crucial in cold chemistry.

\paragraph{External field control}

External electric and magnetic fields distrub the symmetries possessed by the colliding system, and induces couplings between states of the total angular momentum. The projection of total angular momentum of the colliding molecules on space-fixed quantization axis, rather than itself, is a good quantum number in the presence of an external field. Krems and Dalgarno \cite{krems2004quantum} suggest the use of a fully uncoupled space-fixed basis representation of the wave function. The complexity of the collision theory in the fully uncoupled representation does not increase with the number of internal degrees of freedom and the evaluation of the matrix elements of all terms in the Hamiltonian is straightforward. 

The Hamiltonian for two colliding particles A and B in an external field can be simply written as
\begin{equation}
\hat{H} =  \hat{H}_{0}(\rho) + \hat{V}_{field},
\end{equation}
where $\rho$ represents the Jacobi distance between the center of mass of the colliding partners, or the hyperradius describing the collision complex. The field-free Hamiltonian $\hat{H}_{0}(\rho)$ includes the kinetic energy of the relative motion and the potential energy of the collision complex. The term $\hat{V}_{field}$ describes the interaction with an external field assumed to be independent of $\rho$. Expanding the full wavefunction of the collision system in a space-fixed uncoupled basis, typically the field axis, the uncoupled basis functions are written as products of the eigenfunctions of the operator $\mathbf{J}^2$ and $J_Z$,
\begin{equation}
|\Psi \rangle = R^{-1} \sum_{i}F_{i}(R) |JM\rangle,
\end{equation}
where $\mathbf{J}$ is the total angular momentum of the collision system, and $J_Z$ is the component of the total angular momentum along the field axis \cite{krems2004quantum, tscherbul2009magnetic, tscherbul2010quantum, tscherbul2012total, janssen2011cold2}. The full Hamiltonian is diagonal in $M$, but the blocks of states corresponding to $J$ and $J^{\prime} = J\pm 1$ are coupled by the matrix elements of $\hat{V}_{field}$. This basis set expansion, when substituted into the Schr\"odinger equation, leads to a system of coupled differential equations. These close coupled equations at a fixed field strength and total energy $E$ can be written as \cite{krems2004quantum, krems2005molecules, krems2018molecules}
\begin{equation}
\begin{split}
 [\frac{d^2}{d R^2} & + 2\mu E] F_{\alpha l m_{l}}(R) \\
&= 2\mu \sum_{\alpha^{'}l^{'}m_{l}^{'}}H^{\prime}_{\alpha l m_{l}; \alpha^{'}l^{'}m_{l}^{'}}F_{\alpha^{'}l^{'}m_{l}^{'}}(R),
\end{split}
\end{equation}
with the boundary conditions of
\begin{equation}
\begin{split}
F_{\alpha^{'}l^{'}m_{l}^{'}}^{\alpha l m_{l}}&(R\to 0) \to 0,\\
F_{\alpha^{'}l^{'}m_{l}^{'}}^{\alpha l m_{l}}&(R\to \infty) \sim \delta_{\alpha \alpha^{'}}\delta_{ll^{'}}\delta_{m_{l}m_{l}^{'}}exp[-i(k_{\alpha}R-\pi l/2)]\\
&-(\frac{k_{\alpha}}{k_{\alpha^{'}}})^{1/2}S_{\alpha^{'}l^{'}m_{l}^{'}; \alpha l m_{l}}exp[i(k_{\alpha^{'}}R-\pi l^{'}/2)].
\end{split}
\end{equation}
Numerically solve these equations yields the scattering S-matrix, from which the cross-sections for elastic and inelastic collisions are computed as
\begin{equation}
\begin{split}
 \sigma_{\alpha \to \alpha^{'}} &= \frac{\pi}{k_{\alpha}^{2}}\sum_{M}\sum_{l}\sum_{m_l}\sum_{l^{'}}\sum_{m_{l}^{'}}|\delta_{ll^{'}}\delta_{m_{l}m_{l}^{'}}\delta_{\alpha \alpha^{'}} \\
&- S_{\alpha l m_{l}; \alpha^{'}l^{'}m_{l}^{'}}^{M}|^2.
\end{split}
\end{equation}

For polar molecules in an electric field, John L. Bohn have studied their collisional stability in electrostatic traps using a model that emphasizes long-range dipolar forces, whose potential have the threshold form $-C_{3}/R^{3} + \hslash^{2}L(L+1)/mR^{2} -q_{\Lambda} N(N+1)$ \cite{bohn2001inelastic}. The rate constants for collisional losses are shown to vary substantially as a function of molecular parameters, such as mass, the splitting of the molecular $\Lambda$ doublet, dipole moment. The ratio of loss rate to elastic collision rate is shown strongly dependent on the externally applied trapping electric field. Taking OH molecules as an example, the electric field exerts a strong influence on both elastic and state-changing inelastic collision rate constants, nonetheless the scattering properties at ultralow temperatures is independent of the details of the short-range interaction, owing to avoided crossings in the long-range adiabatic potential curves \cite{avdeenkov2002collisional}. In general, evaporative cooling would only be viable in an extremely limited range of temperature and electric field. The situation becomes quite different for $^{2} \Pi_{3/2}$ ground state OH molecules trapped in a magnetic field. Christopher Ticknor and John L. Bohn find that inelastic collisions of weak-field-seeking states can be suppressed by two orders of magnitude with modest fields of few thousand gauss, in contrary to get enhanced in external electric fields \cite{ticknor2005influence}.

For paramagnetic molecules, Alexandr V. Avdeenkov and John L. Bohn has made a detailed analysis for the singlet, triplet, and quintet potential energy surfaces, and then using them to investigate collisional properties of ground-state oxygen molecules at translational energies below $\sim1$ K and in zero magnetic field \cite{avdeenkov2001ultracold}. Both elastic and spin-changing inelastic collision rates for different isotropic molecules are calculated in the framework of a rigid-rotor model. The rotational degrees of freedom is found to be crucial in determining the ratio of elastic and inelastic rate constants. \ce{^{17}O_2} is shown to be a good candidate for evaporative cooling since its elastic collisions strongly dominate spin-changing loss collisions at low temperatures, while \ce{^{16}O_2} is unlikely to survive evaporative cooling. Micha{\l} Hapka et al.
generate new interaction potential curves for the \ce{He(^{3}S)-H2} and \ce{He(^{3}S)-Ar} systems, using both coupled cluster levels of supermolecular theory and symmetry-adapted perturbation theory (SAPT) \cite{hapka2013first}, and then compare the obtained elastic cross sections to the experiment \cite{henson2012observation}.

With an interaction potential of the \ce{He-CaH($^{2}\Sigma^{+}$)} van der Waals complex computed with the partially RCCSD(T), N. Balakrishnan et al. have shown that even small spin-rotational interaction significantly modify the rate coefficients for rotational quenching at low tempertures \cite{balakrishnan2003he}. R. V. Krems et al. have performed a series of calculations on atom-molecule and molecular collisions in a magnetic field, using the same ab initio potential expanded in a Legendre series. They have shown that in low energy collisions, spin-flip relaxation in rotationally ground-state $^{2} \Sigma$ CaH molecules is induced by collisions with \ce{^{3}He} atoms through coupling to rotationally excited molecular levels, and are determined by the spin-rotation interaction in the rotationally excited molecules \cite{krems2003spin}. It is concluded that the collision-induced spin-flip will be least efficient for the diatomic molecules with large rotational constants and small spin-rotational constants $\gamma$.

Similarly, with accurate ab initio calculations of the potential energy surface and bound energy levels of \ce{He-NH(^{3}\Sigma^{-})} van der Waals complex using a supermolecular approach, the effects of the atom-molecule interaction potential and the external magnetic field on the collisionally induced Zeeman relaxation are presented \cite{krems2003low,cybulski2005interaction}, using a formalism for quantum-mechanical close coupling calculations of cross sections \cite{krems2004quantum}. Cross sections for elastic and Zeeman relaxation of NH by \ce{^{3}He} at different magnetic fields are given, and their ratio of rate constants in magnetic fields from 0 to 3 T and collision energies form $10^{-4}$ to 1 K is explored. It is shown that both the rotational constant of the diatomic molecule and the anisotropy of the interaction potential have a strong influence on the rate constants for Zeeman relaxation.

This is confirmed by an experiment where the effect of molecular rotational splitting on Zeeman relaxation rates in collisions of different NH isotopologue with helium atoms in a magnetic field have been determined at low temperatures \cite{campbell2009mechanism}. The measurements support the predicted $1/B_{e} ^{2}$ dependence of the collision-induced Zeeman relaxation rate coefficient on the molecular rotational constant $B_e$, and rate coefficients in collision with \ce{^{3}He} are much larger than the coefficients with \ce{^{4}He}, which is due to a shape resonance in the incoming collision channel.

\subsubsection{Wigner threshold laws}

The essential idea is to divide the configuration space into an inner and an outer region in the three-dimensional scattering calculations, depending on the distrance between collision partners. In the inner regior where there is a strong repulsive potential of the colliding complex, the wave function is described by the Wentzel-Kramers-Brillouin (WKB) form. In the outer region, the wave function is described by the asymptotic form. Generally, for a collision with an effective potential barrier $V_b$, the the scattering process can be described by the Bethe-Wigner threshold laws \cite{bethe1935theory, wigner1948behavior} when collision energy $E_{c}<V_b$. While for $E_{c} \ge E_b$ the collision probability, many quenching rates, the loss rate or chemical reaction rates are well described by the Langevin's classical capture model \cite{langevin1905formule}, which assumes unit probability for chemical reactions and/or hyperfine state-changing collisions if the molecules have surmounted the barrier and become close enough to react.

In quantum threshold model (QT), the molecules will tunnel through a centrifugal barrier with a given probability, which varies with energy according to the Wigner threshold laws. By floating the value of the tunneling probability at the barrier's peak, the QT model is able to describe the chemical reaction probability in an analytic way, even in the presence of an external field \cite{ni2010dipolar}. 

Given a particle in a well-defined initial quantum state $j$ and a collider at very low energy with relative angular momentum $l$, we have two types of processes, elastic collision where the internal state of the particle does not change and inelastic collision where the particle relaxes to a lower quantum level $f$. The dependence of the cross sections on the initial relative velocities $v_j$ (and collision energy $E_k$) for each process is given by
\begin{equation}
\begin{aligned}
\sigma_{j}^{el} &\sim v_{j}^{4l} \sim E_{k}^{2l},\\
\sigma_{j}^{in} &\sim v_{j}^{2l-1} \sim E_{k}^{l-1/2}.
\end{aligned}
\end{equation}
The total quenching cross-section $\sigma^{in}$ is a sum over all the state-to-state cross-sections to all the accessible channels with $E_{f} < E_{j}$,
\begin{equation}
\sigma_{j}(E_{j}) = \sum_{f} \sigma(j \to f, E_{j}).
\end{equation}
When the initial relative velocity goes to zero, all the $l \ne 0$ partial waves stop contributing to the scattering, and the collision enters into s-wave regime, where a complex scattering length $a_{j} = \alpha_{j} - i\beta_{j}$ can be defined. The elastic and total inelastic scattering cross-sections at zero energy are hence given by
\begin{equation}
\begin{aligned}
\sigma_{j}^{el} &= 4\pi |a_{j}|^{2},     \\
\sigma_{j}^{in} &= \frac{4\pi \beta_{j}}{k_j},
\end{aligned}
\end{equation}
where $k_j$ is the wave vector of the collision. 

For a partial wave with $l > 0$, the effective potential is governed at long range by the centrifugal and dispersion terms,
\begin{equation}
V^{l}(R) = \frac{\hbar^{2}l(l+1)}{2\mu R^2} - \frac{C_6}{R^6},
\end{equation}
where $C_6$ is the dispersion coefficient. Then, the total inelastic cross-section and rate coefficient can be given by
\begin{equation}
\begin{aligned}
\sigma_{in}(E) &= 3\pi (\frac{C_6}{4E})^{1/3}, \\
k_{in}(E) &= 3\pi (\frac{C_6}{4E})^{1/3}(\frac{2E}{\mu})^{1/2} = \frac{3\pi C_{6}^{1/3}E^{1/6}}{2^{1/6}\mu^{1/2}}.
\end{aligned}
\end{equation}
Clearly, the reaction rate coefficients can be analytically estimated from the threshold laws. However, they do not yield product resolved reaction rates, or discriminate between multiple product channels if they are present. 

According to Wigner threshold laws, Marko T. Cvita{\v{s}} et al. have performed a series of full quantum dynamics calculation on ultracold \ce{Na + Na_2} \cite{soldan2002quantum,quemener2004sensitivity} and \ce{Li + Li_2} collisions\cite{cvitavs2005ultracold,cvitavs2005ultracold2,cvitavs2007interactions}. Formalism based on body-frame hyperspherical coordinates and the Arthurs-Dalgarno formalism is used in the inner and outer region, respectively. For ultracold collisions between spin-polarized Na atoms and vibrationally excited \ce{Na_2} molecules, the Wigner threshold laws are followed for energies below $10^{-5}$ K, while vibrational relaxation processes dominate elastic processes for temperatures below $10^{-3}-10^{-4}$ K \cite{soldan2002quantum}. In the \ce{Li + Li_2} collision complex consisting of fermionic atoms, there is no systematic suppression of the vibrational relaxation rate coefficient for low-lying vibrational states \cite{cvitavs2005ultracold}. Moreover, it is found that reactive scattering can occur for \ce{^{7}Li + ^{6}Li^{7}Li} and \ce{^{7}Li + ^{6}Li_{2}} systems even when the molecules are in their ground rovibrational states \cite{cvitavs2005ultracold2}. Goulven Qu{\'e}m{\'e}ner et al. also carry out a quantum-dynamical study of vibrational deexcitation and elastic scattering for the bosonic and fermionic \ce{Li + Li_2} system, with \ce{Li2} molecules in high vibrational state \cite{quemener2007ultracold}. The inelastic rates show a strong and irregular dependence on the vibrational state of the molecule.

\begin{figure}
\begin{center}
\includegraphics[width=1.0\linewidth]{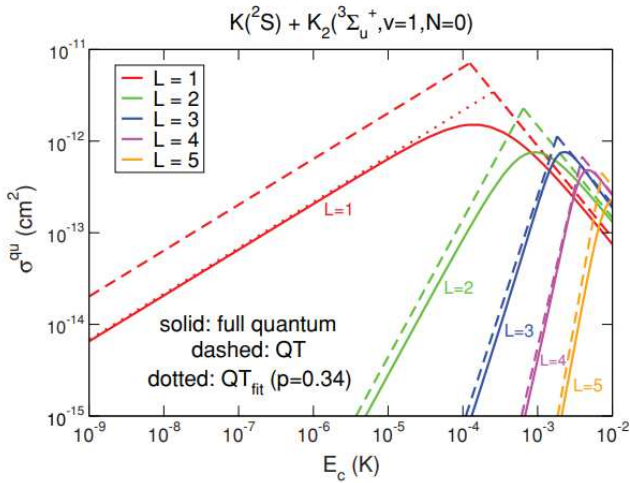}
\caption{Figure adapted from \cite{quemener2010strong}. Quenching cross section of \ce{^{39}K + ^{39}K_2} as a function of the collision energy for the partial wave $L = 1 - 5$. Solid lines represent a full quantum calculation while dashed lines using the quantum threshold model.  }
\label{figure:quenching}
\end{center}
\end{figure}

Qu{\'e}m{\'e}ner and Bohn combine the classical capture theory with the QT laws to give an analytical estimate of the chemical quenching cross sections and rate coefficients of two colliding particles at ultralow tempertures. They apply this method to indistinguishable fermionic KRb molecules in an electric field, and calculate the reaction rates for \ce{K($^{2}S$) + K_{2}($^{3}\Sigma_{u}^{+}, v=1, N=0$)}, as a function of collision energy at low temperatures \cite{quemener2010strong}. Figure ~\ref{figure:quenching} presents quenching cross section of \ce{^{39}K + ^{39}K_2} as a function of the collision energy for the partial wave $L = 1 - 5$.

Universalities in ultracold reactions of alkali-metal polar molecules are proposed based on QT model to explore the dependence of reaction rates on various parameters, which is most of bosonic species of heteronuclear alkali-metal molecules haring a universal reaction rate in a sufficiently large electric field \cite{quemener2011universalities}. Paul S. Julienne et al. also characterize the universal reactive and inelastic relaxation rate constants in 3D and quasi-2D planar geometry for alkali-metal polar molecules in the near-threshold limit of ultracold collisions \cite{julienne2011universal}.

\subsubsection{Quantum defect theory}

Full close-coupling calculations of ultracold collisions including external field effects are a formidable task, and there is intense interest in developing simplified models base on long-range theories. While Langevin type capture models can provide qualitative estimates of reaction rates for systems with strong long-range interaction, their validity is questionable at ultracold energies. Recently, there has been considerable interest in developing formalisms based on quantum defect theory (QDT) \cite{julienne1989collisions}.

The basic idea of QDT is to solve the Schr\"odinger equation accurately for the long-range potential and use an approximate description at short range, by taking advantage of the large disparity in energy scales for the relative motion in the incident channels and the deep potential wells that drive rovibrational transitions and chemical reactions. Compared to the depth of the potential wells which typically range from 100 to 50000 K for many collision/reaction systems, the energy variation in the initial channels are very small for discussed collisions in this review, usually in the range from 1 $\mu$K to 1 K. Therefore, these two regimes can be described differently and the accuracy of this method can be improved by treating the short-range dynamics more accurately.

Especially, a multichannel quantum defect theory (MQDT) approach \cite{burke1998multichannel, gao2005multichannel} successfully replace the short-range physics by suitably parametrized boundary conditions, that acknowledge both a scattering phase shift and the probability of chemical reaction \cite{idziaszek2010universal}. The boundary conditions are matched to highly accurate solutions of the long-range scattering, which are carefully characterized analytically, allowing for simple analytic formulas for scattering observables.

\begin{figure}
\begin{center}
\includegraphics[width=1.0\linewidth]{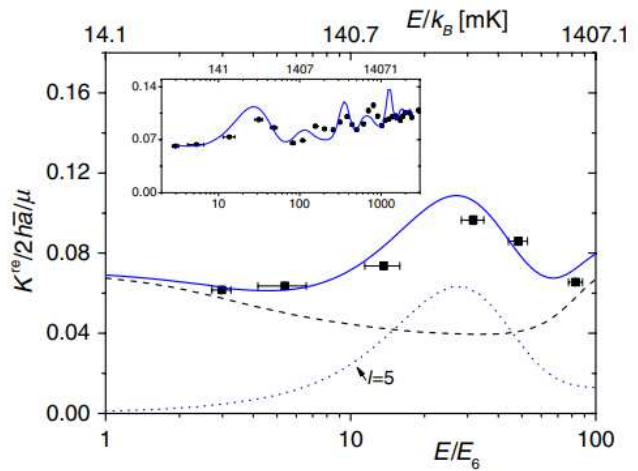}
\caption{Figure adapted from \cite{jachymski2013quantum}. Reactive rate constants calculated using QDT theory (blue line), including contribution from $\emph{l}=5$ (dotted line) and other partial waves (dashed line) for the collision of Ar with \ce{He^*} , are compared to experimental data (black points) taken from \cite{henson2012observation}.  }
\label{figure:PenningQDT}
\end{center}
\end{figure}

Paul S. Julienne and Frederick H. Mies propose a new approach for analysis of quantum threshold behavior, the MQDT \cite{julienne1989collisions}, due to the problem of applying close-coupled scattering approach to ultracold collisions in optical traps.  Zbigniew Idziaszek and Paul S. Julienne propose a general model with a complex potential for threshold molecular collision rates based on MQDT \cite{idziaszek2010universal}. The model is based on the separation of the effects of long- and short-range parts of the intermolecular potential for a given scattering channel \cite{julienne2009ultracold}. The reactive or inelastic part of the short-range collision is characterized by a dimensionless parameter $0\leq y \leq 1$, which depends on dynamics at short range and is related to the probability of irreversible loss. While the long-range potential determines how many partial wave is transmitted to short range. The model gives universal rate constants for \emph{s}- and \emph{p}-wave collisions that are independent of short-range dynamics, if the probability of reaction in the short range is high. It predicts a universal rate constant of $1.1(1)\times 10^{-10} cm^{3} /s$ for the reactive \emph{s}-wave collision of \ce{^{40}K} atoms with \ce{^{40}K^{87}Rb} molecules, close to the measured value of $1.7(3)\times 10^{-10} cm^{3} /s$ for the reaction of \ce{^{40}K^{87}Rb + ^{40}K} \cite{ospelkaus2010quantum}.

Bo Gao also derive from MQDT a universal model of exoergic bimolecular reactions or inelastic processes for a wide range of temperatures \cite{gao2010universal}. The quantum threshold behaviour given by the theory agrees with the classical Gorin model at higher temperatures, connecting between the ultracold chemistry and regular chemistry. In between, the reaction rates first decrease with temperature outside of the Wigner threshold region, before rising after a minimum is reached. Piotr S. {\.Z}uchowski and Jeremy M. Hutson investigate the energetics of trimer formation reactions involving pairs of alkali-metal dimers, such as \ce{KRb + KRb \rightarrow K + KRb_{2}} or \ce{K_{2}Rb + Rb} \cite{zuchowski2010reactions}. They demonstrate that trimer formation reactions are always energetically forbidden for alkali-metal dimers in low-lying singlet states.

In order to explain the fast loss of rovibrational ground state molecules from the trap, which is found not only in chemically reactive KRb \cite{ospelkaus2010quantum,ni2010dipolar,de2019degenerate}, NaLi \cite{rvachov2017long}, triplet \ce{Rb2} \cite{drews2017inelastic}, but in nonreactive NaRb \cite{ye2018collisions} and RbCs \cite{gregory2019sticky} due to unfavourable energetics, Michael Mayle et al. propose a sticky collision hypothesis by formulating a theory within the framework of MQDT that incorporates all possible rovibrational Fano-Feshbach resonances in a statistical manner \cite{mayle2012statistical,mayle2013scattering}. They argue that for nonreactive molecule, the large number of available rovibrational states of the molecule involved in the ultracold collision support a dense manifold of Feshbach resonances and long-lived two-molecule collision complex can be formed by resonant collisions. A further collision between the collision complex and a molecule then lead to the ejection of all three molecules from the trap.

A universal behaviour charactering ultracold collisions or reactions involving ultracold molecules has also attracted considerable interest. Zbigniew Idziaszek et al. present a unified formalism of MQDT and QT, in order to describe chemical reaction rates for trapped ultracold molecules, which reduces the scattering to a propagation of the reactant molecules along the long-range potential surface and ultimately react with a probability \cite{idziaszek2010simple}. Krzysztof Jachymski et al. \cite{jachymski2013quantum} develop a general quantum theory for reactive collisions from the ultracold to the high-temperature limit using the formulation of QDT by Mies \cite{mies1984multichannel, mies1984multichannelb}, and applies it to the case of Penning ionization of argon by metastable helium measured using merged molecular beams \cite{henson2012observation}. Good agreement with the the experimental data has been achieved by fitting at energies below ~1.4 K, as can be seen in Figure ~\ref{figure:PenningQDT}.

\begin{figure}
\begin{center}
\includegraphics[width=1.0\linewidth]{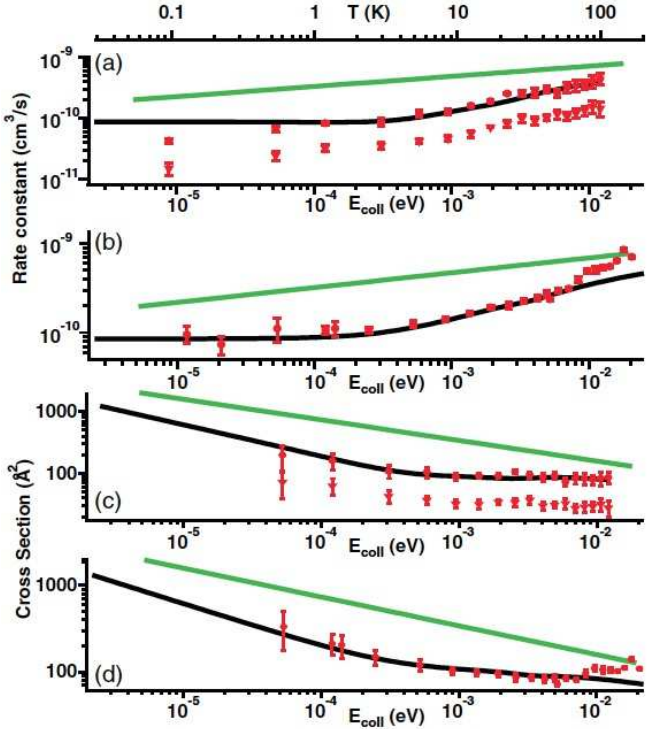}
\caption{Figure adapted from  \cite{jankunas2014dynamics}. Rate constants for (a) the \ce{Ne^* + NH3} reaction and (b) the \ce{Ne^* + ND3} reaction. Panels (c) and (d) show the cross sections for \ce{Ne^* + NH3} and \ce{Ne^* + ND3}, respectively.  The red symbols are the experimental data and the black lines are the results from the MQDT calculations. The green lines show the rate predicted on the basis of a Langevin model with unit reaction probability at short range. }
\label{figure:ND3Merged}
\end{center}
\end{figure}

In another merged molecular beam experiment, Justin Jankunas et al. use MQDT to obtain both the reaction rates and cross sections for reactions of the \ce{Ne^* + NH3} and \ce{Ne^* + ND3} from 0.1 to 100 K, all of which agree well with experimental results \cite{jankunas2014dynamics}, as shown in Figure ~\ref{figure:ND3Merged}.  The application of MQDT formalism to \ce{He(^3S_1) + NH3} Penning ionization with the same experimental setup has also achieved good agreement with experimental data \cite{jankunas2015observation}. However, when applying QDT to the reaction of \ce{He(^3S_1) + CHF3} \cite{jankunas2016communication}, there is a large deviation from experimental data when the collision energy is smaller than 2 K. From the comparison, the effect of inelastic collisions on rate coefficient is also evident. The rotationally inelastic collisions suppress the reaction rate for \ce{He(^3S_1) + CHF_3} from 10 to 100 K collision energy, while \ce{Ne(^3P_2) + NH3} reaction does not show this signature \cite{jankunas2014dynamics}, which indicates that the magnitude of the rotational constant may be critical for this effect since large number of inelastic collision channels may be open when the rotational constant is low.

\section{Experiment: Near cold collision}

In the dense molecular clouds of the interstellar medium, which are at the origin of star formation, gas phase chemistry are characterized by ultralow gas density and low temperatures, typically the temperature of the cold cores could reach as low as 5-10 Kelvin \cite{herbst2013introduction}. The Boomerang Nebula, one of the Universe's peculiar places, has a temperature of 1 K \cite{sahai1997boomerang}, falling below the temperature of the cosmic background radiation T=2.8K. Typically, gas-phase reactions involving two neutral species possess activation energy barriers, and reaction rate coefficients k are giving by the Arrhenius expression:
\begin{equation}
\displaystyle k(T)=A(T)\exp(-E_{a}/T)
\end{equation}
where T is the temperature, $E_a$ is the activation energy in K, and A(T) is a weakly temperature-dependent term known as the pre-exponential factor. However, atom-radical and radical-radical processes occur rapidly with $E_{a}=0$ and pre-exponential factor approaches the so-called collision limit. It has often been assumed the rate coefficient k for these reactions can be approximated by the hard-sphere relation
\begin{equation}
\displaystyle k(T)=A(T)=k(300K)(T/300)^{0.5}
\end{equation}
where typical value for k(300K) is 1.0-3.0$\times 10^{-11} cm^3 s^{-1}$. With this assumption, the rate coefficients are considerably small. However, many crossed-beam experiments with variable low collision energies in conjunction with ab initio quantum chemical calculations have shown that the neutral-neutral reactions have zero threshold energy and proceed via strongly bound energized complexes, which have drastically altered our understanding of the low temperature interstellar chemistry over the last two decades \cite{herschbach2009molecular}. Many reactions have been found to have little temperature dependence or become more rapid as the temperature is decreased, and their rate coefficients can often be fit to a weak inverse power law in temperature. Although the temperature is still high compared to cold regime and collision energy are far above pure s-wave scattering, this near cold regime will open a restricted number of partial waves with which the resonance signatures are not completely blurred by partial wave averaging.

For rotational excitation of simple molecules of astrophysical interest such as CO, \ce{O2}, OH, NO by inelastic collisions with \ce{H2} molecules or He atoms, quantum-mechanical studies of these fundamental collision processes occurred in dense molecular clouds could establish the existence of shape and Feshbach resonances \cite{chefdeville2012appearance,chefdeville2013observation,xiao2011experimental}.

The exploration on such molecular collisions, both experimental and theoretical, can be separated into a few branches, the first is in supersonic beam condition, such as crossed molecular beam technique pioneered by Lee and Herschbach in late 1960s \cite{lee1969molecular}. The second takes place inside a beam, where both collision partners are seeded and collide along the propagation following appropriate initial preparation. The third is under in-situ condition, which is typically in an isolated manner and use cryogenic technique, such as collisional cooling \cite{willey1988very}, matrix isolation \cite{nandi2002polarized} and helium nanodroplets \cite{morrison2012infrared}. The last is target collision, where the trapped samples are bombarded by the incoming beam. In the following, we will focus on some main subbranches for each branch instead of give a detailed review of all the branches.

\subsection{Crossed molecular beam}

Since the invention of crossed-molecular beam technique, it has become the standard tool for study inelastic collisions and reactions in both atom-molecule \cite{schnieder1995experimental} and molecule-molecule systems \cite{xiao2011experimental}.  Exquisite agreement between theory and experiment in the benchmark three-atom \ce{F + HD \rightarrow HF + D} \cite{skodje2000resonance,dong2010transition,wang2013dynamical} and \ce{F + H2 \rightarrow HF + H} \cite{qiu2006observation,kim2015spectroscopic} reactions has been achieved, where scattering resonance features have been fully rationalized. Agreement has also been obtained in high-resolution crossed-beam experiments for prototypical four-atom \ce{HD + OH \rightarrow H2O +D} \cite{xiao2011experimental} and six-atom \ce{Cl + CHD3 \rightarrow HCl + CD3} systems \cite{wang2011steric}. However, for many years the studies have been performed at high temperature regime with contributions from many partial waves and collision energy large enough to surmount the classical energy barriers.

\begin{figure}
\begin{center}
\includegraphics[width=1.0\linewidth]{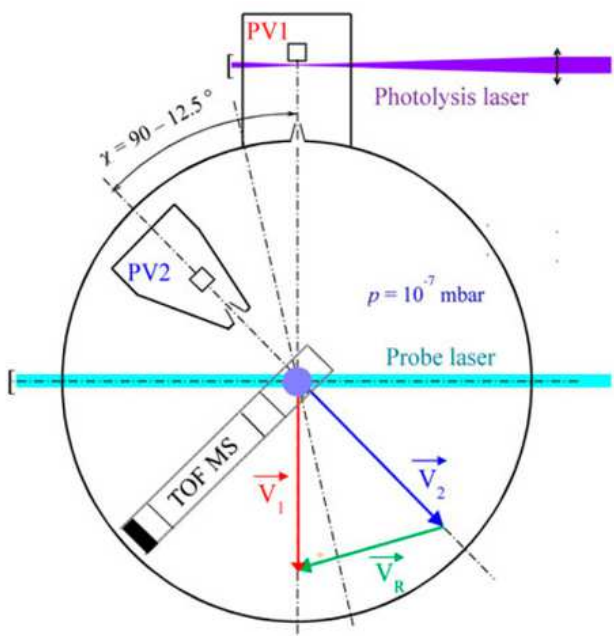}
\caption{Figure adapted from \cite{naulin2014experimental}.  Experimental crossed molecular beam setup for study of reaction dynamics with tunable angle. }
\label{figure:crossed}
\end{center}
\end{figure}

Recently, motivated by the interest in the collision processes in astrophysical environments, significant progress has been made in crossed molecular beam studies of inelastic collisions and chemical reactions. In crossed molecular beam experiments, collision parter A with velocity $v_1$ and parter B with velocity $v_2$ in laboratory frame, will have a collision energy of $E_c=1/2\mu v_{r}^2=1/2\mu (v_1^2+v_2^2-2v_1v_2cos\chi)$, where $\mu$ is the reduced mass, $v_r$ is the relative velocity, $\chi$ is the beam intersection angle. The minimal relativ ecollision energy is $E_c=\mu v_1^2 (1-cos\chi)$ when $v_1=v_2$. Therefore, both small beam intersection angle and small relative velocity are required to attain low collision energy in crossed molecular beam experiments. It has been reduced to nearly 4 Kelvin \cite{berteloite2010kinetics,bergeat2015quantum}, as can be represented by Figure ~\ref{figure:crossed}. These experiments have demonstrated observation of quantum effects in molecular collisions, such as orbiting resonance \cite{vogels2018scattering} and interferences \cite{von2014state,onvlee2017imaging}, entering the near cold regime.

Chefdeville et al. reported experimental and quantum-mechanical studies on the \ce{CO(j=0) + H2(j=0) \rightarrow CO(j=1) + H2(j=0)} inelastic collisions \cite{chefdeville2012appearance}. In the experiment, both beams were from cryogenically cooled Even-Lavie fast-pulsed valves, which enable not only good control over velocity distribution but also quench of most internal state population to the ground rotational state. The beam intersection angle was also reduced to 12.5 $^\circ$. The integral cross section were measured in the threshold region of the \ce{CO(j=0 \rightarrow j=1)} transition between 3 and 23 cm$^{-1}$, corresponding to 4.32 and 33.09 K, which show resonance structures but is only in qualitative agreement from close-coupling calculations within the coupled channel methodology (section II.B.1), as is discussed in section 2. This indicated a more stringent requirement on the potential energy surface for completely describing the experimental measurements.

In the study of \ce{S(^1D_2) + H_2 \rightarrow SH + H} reaction \cite{berteloite2010kinetics}, which is consider as a prototype insertion reaction and has no barrier, Berteloite et al. measured absolute rate coefficients and relative integral cross sections down to $\sim5$ K collision energy with normal hydrogen molecule. The theoretical calculation with adiabatical treatment (section II. B. 1) and single potential energy surface agreed well with measurements. Later in a similar study of the \ce{S(^{1}D_2) + H_2(j=0)} reaction at energies from 0.820 down to 0.078 kJ/mol (i.e. from 98.62 to 9.38 K) \cite{lara2011observation}, normal hydrogen molecule is replaced with neat para hydrogen molecule in order to facilitate the observation of structures in integral cross sections. Explicit measurements of integral cross sections clearly show undulation structures. It is believed that the structures observed at low energy where only a few partial waves contribute, are a signature of the sequential opening of individual channels. However, it also show obvious mismatch between theoretical calculation and experimental measurement, which is mainly attributed to the accuracy of effective potential in the calculation and the collision energy spread.

\begin{figure}
\begin{center}
\includegraphics[width=1.0\linewidth]{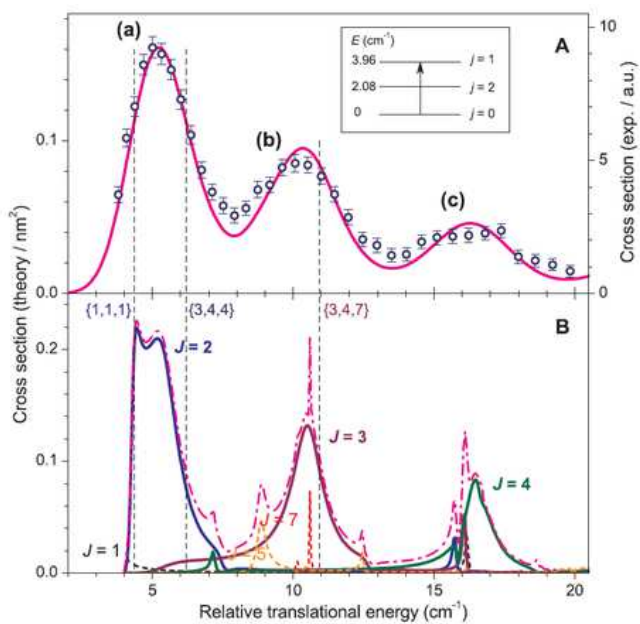}
\caption{Figure adapted from  \cite{chefdeville2013observation}.  Collisional energy dependence of the integral cross sections for \ce{O_2} excitation $|N=1,j=0\rangle \rightarrow |N=1,j=1\rangle$. (A) Experimental data with para-\ce{H2}. (B) Theoretical results with partial waves $J=2$,3,4 (solid lines), partial waves $J=1$ and $J=5$ to 7 (dashed lines), integral cross section (dashed-dotted line). }
\label{figure:nearcold_partial}
\end{center}
\end{figure}

In a following study of the \ce{S(^{1}D_2) + HD(j=0)} reaction at energies from 54.2 meV down to 0.46 meV (i.e. from 628.97 to 5.34 K) \cite{lara2012dynamics}, explicit measurements of integral cross sections for the \ce{HS + D} and \ce{DS + H} products and their branching ratios were compared to adiabatic time independent quantum mechanical calculations (section II. B). Neither the product channel branching ratio, nor the low energy resonance features between 1.5 and 5 meV in the HS + D channel can be reproduced by theoretical calculations. This discrepancy was attributed to the restriction of a single adiabatic electronic potential energy surface instead of a full time-dependent quantum mechanical treatment considering all four potential energy surface and their nonadiabatic couplings for the scattering calculations. Soon for the \ce{O2-H2} inelastic collisions performed with the same experimental apparatus \cite{chefdeville2013observation}, with a full quantum close-coupling scattering calculations (section II. B. 1) by using a four-dimensional ab initio potential energy surface obtained by the coupled cluster method while using single and double excitation with perturbative contributions from connected triple excitations and large sets of atomic basis orbitals, complete characterization of fully resolved partial wave resonances and excellent agreement with experimental integral cross sections has been obtained in 3.7-20 cm$^{-1}$ energy range, as illustrated in Figure ~\ref{figure:nearcold_partial}.

In the last decade, there is another direction which replace one of the crossed beam with Stark-decelerated molecular beam and have achieved many successes. The operation principles of Stark deceleration have been extensively described and reviewed \cite{van2012manipulation}. Its basic concepts, similar to the linear accelerator for charged particles, is that decelerating a molecular beam by using the interaction of neutral polar molecules with time-varying electric fields. A narrow velocity distribution of the original molecular beam will be selected and decelerated to any desired final velocity while maitain its narrow velocity distribution and particle density, with specific timing sequence of the electric field. Because the process is quantum state specific, the collision process can be studied with very high quantum state purities.  Thus the main advantage of this approach is the exquisite level of control over both velocity and the quantum state purity.

\begin{figure}
\begin{center}
\includegraphics[width=1.0\linewidth]{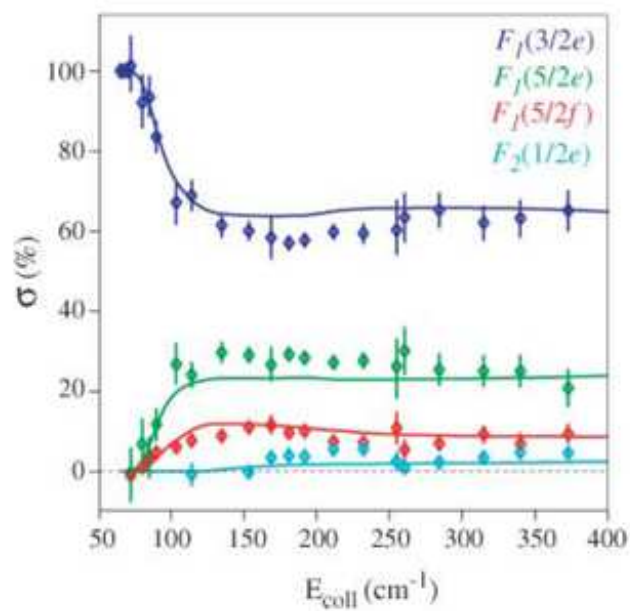}
\caption{Figure adapted from  \cite{gilijamse2006near}. Comparison of the measured (data points) and calculated (solid curves) relative cross-sections dependence on the collision energy. }
\label{figure:StarkDecScat}
\end{center}
\end{figure}

In the first scattering experiment with the Stark-decelerated molecules, a packet of OH ($X ^{2}\Pi_{3/2}, J=3/2, f$) radicals was decelerated and its velocity was tuned between 33 and 700 m/s, then scattered with a pulsed beam of Xe in a crossed-beam geometry with a 90$^\circ$ intersection angle \cite{gilijamse2006near}. The state-to-state integral scattering cross-sections were measured and calculated as a function of the collision energy between 50 and 400 cm$^{-1}$ with an energy resolution of 13 cm$^{-1}$. Figure ~\ref{figure:StarkDecScat} shows a resulting comparison of the collision energy dependence of the measured and calculated relative cross-sections. Clear threshold behavior in the cross-sections was observed where collision energies were resonant with the rotational energies of a specific rotational state of the OH radicals. Moreover, the measured relative cross-sections were in excellent agreement with that derived from coupled channels calculations (section II. B. 1) of ab initio calculated OH-Xe potential energy surfaces. Soon the measurement has been performed with other molecule-rare gas systems \cite{scharfenberg2010state, scharfenberg2011scattering}, and even with molecule-molecule systems such as \ce{OH-D_2} \cite{kirste2010low} and OH-NO \cite{kirste2012quantum}. These measurements provide a very sensitive probe for the theoretical potential energy surfaces, from which a deep understanding of the collision dynamics in these systems have been achieved.

\begin{figure}
\begin{center}
\includegraphics[width=1.0\linewidth]{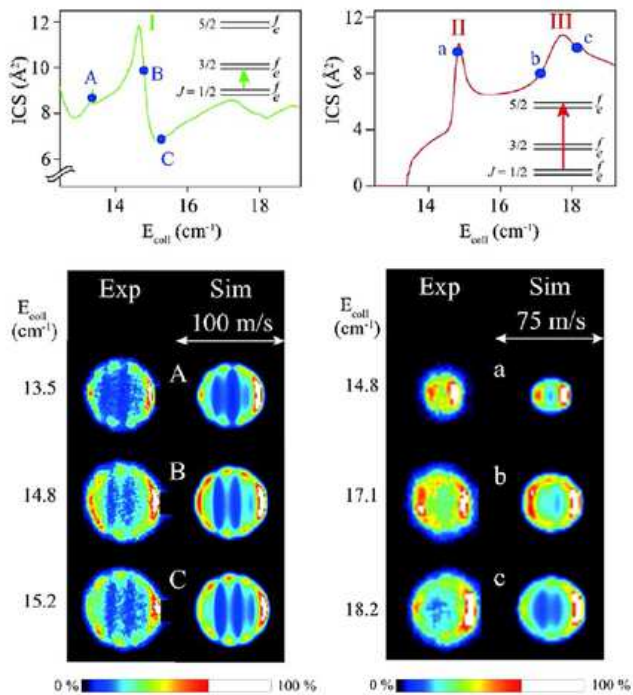}
\caption{Figure adapted from \cite{vogels2015imaging}. Experimental and simulated ion images at selected collision energies for channels of $|J=1/2, f\rangle \rightarrow |J=3/2, e\rangle$ (left) and $|J=1/2, f\rangle \rightarrow |J=5/2, f\rangle$ (right) in inelastic collisions between NO and He. }
\label{figure:IonImaging}
\end{center}
\end{figure}

Besides, this approach is also a useful tool for imaging the scattering resonances in atom-molecule collisions. Using a combination of Stark deceleration and velocity-mapped imaging technique, the fingerprints of scattering resonances in differential cross-sections (DCSs) for inelastic collisions between NO radicals and He atoms are observed \cite{vogels2015imaging}. In this experiment, the slow He beams from a cryogenically cooled Even-Lavie valve crossed with the Stark-decelerated NO beam at an angle of 45$^\circ$, resulting in collision energies as low as 12.5cm$^{-1}$, equivalent to a temperature of 12.0 K. The relative velocity vector was chosen amost perpendicular to the He propagation direction, which resulted in an optimal energy resolution of $\sim0.3$ cm$^{-1}$ at collision energies near the resonances. With these experimental configurations, the scattering images were measured at collision energies between 13 and 19 cm$^{-1}$, which has been pridicted as where scattering resonances take place by high-level quantum-mechanical scattering calculations. As shown in Figure ~\ref{figure:IonImaging}, striking variations in the angular distributions were observed when varying the energy in small steps over the resonances. As seen from the Figure ~\ref{figure:IonImaging}, the diameter of the images for the $|J=1/2, f\rangle \rightarrow |J=5/2, f\rangle$ scattering channel is approaching zero when the collision energy reach the thermodynamic threshold of the channel, which reveals a clear threshold behavior. In general, the measured DCSs agree very well with that obtained from the ab initio calculations. Since scattering resonance is very sensitive to the potential energy surface \cite{chefdeville2012appearance}, this could be a promising method to provide key information for the theoretical models.

To conclude, crossed-beam experiment is an excellent tool to study the collision/reaction dynamics, obtaining integral cross sections, product angular distributions and product translational energy distributions. But it is difficult to identify each of the peaks in the structure of cross section and assign them to the contributions of specific partial waves at near cold regime where almost all partial wave channels contribute to the total integral cross sections. Besides, the theoretical calculation of integral cross section at few Kelvin range becomes highly sensitive to even small inaccuracies in the potential energy surface, thus challenge and push the theorists to further improve the accuracy of the ab initio calculation on the potential energy surface and of scattering calculation methods. Nevertheless, by further reducing the crossed beam angle to zero degree and precisely control the relative velocity between both beams, the collision energy can be tuned to cold regime, thus only few partial waves come into play in the collisions. We will discuss it in the subsection of merged molecular beam.

\subsection{Intra-beam collision}

\paragraph{CRESU}

\begin{figure}
\begin{center}
\includegraphics[width=1.0\linewidth]{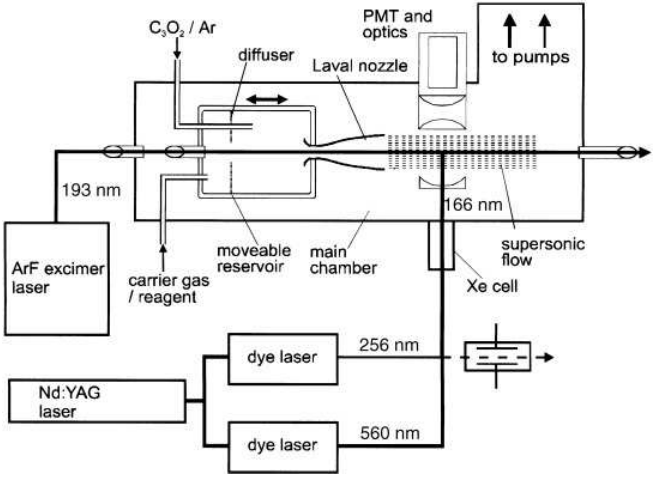}
\caption{Figure adapted from \cite{chastaing1999neutral}. Schematic diagram of the CRESU apparatus for the study of C($^{3}P$) reaction kinetics. The C($^{3}P$) atoms were created by photolysis of \ce{C3O2} at 193nm and observed by laser induced fluorescence in the vacuum ultraviolet.  }
\label{figure:cresu}
\end{center}
\end{figure}

There are a few types of collision studies can be viewed as intra-beam collisions, where collisions between different collision partners can be studied inside single molecular beam. The first one is CRESU, which is the abbreviation of the Cin\'etique de R\'eaction en Ecoulement Supersonique Uniforme. In the CRESU apparatus, a continuous, uniform supersonic flow of constant density and temperature is generated by expansion of gas from a reservoir at moderate pressure through a convergent-divergent Laval nozzle \cite{wm2008low}. Its schematics is shown in Figure ~\ref{figure:cresu}. At the exit of the Laval nozzle, as there is no further expansion downstream, the flow parameters (i.e. temperature, density, pressure and velocity) do not exhibit any axial and radial variations in the centre of the jet, where the flow is isentropic for several tens of centimeters. The diffusion velocity is always negligible with respect to the bulk velocity therefore avoiding the condensation problem with the use of cryogenically cooled cells. As a consequence, in such expansions, heavily supersaturated conditions prevail. 

Typical gas density as large as $10^{16}$-$10^{17}$ cm$^{-3}$ in the uniform supersonic flow ensures that frequent collisions take place during the expansion and subsequent flow, maintaining thermal equilibrium. Therefore, such uniform supersonic flows provide an ideal flow reactor for the study of chemical reactivity at low and very low temperature, where the kinematics, most obviously the rate coefficients can be measured. Notice, in the crossed-beam experiment, the concept of temperature is not really valid, and the supersonic flow is inhomogenius, more importantly the time scale for reaction to happen does not exist as single collision event dominates. Consequently, only dynamics instead of kinematics exists in the crossed-beam studies. 

The first measurements of rate coefficients down to 20 K were reported in 1984 \cite{rowe1984study,dupeyrat1985design}. The CRESU apparatus provides a versatile method for measuring rate constants and have been applied to many neutral atom-molecule and molecule-molecule reactions involving radical species in the 1-100K range \cite{sims1995gas,smith2000reaction,smith2006reactions}. The lowest temperature in a CRESU experiment have achieved 7 K so far \cite{james1998combined} with a precooled reservoir by liquid \ce{N2}. Despite great success, it also have few disadvantages: the supersonic flow is uniform for only tens of centimeters, which restricts the available hydrodynamic time (i.e. the observation time scale of change in concentration for a reagent) to few hundreds $\mu s$, and new nozzle for each set of conditions, requirement of large pumping capacities, as well as limited range of total gas densities \cite{smith2000reaction}.

\paragraph{SARP}

\begin{figure}
\begin{center}
\includegraphics[width=1.0\linewidth]{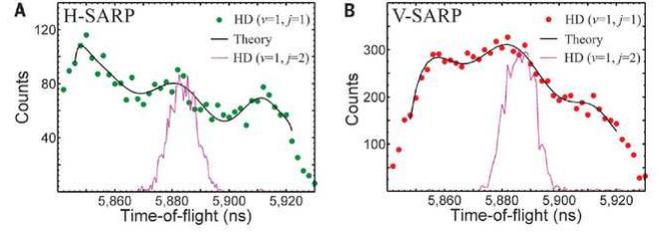}
\caption{Figure adapted from \cite{perreault2017quantum}. Time-of-flight distribution of rotationally relaxed HD ($v=1, j=1$).  }
\label{figure:scatteredHD}
\end{center}
\end{figure}

The Second is coexpanding gases in a mixed molecular beam following Stark-induced Adiabatic Raman Passage (SARP) \cite{perreault2017quantum,perreault2018cold,perreault2019hd}. This method prepare both colliding partners in a single coexpanding supersonic beam, thus bring the collision temperature down to 1 Kelvin, restricting scattering to s and p partial waves. With coherently transfer the population of one partner to an excited state and alignment of the molecular bond axis parallel/perpendicular to the relative velocity of the colliding partners using SARP, HD molecules are prepared in specific $m$ states of a rovibrationally excited energy eigenstate. Thus, full stereodynamic control of the scattering event can be achieved. In Figure ~\ref{figure:scatteredHD}, the time-of-flight spectrum of scattered HD ($v=1, j=1$) demonstrates remarkably stereodynamic effects. When the molecular bond axis is aligned perpendicular to the relative velocity of the colliding partners (V-SARP), the time-of-flight distribution of HD ($v=1, j=1$) is nearly three times strong compared to that when the molecular bond axis aligned parallel to the relative velocity of the colliding partners (H-SARP). This can be understood by the partial-wave analysis of the time-of-flight distribution. 

Assuming the quantization $z$ axis oriented along the relative velocity of the colliding partners, which coincides with the lab-frame molecular beam axis as well as the time-of-flight axis, then the angular distribution $\frac{d\sigma}{d\theta}$ of inelastically scattered HD ($v=1, j=1$) as a function of the polar angle $\theta$ is obtained by integrating the differential scattering cross section $\frac{d\sigma}{d\Omega}$, per unit solid angle over all azimuthal angle $\phi$,
\begin{equation}
\begin{split}
&\frac{d\sigma}{d\theta}(j=2 \to j=1) = \sin{\theta}\int_{0}^{2\pi}[(\frac{d\sigma}{d\Omega})_{j=1,m_{f}=0}\\
&+(\frac{d\sigma}{d\Omega})_{j=1,m_{f}=+1}+(\frac{d\sigma}{d\Omega})_{j=1,m_{f}=-1}]d\phi,
\end{split}
\end{equation}
where $d\Omega$ gives the differential solid angle, $m_f$ refers to the the $z$-axis projectio of the angular momentum vector $\textbf{j}$ of the scattered HD ($v=1, j=1$). For each outgoing channel(designated by a given value of $m_f$), the scattered angular distribution for V-SARP results from the interference of the scattering amplitudes associated with three input channels,
\begin{equation}
\begin{split}
&(\frac{d\sigma}{d\Omega})^{V}_{j=1,m_{f}}=|\sqrt{3/8}q_{j=2,m=+2\to j=1,m_{f}}\\
&-1/2 q_{j=2,m=0\to j=1,m_{f}}+\sqrt{3/8}q_{j=2,m=-2\to j=1,m_{f}}|^{2},
\end{split}
\end{equation}
while for H-SARP, the differential scattering cross section is given by
\begin{equation}
\displaystyle (\frac{d\sigma}{d\Omega})^{H}_{j=1,m_{f}}=|q_{j=2,m=0\to j=1,m_{f}}|^{2},
\end{equation}

\begin{figure}
\begin{center}
\includegraphics[width=1.0\linewidth]{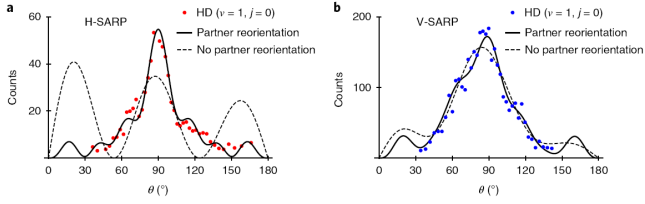}
\caption{Figure adapted from \cite{perreault2018cold}. The angular distribution of the scattered HD ($v=1, j=0$) derived from time-of-flight measurements of the REMPI-generated ions.  }
\label{figure:angular}
\end{center}
\end{figure}

The angular distributions can be extracted from experimentally measured time-of-flight distributions of scattered HD molecules. Figure ~\ref{figure:angular} show the angular distribution of the scattered HD ($v=1, j=0$) for HD molecular bond axis aligned parallel and perpendicular to the relative velocity between HD and \ce{D2}. Therefore, the full stereodynamics in molecular collisions are well studied for \ce{HD-H_2} and \ce{HD-D_2} systems using measured four-vector correlations \cite{barnwell83} and partial wave analysis. 

\paragraph{Cryofuge}

\begin{figure}
\begin{center}
\includegraphics[width=1.0\linewidth]{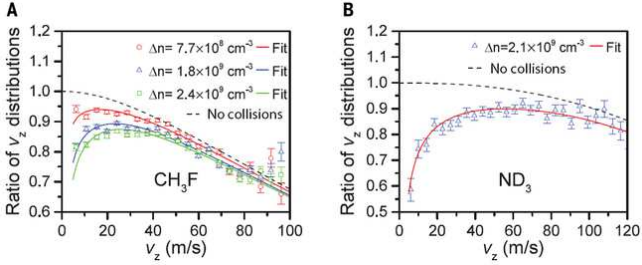}
\caption{Figure adapted from \cite{wu2017cryofuge}. Depletiion from molecular collisions in the centrifuge-decelerated beam for both (a) \ce{CH3F} and (b) \ce{ND3}.  }
\label{figure:cryofuge}
\end{center}
\end{figure}

The third method is the cryofuge \cite{wu2017cryofuge}, which uses centrifugal force to bring cryogenically cooled molecules to kinetic energies below 1 K in the laboratory frame. In the experiment, clear collisional losses are observed due to attained density higher than $10^{9} cm^{-3}$ and interaction times longer than 25 ms for samples of \ce{CH3F} and \ce{ND3}. In Figure ~\ref{figure:cryofuge}, the collisional loss dependence on the molecular density $n$ and the longitudinal velocity $v_z$ of the guided beam are investigated. In a guided beam, the collisional loss modifies longitudinal velocity distribution according to $P=P_{0}exp(\frac{-k_{loss}nL}{v_{z}})$, where $k_{loss}=\sigma_{loss}v_{rel}$ is the loss rate coefficient. Thus, the ratio for two longitudinal velocity distributions can be explained by an $exp(\frac{-\alpha k_{th}\Delta n L}{v_z})$ model, where $k_{th}$ is the theoretically predicted loss rate coefficient and density difference $\Delta n = n_{high}-n_{low}$. Using the semiclassical eikonal approximation and the Langevin capture model (section II. B. 1), the theoretical loss rates are estimated to be $k_{th}^{\ce{CH_3F}}=7.7\times 10^{-10} cm^{3}s^{-1}$ and $k_{th}^{\ce{ND_3}}=1.3\times 10^{-9} cm^{3}s^{-1}$ at 0.8 K and 1.1 K collision energy, respectively. The values are about $40\%$ and $60\%$ larger than theory, which reveals the imprecision of the Langevin model.  Additionally, the ratio of the loss rate coefficient agrees well with the semiclassical calculations \cite{cavagnero2009}, which predicts a rate coefficient $k\propto d^2$, where $d$ is the dipole moment of the molecule. 

\subsection{ Collisional cooling }

Collisional cooling, typically constrain the candidate gas in a vapor cell with preferentially cryogenically cooled helium gas, making it possible to study gas phase collisional processes under equilibrium conditions in a temperature-variable environment. Frank C. De lucia et al. studied pressure broadening coefficients and rotational inelastic cross sections of various molecular collisions like CO-He and \ce{CH3F}-He with millimeter wave spectroscopy \cite{willey1988very, willey1988very2,ball1998direct,ball1999direct} in the late 1980s and 1990s. This enables many molecular collision processes, especially of astronomy importance to be studied at temperatures far below the limits ordinarily imposed by their vapor pressures. Recently, David Patterson group studied conformational dynamics of large polyatomic molecules  \cite{drayna2016direct}. In the experiment, the conformational relaxation of 1,2-propanediol in collisions with helium in 6 K cryogenic environment is observed directly using microwave spectroscopy.

However, due to the limitation imposed by the current cryogenic technology, the lowest temperature  is minimally ~300mK and it suffers from the disadvantage that any gas would condense on the cold walls of the refrigerated-cooled cell. As we can see in section IV. A(buffer gas cooling), with the capability of in-situ magnetic trapping, the collision energy has directly been brought down to cold regime, opening many new possibilities in studying molecular collisions.

\subsection{ Target collision }

Early Target collision studies was made immediately after capability of trapping molecular samples was realized. An example of the very first measurement is \ce{OH-N_2} total scattering cross section at ambient 295 K in a magnetic trap, where OH molecules are loaded from a Stark decelerator \cite{sawyer2007magnetoelectrostatic}. This is recently followed by a collision study of magnetically trapped \ce{CH3} with ambient \ce{H2} molecules \cite{liu2017magnetic}. In such experiments, the measured corss-section is an average over background gas with broad Boltzmann velocity distribution, since they are at room temperature.

\begin{figure}
\begin{center}
\includegraphics[width=1.0\linewidth]{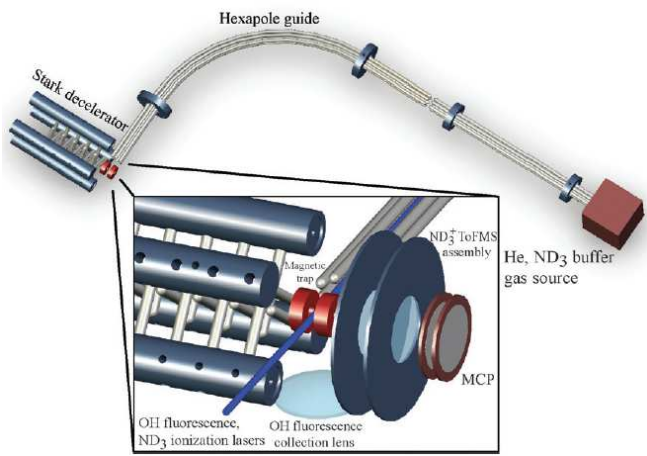}
\caption{Figure adapted from \cite{sawyer2011cold}. Illustration of the combined experimental assembly for study of low energy collision of \ce{ND3} and OH.  }
\label{figure:target}
\end{center}
\end{figure}

In order to reach lower collision energies, Sawyer et al. have measured OH-He and \ce{OH-D_2} cross sections by using a liquid-nitrogen cooled supersonic beam of He or \ce{D_2} to bombard the target, i.e. magnetically-trapped OH molecules \cite{sawyer2008molecular}. The experiment reveals a striking feature in the \ce{OH-D_2} collisions, a pronounced peak in the cross section at ~305 cm$^{-1}$ emerges, which suggests resonant energy transfer possibly contributing from $J=3 \leftarrow J=1$ transition or collision-induced decay of the \ce{^{1}\Sigma_{g}^{+} D_2} molecule. Following this, the collision energy is further lowered using a slow \ce{ND3} beam, which was buffer-gas-cooled and subsequently velocity filtered by a 90$^\circ$ bent hexapole guide, colliding with magnetically trapped OH molecules \cite{sawyer2011cold}. The experimental assembly is illustrated in Figure ~\ref{figure:target}. The absolute trap loss cross sections are measured and electric-field enhancement of collision cross sections have been observed.

In a similar fashion, elastic scattering between Li atoms and \ce{SF6} molecules is studied \cite{strebel2012quantum}. In this study, the Li atoms instead of molecules are confined in a magneto-optical trap (MOT), while \ce{SF6} molecules are released from a nozzle mounted at the tip of a fast-spinning rotor \cite{gupta2001slowing, strebel2010improved}, which enable the tuning of velocity through varying the rotating frequency of the rotor. By detecting the trap loss of the Li atoms out of the MOT and scattering products in the direction of the molecular beam, multiple rainbow features are resolved, revealing a rich structure of orbiting resonances.

Such target collision experiment has an inherent advantage: it can directly measure the absolute collision cross section by using trapped molecules as the scattering target \cite{sawyer2008molecular}. Furthermore, the technique has a high sensitivity to the trap loss and to elastic differential cross sections, allowing for direct measurement of the differential cross section in a trap of variable depth. However, because the trapping potential requires a minimum energy tranfer for an elastic collision to eject a molecule from the trap, the observed trap loss cross section is the sum of the inelastic cross section and some unknown fraction o fthe elastic cross section.

\section{Experiment: Cold collision}

In the community of atomic, molecular and optical physics, chemical physics and astrochemistry, there is always a great desire of studying cold molecular collisions, because they are ideal to observe the quantum nature of molecular collisions since the angular momentum becomes severely constrained, eventually reaching the limit $l=0$ where only s-wave scattering could take place and the de Broglie wavelength associated with relative velocity becomes comparable to or even greater than the size of the reactants \cite{thomas2004imaging}. It is also an ideal platform to obtain precise information of interatomic interaction potential and to study cold chemistry. Moreover, it can be used to test various state-of-art ab initio calculations on molecular potential energy surfaces, and different scattering theories in obtaining integral and differential cross section.

Despite of its importance, it have been out of reach for many decades in atomic and molecular collision studies. The success of laser cooling and trapping render full ability to studies of cold collisions, however the researches have since been mainly focused on ultracold atomic collisions where collisions can be studied in complete quantum regime with only the lowest partial wave exist, reducing the complexity of the related researches. While most of chemically interested species cannot be cooled by the laser, since laser cooling technique is limited to alkali and alkali-earth metal, or atomic species whose electronic configuration bear striking resemblance. Nonetheless, a few techniques have been developed so far, and in the following we will discuss three main techniques, which are buffer gas cooling, merged molecular beam and molecular synchrotron.

\subsection{Buffer gas cooling}

This method relies on the elastic collisions between the target molecules and cryogenically precooled helium/neon gas. It is a universal technique and in principle can be applied to any kind of molecule. In late 1990s, coupling in situ magnetic trapping enabled by superconducting magnets with cryogenic cooling, group of John Doyle and Jonathan Weinstein are able to study Zeeman relaxation rate and cross section in cold collisions of many molecules with helium, including CaH \cite{weinstein1998magnetic}, CaF \cite{maussang2005zeeman}, and NH \cite{campbell2007magnetic,campbell2009mechanism,tsikata2010magnetic}.

\begin{figure}
\begin{center}
\includegraphics[width=1.0\linewidth]{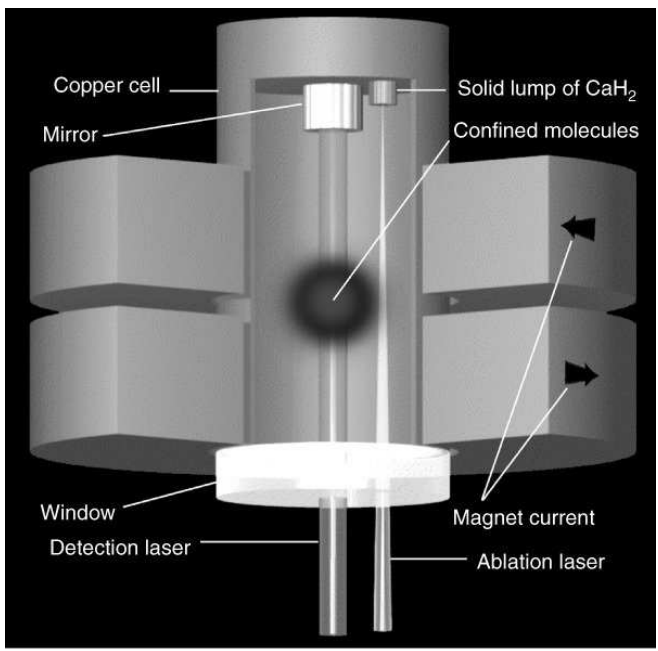}
\caption{Figure adapted from \cite{weinstein1998magnetic}. Schematic of magnetic trapping of CaH molecules collisionally cooled by He, which in term cryogenically cooled by dilution refrigerator.  }
\label{figure:cryogenics}
\end{center}
\end{figure}

In Figure ~\ref{figure:cryogenics}, the CaH molecule is produced from laser ablation of a solid lump of \ce{CaH2} inside the cryogenic cell with temperature up to $\sim1000 K$.  Successively the molecules exchange energy with the cold bath of \ce{^{4}He} at a temperature of 800 mK or \ce{^{3}He} at 240 mK. Typical gas density of \ce{^{4}He} is about $10^{16}cm^{-3}$ and cross section for elastic scattering is assumed to be on the order of $10^{-14}cm^{2}$. After about 100 collisions, the molecules are cooled to the same temperature of these buffer gas and subsequently be trapped by a superconducting magnetic field positioned outside the cell and arranged in an anti-Helmoltz configuration, whose strength is about 3 T, corresponding to a trap depth of 2 K for CaH molecules. The molecules are excited through  $|B^{2}\Sigma^{+}, v'=0\rangle \leftarrow |X^{2}\Sigma^{+}, v" =0\rangle$ transition and detected using rotational transition $|N'=0, J'=3/2\rangle \leftarrow |N"=0, J" =1/2\rangle$ from the radiative decay of $|B^{2}\Sigma^{+}, v'=0\rangle \rightarrow |X^{2}\Sigma^{+}, v" =1\rangle$. By ovserving the time evolution of the molecules, the dynamics of the loading and trapping process can be studied.

\begin{figure}
\begin{center}
\includegraphics[width=1.0\linewidth]{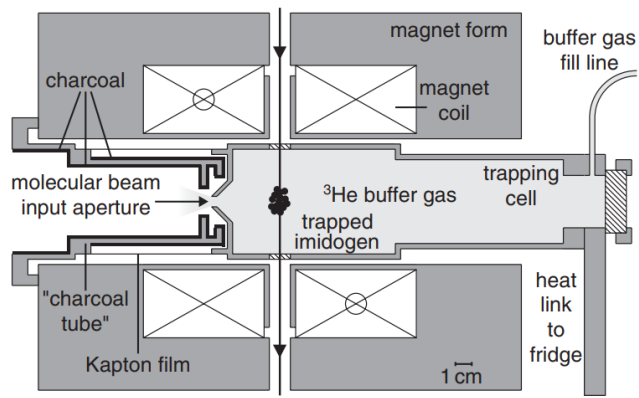}
\caption{Figure adapted from  \cite{campbell2007magnetic}. Schematic diagram loading NH beam into the magnetic trap.  }
\label{figure:beam_loading}
\end{center}
\end{figure}

Except for the in-situ loading by the laser ablation, the molecules can also be loaded using a molecular beam \cite{egorov2004buffer,campbell2007magnetic}.  As in Figure ~\ref{figure:beam_loading}, the NH radical beam, which is converted from ammonia by room-temperature glow discharge, enter the cryogenic cell through an orifice. The molecules are detected by laser induced fluorescence and absorption spectroscopy on the $|A^{3}\Pi, v'=0\rangle \leftarrow |X^{3}\Sigma^{-}, v" =0\rangle$ transition. They are trapped with $1/e$ lifetimes exceeding 20 s \cite{tsikata2010magnetic}. Cold N-NH collisions at a temperature of $\sim600$ mK is also studied \cite{hummon2011cold}. The \ce{N + NH} trap loss rate coefficient attributing to inelastic relaxation is measured, and the ratio of their elastic-to-inelastic collisions is large enough over the temperature range $\sim10$ mK - 1 K from ab initio quantum scattering calculations.

Moreover, the formation and dynamics of van der Waals molecules \cite{brahms2011formation}, including \ce{Ag^{3}He(^{2} \Sigma ^+)} \cite{brahms2010formation}, \ce{^7Li^4He} \cite{tariq2013spectroscopic}, and \ce{^{48}Ti^4He} \cite{quiros2017cold}, have also been studied. In these experiments, the formation of weakly bound He-containing van der Waals molecules through three-body recombination is confirmed either by their dynamics such as spin relaxation in the magnetic trap \cite{brahms2011formation}, or by laser spectroscopy \cite{tariq2013spectroscopic}.

Currently, buffer gas cooling is not only used for the collisional studies, it is a widely used technique in many aspects of the molecular physics and chemistry, where molecules cooled by the buffer gas cooling are used as ideal samples for deceleration of the molecular beam \cite{fabrikant2014, wu2017cryofuge, petzold2018}, laser cooling \cite{shuman2009radiative, shuman2010laser}, opto-electrical (or Sisyphus) cooling \cite{zeppenfeld2012sisyphus, prehn2016}, precision microwave and optical spectroscopy \cite{patterson2013enantiomer, eibenberger2017enantiomer, spaun2016continuous, changala2019rovibrational}, measurement of the electric dipole moment of the electron \cite{baron2014, ACME2018}, and cold ion-molecule reactions \cite{chen2019new, greenberg2020cold}.

\subsection{Merged molecular beam}

\begin{figure}
\begin{center}
\includegraphics[width=1.0\linewidth]{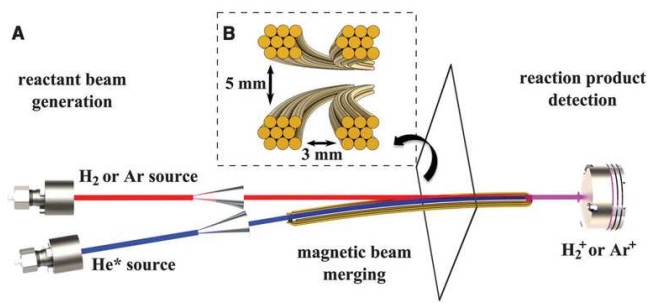}
\caption{Figure adapted from \cite{henson2012observation}. Schematic of Penning ionization reaction setup with merged atomic and molecular beam. The supersonic beam of metastable helium (blue) moving along the curved magnetic quadrupole guide merges with another beam of \ce{H2} or Ar (red) at the exit of the guide.  }
\label{figure:merge}
\end{center}
\end{figure}

As we have seen in the previous section, the relative collision energy is in near cold regime or even high-temperature regime in typical crossed molecular beam experiments. In order to reduce the relative collision energy to cold regime, different proposals using supersonic expansion technique have been made. Qi Wei et al. proposed merging slowed beam from counter-rotating nozzle with pulsed molecular beam such that the scattering angle approaching zero, and relative velocity can be smoothly tuned to ultracold collision regime by adjusting the repetition frequency of the counter-rotating nozzle and the condition of supersonic expansion for pulsed molecular beam \cite{wei2012merged}. This idea of beam-merging has been first realized in a typical \ce{He^* + H2 \rightarrow He + H2^+} Penning ionization experiment \cite{henson2012observation}, where metastable helium generated by dielectric barrier discharge is guided by a curved magnetic quadrupole guide with a small angle and large curvature, and merge with a pulsed beam of molecular hydrogen at zero degree.

\begin{figure}
\begin{center}
\includegraphics[width=1.0\linewidth]{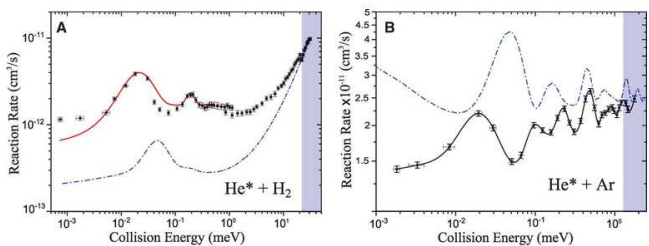}
\caption{Figure adapted from \cite{henson2012observation}. Reaction rate measurement for (A) \ce{(^{3}S)He^*} and \ce{H2}, (B) \ce{(^{3}S)He^*} and Ar. In panel A, red solid line is the calculated reaction rate by using the Tang-Toennies potential. In panel B, the solid black line is a guide to the eye. }
\label{figure:reaction_rate}
\end{center}
\end{figure}

The schematics is shown in Figure ~\ref{figure:merge}. Their relative velocity can be finely tuned from 350K to the lowest collision energy of $\sim10$ mK. With this ability, they have observed orbiting resonance features in the reaction rate. More importantly, this method enables the investigation of cold collisions of atomic/molecular beams with collision energy spanning up to 6 order of magnitude, as is shown in Figure ~\ref{figure:reaction_rate}. It thus bridges an enormous energy gap between conventional studies of crossed molecular beam collisions and ultracold collision studies, which have been beyond of reach for many decades in collision studies. 

\begin{figure}
\begin{center}
\includegraphics[width=1.0\linewidth]{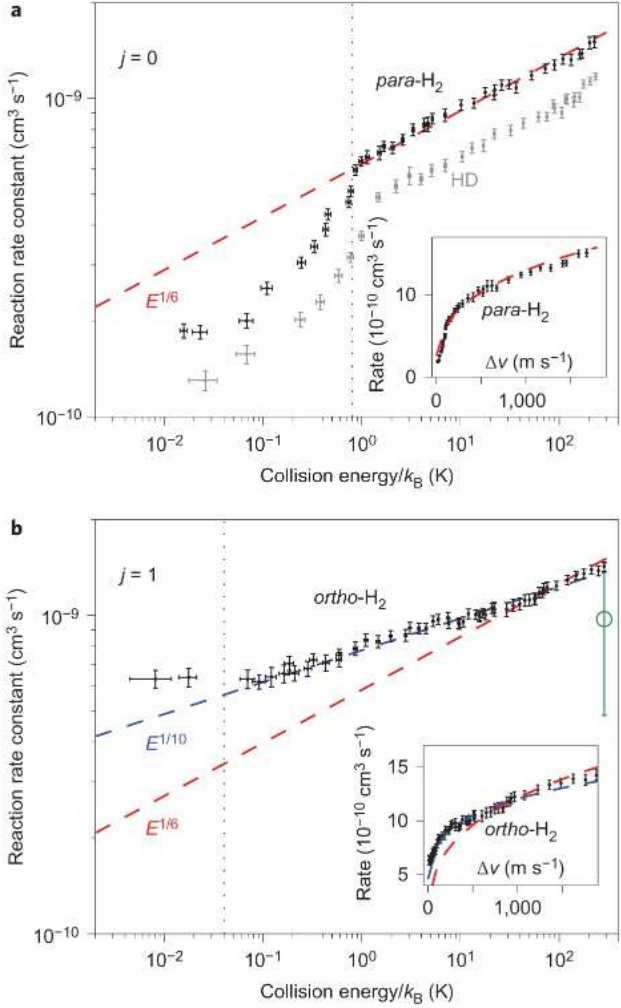}
\caption{Figure adapted from \cite{Shagam2015}. Reaction rate measurement for hydrogen molecules in (a) ground and (b) excited rotational states with \ce{He(2^3P_2)} from 10 mK to 300 K.   }
\label{figure:para2ortho}
\end{center}
\end{figure}

With this technique, the isotope effect on the reaction rate has been studied \cite{lavert2014observation}. Partial waves of collision with different isotopologues causes scattering resonances at different collision energies, leading to clearly different reaction rates for different isotopes in the collision temperature range from 10 mK to 100 K. For the same isotope of hydrogen molecule reacting with helium, the reaction rate constant with \ce{He(2^3P_2)} is more than two order of magnitude than reaction with \ce{He(2^3P_2)}. It also show significant dependence on the rotational state \cite{Shagam2015}, as is shown in Figure ~\ref{figure:para2ortho}. Reaction rate constant of para-\ce{H2} molecules with \ce{He(2^3P_2)} follows theoretical universal Langevin prediction (section II. B. 1) with the $E^{1/6}$ scaling law down to 0.8 K, below which the rate constant falls below $E^{1/6}$ scaling and eventually slow by more than $50\%$ at 10mK lowest collision energy. While the reaction rate constant with rotationally excited ortho-\ce{H2} is consistent with the $E^{1/6}$ scaling law down to $\sim30$ K,  below which follows scales with $E^{1/10}$ until becoming a constant value when the collision energy is less than 0.04 K, which is consistent with the Wigner threshold laws (section II. B. 2) . By comparison, the reaction of ortho-\ce{H2} with \ce{He(2^3P_2)} is faster than the reaction of para-\ce{H2} with \ce{He(2^3P_2)} due to emerging anisotropic quadrupole-quadrupole interaction in the effective long-range potential for rotationally excited molecules.

\begin{figure}
\begin{center}
\includegraphics[width=1.0\linewidth]{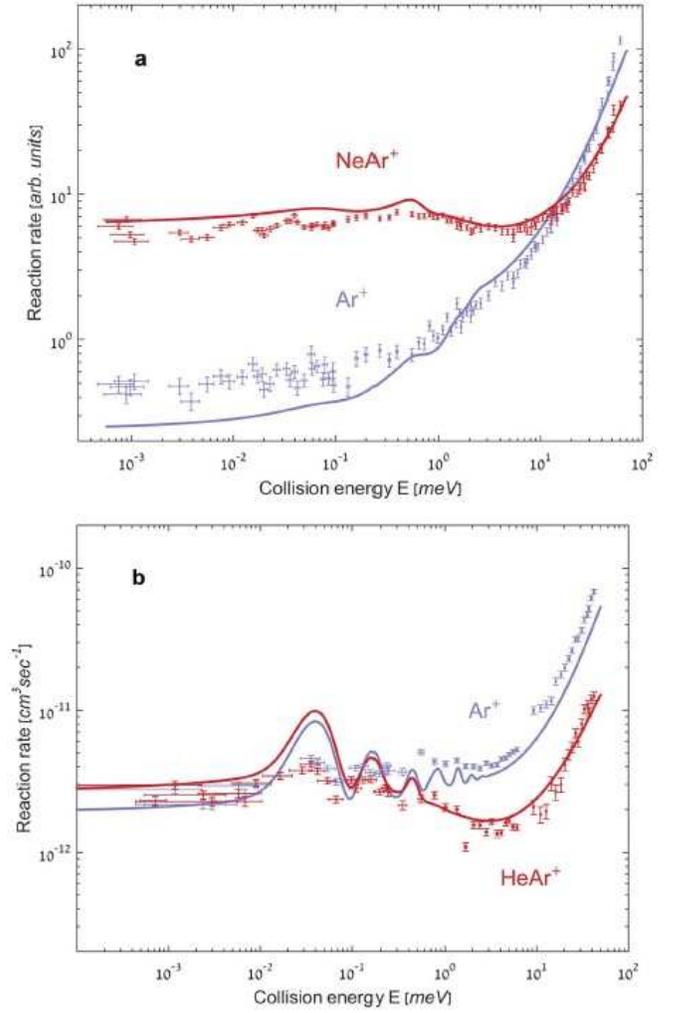}
\caption{Figure adapted from \cite{bibelnik2019cold}. Reaction rate measurement of associative ionization (red) and Penning ionization (blue) in reaction \ce{Ne(^3P) + Ar} (a) and \ce{He(^3S) + Ar} (b). The solid lines denote numerical simulations. }
\label{figure:branching}
\end{center}
\end{figure}

Following this research, the effect of anisotropy on shape resonances in cold Penning ionization reactions of \ce{He(2^3S_1)} with molecular hydrogen in the rotational ground and first excited states have been demonstrated \cite{klein2017directly}. Only when the molecule is rotationally excited does the scattering resonance at the collision energy of $k_B \times 270$ mK appear in the Penning ionization. The isotropic potential dominates the reaction dynamics when molecular hydrogen is in the ground rotational state, while for \ce{He(2^3S_1)} + ortho-\ce{H2} the contribution of the anisotropic part of the potential reduces the well depth with total angular momentum $J=3$, thus raises the bound state above the dissociation threshold by 270 mK which is accessible via tunnelling in scattering experiments. Except the effects mentioned above, the possible outcome of a chemical reaction, namely the branching ratio, could also be affected by the collision energy \cite{bibelnik2019cold}. In barrierless systems such as Penning ionization system \ce{Rg^* + Ar}(here Rg represents rare gas), the reaction is governed by long-range interactions. The branching ratio is sensitive to collision energy and its dependence on collision energy can be used for better understanding the reaction dynamics and reconstructing interaction potentials. In the measured collision energy range, the change of two orders of magnitude in the branching ratio is observed for both \ce{Ne(^3P) + Ar} and \ce{He(^3S) + Ar} systems, as Figure ~\ref{figure:branching} shows. Similar behaviour can be expected in chemical reactions involving molecules.

The merged molecular beam technique has also been extended to study cold collisions with polyatomic molecules. The Ecole Polytechnique F\'ed\'erale de Lausanne (EPFL) group have studied a set of Penning ionizations, such as \ce{Ne^* +NH3}, \ce{Ne^* +ND3} \cite{jankunas2014dynamics, bertsche2014low}, \ce{He(^3S_1) +NH3} \cite{jankunas2015observation} and \ce{He(^3S_1) +CHF3} \cite{jankunas2016communication} from 0.1 to 100 K. In their experiments, the supersonic beam of ammonia molecule and metastable noble gas is injected into curved electrostatic guide and curved magnetic guide, respectively. Each beam perfectly overlap with each other in space and time when they exit the guide.

\subsection{Molecular synchrotron}

Molecular synchrotron can be viewed as a variation of crossed-beam apparatus with 0$^{\circ}$ (i.e. merged beam) or 180$^{\circ}$ angle \cite{van2009collision}, and is similar to the fashion of a slow molecular beam colliding with particles in a trap, which is discussed in section 3.4. One difference is that molecular packets can be continuously loaded, producing large number density of molecules. As an arena for low-energy molecular collision studies, molecular synchrotron offers multiple desirable advantages. First, collision partners can move in the same direction, resulting a small relatvie velocity and thus a low collision energy. If we choose beams that move with the same speed as the stored molecules, in principle, zero collision energy can be attained. 

Second, since the cross sections of collisions involving neutral molecules or atoms are small (typically below 500 {\AA}$^2$) and molecular densities are low (typically $10^8$ molecules/cm$^3$) from currently available deceleration or cooling technique,  a sufficiently large sensitivity is highly demanded for studying cold molecular collisions. Since molecular synchrotron can store molecular packets for many round-trips, they can interact repeatedly with their collision partners at well defined times and at distinct positions. The longer the packets are stored before detection, the more molecules are lost due to (in)elastic collsions accumulate over the collision events, thus collision signals are accumulated and the sensitivity to collisions will be strongly enhanced.  It increases linearly with the number of collision events. Considering there are $n$ packets revolving clockwise with the same speed in the synchrotron consisting of $N$ segments, each packet of molecules collide with copropagating injection packet once every round trip. After $m$ round trips, it would have interacted $mn$ times. With an additional factor $\sqrt{n}$ due to summed intensity of the $n$ packets, the total gain in sensitivity is $mn\sqrt{n}$, which could amount to a factor of 100 - 1000 \cite{van2018detailed}. 

\begin{figure}
\begin{center}
\includegraphics[width=1.0\linewidth]{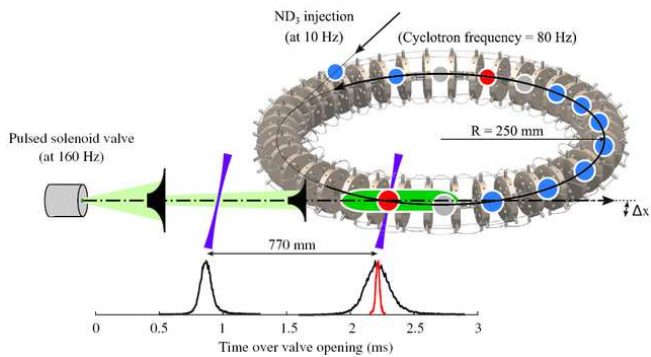}
\caption{Figure adapted from \cite{van2018cold}. Schematic view of the synchrotron and beam line. Supersonic packets of Ar (green) from a cooled Even-Lavie valve encounter copropagating probe \ce{ND3} packets (red) stored in the synchrotron on every round-trap, while reference \ce{ND3} packets (blue) provide a simultaneous measurement of the background loss. }
\label{figure:synchrotron2}
\end{center}
\end{figure}

The molecular synchrotron was demonstrated by Heiner et al. \cite{heiner2007molecular, heiner2009synchrotron}, which breaks a molecular storage ring \cite{crompvoets2001prototype, crompvoets2004dynamics} into two half-circles and switches the voltages on the hexapole electrodes such that molecules are bunched together as they fly through the gap between the two half rings. In this way, the molecular packets are both transversely and longitudinally confined, thus avoiding spreading out along the flight direction and maintaining sufficient density to perform collision experiments. In this prototype synchrotron, the molecules were kept together for over 100 round trips, and injection of multiple molecular packets was demonstrated. With an improved synchrotron consisting of 40 straight hexapoles placed in a circle \cite{zieger2010multiple}, molecular packets of \ce{ND3} possessing both small velocity spread ($\sim10$ mK) and widely tunable velocities (100-150 m/s), were stored over 150 round trips (over 13 s storage time) and up to 19 packets were hold simultaneously. 

\begin{figure}
\begin{center}
\includegraphics[width=1.0\linewidth]{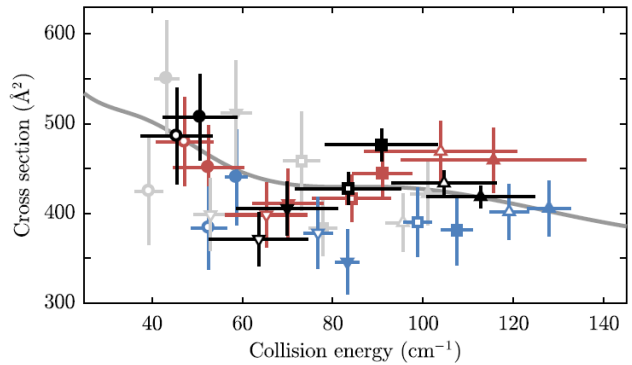}
\caption{Figure adapted from \cite{van2018cold}. Integrated collision cross section for \ce{ND3 + Ar} versus collision energy. The measurements are fit to a theoretical calculation \cite{loreau2015scattering}. }
\label{figure:synchrotron}
\end{center}
\end{figure}

Recently, the collisions between \ce{ND3} stored in a 50 cm diameter synchrotron and Ar atoms in copropagating supersonic beams are studied \cite{van2018cold}. The schematics of the experiment is shown in Figure ~\ref{figure:synchrotron2}. In this study, the collision energy are tuned by varying the velocity of the stored \ce{ND3} packets, by varying the temperature of the pulsed valve for Ar atoms, and by varying the timing between the Ar beam and \ce{ND3} packet. Figure ~\ref{figure:synchrotron} shows the total cross section obtained from the measured loss rates and agrees well with the quantum close-coupling calculations \cite{loreau2015scattering}.

\section{Experiment: Ultracold collisions}

Understanding collisions involving molecules at the quantum level has been a long-standing goal in chemical physics. As elastic, inelastic and reactive collision cross-section are sensitive to the details of both short-range and long-range part of the molecular interaction potential, ultracold molecular collision offers an ideal probe of the potential energy surface, in other words, the measurement of scattering resonances provides an accurate probe of the potential energy surfaces. In this perspective, ultracold molecules offer great opportunities to study molecular collisions in the quantum regime, because the de Broglie wavelength of the collision partners is much larger than the range of molecular interaction potential at ultracold temperature and thus only the lowest partial wave of relative orbital angular momentum dominates the collision process. Since the first association of ultracold atoms \cite{thorsheim1987laser}, the collisions between the molecules and the atoms have been studied extensively.

All of the ultracold collision studies involving molecules can be divided into second categories, first is collisions with unstable highly-lying vibrational excited molecules such as Feshbach molecules, second is collisions with rovibrational ground state molecules. Until creation of rovibrational ground state molecules \cite{deiglmayr2008formation,lang2008ultracold,ni2008high}, studies on ultracold collisions belong to the first and the last one. The molecules are prepared by association of two ultracold atoms \cite{hutson2006molecule}, e.g. using magnetically tuned Feshbach resonances \cite{kohler2006production}. Such a scattering resonance occurs when a free colliding atom pair energetically coincides with a bound molecular state. The formation of molecules near Feshbach resonances in ultracold gases has been reported for bosons \cite{donley2002atom,chin2003sensitive,herbig2003preparation,durr2004observation} and fermions \cite{regal2003creation,strecker2003conversion,cubizolles2003production}.

\subsection{Vibrational excited molecule}

\subsubsection{Homonuclear diatomic molecule}

Dimers of fermions have been observed with lifetimes far longer than the times scales for elastic collisions and thermalization, exhibiting remarkable stability against the atom-molecule and molecule-molecule collisions. \ce{^{40}K_2} molecules are created by scan the nearly equal mixture of $|f=9/2,m_{f}=-5/2\rangle$ and $|9/2,-9/2\rangle$ spin state over a Feshbach resonance at 224.21 G adiabatically at a ramping rate of $40\mu s G^{-1}$ \cite{regal2003creation}. The measured lifetime of the molecules have been only 1 ms owing to vibrational quenching, however it dramatically as the Feshbach resonance it approached, which is demonstrated in a following experiment \cite{regal2004lifetime}. The molecules are created from degenerate \ce{^{40}K} atoms in equal mixture of $|9/2,-7/2\rangle$ and $|9/2,-9/2\rangle$ state, and the lifetime increases from $\sim1$ ms to greater than 100 ms when the magnetic field is tuned to the $|9/2,-7/2\rangle$ and $|9/2,-9/2\rangle$ resonance at 224.21 G.

The \ce{Li2} molecules show a similar collisional stability near the Feshbach resonance. Kevin E. Strecker et al. create \ce{Li2} molecules in the least-bound $|v=38\rangle$ vibrational level of the $X^{1}\Sigma_{g}^{+}$ singlet state from a degenerate two component Fermi gas $|1/2,-1/2\rangle$ and $|1/2,1/2\rangle$ by adiabatic passage through a narrow Feshbach resonance at $\sim543$ G \cite{strecker2003conversion}. The molecules are produced with an efficiency of 50\% and remain confined in an optical trap for up to 1 s. Similarly, J. Cubizolles et al. create weakly weakly bound \ce{Li2} molecules by sweeping a magnetic field across a broad Feshbach resonance at 810 G with 85\% efficiency \cite{cubizolles2003production}. The decay of \ce{Li2} molecules show a lifetime $\simeq 500 ms$ at 689 G, while only 20 ms at 636 G, exhibiting strikingly different collisional relaxation. S. Jochim et al. create pure sample of weakly bound \ce{Li2} molecules from three-body recombination in a 50-50 spin mixture in the lowest two spin states at a field of 690 G, and observe a strong coupling between atomic gas and the molecules \cite{jochim2003pure}. The long lifetime of the molecules represents a large elastic collision rate between the particles. Their collisional stability is explained by a fermionic suppression of vibrational quenching in molecule collisions \cite{petrov2004weakly}, and leads to the immediate formation of molecular Bose-Einstein condensation following evaporative cooling. \cite{greiner2003emergence,jochim2003bose,zwierlein2003observation}. Tout T. Wang et al. have also observed different collisional behaviour in collisions of \ce{Li_2} in the highest vibrational state with free Li atoms, compared to \ce{Li_2 + Li_2} and \ce{Li_2 + Na} collisions \cite{wang2013deviation}. The two-body inelastic decay coefficients for latter two collisions match with predictions from the quantum Langevin model (section II. B. 1) for universal collisions of ultracold molecules, which assumes unit probability of loss at short range and the decay rate dependent only on the long-range van der Waals interaction between collision partners. However, the two-body decay coefficient for former collision is exceptionally small, more than 10 times smaller than the universal predictions, which is explained by the low density of available decay sates in systems of light atoms \cite{quemener2007ultracold}.

In contrast, the dimers of bosons show a quick decay via inelastic collisions, so that quantum degenerate molecular clouds can only survive few milliseconds \cite{xu2003formation}. Roahn Wynar et al. report the creation of state-selected \ce{Rb2} molecules in the Rb Bose-Einstein condensation with coherent free-bound stimulated Raman transitions \cite{wynar2000molecules}, whose line shapes reflect the collisional behaviour of molecules and atoms. The inelastic atom-molecule collisions result in a finite lifetime of the molecules, while the elastic interactions between the molecule and atomic condensate give rise to a mean-field contribution to the chemical potential of an atom or molecule, which causes the shift of the resonance Raman transition. Both of these interaction are responsible of the broadening of the Raman transition line shapes.

\begin{figure}
\begin{center}
\includegraphics[width=1.0\linewidth]{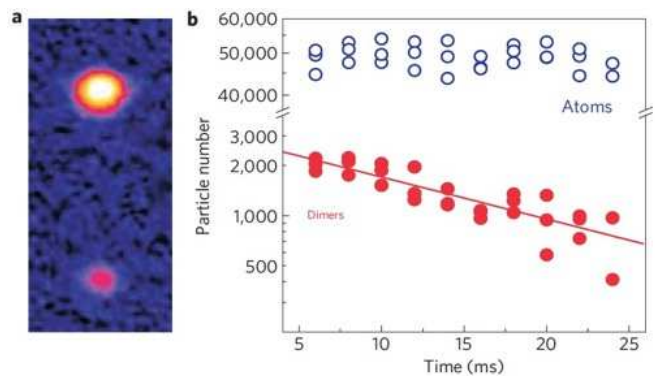}
\caption{Figure adapted from \cite{knoop2009observation}. Measuring the loss rate of the \ce{Cs2} dimers. (a) Absorption image of the atom-dimer mixture after release from the trap and Stern-Gerlach separation. (b) Time evolution of the number of atoms and dimers at 35 G. The loss of dimers is fitted with an exponential decay curve. }
\label{figure:Cs2relaxation}
\end{center}
\end{figure}

Takashi Mukaiyama et al. study the inelastic collisions between sodium condensates in $|F,m_{F}\rangle =|1,1\rangle$ state and Feshbach sodium molecules in an optical dipole trap \cite{mukaiyama2004dissociation}. N. Zahzam et al. investigate the cold inelastic collisions between Cs atoms and \ce{Cs2} dimers in an optical dipole trap \cite{zahzam2006atom}. The atom-molecule collision is the dominant mechanism for the loss of the molecules with a rate coefficient, which shows no dependence on the photoassociated state. In a similar experiment \cite{staanum2006experimental}, the dimers are prepared in high-lying vibrational levels ($v'=32-47$) or low-lying levels ($v'=4-6$) in the $a^{3}\Sigma^+$ triplet ground state by photoassociation through $v=6$ or $v=79$ state of the outer well of the $0_{g}^{-} (^{2}P_{3/2})$ potential. The storage time of \ce{Cs2} molecules at different atom density are measured, and corresponding loss rate coefficients are equally large for both molecular states in inelastic collisions with Cs(F=3). S. Knoop et al. observe a trimer state in ultracold \ce{Cs-Cs_{2}} scattering, in the positive scattering length regime where a dimer state exists \cite{knoop2009observation}. Figure ~\ref{figure:Cs2relaxation} presents the decay of \ce{Cs2} dimers measured from the absorption image. The trimer state manifests itself in a resonant enhancement of inelastic collisions in the mixture of Cs atoms and \ce{Cs2} dimers, with loss rate coefficient $\beta$ of dimers as a function of the scattering length and the magnetic field. They also observe an elementary exchange process \ce{A_{2} + B \rightarrow A + AB} in the mixture \cite{knoop2010magnetically}, where A represents the lowest hyperfine sublevel $|F=3, m_{F} =3\rangle$, and B one of the upper $|F=4, m_{F}\rangle$ hyperfine sublevels with $m_{F} =2$,3,4. The energetics of the process can be magnetically controlled and be tuned from endoergic to exoergic, enabling the observation of a pronounced threshold behaviour in the rate coefficient $\beta$ for inelastic atom-molecule collisions. A. Zenesini et al. study the collisional properties of ultracold Cs atoms and dimers close to a Feshbach resonance near 550 G and observe an atom-dimer loss resonance that is related to Efimov's scenario of trimer states \cite{zenesini2014resonant}.

\subsubsection{Heteronuclear diatomic molecule}

Eric R. Hudson et al. observe strong inelastic collisions of ultracold RbCs with background Cs or Rb atoms, which lead to dramatic reduction of the molecular lifetime \cite{hudson2008inelastic}. The vibrationally excited RbCs molecules in their $a^{3}\Sigma^+$ ground electronic state is produced via photoassociation of laser-cooled \ce{^{85}Rb} and \ce{^{133}Cs} atoms. The inelastic scattering rate constant show a lack of dependence on molecular binding energy, indicating the details of the short-range interaction potential are unimportant, consistent with the Langevin model (section II. B. 1) . J. J. Zirbel et al. create ultracold weakly bound fermionic \ce{^{40}K^{87}Rb} molecules using a Feshbach resonance between Rb $|1,1\rangle$ and K $|9.2,-9/2\rangle$ atoms at $B_{0} =546.7 G$, and investigate their stability in collision with bosonic \ce{^{87}Rb}, fermionic \ce{^{40}K}, or distinguishable \ce{^{40}K} $|9/2,-7/2\rangle$ atoms \cite{zirbel2008collisional}. It is shown that inelastic loss of the molecules near the Feshbach resonance is dramatically affected by the quantum statistics of the colliding particles and the scattering length. A molecule lifetime as long as 100 ms suggests that collisions between identical fermionic molecules are suppressed.

However, enhanced decay rates for $a \geq1000a_0$ are observed when colliding with bosonic \ce{^{87}Rb} atoms. Johannes Deiglmayr et al. investigate collisions of ultracold polar LiCs molecules and ultracold caesium atoms in an optical dipole trap\cite{deiglmayr2011inelastic}. The LiCs molecules are created by photoassociation of caesium and lithium atoms the $B^1\Pi$ excited state followed by spontaneous emission to the $X^1\Sigma^+$ ground state and the lowest triplet sate $a^3\Sigma^+$. The measured lifetime is identical for molecular ensembles produced by photoassociation via $|B^1\Pi, v'=26, J'=1\rangle$ and via $|B^{1}\Pi, v'=26, J'=1\rangle$.

Bo Zhao group recently have observed state-to-state atom-exchange reaction \ce{AB + C \rightarrow AC + B} in an ultracold \ce{^{40}K} atom and \ce{^{23}Na^{40}K} Feshbach molecule mixture near magnetic field of 130.24 G \cite{rui2017controlled}. The \ce{^{23}Na^{40}K} molecules are associated from ultracold \ce{^{23}Na} and \ce{^{40}K} mixture by a Blackman rf pulse, and is detected by rf dissociation. Both the atom and molecule product, as well as the reaction rate coefficient are measured for different magnetic fields. The dependence of rate coefficient on magnetic field indicates the reaction can be switched on or off by fine tuning the magnetic field.

\subsubsection{Collision between Feshbach molecules}

There are also studies on ultracold molecule-molecule collisions. C. Chin et al. observe magnetically tuned collision resonances for ultracold \ce{Cs2} molecules stored in a \ce{CO2}-laser trap \cite{chin2005observation}. The observed "double peak" structure strongly suggests that the existence of bound states of two \ce{Cs2} molecules or \ce{Cs4} tetramer states in collisions between ultracold \ce{Cs2} molecules, inducing Feshbach-like couplings to inelastic decay channels. F. Ferlaino et al. study the loss of \ce{Cs2} halo dimers due to inelastic dimer-dimer collisions in a pure trapped sample of Feshbach \ce{Cs2} molecules \cite{ferlaino2008collisions}. The dependences of relaxation rate coefficient on scattering length and temperature are measured. Fudong Wang et al. recently report the observation of a dimer-dimer inelastic collision resonance for ultracold \ce{^{23}Na^{87}Rb} Feshbach molecules \cite{wang2019observation}. The resonance manifests itself as a pronounced inelastic loss peak of dimers when the interspecies scattering length is tuned and is attributed to the crossing of the dimer-dimer threshold with a tetramer state. The loss rate coefficients also show a temperature dependence. When $a=301a_0$, the dimer is deeply bound with a binding energy of $\sim1$ MHz, $\beta_{dd}$ is nearly temperature independent, which agrees with the Wigner threshold law. At $a=596a_0$ and $1599a_0$, $\beta_{dd}$ decreases with increasing temperature according to the power law $\beta_{dd} \propto T^{-0.42}$ and $\beta_{dd} \propto T^{-0.21}$, respectively.

The dimer-dimer collision can also be studied in a reversible fashion and reactions proceed along a single channel. Daniel K. Hoffmann et al. investigate the reaction of \ce{2AB <=> AB + A + B} with \ce{Li2} Feshbach molecules \cite{hoffmann2018reaction}. Pairs of $|F=1/2,m_{F} = -1/2\rangle$, $|1/2,1/2\rangle$ \ce{^6Li} atoms in the two lowest hyperfine states of the electronic ground state are converted into weakly bound Feshbach molecules by exothermic three-body recombination. Then the temperature is suddenly raised using an excitation pulse of parametric heating, which is realized by frequency modulation of the dipole trap depth during a short period and shift the atom-molecule mixture out of equilibrium. The evolution of the system is measured until it reaches a new equilibrium. A strong dependence on the temperature of the reaction dynamics is observed and is consistent with the Arrhenius law. The increasing of association rate constant with scattering length \emph{a} agrees with the low-energy threshold law (section II. B. 2) $R\propto \hbar a^{4}/m$, and the data for the dependence of dissociation rate constant on \emph{a} are raw but in reasonably agreement with theoretical prediction $C\propto T^{3/2} a^{4} e^{-E_{b}/k_{B}T}$.

\subsection{Ground state molecule}

\subsubsection{Field-free}

\begin{figure}
\begin{center}
\includegraphics[width=1.0\linewidth]{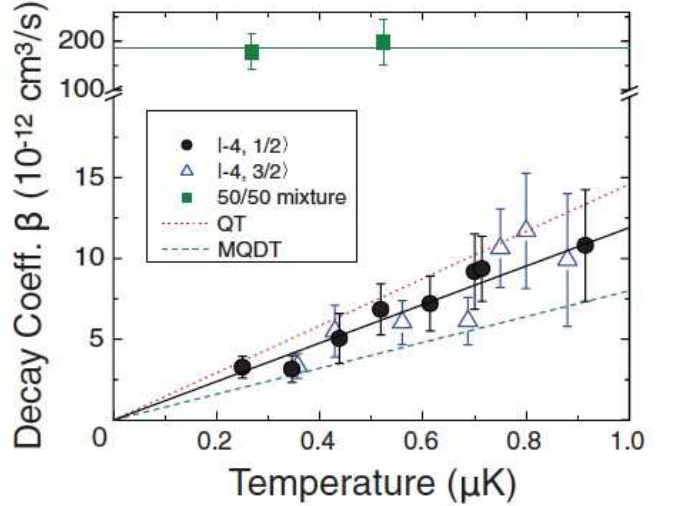}
\caption{Figure adapted from  \cite{ospelkaus2010quantum}. Loss rate coefficient of \ce{^{40}K^{87}Rb} versus temperature. Experimental data for a single component gas of molecules in spin states $| m_{\textit{I}}^{K} =-4, m_{\textit{I}}^{Rb} =1/2\rangle$ (solid circles), $| m_{\textit{I}}^{K} =-4, m_{\textit{I}}^{Rb} =3/2\rangle$ (open triangles), and for 50/50 mixture of these two spin states (solid squares), are compared with prediction of QT (dotted line) and MQDT (dashed line) model. A linear fit (solid line) to the $| -4, 1/2\rangle$ yields the temperature-dependent loss rate to be $1.2 (\pm0.3) \times 10^{-5} cm^{3} s^{-1} K^{-1}$. }
\label{figure:KRbReaction}
\end{center}
\end{figure}

The first ultracold collisions between atoms and rovibrational ground state molecules are first studied by D. S. Jin and Jun Ye's group \cite{ospelkaus2010quantum}, where the quantum nature of the relative motion of the reactants is reflected in the quantum threshold behavior,  and the quantum statistics become unequivocally important. The \ce{^{40}K^{87}Rb} molecules are prepared in $| m_{\textit{I}}^{K} =-4, m_{\textit{I}}^{Rb} =1/2\rangle$ or $| m_{\textit{I}}^{K} =-4, m_{\textit{I}}^{Rb} =3/2\rangle$ hyperfine state of rovibronic ground state $| X^{1}\Sigma ^{+}, v=0, N=0\rangle$. The molecules are produced using single step of two-photon coherent Raman transfer from weakly bound molecules with an averaged density of $10^{11}$ to $10^{12}$ cm$^{-3}$ and translational temperature of a few hundred nanoKelvin. When KRb molecules are prepared in either $| m_{\textit{I}}^{K} =-4, m_{\textit{I}}^{Rb} =1/2\rangle$ or $| m_{\textit{I}}^{K} =-4, m_{\textit{I}}^{Rb} =3/2\rangle$, the collisions of the KRb-KRb is strongly suppressed by the Pauli principle and proceed predominantly by tunnelling through p-wave centrifugal barrier with height of $k_{B} \times 24 \mu$ K, and loss rate coefficient dependence on temperature is shown in Figure ~\ref{figure:KRbReaction}, which yields a temperature-dependent loss rate to be $1.2 (\pm0.3) \times 10^{-5} cm^{3} s^{-1} K^{-1}$. Similar dependence is also found in a degenerate Fermi gas of KRb molecules \cite{de2019degenerate}. When the molecules are prepared in a equal mixture of the two hyperfine states, the reaction proceeds very fast and a temperature-independent loss rate is observed, where the loss of the molecules is found due to barrierless chemical reaction with potassium atoms.

\begin{figure}
\begin{center}
\includegraphics[width=1.0\linewidth]{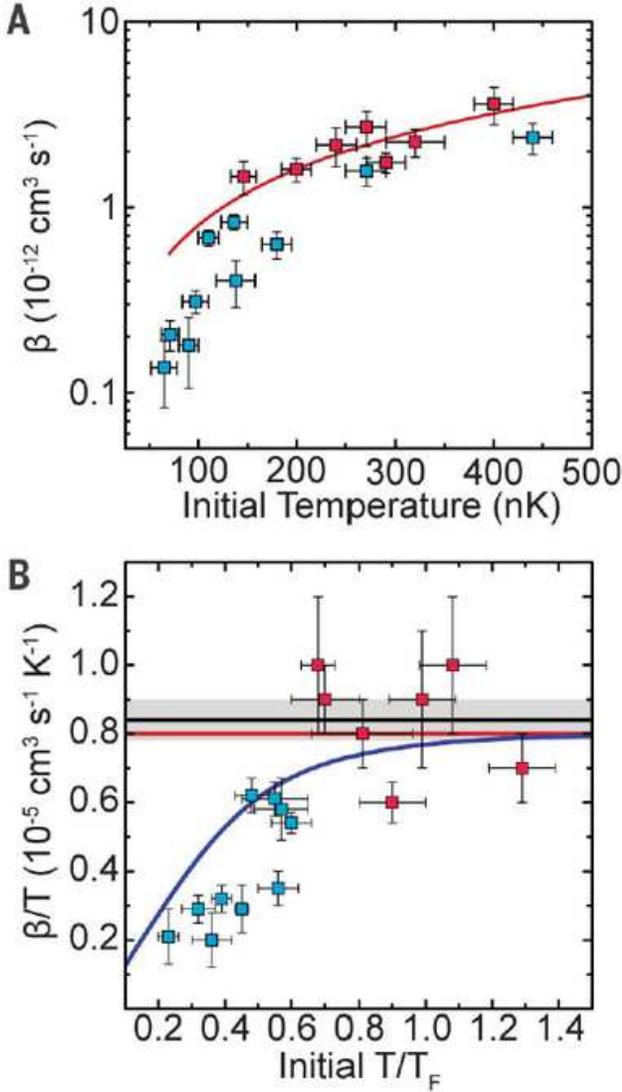}
\caption{Figure adapted from \cite{de2019degenerate}. Temperature dependence of reaction rate constants. (A) The reaction rate constant $\beta$ at various initial temperatures. (B) Temperature normalized reaction rate constants as a function of degeneracy. Soldi bule points correspond to $T/T_{F}\leq 0.6$, and solid red points $T/T_{F}>0.6$. Solid black, red and blue line represent average $\beta/T$ for $T/T_{F}>0.6$, MQDT calculations and average relative density fluctuations, respectively.  }
\label{figure:degenerateKRb}
\end{center}
\end{figure}

To explain the experimental results for reaction of ultracold KRb molecules, S. Ospelkaus et al. have used both QT (section II. B. 2) and MQDT model (section II. B. 3) to analyze the loss rate of rovibronic ground state \ce{^{40}K^{87}Rb} molecules. The comparison of the calculation results based on each model to the experimental measurements is presented in Figure ~\ref{figure:KRbReaction}. The experimental data is consistent with the Wigner threshold laws, where the reaction rate slows down at low temperatures since the tunnelling rate is proportional to the temperature. In a recent experiment \cite{de2019degenerate}, the two-body loss coefficients of KRb molecules follow the predicted MQDT trend closely with temperature $T/T_{F} > 0.6$, while show strong deviations with $T/T_{F}\leq 0.6$ at all temperatures. Moreover, temperature normalized reaction rate constants $\beta/T$ is constant above $T/T_{F} =0.6$, in excellent agreement with predicted MQDT value of $\beta/T=0.8\times10^{-5}cm^{3}s^{-1}K^{-1}$, whereas $\beta/T$ shows a strong deviation from the Wigner threshold law and MQDT trend at $T/T_{F}\leq 0.6$, as can be seen in Figure ~\ref{figure:degenerateKRb}. The observed reduction of $\beta/T$, manifesting suppressed collision rate beyond the prediction by the Wigner threshold law, is attributed to the reduced density fluctuation, i.e. reduced probability of finding two molecules within a short distance of each other as $T/T_{F}$ is lowered because of anti-bunching, which is the same physical phenomenon giving rise to the Pauli pressure and reduced compressibility of a Fermi gas.

Timur M. Rvachov et al. create fermionic dipolar \ce{^{23}Na^{6}Li} molecules in their triplet ground state from an ultracold mixture of \ce{^{23}Na} and \ce{^{6}Li} \cite{rvachov2017long}. The isolated molecular sample has a lifetime of 4.6 s and two-body loss coefficients for NaLi with Li, Na, and NaLi are measured. Collisions of the ground state molecule with Na $|1,1\rangle$ or Li $|1/2,1/2\rangle$ yield lifetime $\tau \approx 2 ms$. The molecule-molecule proceed through p-wave scattering due to their fermionic nature and give a low loss rate $K\approx 10^{-11} cm^3/s$.

Dajun Wang group recently presents detailed studies on the ultracold bimolecular reaction \ce{NaRb + NaRb \rightarrow Na_{2} + Rb_{2}} for both ro-vibrational ground state and the first vibrational excited state molecule\cite{ye2018collisions}. For ultracold \ce{^{23}Na^{87}Rb} molecules, the reaction \ce{NaRb(v=0,J=0) + NaRb(v=0,J=0) \rightarrow Na_{2} (v=0,J=0) + Rb_{2} (v=0,J=0)} is endothermic by 47 cm$^{-1}$, while the reaction \ce{NaRb(v=0,J=0) + NaRb (v=0,J=0) \rightarrow Na_{2} (v=0,J=0) + Rb_{2} (v=0,J=0)} is exothermic by 164 cm$^{-1}$. However, they observe very similar loss and heating for both cases. In addition, increasing temperatures give reducing loss rate constants for each case, and loss rate constant is larger for $|v=0,J=0\rangle$ molecules than $|v=1,J=0\rangle$.

In order to test the sticky collision hypothesis \cite{mayle2012statistical,mayle2013scattering} (section II. B. 3) , which argues that pairs of molecules form long-lived collision complexes and is considered to be the dominant mechanism for the loss of ground state molecules from the trap, Philip D. Gregory et al. experimentally investigate collisional losses of optically trapped sample of ground state \ce{^{87}Rb^{133}Cs} molecules \cite{gregory2019sticky}. The molecules are prepared in a single hyperfine level of the $X^{1}\Sigma^+$ rovibrational ground state via stimulated Raman adiabatic passage. The collisional losses of these nonreactive molecules are measured. The loss of molecules is found consistent with the expectation for complex-mediated collisions, but the loss rate is lower than the limit of universal loss.

\subsubsection{External field}

\begin{figure}
\begin{center}
\includegraphics[width=1.0\linewidth]{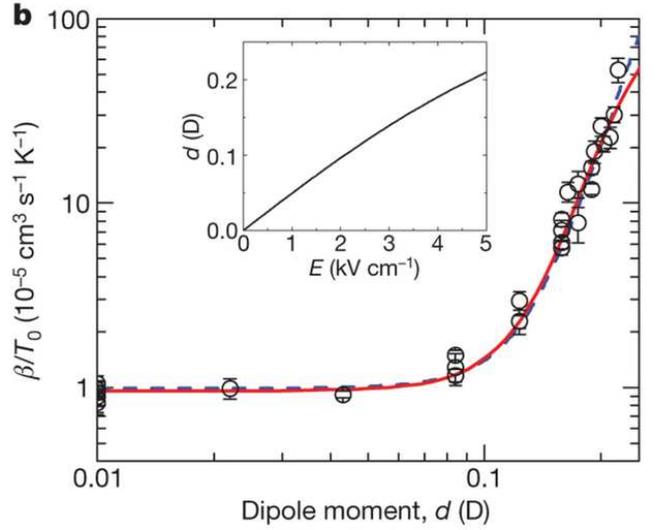}
\caption{Figure adapted from \cite{ni2010dipolar}. Two-body inelastic loss for fermionic KRb molecules. The dashed line and solid line show the fit to simple quantum threshold model and complete quantum scattering model, respectively. The inset show the calculated dependence of d on the applied electric field. }
\label{figure:ElectricFieldLoss}
\end{center}
\end{figure}

The tunability of the chemical reactivity of polar molecules have also been studied under external field. Tetsu Takekoshi et al. observe the onset of hyperfine-changing collisions when the magnetic field is ramped so that the \ce{^{87}Rb^{133}Cs} molecules are no longer in the hyperfine ground state $|v"=0,J"=0,M_{tot}=5\rangle$, and demonstrate that the two-body loss rate coefficient reduce by an order of magnitude when magnetic field is increased to more than 90 G \cite{takekoshi2014ultracold}.  D. S. Jin and Jun Ye's group demonstrate modest external electric fields can drastically alter the dipole interactions between fermionic KRb molecules. In the ultracold dipolar collisions, the loss of fermionic KRb molecules due to ultracold chemical reactions \ce{KRb + KRb \rightarrow K_{2} + Rb_{2}}, and a power law dependence of the loss rate on the induced electric dipole moment have been observed\cite{ni2010dipolar}, as shown in Figure ~\ref{figure:ElectricFieldLoss}. At low electric field where $d<0.1 D$, no significant change to the loss rate have been observed. For higher electric fields, loss rate $\beta /T_{0}$ rapidly increase well over an order of magnitude by $d=0.2 D$, and the rate coefficient can be fitted to $\beta /T_{0} \propto d^{p}$, where $p=6.1\pm 0.8$. The normalized fractional heating rate \'{T}$/T_{0}^{2}n$ follows the same trend with the loss rate, agreeing with a simple thermodynamic model assuming that heating is directly caused by the inelastic loss. The dipolar interaction also show the spatial anisotropy nature from parametric driving measurements under external electric fields. For the case in which initially $T_{z} > T_{x}$, $T_{z} $ and $T_{x}$ approach each other and the timescale for the rethermalization decreases steeply as d increases. While $T_{z} $ and $T_{x}$ do not equilibrate when the gas initially has $T_{z} < T_{x}$. This suggests that it is necessary to protect the molecules from strong inelastic loss and heating and confining them in an array of pancake-shaped traps in a one-dimensional optical lattice configuration would be a promising route to realizing a long-lived quantum gas of polar molecules.

Dajun Wang group also have studied the dipolar collisions of ultracold ground state NaRb molecules in electric fields with induced electric dipole moment up to 0.7 D \cite{guo2018dipolar}. They observe stepwise enhancement of losses with increasing induced dipole moment due to the coupling between s-wave and d-wave induced by stronger anisotropic dipolar interaction.  Bo Zhao group recently observes atom-molecule Feshbach resonances in ultracold collisions between \ce{40K} atoms and \ce{^{23}Na^{40}K} molecules in the rovibrational ground state \cite{yang2019observation}. By preparing atoms and molecules in various hyperfine levels of their ground states, 11 resonances representing enhanced loss of molecules as a function of the magnetic field are observed in the range of 43 to 120 G, in 4 collision channels $|0,0,-3/2,-2\rangle +|9/2,-3/2\rangle$, $|0,0,-3/2,-2\rangle +|9/2,-5/2\rangle$, $|0,0,-3/2,-2\rangle +|9/2,-7/2\rangle$, and $|0,0,-1/2,-3\rangle +|9/2,-7/2\rangle$.

D. S. Jin and Jun Ye's group have also demonstrated that two-dimensional KRb molecules with sufficiently tight confinement can strongly suppress the chemical reactions between molecules \cite{de2011controlling}. For a dipole moment greater than 0.1 D, the 3D loss rate constant increases markedly as $d^6$, whereas the rate constant scaled for the quasi-2D case using $\beta_{3D} = \sqrt{\pi} a \beta_{2D}$ remains close to the value at zero electric field. At a dipole moment of 0.174 D, the measured suppression in quasi-two dimensions is a factor of 60 and the lifetime is approximately 1 s at temperature of 800 nK and density of $10^7 cm^{-2}$.

By loading ground state molecules into a 3D optical lattice, inelastic collisional loss can be strongly suppressed and thus realize long-lived molecules. For homonuclear \ce{Cs2} molecules, lifetime of 8 s has been achieved in a 3D lattice \cite{danzl2010ultracold}. Similarly, long-lived ground state KRb molecules with lifetime of up to 25 s is realized\cite{chotia2012long}. Bo Yan et al. have studied spin dynamics in a 3D lattice using $|N=0, m_{N}=0\rangle$ and $|1, -1\rangle$ of KRb molecule as two spin states \cite{yan2013observation}. By employing coherent microwave spectroscopy and spin echo technique, the spin-exchange interaction has been realized by resonant exchange of rotational angular momentum. The suppression of loss in weak lattices due to a continuous quantum Zeno mechanism has also been demonstrated, which is explained by a following theoretical description of the dissipative dynamics that nonperturbatively includes 3D multiband effects \cite{zhu2014suppressing}.

Similarly, Johannes Hecker Denschlag group investigate the collisional properties of \ce{Rb2} molecules in both 2D lattice and 3D lattice. Pure sample of $\sim1.5\times 10^{4}$ rovibrational ground state $|v=0, N=0, m_{N} = 0\rangle$ \ce{87Rb2} molecules are prepared in the rovibrational ground state of the $a^{3}\Sigma_{u}^+$ potential through stimulated Raman adiabatic passage, and are trapped in the lowest Bloch band of a 3D optical lattices with no more than a molecule per lattice site \cite{deiss2015polarizability}. The lifetime of more than 2 s is observed and for large lattice depth, the lifetime is affected by photon scattering. By quickly ramping down one direction of the 3D optical lattice, it is converted into an array of quasi-1D potential tubes, and molecules within the same tube collide with relative energies on the order of $\mu K \times k_B$ \cite{drews2017inelastic}. Each molecule is considered as a wave packet and collision probability between two molecules is taken as the spatial overlap of their wave packets. The lifetime of \ce{Rb2} Feshbach molecule, $|v=0, R=2\rangle$ and $|v=0, R=0\rangle$ molecules are measured. Remarkably their decay follow a similar timescale, which indicate no suppression of the loss rate by coherent population transferring to rovibrational ground state of a triplet electronic state.

\section{Application: Creating ultracold molecules}

Currently the research of ultracold molecules has emerged as a fast growing field due to their potential application in many different subjects ranging from precision measurement of fundamental physical constants to quantum calculation and simulation, to the fundamental test of Charge-Parity-Time symmetry and search of physics beyond the Standard Model, and even to ultracold chemistry \cite{carr2009cold,wm2008low,krems2009cold,dulieu2017cold}. However cooling molecules directly to the ultracold regime from room temperature or higher temperature remains a major challenge because of the complicated molecular internal degrees of freedom. So far, ultracold molecules can be created by association of ultracold atoms \cite{deiglmayr2008formation,lang2008ultracold,ni2008high}, laser cooling \cite{shuman2009radiative,shuman2010laser,barry2014magneto,norrgard2016submillikelvin,mccarron2018magnetic}, electro-optic Sisyphus cooling \cite{zeppenfeld2012sisyphus}.

Although those methods above mentioned have achieved significant success, they are limited to specific class of molecules due to the features of each method. Associated diatomic molecules are commonly through Feshbach resonance followed by rather complicated scheme of Stimulated Raman Adiabatic Passage transfer and limited to alkali atoms and alkaline-earth atoms which can be laser cooled. Laser cooling so far are limited to some fluoride compound of alkaline-earth atom or molecules whose energy level structure and Franck-Condon factor between vibrational energy levels of electronic ground state and vibrational levels of first electronic excited state, rendering the closed optical cycling transition possible, such as SrF \cite{shuman2009radiative,shuman2010laser,barry2014magneto,norrgard2016submillikelvin,mccarron2018magnetic}, YO \cite{hummon20132d,yeo2015rotational,collopy20183d}, CaF \cite{zhelyazkova2014laser,williams2018magnetic,anderegg2018laser,caldwell2019deep}, YbF \cite{lim2018laser}, BaF \cite{chen2017radiative}, and MgF \cite{xu2019determination}. While electro-optic Sisyphus cooling \cite{zeppenfeld2012sisyphus} is limited to symmetric-top molecules whose rotational energy levels having appropriate Stark shift. Thus evaporative cooling or sympathetic cooling of molecules utilizing ultracold collisional properties of particles appears to be a universal method for producing ultracold molecules.

\subsection{Evaporative cooling}

Evaporative cooling of a thermal distribution is in principle reducing the ensemble temperature by selectively removing particles with energies much greater than the average total energy per particle. In the presence of elastic collisions, the high-energy tail is repeatedly repopulated and be trimmed, realizing removal of energy at cost in particle number. Evaporative cooling has achieved great success in realization of quantum degenerate gases \cite{ketterle1996evaporative}, and the applicability to molecules has also been theoretically discussed. Cold and ultracold collisions of NH molecules has been investigated with or without external magnetic field, demonstrating large ratio of elastic to inelastic cross section and prospects for evaporative cooling \cite{janssen2011cold,janssen2011cold2,suleimanov2012efficient}. However, it is found the ultracold reaction of magnetically trapped \ce{NH(X^{3} \Sigma^{-}} radicals is driven by a short-ranged collisional mechanism and the magnitude of the reactive cross section is weakly dependent on magnetic field strength, which indicate chemical reactions may cause more trap loss than inelastic spin-changing collisions, making evaporative cooling much more difficult \cite{janssen2013quantum}. Previous experimental work on both KRb \cite{ospelkaus2010quantum,ni2010dipolar} and \ce{ND3} \cite{parazzoli2011large} molecules suggest that evaporative cooling is unfavourable for these molecules. For OH molecules, despite the open shell configuration and large anisotropy in the OH-OH interaction potential, theoretical work find strong suppression of inelastic collision rate constant between OH molecules in a magnetic field\cite{ticknor2005influence}. Benjamin K. Stuhl et al. experimentally realize the evaporative cooling of the OH molecules in the ground-state $\Lambda$-doublet structure using so called radio frequency knife to successively lower the trap depth \cite{stuhl2012evaporative}.

\subsection{Sympathetic cooling}

Sympathetic cooling is the thermalization of candidate species in a heat bath of ultracold particles. It has seen great success in cooling both ions \cite{larson1986sympathetic} and atoms \cite{myatt1997production,schreck2001sympathetic}. Due to this reason, its application to molecules was naturally expected and thus a lot of efforts have been devoted for many different atom-molecule systems by theorists in the last two decades, such as \ce{NH3-Rb} \cite{zuchowski2008prospects,zuchowski2009low}, \ce{H2}-Rare gas \cite{barletta2008creating,barker2009sympathetic,barletta2010direct}, OH-Rb \cite{lara2006ultracold,lara2007cold}, NH-Alkali(or Alkali earth) atom \cite{tacconi2007collisions,soldan2009prospects,wallis2009production,gonzalez2011effect,wallis2011prospects}, N-NH \cite{zuchowski2011cold,hummon2011cold}, OH-H and NH-H \cite{gonzalez2013ultracold}, LiH-Li \cite{tokunaga2011prospects}, CaH-Li and CaH-Mg \cite{tscherbul2011ultracold}, YbF-He \cite{tscherbul2007fine}, CaF-Li and CaF-Rb \cite{lim2015modeling}, SrF-Rb \cite{morita2018atom,morita2019universal}, SrOH-Li \cite{morita2017cold,morita2019universal} et al.

\begin{figure}
\begin{center}
\includegraphics[width=1.0\linewidth]{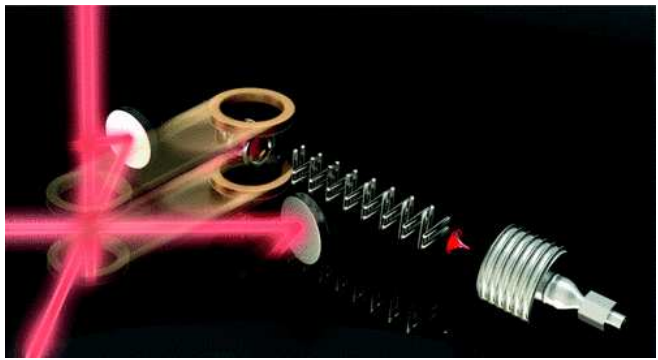}
\caption{Figure adapted from \cite{parazzoli2011large}. Rb atoms are cooled and trapped at the intersection of laser beams, while supersonic beam of \ce{ND3} molecules is created by a pulsed valve, slowed and trapped using inhomogeneous time-varying electric fields created by array of high voltage electrodes. A pair of anti-Helmholtz coils composing atomic trap are translated along a linear translation stage such that the atomic trap overlap with the electrostatic molecular trap.  }
\label{figure:overlap}
\end{center}
\end{figure}

Typical approach of sympathetic cooling of molecules using ultracold atoms would be overlap magneto-optical trap of atoms with molecular trap assuming favorable ratio of rate constants for elastic scattering and spin relaxation between atoms and molecules. However, large optical access of the magneto-optical trap and enclosure configuration of electrostatic or magneto-static trap for molecules, as well as small volume of these molecular trap, poses an significant technical challenge for experimental realization. So far, only inelastic collisions between \ce{ND3} and Rb has been experimentally investigated \cite{parazzoli2011large}. Ultracold Rb atoms are prepared in a separate chamber mainly due to the need for large access. Rubidium atoms are first laser cooled and confined in a magneto-optical trap, then they are loaded into a magnetic trap and transferred to the trap region of \ce{ND3} by moving the magnetic coils along a linear translation stage, as is shown in Figure ~\ref{figure:overlap}. Good overlap between \ce{ND3} and Rb is an experimental problem, although approximately 100$\mu m$ uncertainty have been achieved using tricky operations. Besides, the strong electric fields created by the electrostatic trap would pull Rb atoms towards electrodes and eventually cause them fall out of the trap.

Recently, Narevicius group used a different technique - moving trap decelerator \cite{lavert2011moving,lavert2011stopping} - to overlap molecules with alkali atoms. Similar scheme with different coil configuration have also been demonstrated \cite{trimeche2011trapping}. The principle has previously been applied to decelerate polar molecules with travelling electric potential well both on a chip \cite{meek2008trapping,meek2009trapping,meek2009stark} and in free space \cite{osterwalder2010deceleration,meek2011traveling}. They simultaneous decelerate supersonic beams of \ce{O2} and Li with moving magnetic trap and finally load them into a static magnetic trap \cite{akerman2017trapping}, thus circumvent the difficulty of spatially overlap atomic trap with molecular trap with reduced complexity, facilitating the studies of cold collisions between molecular and atomic species.

\begin{figure}
\begin{center}
\includegraphics[width=1.0\linewidth]{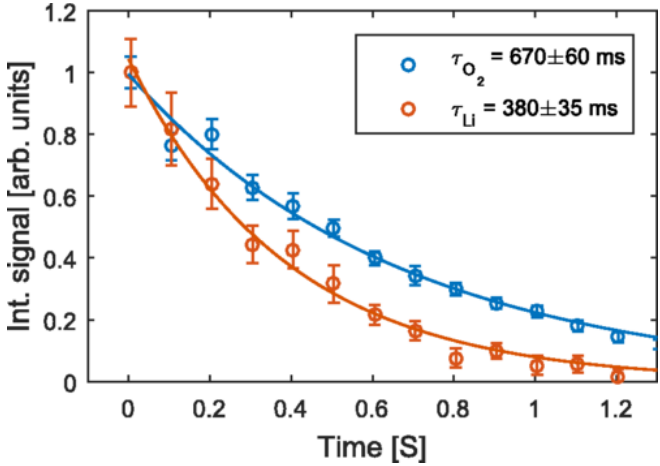}
\caption{Figure adapted from \cite{segev2019collisions}. Measured lifetime of the trapped \ce{O2} molecules in the magnetic trap for different trap depths.  }
\label{figure:lifetime}
\end{center}
\end{figure}

\begin{figure}
\begin{center}
\includegraphics[width=1.0\linewidth]{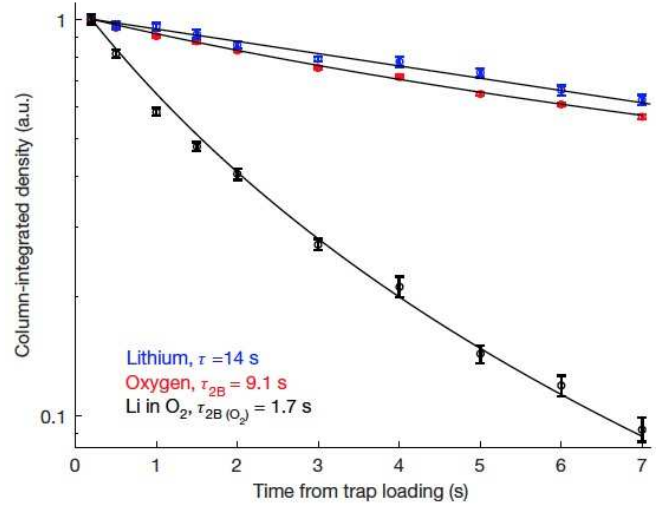}
\caption{Figure adapted from \cite{segev2019collisions}. Measured lifetime of the trapped \ce{O2} molecules and Li atoms in the magnetic trap.  }
\label{figure:Cotrap}
\end{center}
\end{figure}

Very recently, the group studies the collision properties for oxygen molecule alone and for oxygen molecule with lithium atom inside a superconducting magnetic trap by measuring their lifetimes \cite{segev2019collisions}, as shown in Figure ~\ref{figure:lifetime}. The \ce{O2} molecules alone can be trapped more than 90 s with lifetime $\tau \approx 52 s$ when they are loaded into a shallow trap of 50 mK, consistent with estimated background collision rate, indicating the absence of intermolecular collisions. With increasing trap depth, a clear deviation from an exponential decay has been observed and \ce{O2-O2} two-body collisions are more and more pronounced. Approximately 9 s two-body lifetime is estimated at 800 mK trap depth. In Figure ~\ref{figure:Cotrap}, trapped lithium atoms also displays an exponential decay with a lifetime of $\tau \approx 14 s$ when loading into the trap without oxygen molecules. When lithium is co-loaded with \ce{O2} the Li signal shows a fast non-exponential decay with 1.7 s lifetime, which can be well described by \ce{Li-O2} two-body loss-rate equation. The \ce{O2} lifetime does not show any change due to the addition of lithium, confirming trapped lithium density is much lower than that of oxygen.

\section{Conclusions}

This article has given a comprehensive overview of experimental and theoretical investigations of molecular collisions covering a wide range of temperature from few Kelvins to nano-Kelvin. We have reviewed a variety of techniques exploring collision dynamics in these temperature regimes, especially emphasizing techniques such as merged molecular beam, in which collision energy can be tuned over multiple regimes, and external field control for collisions with ultracold ground state diatomic molecules, etc. These techniques provide a unique testbed of the universality for a variety of collision theories. We also discuss interesting features such as scattering resonances observed in crossed molecular beam and merged molecular beam experiment, chemical reactions in cold and ultracold temperatures, etc. Currently, the fundamental physics underlying molecular collisions for a wide range of systems have been reasonably well understood. However, many important issues remain to be explored and addressed,  such as many-body physics in molecular collisions, collision dynamics involving polyatomic and complex molecules, demanding of higher precision in molecular scattering calculations, etc.

Fast improving techniques, such as superconducting magnetic trap \cite{segev2019collisions}, laser cooling of polyatomic molecules \cite{morita2019universal}, and ability of simulate solid state physics with ultracold molecules loaded into optical lattices, provide us a great opportunity to get deeper insight into the fundamental physics related to molecular collisions. Various newly modified or developed theoretical approaches based on quantum defect theory or quantum capture theory or ab initio calculations,  and rapidly increasing computational capability, greatly improve the calculation precision to a new level, reaching much better agreements with various experimental measurements. We believe, in this rapidly developing field, deep and diversified physics and chemistry will be advanced in the near future.

\section{Acknowlegements}

Yang Liu acknowledges the financial support from National Natural Science Foundation of China(NSFC) under Grant No. 11974434, Fundamental Research Funds for the Central Universities of Education of China under Grant No. 191gpy276, Natural Science Foundation of Guangdong Province under Grant 2020A1515011159. Le Luo thanks for supports from NSFC under Grant No.11774436, Guangdong Province Youth Talent Program under Grant No.2017GC010656, Sun Yat-sen University Core Technology Development Fund, and the Key-Area Research and Development Program of GuangDong Province under Grant No.2019B030330001.

\bibliography{collisions}

\begin{thebibliography}{330}%
\makeatletter
\providecommand \@ifxundefined [1]{%
 \@ifx{#1\undefined}
}%
\providecommand \@ifnum [1]{%
 \ifnum #1\expandafter \@firstoftwo
 \else \expandafter \@secondoftwo
 \fi
}%
\providecommand \@ifx [1]{%
 \ifx #1\expandafter \@firstoftwo
 \else \expandafter \@secondoftwo
 \fi
}%
\providecommand \natexlab [1]{#1}%
\providecommand \enquote  [1]{``#1''}%
\providecommand \bibnamefont  [1]{#1}%
\providecommand \bibfnamefont [1]{#1}%
\providecommand \citenamefont [1]{#1}%
\providecommand \href@noop [0]{\@secondoftwo}%
\providecommand \href [0]{\begingroup \@sanitize@url \@href}%
\providecommand \@href[1]{\@@startlink{#1}\@@href}%
\providecommand \@@href[1]{\endgroup#1\@@endlink}%
\providecommand \@sanitize@url [0]{\catcode `\\12\catcode `\$12\catcode
  `\&12\catcode `\#12\catcode `\^12\catcode `\_12\catcode `\%12\relax}%
\providecommand \@@startlink[1]{}%
\providecommand \@@endlink[0]{}%
\providecommand \url  [0]{\begingroup\@sanitize@url \@url }%
\providecommand \@url [1]{\endgroup\@href {#1}{\urlprefix }}%
\providecommand \urlprefix  [0]{URL }%
\providecommand \Eprint [0]{\href }%
\providecommand \doibase [0]{http://dx.doi.org/}%
\providecommand \selectlanguage [0]{\@gobble}%
\providecommand \bibinfo  [0]{\@secondoftwo}%
\providecommand \bibfield  [0]{\@secondoftwo}%
\providecommand \translation [1]{[#1]}%
\providecommand \BibitemOpen [0]{}%
\providecommand \bibitemStop [0]{}%
\providecommand \bibitemNoStop [0]{.\EOS\space}%
\providecommand \EOS [0]{\spacefactor3000\relax}%
\providecommand \BibitemShut  [1]{\csname bibitem#1\endcsname}%
\let\auto@bib@innerbib\@empty
\bibitem [{\citenamefont {Lee}\ \emph {et~al.}(1969)\citenamefont {Lee},
  \citenamefont {McDonald}, \citenamefont {LeBreton},\ and\ \citenamefont
  {Herschbach}}]{lee1969molecular}%
  \BibitemOpen
  \bibfield  {author} {\bibinfo {author} {\bibfnamefont {Y.-T.}\ \bibnamefont
  {Lee}}, \bibinfo {author} {\bibfnamefont {J.}~\bibnamefont {McDonald}},
  \bibinfo {author} {\bibfnamefont {P.}~\bibnamefont {LeBreton}}, \ and\
  \bibinfo {author} {\bibfnamefont {D.}~\bibnamefont {Herschbach}},\
  }\href@noop {} {\bibfield  {journal} {\bibinfo  {journal} {Review of
  Scientific Instruments}\ }\textbf {\bibinfo {volume} {40}},\ \bibinfo {pages}
  {1402} (\bibinfo {year} {1969})}\BibitemShut {NoStop}%
\bibitem [{\citenamefont {Herschbach}(1987)}]{herschbach1987molecular}%
  \BibitemOpen
  \bibfield  {author} {\bibinfo {author} {\bibfnamefont {D.~R.}\ \bibnamefont
  {Herschbach}},\ }\href@noop {} {\bibfield  {journal} {\bibinfo  {journal}
  {Angewandte Chemie International Edition in English}\ }\textbf {\bibinfo
  {volume} {26}},\ \bibinfo {pages} {1221} (\bibinfo {year}
  {1987})}\BibitemShut {NoStop}%
\bibitem [{\citenamefont {Weiner}\ \emph {et~al.}(1999)\citenamefont {Weiner},
  \citenamefont {Bagnato}, \citenamefont {Zilio},\ and\ \citenamefont
  {Julienne}}]{weiner1999experiments}%
  \BibitemOpen
  \bibfield  {author} {\bibinfo {author} {\bibfnamefont {J.}~\bibnamefont
  {Weiner}}, \bibinfo {author} {\bibfnamefont {V.~S.}\ \bibnamefont {Bagnato}},
  \bibinfo {author} {\bibfnamefont {S.}~\bibnamefont {Zilio}}, \ and\ \bibinfo
  {author} {\bibfnamefont {P.~S.}\ \bibnamefont {Julienne}},\ }\href@noop {}
  {\bibfield  {journal} {\bibinfo  {journal} {Reviews of Modern Physics}\
  }\textbf {\bibinfo {volume} {71}},\ \bibinfo {pages} {1} (\bibinfo {year}
  {1999})}\BibitemShut {NoStop}%
\bibitem [{che()}]{chem86}%
  \BibitemOpen
  \href@noop {} {}\bibinfo {howpublished}
  {\url{https://www.nobelprize.org/prizes/chemistry/1986/}}\BibitemShut
  {NoStop}%
\bibitem [{phy()}]{phys97}%
  \BibitemOpen
  \href@noop {} {}\bibinfo {howpublished}
  {\url{https://www.nobelprize.org/prizes/physics/1997/}}\BibitemShut {NoStop}%
\bibitem [{\citenamefont {Hudson}\ \emph {et~al.}(2006)\citenamefont {Hudson},
  \citenamefont {Lewandowski}, \citenamefont {Sawyer},\ and\ \citenamefont
  {Ye}}]{hudson2006cold}%
  \BibitemOpen
  \bibfield  {author} {\bibinfo {author} {\bibfnamefont {E.~R.}\ \bibnamefont
  {Hudson}}, \bibinfo {author} {\bibfnamefont {H.}~\bibnamefont {Lewandowski}},
  \bibinfo {author} {\bibfnamefont {B.~C.}\ \bibnamefont {Sawyer}}, \ and\
  \bibinfo {author} {\bibfnamefont {J.}~\bibnamefont {Ye}},\ }\href@noop {}
  {\bibfield  {journal} {\bibinfo  {journal} {Physical review letters}\
  }\textbf {\bibinfo {volume} {96}},\ \bibinfo {pages} {143004} (\bibinfo
  {year} {2006})}\BibitemShut {NoStop}%
\bibitem [{\citenamefont {Zelevinsky}\ \emph {et~al.}(2008)\citenamefont
  {Zelevinsky}, \citenamefont {Kotochigova},\ and\ \citenamefont
  {Ye}}]{zelevinsky2008precision}%
  \BibitemOpen
  \bibfield  {author} {\bibinfo {author} {\bibfnamefont {T.}~\bibnamefont
  {Zelevinsky}}, \bibinfo {author} {\bibfnamefont {S.}~\bibnamefont
  {Kotochigova}}, \ and\ \bibinfo {author} {\bibfnamefont {J.}~\bibnamefont
  {Ye}},\ }\href@noop {} {\bibfield  {journal} {\bibinfo  {journal} {Physical
  review letters}\ }\textbf {\bibinfo {volume} {100}},\ \bibinfo {pages}
  {043201} (\bibinfo {year} {2008})}\BibitemShut {NoStop}%
\bibitem [{\citenamefont {Chin}\ \emph {et~al.}(2009)\citenamefont {Chin},
  \citenamefont {Flambaum},\ and\ \citenamefont {Kozlov}}]{chin2009ultracold}%
  \BibitemOpen
  \bibfield  {author} {\bibinfo {author} {\bibfnamefont {C.}~\bibnamefont
  {Chin}}, \bibinfo {author} {\bibfnamefont {V.}~\bibnamefont {Flambaum}}, \
  and\ \bibinfo {author} {\bibfnamefont {M.}~\bibnamefont {Kozlov}},\
  }\href@noop {} {\bibfield  {journal} {\bibinfo  {journal} {New Journal of
  Physics}\ }\textbf {\bibinfo {volume} {11}},\ \bibinfo {pages} {055048}
  (\bibinfo {year} {2009})}\BibitemShut {NoStop}%
\bibitem [{\citenamefont {Baron}\ \emph {et~al.}(2014)\citenamefont {Baron},
  \citenamefont {Campbell}, \citenamefont {DeMille}, \citenamefont {Doyle},
  \citenamefont {Gabrielse}, \citenamefont {Gurevich}, \citenamefont {Hess},
  \citenamefont {Hutzler}, \citenamefont {Kirilov}, \citenamefont {Kozyryev}
  \emph {et~al.}}]{baron2014order}%
  \BibitemOpen
  \bibfield  {author} {\bibinfo {author} {\bibfnamefont {J.}~\bibnamefont
  {Baron}}, \bibinfo {author} {\bibfnamefont {W.~C.}\ \bibnamefont {Campbell}},
  \bibinfo {author} {\bibfnamefont {D.}~\bibnamefont {DeMille}}, \bibinfo
  {author} {\bibfnamefont {J.~M.}\ \bibnamefont {Doyle}}, \bibinfo {author}
  {\bibfnamefont {G.}~\bibnamefont {Gabrielse}}, \bibinfo {author}
  {\bibfnamefont {Y.~V.}\ \bibnamefont {Gurevich}}, \bibinfo {author}
  {\bibfnamefont {P.~W.}\ \bibnamefont {Hess}}, \bibinfo {author}
  {\bibfnamefont {N.~R.}\ \bibnamefont {Hutzler}}, \bibinfo {author}
  {\bibfnamefont {E.}~\bibnamefont {Kirilov}}, \bibinfo {author} {\bibfnamefont
  {I.}~\bibnamefont {Kozyryev}},  \emph {et~al.},\ }\href@noop {} {\bibfield
  {journal} {\bibinfo  {journal} {Science}\ }\textbf {\bibinfo {volume}
  {343}},\ \bibinfo {pages} {269} (\bibinfo {year} {2014})}\BibitemShut
  {NoStop}%
\bibitem [{\citenamefont {Kobayashi}\ \emph {et~al.}(2019)\citenamefont
  {Kobayashi}, \citenamefont {Ogino},\ and\ \citenamefont
  {Inouye}}]{kobayashi2019measurement}%
  \BibitemOpen
  \bibfield  {author} {\bibinfo {author} {\bibfnamefont {J.}~\bibnamefont
  {Kobayashi}}, \bibinfo {author} {\bibfnamefont {A.}~\bibnamefont {Ogino}}, \
  and\ \bibinfo {author} {\bibfnamefont {S.}~\bibnamefont {Inouye}},\
  }\href@noop {} {\bibfield  {journal} {\bibinfo  {journal} {Nature
  communications}\ }\textbf {\bibinfo {volume} {10}},\ \bibinfo {pages} {1}
  (\bibinfo {year} {2019})}\BibitemShut {NoStop}%
\bibitem [{\citenamefont {Baranov}(2008)}]{baranov2008theoretical}%
  \BibitemOpen
  \bibfield  {author} {\bibinfo {author} {\bibfnamefont {M.~A.}\ \bibnamefont
  {Baranov}},\ }\href@noop {} {\bibfield  {journal} {\bibinfo  {journal}
  {Physics Reports}\ }\textbf {\bibinfo {volume} {464}},\ \bibinfo {pages} {71}
  (\bibinfo {year} {2008})}\BibitemShut {NoStop}%
\bibitem [{\citenamefont {Eisert}\ \emph {et~al.}(2015)\citenamefont {Eisert},
  \citenamefont {Friesdorf},\ and\ \citenamefont
  {Gogolin}}]{eisert2015quantum}%
  \BibitemOpen
  \bibfield  {author} {\bibinfo {author} {\bibfnamefont {J.}~\bibnamefont
  {Eisert}}, \bibinfo {author} {\bibfnamefont {M.}~\bibnamefont {Friesdorf}}, \
  and\ \bibinfo {author} {\bibfnamefont {C.}~\bibnamefont {Gogolin}},\
  }\href@noop {} {\bibfield  {journal} {\bibinfo  {journal} {Nature Physics}\
  }\textbf {\bibinfo {volume} {11}},\ \bibinfo {pages} {124} (\bibinfo {year}
  {2015})}\BibitemShut {NoStop}%
\bibitem [{\citenamefont {DeMille}(2002)}]{demille2002quantum}%
  \BibitemOpen
  \bibfield  {author} {\bibinfo {author} {\bibfnamefont {D.}~\bibnamefont
  {DeMille}},\ }\href@noop {} {\bibfield  {journal} {\bibinfo  {journal}
  {Physical Review Letters}\ }\textbf {\bibinfo {volume} {88}},\ \bibinfo
  {pages} {067901} (\bibinfo {year} {2002})}\BibitemShut {NoStop}%
\bibitem [{\citenamefont {Rabl}\ \emph {et~al.}(2006)\citenamefont {Rabl},
  \citenamefont {DeMille}, \citenamefont {Doyle}, \citenamefont {Lukin},
  \citenamefont {Schoelkopf},\ and\ \citenamefont {Zoller}}]{rabl2006hybrid}%
  \BibitemOpen
  \bibfield  {author} {\bibinfo {author} {\bibfnamefont {P.}~\bibnamefont
  {Rabl}}, \bibinfo {author} {\bibfnamefont {D.}~\bibnamefont {DeMille}},
  \bibinfo {author} {\bibfnamefont {J.~M.}\ \bibnamefont {Doyle}}, \bibinfo
  {author} {\bibfnamefont {M.~D.}\ \bibnamefont {Lukin}}, \bibinfo {author}
  {\bibfnamefont {R.}~\bibnamefont {Schoelkopf}}, \ and\ \bibinfo {author}
  {\bibfnamefont {P.}~\bibnamefont {Zoller}},\ }\href@noop {} {\bibfield
  {journal} {\bibinfo  {journal} {Physical review letters}\ }\textbf {\bibinfo
  {volume} {97}},\ \bibinfo {pages} {033003} (\bibinfo {year}
  {2006})}\BibitemShut {NoStop}%
\bibitem [{\citenamefont {Micheli}\ \emph {et~al.}(2006)\citenamefont
  {Micheli}, \citenamefont {Brennen},\ and\ \citenamefont
  {Zoller}}]{micheli2006toolbox}%
  \BibitemOpen
  \bibfield  {author} {\bibinfo {author} {\bibfnamefont {A.}~\bibnamefont
  {Micheli}}, \bibinfo {author} {\bibfnamefont {G.}~\bibnamefont {Brennen}}, \
  and\ \bibinfo {author} {\bibfnamefont {P.}~\bibnamefont {Zoller}},\
  }\href@noop {} {\bibfield  {journal} {\bibinfo  {journal} {Nature Physics}\
  }\textbf {\bibinfo {volume} {2}},\ \bibinfo {pages} {341} (\bibinfo {year}
  {2006})}\BibitemShut {NoStop}%
\bibitem [{\citenamefont {Gorshkov}\ \emph {et~al.}(2011)\citenamefont
  {Gorshkov}, \citenamefont {Manmana}, \citenamefont {Chen}, \citenamefont
  {Ye}, \citenamefont {Demler}, \citenamefont {Lukin},\ and\ \citenamefont
  {Rey}}]{gorshkov2011tunable}%
  \BibitemOpen
  \bibfield  {author} {\bibinfo {author} {\bibfnamefont {A.~V.}\ \bibnamefont
  {Gorshkov}}, \bibinfo {author} {\bibfnamefont {S.~R.}\ \bibnamefont
  {Manmana}}, \bibinfo {author} {\bibfnamefont {G.}~\bibnamefont {Chen}},
  \bibinfo {author} {\bibfnamefont {J.}~\bibnamefont {Ye}}, \bibinfo {author}
  {\bibfnamefont {E.}~\bibnamefont {Demler}}, \bibinfo {author} {\bibfnamefont
  {M.~D.}\ \bibnamefont {Lukin}}, \ and\ \bibinfo {author} {\bibfnamefont
  {A.~M.}\ \bibnamefont {Rey}},\ }\href@noop {} {\bibfield  {journal} {\bibinfo
   {journal} {Physical review letters}\ }\textbf {\bibinfo {volume} {107}},\
  \bibinfo {pages} {115301} (\bibinfo {year} {2011})}\BibitemShut {NoStop}%
\bibitem [{\citenamefont {Yan}\ \emph {et~al.}(2013)\citenamefont {Yan},
  \citenamefont {Moses}, \citenamefont {Gadway}, \citenamefont {Covey},
  \citenamefont {Hazzard}, \citenamefont {Rey}, \citenamefont {Jin},\ and\
  \citenamefont {Ye}}]{yan2013observation}%
  \BibitemOpen
  \bibfield  {author} {\bibinfo {author} {\bibfnamefont {B.}~\bibnamefont
  {Yan}}, \bibinfo {author} {\bibfnamefont {S.~A.}\ \bibnamefont {Moses}},
  \bibinfo {author} {\bibfnamefont {B.}~\bibnamefont {Gadway}}, \bibinfo
  {author} {\bibfnamefont {J.~P.}\ \bibnamefont {Covey}}, \bibinfo {author}
  {\bibfnamefont {K.~R.}\ \bibnamefont {Hazzard}}, \bibinfo {author}
  {\bibfnamefont {A.~M.}\ \bibnamefont {Rey}}, \bibinfo {author} {\bibfnamefont
  {D.~S.}\ \bibnamefont {Jin}}, \ and\ \bibinfo {author} {\bibfnamefont
  {J.}~\bibnamefont {Ye}},\ }\href@noop {} {\bibfield  {journal} {\bibinfo
  {journal} {Nature}\ }\textbf {\bibinfo {volume} {501}},\ \bibinfo {pages}
  {521} (\bibinfo {year} {2013})}\BibitemShut {NoStop}%
\bibitem [{\citenamefont {Balakrishnan}\ and\ \citenamefont
  {Dalgarno}(2001)}]{balakrishnan2001chemistry}%
  \BibitemOpen
  \bibfield  {author} {\bibinfo {author} {\bibfnamefont {N.}~\bibnamefont
  {Balakrishnan}}\ and\ \bibinfo {author} {\bibfnamefont {A.}~\bibnamefont
  {Dalgarno}},\ }\href@noop {} {\bibfield  {journal} {\bibinfo  {journal}
  {Chemical physics letters}\ }\textbf {\bibinfo {volume} {341}},\ \bibinfo
  {pages} {652} (\bibinfo {year} {2001})}\BibitemShut {NoStop}%
\bibitem [{\citenamefont {Krems}(2008)}]{krems2008cold}%
  \BibitemOpen
  \bibfield  {author} {\bibinfo {author} {\bibfnamefont {R.~V.}\ \bibnamefont
  {Krems}},\ }\href@noop {} {\bibfield  {journal} {\bibinfo  {journal}
  {Physical Chemistry Chemical Physics}\ }\textbf {\bibinfo {volume} {10}},\
  \bibinfo {pages} {4079} (\bibinfo {year} {2008})}\BibitemShut {NoStop}%
\bibitem [{\citenamefont {Bell}\ and\ \citenamefont
  {P.~Softley}(2009)}]{bell2009ultracold}%
  \BibitemOpen
  \bibfield  {author} {\bibinfo {author} {\bibfnamefont {M.~T.}\ \bibnamefont
  {Bell}}\ and\ \bibinfo {author} {\bibfnamefont {T.}~\bibnamefont
  {P.~Softley}},\ }\href@noop {} {\bibfield  {journal} {\bibinfo  {journal}
  {Molecular Physics}\ }\textbf {\bibinfo {volume} {107}},\ \bibinfo {pages}
  {99} (\bibinfo {year} {2009})}\BibitemShut {NoStop}%
\bibitem [{\citenamefont {Ni}\ \emph {et~al.}(2010)\citenamefont {Ni},
  \citenamefont {Ospelkaus}, \citenamefont {Wang}, \citenamefont
  {Qu{\'e}m{\'e}ner}, \citenamefont {Neyenhuis}, \citenamefont {De~Miranda},
  \citenamefont {Bohn}, \citenamefont {Ye},\ and\ \citenamefont
  {Jin}}]{ni2010dipolar}%
  \BibitemOpen
  \bibfield  {author} {\bibinfo {author} {\bibfnamefont {K.-K.}\ \bibnamefont
  {Ni}}, \bibinfo {author} {\bibfnamefont {S.}~\bibnamefont {Ospelkaus}},
  \bibinfo {author} {\bibfnamefont {D.}~\bibnamefont {Wang}}, \bibinfo {author}
  {\bibfnamefont {G.}~\bibnamefont {Qu{\'e}m{\'e}ner}}, \bibinfo {author}
  {\bibfnamefont {B.}~\bibnamefont {Neyenhuis}}, \bibinfo {author}
  {\bibfnamefont {M.}~\bibnamefont {De~Miranda}}, \bibinfo {author}
  {\bibfnamefont {J.}~\bibnamefont {Bohn}}, \bibinfo {author} {\bibfnamefont
  {J.}~\bibnamefont {Ye}}, \ and\ \bibinfo {author} {\bibfnamefont
  {D.}~\bibnamefont {Jin}},\ }\href@noop {} {\bibfield  {journal} {\bibinfo
  {journal} {Nature}\ }\textbf {\bibinfo {volume} {464}},\ \bibinfo {pages}
  {1324} (\bibinfo {year} {2010})}\BibitemShut {NoStop}%
\bibitem [{\citenamefont {Ospelkaus}\ \emph {et~al.}(2010)\citenamefont
  {Ospelkaus}, \citenamefont {Ni}, \citenamefont {Wang}, \citenamefont
  {De~Miranda}, \citenamefont {Neyenhuis}, \citenamefont {Qu{\'e}m{\'e}ner},
  \citenamefont {Julienne}, \citenamefont {Bohn}, \citenamefont {Jin},\ and\
  \citenamefont {Ye}}]{ospelkaus2010quantum}%
  \BibitemOpen
  \bibfield  {author} {\bibinfo {author} {\bibfnamefont {S.}~\bibnamefont
  {Ospelkaus}}, \bibinfo {author} {\bibfnamefont {K.-K.}\ \bibnamefont {Ni}},
  \bibinfo {author} {\bibfnamefont {D.}~\bibnamefont {Wang}}, \bibinfo {author}
  {\bibfnamefont {M.}~\bibnamefont {De~Miranda}}, \bibinfo {author}
  {\bibfnamefont {B.}~\bibnamefont {Neyenhuis}}, \bibinfo {author}
  {\bibfnamefont {G.}~\bibnamefont {Qu{\'e}m{\'e}ner}}, \bibinfo {author}
  {\bibfnamefont {P.}~\bibnamefont {Julienne}}, \bibinfo {author}
  {\bibfnamefont {J.}~\bibnamefont {Bohn}}, \bibinfo {author} {\bibfnamefont
  {D.}~\bibnamefont {Jin}}, \ and\ \bibinfo {author} {\bibfnamefont
  {J.}~\bibnamefont {Ye}},\ }\href@noop {} {\bibfield  {journal} {\bibinfo
  {journal} {Science}\ }\textbf {\bibinfo {volume} {327}},\ \bibinfo {pages}
  {853} (\bibinfo {year} {2010})}\BibitemShut {NoStop}%
\bibitem [{\citenamefont {Stuhl}\ \emph {et~al.}(2014)\citenamefont {Stuhl},
  \citenamefont {Hummon},\ and\ \citenamefont {Ye}}]{stuhl2014cold}%
  \BibitemOpen
  \bibfield  {author} {\bibinfo {author} {\bibfnamefont {B.~K.}\ \bibnamefont
  {Stuhl}}, \bibinfo {author} {\bibfnamefont {M.~T.}\ \bibnamefont {Hummon}}, \
  and\ \bibinfo {author} {\bibfnamefont {J.}~\bibnamefont {Ye}},\ }\href@noop
  {} {\bibfield  {journal} {\bibinfo  {journal} {Annual review of physical
  chemistry}\ }\textbf {\bibinfo {volume} {65}},\ \bibinfo {pages} {501}
  (\bibinfo {year} {2014})}\BibitemShut {NoStop}%
\bibitem [{\citenamefont {Dulieu}\ and\ \citenamefont
  {Osterwalder}(2017)}]{dulieu2017cold}%
  \BibitemOpen
  \bibfield  {author} {\bibinfo {author} {\bibfnamefont {O.}~\bibnamefont
  {Dulieu}}\ and\ \bibinfo {author} {\bibfnamefont {A.}~\bibnamefont
  {Osterwalder}},\ }\href@noop {} {\emph {\bibinfo {title} {Cold chemistry:
  molecular scattering and reactivity near absolute zero}}},\ Vol.~\bibinfo
  {volume} {11}\ (\bibinfo  {publisher} {Royal Society of Chemistry},\ \bibinfo
  {year} {2017})\BibitemShut {NoStop}%
\bibitem [{\citenamefont {Naulin}\ and\ \citenamefont
  {Costes}(2014)}]{naulin2014experimental}%
  \BibitemOpen
  \bibfield  {author} {\bibinfo {author} {\bibfnamefont {C.}~\bibnamefont
  {Naulin}}\ and\ \bibinfo {author} {\bibfnamefont {M.}~\bibnamefont
  {Costes}},\ }\href@noop {} {\bibfield  {journal} {\bibinfo  {journal}
  {International Reviews in Physical Chemistry}\ }\textbf {\bibinfo {volume}
  {33}},\ \bibinfo {pages} {427} (\bibinfo {year} {2014})}\BibitemShut
  {NoStop}%
\bibitem [{\citenamefont {Carr}\ \emph {et~al.}(2009)\citenamefont {Carr},
  \citenamefont {DeMille}, \citenamefont {Krems},\ and\ \citenamefont
  {Ye}}]{carr2009cold}%
  \BibitemOpen
  \bibfield  {author} {\bibinfo {author} {\bibfnamefont {L.~D.}\ \bibnamefont
  {Carr}}, \bibinfo {author} {\bibfnamefont {D.}~\bibnamefont {DeMille}},
  \bibinfo {author} {\bibfnamefont {R.~V.}\ \bibnamefont {Krems}}, \ and\
  \bibinfo {author} {\bibfnamefont {J.}~\bibnamefont {Ye}},\ }\href@noop {}
  {\bibfield  {journal} {\bibinfo  {journal} {New Journal of Physics}\ }\textbf
  {\bibinfo {volume} {11}},\ \bibinfo {pages} {055049} (\bibinfo {year}
  {2009})}\BibitemShut {NoStop}%
\bibitem [{\citenamefont {Harich}\ \emph {et~al.}(2002)\citenamefont {Harich},
  \citenamefont {Dai}, \citenamefont {Wang}, \citenamefont {Yang},
  \citenamefont {Der~Chao},\ and\ \citenamefont {Skodje}}]{harich2002forward}%
  \BibitemOpen
  \bibfield  {author} {\bibinfo {author} {\bibfnamefont {S.~A.}\ \bibnamefont
  {Harich}}, \bibinfo {author} {\bibfnamefont {D.}~\bibnamefont {Dai}},
  \bibinfo {author} {\bibfnamefont {C.~C.}\ \bibnamefont {Wang}}, \bibinfo
  {author} {\bibfnamefont {X.}~\bibnamefont {Yang}}, \bibinfo {author}
  {\bibfnamefont {S.}~\bibnamefont {Der~Chao}}, \ and\ \bibinfo {author}
  {\bibfnamefont {R.~T.}\ \bibnamefont {Skodje}},\ }\href@noop {} {\bibfield
  {journal} {\bibinfo  {journal} {Nature}\ }\textbf {\bibinfo {volume} {419}},\
  \bibinfo {pages} {281} (\bibinfo {year} {2002})}\BibitemShut {NoStop}%
\bibitem [{\citenamefont {Yang}\ and\ \citenamefont
  {Zhang}(2008)}]{yang2008dynamical}%
  \BibitemOpen
  \bibfield  {author} {\bibinfo {author} {\bibfnamefont {X.}~\bibnamefont
  {Yang}}\ and\ \bibinfo {author} {\bibfnamefont {D.~H.}\ \bibnamefont
  {Zhang}},\ }\href@noop {} {\bibfield  {journal} {\bibinfo  {journal}
  {Accounts of chemical research}\ }\textbf {\bibinfo {volume} {41}},\ \bibinfo
  {pages} {981} (\bibinfo {year} {2008})}\BibitemShut {NoStop}%
\bibitem [{\citenamefont {Berteloite}\ \emph {et~al.}(2010)\citenamefont
  {Berteloite}, \citenamefont {Lara}, \citenamefont {Bergeat}, \citenamefont
  {Le~Picard}, \citenamefont {Dayou}, \citenamefont {Hickson}, \citenamefont
  {Canosa}, \citenamefont {Naulin}, \citenamefont {Launay}, \citenamefont
  {Sims} \emph {et~al.}}]{berteloite2010kinetics}%
  \BibitemOpen
  \bibfield  {author} {\bibinfo {author} {\bibfnamefont {C.}~\bibnamefont
  {Berteloite}}, \bibinfo {author} {\bibfnamefont {M.}~\bibnamefont {Lara}},
  \bibinfo {author} {\bibfnamefont {A.}~\bibnamefont {Bergeat}}, \bibinfo
  {author} {\bibfnamefont {S.~D.}\ \bibnamefont {Le~Picard}}, \bibinfo {author}
  {\bibfnamefont {F.}~\bibnamefont {Dayou}}, \bibinfo {author} {\bibfnamefont
  {K.~M.}\ \bibnamefont {Hickson}}, \bibinfo {author} {\bibfnamefont
  {A.}~\bibnamefont {Canosa}}, \bibinfo {author} {\bibfnamefont
  {C.}~\bibnamefont {Naulin}}, \bibinfo {author} {\bibfnamefont {J.-M.}\
  \bibnamefont {Launay}}, \bibinfo {author} {\bibfnamefont {I.~R.}\
  \bibnamefont {Sims}},  \emph {et~al.},\ }\href@noop {} {\bibfield  {journal}
  {\bibinfo  {journal} {Physical Review Letters}\ }\textbf {\bibinfo {volume}
  {105}},\ \bibinfo {pages} {203201} (\bibinfo {year} {2010})}\BibitemShut
  {NoStop}%
\bibitem [{\citenamefont {Bergeat}\ \emph {et~al.}(2015)\citenamefont
  {Bergeat}, \citenamefont {Onvlee}, \citenamefont {Naulin}, \citenamefont {Van
  Der~Avoird},\ and\ \citenamefont {Costes}}]{bergeat2015quantum}%
  \BibitemOpen
  \bibfield  {author} {\bibinfo {author} {\bibfnamefont {A.}~\bibnamefont
  {Bergeat}}, \bibinfo {author} {\bibfnamefont {J.}~\bibnamefont {Onvlee}},
  \bibinfo {author} {\bibfnamefont {C.}~\bibnamefont {Naulin}}, \bibinfo
  {author} {\bibfnamefont {A.}~\bibnamefont {Van Der~Avoird}}, \ and\ \bibinfo
  {author} {\bibfnamefont {M.}~\bibnamefont {Costes}},\ }\href@noop {}
  {\bibfield  {journal} {\bibinfo  {journal} {Nature chemistry}\ }\textbf
  {\bibinfo {volume} {7}},\ \bibinfo {pages} {349} (\bibinfo {year}
  {2015})}\BibitemShut {NoStop}%
\bibitem [{\citenamefont {Gilijamse}\ \emph {et~al.}(2006)\citenamefont
  {Gilijamse}, \citenamefont {Hoekstra}, \citenamefont {van~de Meerakker},
  \citenamefont {Groenenboom},\ and\ \citenamefont
  {Meijer}}]{gilijamse2006near}%
  \BibitemOpen
  \bibfield  {author} {\bibinfo {author} {\bibfnamefont {J.~J.}\ \bibnamefont
  {Gilijamse}}, \bibinfo {author} {\bibfnamefont {S.}~\bibnamefont {Hoekstra}},
  \bibinfo {author} {\bibfnamefont {S.~Y.}\ \bibnamefont {van~de Meerakker}},
  \bibinfo {author} {\bibfnamefont {G.~C.}\ \bibnamefont {Groenenboom}}, \ and\
  \bibinfo {author} {\bibfnamefont {G.}~\bibnamefont {Meijer}},\ }\href@noop {}
  {\bibfield  {journal} {\bibinfo  {journal} {Science}\ }\textbf {\bibinfo
  {volume} {313}},\ \bibinfo {pages} {1617} (\bibinfo {year}
  {2006})}\BibitemShut {NoStop}%
\bibitem [{\citenamefont {Bethlem}\ \emph {et~al.}(1999)\citenamefont
  {Bethlem}, \citenamefont {Berden},\ and\ \citenamefont
  {Meijer}}]{bethlem1999decelerating}%
  \BibitemOpen
  \bibfield  {author} {\bibinfo {author} {\bibfnamefont {H.~L.}\ \bibnamefont
  {Bethlem}}, \bibinfo {author} {\bibfnamefont {G.}~\bibnamefont {Berden}}, \
  and\ \bibinfo {author} {\bibfnamefont {G.}~\bibnamefont {Meijer}},\
  }\href@noop {} {\bibfield  {journal} {\bibinfo  {journal} {Physical Review
  Letters}\ }\textbf {\bibinfo {volume} {83}},\ \bibinfo {pages} {1558}
  (\bibinfo {year} {1999})}\BibitemShut {NoStop}%
\bibitem [{\citenamefont {Kirste}\ \emph {et~al.}(2012)\citenamefont {Kirste},
  \citenamefont {Wang}, \citenamefont {Schewe}, \citenamefont {Meijer},
  \citenamefont {Liu}, \citenamefont {van~der Avoird}, \citenamefont {Janssen},
  \citenamefont {Gubbels}, \citenamefont {Groenenboom},\ and\ \citenamefont
  {van~de Meerakker}}]{kirste2012quantum}%
  \BibitemOpen
  \bibfield  {author} {\bibinfo {author} {\bibfnamefont {M.}~\bibnamefont
  {Kirste}}, \bibinfo {author} {\bibfnamefont {X.}~\bibnamefont {Wang}},
  \bibinfo {author} {\bibfnamefont {H.~C.}\ \bibnamefont {Schewe}}, \bibinfo
  {author} {\bibfnamefont {G.}~\bibnamefont {Meijer}}, \bibinfo {author}
  {\bibfnamefont {K.}~\bibnamefont {Liu}}, \bibinfo {author} {\bibfnamefont
  {A.}~\bibnamefont {van~der Avoird}}, \bibinfo {author} {\bibfnamefont
  {L.~M.}\ \bibnamefont {Janssen}}, \bibinfo {author} {\bibfnamefont {K.~B.}\
  \bibnamefont {Gubbels}}, \bibinfo {author} {\bibfnamefont {G.~C.}\
  \bibnamefont {Groenenboom}}, \ and\ \bibinfo {author} {\bibfnamefont {S.~Y.}\
  \bibnamefont {van~de Meerakker}},\ }\href@noop {} {\bibfield  {journal}
  {\bibinfo  {journal} {Science}\ }\textbf {\bibinfo {volume} {338}},\ \bibinfo
  {pages} {1060} (\bibinfo {year} {2012})}\BibitemShut {NoStop}%
\bibitem [{\citenamefont {Vogels}\ \emph {et~al.}(2015)\citenamefont {Vogels},
  \citenamefont {Onvlee}, \citenamefont {Chefdeville}, \citenamefont {van~der
  Avoird}, \citenamefont {Groenenboom},\ and\ \citenamefont {van~de
  Meerakker}}]{vogels2015imaging}%
  \BibitemOpen
  \bibfield  {author} {\bibinfo {author} {\bibfnamefont {S.~N.}\ \bibnamefont
  {Vogels}}, \bibinfo {author} {\bibfnamefont {J.}~\bibnamefont {Onvlee}},
  \bibinfo {author} {\bibfnamefont {S.}~\bibnamefont {Chefdeville}}, \bibinfo
  {author} {\bibfnamefont {A.}~\bibnamefont {van~der Avoird}}, \bibinfo
  {author} {\bibfnamefont {G.~C.}\ \bibnamefont {Groenenboom}}, \ and\ \bibinfo
  {author} {\bibfnamefont {S.~Y.}\ \bibnamefont {van~de Meerakker}},\
  }\href@noop {} {\bibfield  {journal} {\bibinfo  {journal} {Science}\ }\textbf
  {\bibinfo {volume} {350}},\ \bibinfo {pages} {787} (\bibinfo {year}
  {2015})}\BibitemShut {NoStop}%
\bibitem [{\citenamefont {Liu}(2001)}]{liu2001crossed}%
  \BibitemOpen
  \bibfield  {author} {\bibinfo {author} {\bibfnamefont {K.}~\bibnamefont
  {Liu}},\ }\href@noop {} {\bibfield  {journal} {\bibinfo  {journal} {Annual
  Review of Physical Chemistry}\ }\textbf {\bibinfo {volume} {52}},\ \bibinfo
  {pages} {139} (\bibinfo {year} {2001})}\BibitemShut {NoStop}%
\bibitem [{\citenamefont {Liu}(2016)}]{liu2016vibrational}%
  \BibitemOpen
  \bibfield  {author} {\bibinfo {author} {\bibfnamefont {K.}~\bibnamefont
  {Liu}},\ }\href@noop {} {\bibfield  {journal} {\bibinfo  {journal} {Annual
  review of physical chemistry}\ }\textbf {\bibinfo {volume} {67}},\ \bibinfo
  {pages} {91} (\bibinfo {year} {2016})}\BibitemShut {NoStop}%
\bibitem [{\citenamefont {Liu}\ \emph {et~al.}(2014)\citenamefont {Liu},
  \citenamefont {Wang}, \citenamefont {Jiang}, \citenamefont {Czak{\'o}},
  \citenamefont {Yang}, \citenamefont {Liu},\ and\ \citenamefont
  {Guo}}]{liu2014rotational}%
  \BibitemOpen
  \bibfield  {author} {\bibinfo {author} {\bibfnamefont {R.}~\bibnamefont
  {Liu}}, \bibinfo {author} {\bibfnamefont {F.}~\bibnamefont {Wang}}, \bibinfo
  {author} {\bibfnamefont {B.}~\bibnamefont {Jiang}}, \bibinfo {author}
  {\bibfnamefont {G.}~\bibnamefont {Czak{\'o}}}, \bibinfo {author}
  {\bibfnamefont {M.}~\bibnamefont {Yang}}, \bibinfo {author} {\bibfnamefont
  {K.}~\bibnamefont {Liu}}, \ and\ \bibinfo {author} {\bibfnamefont
  {H.}~\bibnamefont {Guo}},\ }\href@noop {} {\bibfield  {journal} {\bibinfo
  {journal} {The Journal of chemical physics}\ }\textbf {\bibinfo {volume}
  {141}},\ \bibinfo {pages} {074310} (\bibinfo {year} {2014})}\BibitemShut
  {NoStop}%
\bibitem [{\citenamefont {Shagam}\ \emph
  {et~al.}(2015{\natexlab{a}})\citenamefont {Shagam}, \citenamefont {Klein},
  \citenamefont {Skomorowski}, \citenamefont {Yun}, \citenamefont {Averbukh},
  \citenamefont {Koch},\ and\ \citenamefont
  {Narevicius}}]{shagam2015molecular}%
  \BibitemOpen
  \bibfield  {author} {\bibinfo {author} {\bibfnamefont {Y.}~\bibnamefont
  {Shagam}}, \bibinfo {author} {\bibfnamefont {A.}~\bibnamefont {Klein}},
  \bibinfo {author} {\bibfnamefont {W.}~\bibnamefont {Skomorowski}}, \bibinfo
  {author} {\bibfnamefont {R.}~\bibnamefont {Yun}}, \bibinfo {author}
  {\bibfnamefont {V.}~\bibnamefont {Averbukh}}, \bibinfo {author}
  {\bibfnamefont {C.~P.}\ \bibnamefont {Koch}}, \ and\ \bibinfo {author}
  {\bibfnamefont {E.}~\bibnamefont {Narevicius}},\ }\href@noop {} {\bibfield
  {journal} {\bibinfo  {journal} {Nature chemistry}\ }\textbf {\bibinfo
  {volume} {7}},\ \bibinfo {pages} {921} (\bibinfo {year}
  {2015}{\natexlab{a}})}\BibitemShut {NoStop}%
\bibitem [{\citenamefont {WM}(2008)}]{wm2008low}%
  \BibitemOpen
  \bibfield  {author} {\bibinfo {author} {\bibfnamefont {S.~I.}\ \bibnamefont
  {WM}},\ }\href@noop {} {\emph {\bibinfo {title} {Low temperatures and cold
  molecules}}}\ (\bibinfo  {publisher} {World Scientific},\ \bibinfo {year}
  {2008})\BibitemShut {NoStop}%
\bibitem [{\citenamefont {Krems}\ \emph {et~al.}(2009)\citenamefont {Krems},
  \citenamefont {Friedrich},\ and\ \citenamefont {Stwalley}}]{krems2009cold}%
  \BibitemOpen
  \bibfield  {author} {\bibinfo {author} {\bibfnamefont {R.}~\bibnamefont
  {Krems}}, \bibinfo {author} {\bibfnamefont {B.}~\bibnamefont {Friedrich}}, \
  and\ \bibinfo {author} {\bibfnamefont {W.~C.}\ \bibnamefont {Stwalley}},\
  }\href@noop {} {\emph {\bibinfo {title} {Cold molecules: theory, experiment,
  applications}}}\ (\bibinfo  {publisher} {CRC press},\ \bibinfo {year}
  {2009})\BibitemShut {NoStop}%
\bibitem [{\citenamefont {Yang}(2007)}]{yang2007state}%
  \BibitemOpen
  \bibfield  {author} {\bibinfo {author} {\bibfnamefont {X.}~\bibnamefont
  {Yang}},\ }\href@noop {} {\bibfield  {journal} {\bibinfo  {journal} {Annu.
  Rev. Phys. Chem.}\ }\textbf {\bibinfo {volume} {58}},\ \bibinfo {pages} {433}
  (\bibinfo {year} {2007})}\BibitemShut {NoStop}%
\bibitem [{\citenamefont {Bernstein}(1979)}]{bernstein1979atom}%
  \BibitemOpen
  \bibfield  {author} {\bibinfo {author} {\bibfnamefont {R.~B.}\ \bibnamefont
  {Bernstein}},\ }\href@noop {} {\emph {\bibinfo {title} {Atom - molecule
  collision theory}}}\ (\bibinfo  {publisher} {Plenum Press},\ \bibinfo {year}
  {1979})\BibitemShut {NoStop}%
\bibitem [{\citenamefont {Child}(1996)}]{child1996molecular}%
  \BibitemOpen
  \bibfield  {author} {\bibinfo {author} {\bibfnamefont {M.~S.}\ \bibnamefont
  {Child}},\ }\href@noop {} {\emph {\bibinfo {title} {Molecular collision
  theory}}}\ (\bibinfo  {publisher} {Courier Corporation},\ \bibinfo {year}
  {1996})\BibitemShut {NoStop}%
\bibitem [{\citenamefont {Levine}(2009)}]{levine2009molecular}%
  \BibitemOpen
  \bibfield  {author} {\bibinfo {author} {\bibfnamefont {R.~D.}\ \bibnamefont
  {Levine}},\ }\href@noop {} {\emph {\bibinfo {title} {Molecular reaction
  dynamics}}}\ (\bibinfo  {publisher} {Cambridge University Press},\ \bibinfo
  {year} {2009})\BibitemShut {NoStop}%
\bibitem [{\citenamefont {Chalasinski}\ and\ \citenamefont
  {Szczesniak}(1994)}]{chalasinski1994origins}%
  \BibitemOpen
  \bibfield  {author} {\bibinfo {author} {\bibfnamefont {G.}~\bibnamefont
  {Chalasinski}}\ and\ \bibinfo {author} {\bibfnamefont {M.~M.}\ \bibnamefont
  {Szczesniak}},\ }\href@noop {} {\bibfield  {journal} {\bibinfo  {journal}
  {Chemical Reviews}\ }\textbf {\bibinfo {volume} {94}},\ \bibinfo {pages}
  {1723} (\bibinfo {year} {1994})}\BibitemShut {NoStop}%
\bibitem [{\citenamefont {Chalasinski}\ and\ \citenamefont
  {Szczesniak}(2000)}]{chalasinski2000state}%
  \BibitemOpen
  \bibfield  {author} {\bibinfo {author} {\bibfnamefont {G.}~\bibnamefont
  {Chalasinski}}\ and\ \bibinfo {author} {\bibfnamefont {M.~M.}\ \bibnamefont
  {Szczesniak}},\ }\href@noop {} {\bibfield  {journal} {\bibinfo  {journal}
  {Chemical reviews}\ }\textbf {\bibinfo {volume} {100}},\ \bibinfo {pages}
  {4227} (\bibinfo {year} {2000})}\BibitemShut {NoStop}%
\bibitem [{\citenamefont {Raghavachari}\ \emph {et~al.}(1989)\citenamefont
  {Raghavachari}, \citenamefont {Trucks}, \citenamefont {Pople},\ and\
  \citenamefont {Head-Gordon}}]{raghavachari1989fifth}%
  \BibitemOpen
  \bibfield  {author} {\bibinfo {author} {\bibfnamefont {K.}~\bibnamefont
  {Raghavachari}}, \bibinfo {author} {\bibfnamefont {G.~W.}\ \bibnamefont
  {Trucks}}, \bibinfo {author} {\bibfnamefont {J.~A.}\ \bibnamefont {Pople}}, \
  and\ \bibinfo {author} {\bibfnamefont {M.}~\bibnamefont {Head-Gordon}},\
  }\href@noop {} {\bibfield  {journal} {\bibinfo  {journal} {Chemical Physics
  Letters}\ }\textbf {\bibinfo {volume} {157}},\ \bibinfo {pages} {479}
  (\bibinfo {year} {1989})}\BibitemShut {NoStop}%
\bibitem [{\citenamefont {Roos}(1987)}]{roos1987complete}%
  \BibitemOpen
  \bibfield  {author} {\bibinfo {author} {\bibfnamefont {B.~O.}\ \bibnamefont
  {Roos}},\ }\href@noop {} {\bibfield  {journal} {\bibinfo  {journal} {Advances
  in chemical physics}\ }\textbf {\bibinfo {volume} {69}},\ \bibinfo {pages}
  {399} (\bibinfo {year} {1987})}\BibitemShut {NoStop}%
\bibitem [{\citenamefont {Werner}\ and\ \citenamefont
  {Knowles}(1985)}]{werner1985second}%
  \BibitemOpen
  \bibfield  {author} {\bibinfo {author} {\bibfnamefont {H.-J.}\ \bibnamefont
  {Werner}}\ and\ \bibinfo {author} {\bibfnamefont {P.~J.}\ \bibnamefont
  {Knowles}},\ }\href@noop {} {\bibfield  {journal} {\bibinfo  {journal} {The
  Journal of chemical physics}\ }\textbf {\bibinfo {volume} {82}},\ \bibinfo
  {pages} {5053} (\bibinfo {year} {1985})}\BibitemShut {NoStop}%
\bibitem [{\citenamefont {Dunning~Jr}(1989)}]{dunning1989gaussian}%
  \BibitemOpen
  \bibfield  {author} {\bibinfo {author} {\bibfnamefont {T.~H.}\ \bibnamefont
  {Dunning~Jr}},\ }\href@noop {} {\bibfield  {journal} {\bibinfo  {journal}
  {The Journal of chemical physics}\ }\textbf {\bibinfo {volume} {90}},\
  \bibinfo {pages} {1007} (\bibinfo {year} {1989})}\BibitemShut {NoStop}%
\bibitem [{\citenamefont {Sachs}\ \emph {et~al.}(1975)\citenamefont {Sachs},
  \citenamefont {Hinze},\ and\ \citenamefont {Sabelli}}]{sachs1975frozen}%
  \BibitemOpen
  \bibfield  {author} {\bibinfo {author} {\bibfnamefont {E.~S.}\ \bibnamefont
  {Sachs}}, \bibinfo {author} {\bibfnamefont {J.}~\bibnamefont {Hinze}}, \ and\
  \bibinfo {author} {\bibfnamefont {N.~H.}\ \bibnamefont {Sabelli}},\
  }\href@noop {} {\bibfield  {journal} {\bibinfo  {journal} {The Journal of
  Chemical Physics}\ }\textbf {\bibinfo {volume} {62}},\ \bibinfo {pages}
  {3393} (\bibinfo {year} {1975})}\BibitemShut {NoStop}%
\bibitem [{\citenamefont {Werner}\ \emph {et~al.}(2012)\citenamefont {Werner},
  \citenamefont {Knowles}, \citenamefont {Knizia}, \citenamefont {Manby},\ and\
  \citenamefont {Sch{\"u}tz}}]{werner2012molpro}%
  \BibitemOpen
  \bibfield  {author} {\bibinfo {author} {\bibfnamefont {H.-J.}\ \bibnamefont
  {Werner}}, \bibinfo {author} {\bibfnamefont {P.~J.}\ \bibnamefont {Knowles}},
  \bibinfo {author} {\bibfnamefont {G.}~\bibnamefont {Knizia}}, \bibinfo
  {author} {\bibfnamefont {F.~R.}\ \bibnamefont {Manby}}, \ and\ \bibinfo
  {author} {\bibfnamefont {M.}~\bibnamefont {Sch{\"u}tz}},\ }\href@noop {}
  {\bibfield  {journal} {\bibinfo  {journal} {Wiley Interdisciplinary Reviews:
  Computational Molecular Science}\ }\textbf {\bibinfo {volume} {2}},\ \bibinfo
  {pages} {242} (\bibinfo {year} {2012})}\BibitemShut {NoStop}%
\bibitem [{\citenamefont {Schatz}(1989)}]{schatz1989analytical}%
  \BibitemOpen
  \bibfield  {author} {\bibinfo {author} {\bibfnamefont {G.~C.}\ \bibnamefont
  {Schatz}},\ }\href@noop {} {\bibfield  {journal} {\bibinfo  {journal}
  {Reviews of Modern Physics}\ }\textbf {\bibinfo {volume} {61}},\ \bibinfo
  {pages} {669} (\bibinfo {year} {1989})}\BibitemShut {NoStop}%
\bibitem [{\citenamefont {Hollebeek}\ \emph {et~al.}(1999)\citenamefont
  {Hollebeek}, \citenamefont {Ho},\ and\ \citenamefont
  {Rabitz}}]{hollebeek1999constructing}%
  \BibitemOpen
  \bibfield  {author} {\bibinfo {author} {\bibfnamefont {T.}~\bibnamefont
  {Hollebeek}}, \bibinfo {author} {\bibfnamefont {T.-S.}\ \bibnamefont {Ho}}, \
  and\ \bibinfo {author} {\bibfnamefont {H.}~\bibnamefont {Rabitz}},\
  }\href@noop {} {\bibfield  {journal} {\bibinfo  {journal} {Annual review of
  physical chemistry}\ }\textbf {\bibinfo {volume} {50}},\ \bibinfo {pages}
  {537} (\bibinfo {year} {1999})}\BibitemShut {NoStop}%
\bibitem [{\citenamefont {Kuntz}\ and\ \citenamefont
  {Roach}(1972)}]{kuntz1972ion}%
  \BibitemOpen
  \bibfield  {author} {\bibinfo {author} {\bibfnamefont {P.}~\bibnamefont
  {Kuntz}}\ and\ \bibinfo {author} {\bibfnamefont {A.}~\bibnamefont {Roach}},\
  }\href@noop {} {\bibfield  {journal} {\bibinfo  {journal} {Journal of the
  Chemical Society, Faraday Transactions 2: Molecular and Chemical Physics}\
  }\textbf {\bibinfo {volume} {68}},\ \bibinfo {pages} {259} (\bibinfo {year}
  {1972})}\BibitemShut {NoStop}%
\bibitem [{\citenamefont {Ho}\ and\ \citenamefont
  {Rabitz}(2000)}]{ho2000proper}%
  \BibitemOpen
  \bibfield  {author} {\bibinfo {author} {\bibfnamefont {T.-S.}\ \bibnamefont
  {Ho}}\ and\ \bibinfo {author} {\bibfnamefont {H.}~\bibnamefont {Rabitz}},\
  }\href@noop {} {\bibfield  {journal} {\bibinfo  {journal} {The Journal of
  Chemical Physics}\ }\textbf {\bibinfo {volume} {113}},\ \bibinfo {pages}
  {3960} (\bibinfo {year} {2000})}\BibitemShut {NoStop}%
\bibitem [{\citenamefont {Frishman}\ \emph {et~al.}(1997)\citenamefont
  {Frishman}, \citenamefont {Hoffman},\ and\ \citenamefont
  {Kouri}}]{frishman1997distributed}%
  \BibitemOpen
  \bibfield  {author} {\bibinfo {author} {\bibfnamefont {A.}~\bibnamefont
  {Frishman}}, \bibinfo {author} {\bibfnamefont {D.~K.}\ \bibnamefont
  {Hoffman}}, \ and\ \bibinfo {author} {\bibfnamefont {D.~J.}\ \bibnamefont
  {Kouri}},\ }\href@noop {} {\bibfield  {journal} {\bibinfo  {journal} {The
  Journal of chemical physics}\ }\textbf {\bibinfo {volume} {107}},\ \bibinfo
  {pages} {804} (\bibinfo {year} {1997})}\BibitemShut {NoStop}%
\bibitem [{\citenamefont {Garcia}\ and\ \citenamefont
  {Lagana'}(1984)}]{garcia1984fit}%
  \BibitemOpen
  \bibfield  {author} {\bibinfo {author} {\bibfnamefont {E.}~\bibnamefont
  {Garcia}}\ and\ \bibinfo {author} {\bibfnamefont {A.}~\bibnamefont
  {Lagana'}},\ }\href@noop {} {\bibfield  {journal} {\bibinfo  {journal}
  {Molecular Physics}\ }\textbf {\bibinfo {volume} {52}},\ \bibinfo {pages}
  {1115} (\bibinfo {year} {1984})}\BibitemShut {NoStop}%
\bibitem [{\citenamefont {Murrell}(1984)}]{murrell1984molecular}%
  \BibitemOpen
  \bibfield  {author} {\bibinfo {author} {\bibfnamefont {J.~N.}\ \bibnamefont
  {Murrell}},\ }\href@noop {} {\emph {\bibinfo {title} {Molecular potential
  energy functions}}}\ (\bibinfo  {publisher} {J. Wiley},\ \bibinfo {year}
  {1984})\BibitemShut {NoStop}%
\bibitem [{\citenamefont {Dhont}\ \emph {et~al.}(2005)\citenamefont {Dhont},
  \citenamefont {van Lenthe}, \citenamefont {Groenenboom},\ and\ \citenamefont
  {van~der Avoird}}]{dhont2005ab}%
  \BibitemOpen
  \bibfield  {author} {\bibinfo {author} {\bibfnamefont {G.~S.}\ \bibnamefont
  {Dhont}}, \bibinfo {author} {\bibfnamefont {J.~H.}\ \bibnamefont {van
  Lenthe}}, \bibinfo {author} {\bibfnamefont {G.~C.}\ \bibnamefont
  {Groenenboom}}, \ and\ \bibinfo {author} {\bibfnamefont {A.}~\bibnamefont
  {van~der Avoird}},\ }\href@noop {} {\bibfield  {journal} {\bibinfo  {journal}
  {The Journal of chemical physics}\ }\textbf {\bibinfo {volume} {123}},\
  \bibinfo {pages} {184302} (\bibinfo {year} {2005})}\BibitemShut {NoStop}%
\bibitem [{\citenamefont {Janssen}\ \emph {et~al.}(2009)\citenamefont
  {Janssen}, \citenamefont {Groenenboom}, \citenamefont {van~der Avoird},
  \citenamefont {{\.Z}uchowski},\ and\ \citenamefont
  {Podeszwa}}]{janssen2009ab}%
  \BibitemOpen
  \bibfield  {author} {\bibinfo {author} {\bibfnamefont {L.~M.}\ \bibnamefont
  {Janssen}}, \bibinfo {author} {\bibfnamefont {G.~C.}\ \bibnamefont
  {Groenenboom}}, \bibinfo {author} {\bibfnamefont {A.}~\bibnamefont {van~der
  Avoird}}, \bibinfo {author} {\bibfnamefont {P.~S.}\ \bibnamefont
  {{\.Z}uchowski}}, \ and\ \bibinfo {author} {\bibfnamefont {R.}~\bibnamefont
  {Podeszwa}},\ }\href@noop {} {\bibfield  {journal} {\bibinfo  {journal} {The
  Journal of chemical physics}\ }\textbf {\bibinfo {volume} {131}},\ \bibinfo
  {pages} {224314} (\bibinfo {year} {2009})}\BibitemShut {NoStop}%
\bibitem [{\citenamefont {Li}\ \emph {et~al.}(1997)\citenamefont {Li},
  \citenamefont {Apkarian},\ and\ \citenamefont {Harding}}]{li1997theoretical}%
  \BibitemOpen
  \bibfield  {author} {\bibinfo {author} {\bibfnamefont {Z.}~\bibnamefont
  {Li}}, \bibinfo {author} {\bibfnamefont {V.}~\bibnamefont {Apkarian}}, \ and\
  \bibinfo {author} {\bibfnamefont {L.~B.}\ \bibnamefont {Harding}},\
  }\href@noop {} {\bibfield  {journal} {\bibinfo  {journal} {The Journal of
  chemical physics}\ }\textbf {\bibinfo {volume} {106}},\ \bibinfo {pages}
  {942} (\bibinfo {year} {1997})}\BibitemShut {NoStop}%
\bibitem [{\citenamefont {Alexander}(1998)}]{alexander1998theoretical}%
  \BibitemOpen
  \bibfield  {author} {\bibinfo {author} {\bibfnamefont {M.~H.}\ \bibnamefont
  {Alexander}},\ }\href@noop {} {\bibfield  {journal} {\bibinfo  {journal} {The
  Journal of chemical physics}\ }\textbf {\bibinfo {volume} {108}},\ \bibinfo
  {pages} {4467} (\bibinfo {year} {1998})}\BibitemShut {NoStop}%
\bibitem [{\citenamefont {Rogers}\ \emph {et~al.}(2000)\citenamefont {Rogers},
  \citenamefont {Wang}, \citenamefont {Kuppermann},\ and\ \citenamefont
  {Walch}}]{rogers2000chemically}%
  \BibitemOpen
  \bibfield  {author} {\bibinfo {author} {\bibfnamefont {S.}~\bibnamefont
  {Rogers}}, \bibinfo {author} {\bibfnamefont {D.}~\bibnamefont {Wang}},
  \bibinfo {author} {\bibfnamefont {A.}~\bibnamefont {Kuppermann}}, \ and\
  \bibinfo {author} {\bibfnamefont {S.}~\bibnamefont {Walch}},\ }\href@noop {}
  {\bibfield  {journal} {\bibinfo  {journal} {The Journal of Physical Chemistry
  A}\ }\textbf {\bibinfo {volume} {104}},\ \bibinfo {pages} {2308} (\bibinfo
  {year} {2000})}\BibitemShut {NoStop}%
\bibitem [{\citenamefont {Brandao}\ \emph {et~al.}(2004)\citenamefont
  {Brandao}, \citenamefont {Mogo},\ and\ \citenamefont
  {Silva}}]{brandao2004potential}%
  \BibitemOpen
  \bibfield  {author} {\bibinfo {author} {\bibfnamefont {J.}~\bibnamefont
  {Brandao}}, \bibinfo {author} {\bibfnamefont {C.}~\bibnamefont {Mogo}}, \
  and\ \bibinfo {author} {\bibfnamefont {B.~C.}\ \bibnamefont {Silva}},\
  }\href@noop {} {\bibfield  {journal} {\bibinfo  {journal} {The Journal of
  chemical physics}\ }\textbf {\bibinfo {volume} {121}},\ \bibinfo {pages}
  {8861} (\bibinfo {year} {2004})}\BibitemShut {NoStop}%
\bibitem [{\citenamefont {Atahan}\ \emph {et~al.}(2006)\citenamefont {Atahan},
  \citenamefont {K{\l}os}, \citenamefont {{\.Z}uchowski},\ and\ \citenamefont
  {Alexander}}]{atahan2006ab}%
  \BibitemOpen
  \bibfield  {author} {\bibinfo {author} {\bibfnamefont {S.}~\bibnamefont
  {Atahan}}, \bibinfo {author} {\bibfnamefont {J.}~\bibnamefont {K{\l}os}},
  \bibinfo {author} {\bibfnamefont {P.~S.}\ \bibnamefont {{\.Z}uchowski}}, \
  and\ \bibinfo {author} {\bibfnamefont {M.~H.}\ \bibnamefont {Alexander}},\
  }\href@noop {} {\bibfield  {journal} {\bibinfo  {journal} {Physical Chemistry
  Chemical Physics}\ }\textbf {\bibinfo {volume} {8}},\ \bibinfo {pages} {4420}
  (\bibinfo {year} {2006})}\BibitemShut {NoStop}%
\bibitem [{\citenamefont {Knowles}\ and\ \citenamefont
  {Werner}(1988)}]{knowles1988efficient}%
  \BibitemOpen
  \bibfield  {author} {\bibinfo {author} {\bibfnamefont {P.~J.}\ \bibnamefont
  {Knowles}}\ and\ \bibinfo {author} {\bibfnamefont {H.-J.}\ \bibnamefont
  {Werner}},\ }\href@noop {} {\bibfield  {journal} {\bibinfo  {journal}
  {Chemical physics letters}\ }\textbf {\bibinfo {volume} {145}},\ \bibinfo
  {pages} {514} (\bibinfo {year} {1988})}\BibitemShut {NoStop}%
\bibitem [{\citenamefont {Werner}\ and\ \citenamefont
  {Knowles}(1988)}]{werner1988efficient}%
  \BibitemOpen
  \bibfield  {author} {\bibinfo {author} {\bibfnamefont {H.-J.}\ \bibnamefont
  {Werner}}\ and\ \bibinfo {author} {\bibfnamefont {P.~J.}\ \bibnamefont
  {Knowles}},\ }\href@noop {} {\bibfield  {journal} {\bibinfo  {journal} {The
  Journal of chemical physics}\ }\textbf {\bibinfo {volume} {89}},\ \bibinfo
  {pages} {5803} (\bibinfo {year} {1988})}\BibitemShut {NoStop}%
\bibitem [{\citenamefont {Jeziorski}\ \emph {et~al.}(1994)\citenamefont
  {Jeziorski}, \citenamefont {Moszynski},\ and\ \citenamefont
  {Szalewicz}}]{jeziorski1994perturbation}%
  \BibitemOpen
  \bibfield  {author} {\bibinfo {author} {\bibfnamefont {B.}~\bibnamefont
  {Jeziorski}}, \bibinfo {author} {\bibfnamefont {R.}~\bibnamefont
  {Moszynski}}, \ and\ \bibinfo {author} {\bibfnamefont {K.}~\bibnamefont
  {Szalewicz}},\ }\href@noop {} {\bibfield  {journal} {\bibinfo  {journal}
  {Chemical Reviews}\ }\textbf {\bibinfo {volume} {94}},\ \bibinfo {pages}
  {1887} (\bibinfo {year} {1994})}\BibitemShut {NoStop}%
\bibitem [{\citenamefont {Zyubin}\ \emph {et~al.}(2001)\citenamefont {Zyubin},
  \citenamefont {Mebel}, \citenamefont {Der~Chao},\ and\ \citenamefont
  {Skodje}}]{zyubin2001reaction}%
  \BibitemOpen
  \bibfield  {author} {\bibinfo {author} {\bibfnamefont {A.~S.}\ \bibnamefont
  {Zyubin}}, \bibinfo {author} {\bibfnamefont {A.~M.}\ \bibnamefont {Mebel}},
  \bibinfo {author} {\bibfnamefont {S.}~\bibnamefont {Der~Chao}}, \ and\
  \bibinfo {author} {\bibfnamefont {R.~T.}\ \bibnamefont {Skodje}},\
  }\href@noop {} {\bibfield  {journal} {\bibinfo  {journal} {The Journal of
  Chemical Physics}\ }\textbf {\bibinfo {volume} {114}},\ \bibinfo {pages}
  {320} (\bibinfo {year} {2001})}\BibitemShut {NoStop}%
\bibitem [{\citenamefont {Ho}\ \emph {et~al.}(2002)\citenamefont {Ho},
  \citenamefont {Hollebeek}, \citenamefont {Rabitz}, \citenamefont {Der~Chao},
  \citenamefont {Skodje}, \citenamefont {Zyubin},\ and\ \citenamefont
  {Mebel}}]{ho2002globally}%
  \BibitemOpen
  \bibfield  {author} {\bibinfo {author} {\bibfnamefont {T.-S.}\ \bibnamefont
  {Ho}}, \bibinfo {author} {\bibfnamefont {T.}~\bibnamefont {Hollebeek}},
  \bibinfo {author} {\bibfnamefont {H.}~\bibnamefont {Rabitz}}, \bibinfo
  {author} {\bibfnamefont {S.}~\bibnamefont {Der~Chao}}, \bibinfo {author}
  {\bibfnamefont {R.~T.}\ \bibnamefont {Skodje}}, \bibinfo {author}
  {\bibfnamefont {A.~S.}\ \bibnamefont {Zyubin}}, \ and\ \bibinfo {author}
  {\bibfnamefont {A.~M.}\ \bibnamefont {Mebel}},\ }\href@noop {} {\bibfield
  {journal} {\bibinfo  {journal} {The Journal of chemical physics}\ }\textbf
  {\bibinfo {volume} {116}},\ \bibinfo {pages} {4124} (\bibinfo {year}
  {2002})}\BibitemShut {NoStop}%
\bibitem [{\citenamefont {Pack}\ and\ \citenamefont
  {Parker}(1987)}]{pack1987quantum}%
  \BibitemOpen
  \bibfield  {author} {\bibinfo {author} {\bibfnamefont {R.~T.}\ \bibnamefont
  {Pack}}\ and\ \bibinfo {author} {\bibfnamefont {G.~A.}\ \bibnamefont
  {Parker}},\ }\href@noop {} {\bibfield  {journal} {\bibinfo  {journal} {The
  Journal of chemical physics}\ }\textbf {\bibinfo {volume} {87}},\ \bibinfo
  {pages} {3888} (\bibinfo {year} {1987})}\BibitemShut {NoStop}%
\bibitem [{\citenamefont {Makrides}\ \emph {et~al.}(2015)\citenamefont
  {Makrides}, \citenamefont {Hazra}, \citenamefont {Pradhan}, \citenamefont
  {Petrov}, \citenamefont {Kendrick}, \citenamefont {Gonz{\'a}lez-Lezana},
  \citenamefont {Balakrishnan},\ and\ \citenamefont
  {Kotochigova}}]{makrides2015ultracold}%
  \BibitemOpen
  \bibfield  {author} {\bibinfo {author} {\bibfnamefont {C.}~\bibnamefont
  {Makrides}}, \bibinfo {author} {\bibfnamefont {J.}~\bibnamefont {Hazra}},
  \bibinfo {author} {\bibfnamefont {G.}~\bibnamefont {Pradhan}}, \bibinfo
  {author} {\bibfnamefont {A.}~\bibnamefont {Petrov}}, \bibinfo {author}
  {\bibfnamefont {B.~K.}\ \bibnamefont {Kendrick}}, \bibinfo {author}
  {\bibfnamefont {T.}~\bibnamefont {Gonz{\'a}lez-Lezana}}, \bibinfo {author}
  {\bibfnamefont {N.}~\bibnamefont {Balakrishnan}}, \ and\ \bibinfo {author}
  {\bibfnamefont {S.}~\bibnamefont {Kotochigova}},\ }\href@noop {} {\bibfield
  {journal} {\bibinfo  {journal} {Physical Review A}\ }\textbf {\bibinfo
  {volume} {91}},\ \bibinfo {pages} {012708} (\bibinfo {year}
  {2015})}\BibitemShut {NoStop}%
\bibitem [{\citenamefont {Croft}\ \emph {et~al.}(2017)\citenamefont {Croft},
  \citenamefont {Makrides}, \citenamefont {Li}, \citenamefont {Petrov},
  \citenamefont {Kendrick}, \citenamefont {Balakrishnan},\ and\ \citenamefont
  {Kotochigova}}]{croft2017universality}%
  \BibitemOpen
  \bibfield  {author} {\bibinfo {author} {\bibfnamefont {J.}~\bibnamefont
  {Croft}}, \bibinfo {author} {\bibfnamefont {C.}~\bibnamefont {Makrides}},
  \bibinfo {author} {\bibfnamefont {M.}~\bibnamefont {Li}}, \bibinfo {author}
  {\bibfnamefont {A.}~\bibnamefont {Petrov}}, \bibinfo {author} {\bibfnamefont
  {B.}~\bibnamefont {Kendrick}}, \bibinfo {author} {\bibfnamefont
  {N.}~\bibnamefont {Balakrishnan}}, \ and\ \bibinfo {author} {\bibfnamefont
  {S.}~\bibnamefont {Kotochigova}},\ }\href@noop {} {\bibfield  {journal}
  {\bibinfo  {journal} {Nature communications}\ }\textbf {\bibinfo {volume}
  {8}},\ \bibinfo {pages} {15897} (\bibinfo {year} {2017})}\BibitemShut
  {NoStop}%
\bibitem [{\citenamefont {Braams}\ and\ \citenamefont
  {Bowman}(2009)}]{braams2009permutationally}%
  \BibitemOpen
  \bibfield  {author} {\bibinfo {author} {\bibfnamefont {B.~J.}\ \bibnamefont
  {Braams}}\ and\ \bibinfo {author} {\bibfnamefont {J.~M.}\ \bibnamefont
  {Bowman}},\ }\href@noop {} {\bibfield  {journal} {\bibinfo  {journal}
  {International Reviews in Physical Chemistry}\ }\textbf {\bibinfo {volume}
  {28}},\ \bibinfo {pages} {577} (\bibinfo {year} {2009})}\BibitemShut
  {NoStop}%
\bibitem [{\citenamefont {Bowman}\ \emph {et~al.}(2011)\citenamefont {Bowman},
  \citenamefont {Czako},\ and\ \citenamefont {Fu}}]{bowman2011high}%
  \BibitemOpen
  \bibfield  {author} {\bibinfo {author} {\bibfnamefont {J.~M.}\ \bibnamefont
  {Bowman}}, \bibinfo {author} {\bibfnamefont {G.}~\bibnamefont {Czako}}, \
  and\ \bibinfo {author} {\bibfnamefont {B.}~\bibnamefont {Fu}},\ }\href@noop
  {} {\bibfield  {journal} {\bibinfo  {journal} {Physical Chemistry Chemical
  Physics}\ }\textbf {\bibinfo {volume} {13}},\ \bibinfo {pages} {8094}
  (\bibinfo {year} {2011})}\BibitemShut {NoStop}%
\bibitem [{\citenamefont {Czak{\'o}}\ and\ \citenamefont
  {Bowman}(2011)}]{czako2011dynamics}%
  \BibitemOpen
  \bibfield  {author} {\bibinfo {author} {\bibfnamefont {G.}~\bibnamefont
  {Czak{\'o}}}\ and\ \bibinfo {author} {\bibfnamefont {J.~M.}\ \bibnamefont
  {Bowman}},\ }\href@noop {} {\bibfield  {journal} {\bibinfo  {journal}
  {Science}\ }\textbf {\bibinfo {volume} {334}},\ \bibinfo {pages} {343}
  (\bibinfo {year} {2011})}\BibitemShut {NoStop}%
\bibitem [{\citenamefont {Althorpe}\ and\ \citenamefont
  {Clary}(2003)}]{althorpe2003quantum}%
  \BibitemOpen
  \bibfield  {author} {\bibinfo {author} {\bibfnamefont {S.~C.}\ \bibnamefont
  {Althorpe}}\ and\ \bibinfo {author} {\bibfnamefont {D.~C.}\ \bibnamefont
  {Clary}},\ }\href@noop {} {\bibfield  {journal} {\bibinfo  {journal} {Annual
  review of physical chemistry}\ }\textbf {\bibinfo {volume} {54}},\ \bibinfo
  {pages} {493} (\bibinfo {year} {2003})}\BibitemShut {NoStop}%
\bibitem [{\citenamefont {Brouard}\ and\ \citenamefont
  {Vallance}(2015)}]{brouard2015tutorials}%
  \BibitemOpen
  \bibfield  {author} {\bibinfo {author} {\bibfnamefont {M.}~\bibnamefont
  {Brouard}}\ and\ \bibinfo {author} {\bibfnamefont {C.}~\bibnamefont
  {Vallance}},\ }\href@noop {} {\emph {\bibinfo {title} {Tutorials in molecular
  reaction dynamics}}}\ (\bibinfo  {publisher} {Royal Society of Chemistry},\
  \bibinfo {year} {2015})\BibitemShut {NoStop}%
\bibitem [{\citenamefont {Klein}\ \emph {et~al.}(2017)\citenamefont {Klein},
  \citenamefont {Shagam}, \citenamefont {Skomorowski}, \citenamefont
  {{\.Z}uchowski}, \citenamefont {Pawlak}, \citenamefont {Janssen},
  \citenamefont {Moiseyev}, \citenamefont {van~de Meerakker}, \citenamefont
  {van~der Avoird}, \citenamefont {Koch} \emph {et~al.}}]{klein2017directly}%
  \BibitemOpen
  \bibfield  {author} {\bibinfo {author} {\bibfnamefont {A.}~\bibnamefont
  {Klein}}, \bibinfo {author} {\bibfnamefont {Y.}~\bibnamefont {Shagam}},
  \bibinfo {author} {\bibfnamefont {W.}~\bibnamefont {Skomorowski}}, \bibinfo
  {author} {\bibfnamefont {P.~S.}\ \bibnamefont {{\.Z}uchowski}}, \bibinfo
  {author} {\bibfnamefont {M.}~\bibnamefont {Pawlak}}, \bibinfo {author}
  {\bibfnamefont {L.~M.}\ \bibnamefont {Janssen}}, \bibinfo {author}
  {\bibfnamefont {N.}~\bibnamefont {Moiseyev}}, \bibinfo {author}
  {\bibfnamefont {S.~Y.}\ \bibnamefont {van~de Meerakker}}, \bibinfo {author}
  {\bibfnamefont {A.}~\bibnamefont {van~der Avoird}}, \bibinfo {author}
  {\bibfnamefont {C.~P.}\ \bibnamefont {Koch}},  \emph {et~al.},\ }\href@noop
  {} {\bibfield  {journal} {\bibinfo  {journal} {Nature Physics}\ }\textbf
  {\bibinfo {volume} {13}},\ \bibinfo {pages} {35} (\bibinfo {year}
  {2017})}\BibitemShut {NoStop}%
\bibitem [{\citenamefont {Arthurs}\ and\ \citenamefont
  {Dalgarno}(1960)}]{arthurs1960theory}%
  \BibitemOpen
  \bibfield  {author} {\bibinfo {author} {\bibfnamefont {A.}~\bibnamefont
  {Arthurs}}\ and\ \bibinfo {author} {\bibfnamefont {A.}~\bibnamefont
  {Dalgarno}},\ }\href@noop {} {\bibfield  {journal} {\bibinfo  {journal}
  {Proceedings of the Royal Society of London. Series A. Mathematical and
  Physical Sciences}\ }\textbf {\bibinfo {volume} {256}},\ \bibinfo {pages}
  {540} (\bibinfo {year} {1960})}\BibitemShut {NoStop}%
\bibitem [{\citenamefont {Quack}\ and\ \citenamefont
  {Troe}(1974)}]{quack1974specific}%
  \BibitemOpen
  \bibfield  {author} {\bibinfo {author} {\bibfnamefont {M.}~\bibnamefont
  {Quack}}\ and\ \bibinfo {author} {\bibfnamefont {J.}~\bibnamefont {Troe}},\
  }\href@noop {} {\bibfield  {journal} {\bibinfo  {journal} {Berichte der
  Bunsengesellschaft f{\"u}r physikalische Chemie}\ }\textbf {\bibinfo {volume}
  {78}},\ \bibinfo {pages} {240} (\bibinfo {year} {1974})}\BibitemShut
  {NoStop}%
\bibitem [{\citenamefont {Clary}(1985)}]{clary1985calculations}%
  \BibitemOpen
  \bibfield  {author} {\bibinfo {author} {\bibfnamefont {D.}~\bibnamefont
  {Clary}},\ }\href@noop {} {\bibfield  {journal} {\bibinfo  {journal}
  {Molecular Physics}\ }\textbf {\bibinfo {volume} {54}},\ \bibinfo {pages}
  {605} (\bibinfo {year} {1985})}\BibitemShut {NoStop}%
\bibitem [{\citenamefont {Clary}(1987)}]{clary1987rate}%
  \BibitemOpen
  \bibfield  {author} {\bibinfo {author} {\bibfnamefont {D.~C.}\ \bibnamefont
  {Clary}},\ }\href@noop {} {\bibfield  {journal} {\bibinfo  {journal} {Journal
  of the Chemical Society, Faraday Transactions 2: Molecular and Chemical
  Physics}\ }\textbf {\bibinfo {volume} {83}},\ \bibinfo {pages} {139}
  (\bibinfo {year} {1987})}\BibitemShut {NoStop}%
\bibitem [{\citenamefont {Ramillon}\ and\ \citenamefont
  {McCarroll}(1994)}]{ramillon1994adiabatic}%
  \BibitemOpen
  \bibfield  {author} {\bibinfo {author} {\bibfnamefont {M.}~\bibnamefont
  {Ramillon}}\ and\ \bibinfo {author} {\bibfnamefont {R.}~\bibnamefont
  {McCarroll}},\ }\href@noop {} {\bibfield  {journal} {\bibinfo  {journal} {The
  Journal of chemical physics}\ }\textbf {\bibinfo {volume} {101}},\ \bibinfo
  {pages} {8697} (\bibinfo {year} {1994})}\BibitemShut {NoStop}%
\bibitem [{\citenamefont {Clary}(1990)}]{clary1990fast}%
  \BibitemOpen
  \bibfield  {author} {\bibinfo {author} {\bibfnamefont {D.}~\bibnamefont
  {Clary}},\ }\href@noop {} {\bibfield  {journal} {\bibinfo  {journal} {Annual
  Review of Physical Chemistry}\ }\textbf {\bibinfo {volume} {41}},\ \bibinfo
  {pages} {61} (\bibinfo {year} {1990})}\BibitemShut {NoStop}%
\bibitem [{\citenamefont {Frankcombe}\ and\ \citenamefont
  {Nyman}(2007)}]{frankcombe2007adiabatic}%
  \BibitemOpen
  \bibfield  {author} {\bibinfo {author} {\bibfnamefont {T.~J.}\ \bibnamefont
  {Frankcombe}}\ and\ \bibinfo {author} {\bibfnamefont {G.}~\bibnamefont
  {Nyman}},\ }\href@noop {} {\bibfield  {journal} {\bibinfo  {journal} {The
  Journal of Physical Chemistry A}\ }\textbf {\bibinfo {volume} {111}},\
  \bibinfo {pages} {13163} (\bibinfo {year} {2007})}\BibitemShut {NoStop}%
\bibitem [{\citenamefont {Tscherbul}\ and\ \citenamefont
  {Buchachenko}(2015)}]{tscherbul2015adiabatic}%
  \BibitemOpen
  \bibfield  {author} {\bibinfo {author} {\bibfnamefont {T.~V.}\ \bibnamefont
  {Tscherbul}}\ and\ \bibinfo {author} {\bibfnamefont {A.~A.}\ \bibnamefont
  {Buchachenko}},\ }\href@noop {} {\bibfield  {journal} {\bibinfo  {journal}
  {New Journal of Physics}\ }\textbf {\bibinfo {volume} {17}},\ \bibinfo
  {pages} {035010} (\bibinfo {year} {2015})}\BibitemShut {NoStop}%
\bibitem [{\citenamefont {Singh}\ \emph {et~al.}(2012)\citenamefont {Singh},
  \citenamefont {Hardman}, \citenamefont {Tariq}, \citenamefont {Lu},
  \citenamefont {Ellis}, \citenamefont {Morrison},\ and\ \citenamefont
  {Weinstein}}]{singh2012chemical}%
  \BibitemOpen
  \bibfield  {author} {\bibinfo {author} {\bibfnamefont {V.}~\bibnamefont
  {Singh}}, \bibinfo {author} {\bibfnamefont {K.~S.}\ \bibnamefont {Hardman}},
  \bibinfo {author} {\bibfnamefont {N.}~\bibnamefont {Tariq}}, \bibinfo
  {author} {\bibfnamefont {M.-J.}\ \bibnamefont {Lu}}, \bibinfo {author}
  {\bibfnamefont {A.}~\bibnamefont {Ellis}}, \bibinfo {author} {\bibfnamefont
  {M.~J.}\ \bibnamefont {Morrison}}, \ and\ \bibinfo {author} {\bibfnamefont
  {J.~D.}\ \bibnamefont {Weinstein}},\ }\href@noop {} {\bibfield  {journal}
  {\bibinfo  {journal} {Physical review letters}\ }\textbf {\bibinfo {volume}
  {108}},\ \bibinfo {pages} {203201} (\bibinfo {year} {2012})}\BibitemShut
  {NoStop}%
\bibitem [{\citenamefont {Pawlak}\ \emph {et~al.}(2015)\citenamefont {Pawlak},
  \citenamefont {Shagam}, \citenamefont {Narevicius},\ and\ \citenamefont
  {Moiseyev}}]{pawlak2015adiabatic}%
  \BibitemOpen
  \bibfield  {author} {\bibinfo {author} {\bibfnamefont {M.}~\bibnamefont
  {Pawlak}}, \bibinfo {author} {\bibfnamefont {Y.}~\bibnamefont {Shagam}},
  \bibinfo {author} {\bibfnamefont {E.}~\bibnamefont {Narevicius}}, \ and\
  \bibinfo {author} {\bibfnamefont {N.}~\bibnamefont {Moiseyev}},\ }\href@noop
  {} {\bibfield  {journal} {\bibinfo  {journal} {The Journal of chemical
  physics}\ }\textbf {\bibinfo {volume} {143}},\ \bibinfo {pages} {074114}
  (\bibinfo {year} {2015})}\BibitemShut {NoStop}%
\bibitem [{\citenamefont {Pawlak}\ \emph {et~al.}(2017)\citenamefont {Pawlak},
  \citenamefont {Shagam}, \citenamefont {Klein}, \citenamefont {Narevicius},\
  and\ \citenamefont {Moiseyev}}]{pawlak2017adiabatic}%
  \BibitemOpen
  \bibfield  {author} {\bibinfo {author} {\bibfnamefont {M.}~\bibnamefont
  {Pawlak}}, \bibinfo {author} {\bibfnamefont {Y.}~\bibnamefont {Shagam}},
  \bibinfo {author} {\bibfnamefont {A.}~\bibnamefont {Klein}}, \bibinfo
  {author} {\bibfnamefont {E.}~\bibnamefont {Narevicius}}, \ and\ \bibinfo
  {author} {\bibfnamefont {N.}~\bibnamefont {Moiseyev}},\ }\href@noop {}
  {\bibfield  {journal} {\bibinfo  {journal} {The Journal of Physical Chemistry
  A}\ }\textbf {\bibinfo {volume} {121}},\ \bibinfo {pages} {2194} (\bibinfo
  {year} {2017})}\BibitemShut {NoStop}%
\bibitem [{\citenamefont {Bhattacharya}\ \emph {et~al.}(2017)\citenamefont
  {Bhattacharya}, \citenamefont {Ben-Asher}, \citenamefont {Haritan},
  \citenamefont {Pawlak}, \citenamefont {Landau},\ and\ \citenamefont
  {Moiseyev}}]{bhattacharya2017polyatomic}%
  \BibitemOpen
  \bibfield  {author} {\bibinfo {author} {\bibfnamefont {D.}~\bibnamefont
  {Bhattacharya}}, \bibinfo {author} {\bibfnamefont {A.}~\bibnamefont
  {Ben-Asher}}, \bibinfo {author} {\bibfnamefont {I.}~\bibnamefont {Haritan}},
  \bibinfo {author} {\bibfnamefont {M.}~\bibnamefont {Pawlak}}, \bibinfo
  {author} {\bibfnamefont {A.}~\bibnamefont {Landau}}, \ and\ \bibinfo {author}
  {\bibfnamefont {N.}~\bibnamefont {Moiseyev}},\ }\href@noop {} {\bibfield
  {journal} {\bibinfo  {journal} {Journal of chemical theory and computation}\
  }\textbf {\bibinfo {volume} {13}},\ \bibinfo {pages} {1682} (\bibinfo {year}
  {2017})}\BibitemShut {NoStop}%
\bibitem [{\citenamefont {Bhattacharya}\ \emph {et~al.}(2019)\citenamefont
  {Bhattacharya}, \citenamefont {Pawlak}, \citenamefont {Ben-Asher},
  \citenamefont {Landau}, \citenamefont {Haritan}, \citenamefont {Narevicius},\
  and\ \citenamefont {Moiseyev}}]{bhattacharya2019quantum}%
  \BibitemOpen
  \bibfield  {author} {\bibinfo {author} {\bibfnamefont {D.}~\bibnamefont
  {Bhattacharya}}, \bibinfo {author} {\bibfnamefont {M.}~\bibnamefont
  {Pawlak}}, \bibinfo {author} {\bibfnamefont {A.}~\bibnamefont {Ben-Asher}},
  \bibinfo {author} {\bibfnamefont {A.}~\bibnamefont {Landau}}, \bibinfo
  {author} {\bibfnamefont {I.}~\bibnamefont {Haritan}}, \bibinfo {author}
  {\bibfnamefont {E.}~\bibnamefont {Narevicius}}, \ and\ \bibinfo {author}
  {\bibfnamefont {N.}~\bibnamefont {Moiseyev}},\ }\href@noop {} {\bibfield
  {journal} {\bibinfo  {journal} {The journal of physical chemistry letters}\
  }\textbf {\bibinfo {volume} {10}},\ \bibinfo {pages} {855} (\bibinfo {year}
  {2019})}\BibitemShut {NoStop}%
\bibitem [{\citenamefont {Pawlak}\ \emph {et~al.}(2019)\citenamefont {Pawlak},
  \citenamefont {{\.Z}uchowski}, \citenamefont {Moiseyev},\ and\ \citenamefont
  {Jankowski}}]{pawlak2019nonrigidity}%
  \BibitemOpen
  \bibfield  {author} {\bibinfo {author} {\bibfnamefont {M.}~\bibnamefont
  {Pawlak}}, \bibinfo {author} {\bibfnamefont {P.~S.}\ \bibnamefont
  {{\.Z}uchowski}}, \bibinfo {author} {\bibfnamefont {N.}~\bibnamefont
  {Moiseyev}}, \ and\ \bibinfo {author} {\bibfnamefont {P.}~\bibnamefont
  {Jankowski}},\ }\href@noop {} {\bibfield  {journal} {\bibinfo  {journal}
  {arXiv preprint arXiv:1907.05130}\ } (\bibinfo {year} {2019})}\BibitemShut
  {NoStop}%
\bibitem [{\citenamefont {Krems}\ and\ \citenamefont
  {Dalgarno}(2004)}]{krems2004quantum}%
  \BibitemOpen
  \bibfield  {author} {\bibinfo {author} {\bibfnamefont {R.}~\bibnamefont
  {Krems}}\ and\ \bibinfo {author} {\bibfnamefont {A.}~\bibnamefont
  {Dalgarno}},\ }\href@noop {} {\bibfield  {journal} {\bibinfo  {journal} {The
  Journal of chemical physics}\ }\textbf {\bibinfo {volume} {120}},\ \bibinfo
  {pages} {2296} (\bibinfo {year} {2004})}\BibitemShut {NoStop}%
\bibitem [{\citenamefont {Tscherbul}\ \emph {et~al.}(2009)\citenamefont
  {Tscherbul}, \citenamefont {Suleimanov}, \citenamefont {Aquilanti},\ and\
  \citenamefont {Krems}}]{tscherbul2009magnetic}%
  \BibitemOpen
  \bibfield  {author} {\bibinfo {author} {\bibfnamefont {T.}~\bibnamefont
  {Tscherbul}}, \bibinfo {author} {\bibfnamefont {Y.~V.}\ \bibnamefont
  {Suleimanov}}, \bibinfo {author} {\bibfnamefont {V.}~\bibnamefont
  {Aquilanti}}, \ and\ \bibinfo {author} {\bibfnamefont {R.}~\bibnamefont
  {Krems}},\ }\href@noop {} {\bibfield  {journal} {\bibinfo  {journal} {New
  Journal of Physics}\ }\textbf {\bibinfo {volume} {11}},\ \bibinfo {pages}
  {055021} (\bibinfo {year} {2009})}\BibitemShut {NoStop}%
\bibitem [{\citenamefont {Tscherbul}\ and\ \citenamefont
  {Dalgarno}(2010)}]{tscherbul2010quantum}%
  \BibitemOpen
  \bibfield  {author} {\bibinfo {author} {\bibfnamefont {T.~V.}\ \bibnamefont
  {Tscherbul}}\ and\ \bibinfo {author} {\bibfnamefont {A.}~\bibnamefont
  {Dalgarno}},\ }\href@noop {} {\bibfield  {journal} {\bibinfo  {journal} {The
  Journal of chemical physics}\ }\textbf {\bibinfo {volume} {133}},\ \bibinfo
  {pages} {184104} (\bibinfo {year} {2010})}\BibitemShut {NoStop}%
\bibitem [{\citenamefont {Tscherbul}(2012)}]{tscherbul2012total}%
  \BibitemOpen
  \bibfield  {author} {\bibinfo {author} {\bibfnamefont {T.}~\bibnamefont
  {Tscherbul}},\ }\href@noop {} {\bibfield  {journal} {\bibinfo  {journal}
  {Physical Review A}\ }\textbf {\bibinfo {volume} {85}},\ \bibinfo {pages}
  {052710} (\bibinfo {year} {2012})}\BibitemShut {NoStop}%
\bibitem [{\citenamefont {Janssen}\ \emph
  {et~al.}(2011{\natexlab{a}})\citenamefont {Janssen}, \citenamefont
  {{\.Z}uchowski}, \citenamefont {van~der Avoird}, \citenamefont
  {Groenenboom},\ and\ \citenamefont {Hutson}}]{janssen2011cold2}%
  \BibitemOpen
  \bibfield  {author} {\bibinfo {author} {\bibfnamefont {L.~M.}\ \bibnamefont
  {Janssen}}, \bibinfo {author} {\bibfnamefont {P.~S.}\ \bibnamefont
  {{\.Z}uchowski}}, \bibinfo {author} {\bibfnamefont {A.}~\bibnamefont {van~der
  Avoird}}, \bibinfo {author} {\bibfnamefont {G.~C.}\ \bibnamefont
  {Groenenboom}}, \ and\ \bibinfo {author} {\bibfnamefont {J.~M.}\ \bibnamefont
  {Hutson}},\ }\href@noop {} {\bibfield  {journal} {\bibinfo  {journal}
  {Physical Review A}\ }\textbf {\bibinfo {volume} {83}},\ \bibinfo {pages}
  {022713} (\bibinfo {year} {2011}{\natexlab{a}})}\BibitemShut {NoStop}%
\bibitem [{\citenamefont {Krems}(2005)}]{krems2005molecules}%
  \BibitemOpen
  \bibfield  {author} {\bibinfo {author} {\bibfnamefont {R.~V.}\ \bibnamefont
  {Krems}},\ }\href@noop {} {\bibfield  {journal} {\bibinfo  {journal}
  {International Reviews in Physical Chemistry}\ }\textbf {\bibinfo {volume}
  {24}},\ \bibinfo {pages} {99} (\bibinfo {year} {2005})}\BibitemShut {NoStop}%
\bibitem [{\citenamefont {Krems}(2018)}]{krems2018molecules}%
  \BibitemOpen
  \bibfield  {author} {\bibinfo {author} {\bibfnamefont {R.~V.}\ \bibnamefont
  {Krems}},\ }\href@noop {} {\emph {\bibinfo {title} {Molecules in
  electromagnetic fields: from ultracold physics to controlled chemistry}}}\
  (\bibinfo  {publisher} {John Wiley \& Sons},\ \bibinfo {year}
  {2018})\BibitemShut {NoStop}%
\bibitem [{\citenamefont {Bohn}(2001)}]{bohn2001inelastic}%
  \BibitemOpen
  \bibfield  {author} {\bibinfo {author} {\bibfnamefont {J.~L.}\ \bibnamefont
  {Bohn}},\ }\href@noop {} {\bibfield  {journal} {\bibinfo  {journal} {Physical
  Review A}\ }\textbf {\bibinfo {volume} {63}},\ \bibinfo {pages} {052714}
  (\bibinfo {year} {2001})}\BibitemShut {NoStop}%
\bibitem [{\citenamefont {Avdeenkov}\ and\ \citenamefont
  {Bohn}(2002)}]{avdeenkov2002collisional}%
  \BibitemOpen
  \bibfield  {author} {\bibinfo {author} {\bibfnamefont {A.~V.}\ \bibnamefont
  {Avdeenkov}}\ and\ \bibinfo {author} {\bibfnamefont {J.~L.}\ \bibnamefont
  {Bohn}},\ }\href@noop {} {\bibfield  {journal} {\bibinfo  {journal} {Physical
  Review A}\ }\textbf {\bibinfo {volume} {66}},\ \bibinfo {pages} {052718}
  (\bibinfo {year} {2002})}\BibitemShut {NoStop}%
\bibitem [{\citenamefont {Ticknor}\ and\ \citenamefont
  {Bohn}(2005)}]{ticknor2005influence}%
  \BibitemOpen
  \bibfield  {author} {\bibinfo {author} {\bibfnamefont {C.}~\bibnamefont
  {Ticknor}}\ and\ \bibinfo {author} {\bibfnamefont {J.~L.}\ \bibnamefont
  {Bohn}},\ }\href@noop {} {\bibfield  {journal} {\bibinfo  {journal} {Physical
  Review A}\ }\textbf {\bibinfo {volume} {71}},\ \bibinfo {pages} {022709}
  (\bibinfo {year} {2005})}\BibitemShut {NoStop}%
\bibitem [{\citenamefont {Avdeenkov}\ and\ \citenamefont
  {Bohn}(2001)}]{avdeenkov2001ultracold}%
  \BibitemOpen
  \bibfield  {author} {\bibinfo {author} {\bibfnamefont {A.~V.}\ \bibnamefont
  {Avdeenkov}}\ and\ \bibinfo {author} {\bibfnamefont {J.~L.}\ \bibnamefont
  {Bohn}},\ }\href@noop {} {\bibfield  {journal} {\bibinfo  {journal} {Physical
  Review A}\ }\textbf {\bibinfo {volume} {64}},\ \bibinfo {pages} {052703}
  (\bibinfo {year} {2001})}\BibitemShut {NoStop}%
\bibitem [{\citenamefont {Hapka}\ \emph {et~al.}(2013)\citenamefont {Hapka},
  \citenamefont {Cha{\l}asi{\'n}ski}, \citenamefont {K{\l}os},\ and\
  \citenamefont {{\.Z}uchowski}}]{hapka2013first}%
  \BibitemOpen
  \bibfield  {author} {\bibinfo {author} {\bibfnamefont {M.}~\bibnamefont
  {Hapka}}, \bibinfo {author} {\bibfnamefont {G.}~\bibnamefont
  {Cha{\l}asi{\'n}ski}}, \bibinfo {author} {\bibfnamefont {J.}~\bibnamefont
  {K{\l}os}}, \ and\ \bibinfo {author} {\bibfnamefont {P.~S.}\ \bibnamefont
  {{\.Z}uchowski}},\ }\href@noop {} {\bibfield  {journal} {\bibinfo  {journal}
  {The Journal of chemical physics}\ }\textbf {\bibinfo {volume} {139}},\
  \bibinfo {pages} {014307} (\bibinfo {year} {2013})}\BibitemShut {NoStop}%
\bibitem [{\citenamefont {Henson}\ \emph {et~al.}(2012)\citenamefont {Henson},
  \citenamefont {Gersten}, \citenamefont {Shagam}, \citenamefont {Narevicius},\
  and\ \citenamefont {Narevicius}}]{henson2012observation}%
  \BibitemOpen
  \bibfield  {author} {\bibinfo {author} {\bibfnamefont {A.~B.}\ \bibnamefont
  {Henson}}, \bibinfo {author} {\bibfnamefont {S.}~\bibnamefont {Gersten}},
  \bibinfo {author} {\bibfnamefont {Y.}~\bibnamefont {Shagam}}, \bibinfo
  {author} {\bibfnamefont {J.}~\bibnamefont {Narevicius}}, \ and\ \bibinfo
  {author} {\bibfnamefont {E.}~\bibnamefont {Narevicius}},\ }\href@noop {}
  {\bibfield  {journal} {\bibinfo  {journal} {Science}\ }\textbf {\bibinfo
  {volume} {338}},\ \bibinfo {pages} {234} (\bibinfo {year}
  {2012})}\BibitemShut {NoStop}%
\bibitem [{\citenamefont {Balakrishnan}\ \emph {et~al.}(2003)\citenamefont
  {Balakrishnan}, \citenamefont {Groenenboom}, \citenamefont {Krems},\ and\
  \citenamefont {Dalgarno}}]{balakrishnan2003he}%
  \BibitemOpen
  \bibfield  {author} {\bibinfo {author} {\bibfnamefont {N.}~\bibnamefont
  {Balakrishnan}}, \bibinfo {author} {\bibfnamefont {G.~C.}\ \bibnamefont
  {Groenenboom}}, \bibinfo {author} {\bibfnamefont {R.}~\bibnamefont {Krems}},
  \ and\ \bibinfo {author} {\bibfnamefont {A.}~\bibnamefont {Dalgarno}},\
  }\href@noop {} {\bibfield  {journal} {\bibinfo  {journal} {The Journal of
  chemical physics}\ }\textbf {\bibinfo {volume} {118}},\ \bibinfo {pages}
  {7386} (\bibinfo {year} {2003})}\BibitemShut {NoStop}%
\bibitem [{\citenamefont {Krems}\ \emph
  {et~al.}(2003{\natexlab{a}})\citenamefont {Krems}, \citenamefont {Dalgarno},
  \citenamefont {Balakrishnan},\ and\ \citenamefont
  {Groenenboom}}]{krems2003spin}%
  \BibitemOpen
  \bibfield  {author} {\bibinfo {author} {\bibfnamefont {R.}~\bibnamefont
  {Krems}}, \bibinfo {author} {\bibfnamefont {A.}~\bibnamefont {Dalgarno}},
  \bibinfo {author} {\bibfnamefont {N.}~\bibnamefont {Balakrishnan}}, \ and\
  \bibinfo {author} {\bibfnamefont {G.}~\bibnamefont {Groenenboom}},\
  }\href@noop {} {\bibfield  {journal} {\bibinfo  {journal} {Physical Review
  A}\ }\textbf {\bibinfo {volume} {67}},\ \bibinfo {pages} {060703} (\bibinfo
  {year} {2003}{\natexlab{a}})}\BibitemShut {NoStop}%
\bibitem [{\citenamefont {Krems}\ \emph
  {et~al.}(2003{\natexlab{b}})\citenamefont {Krems}, \citenamefont
  {Sadeghpour}, \citenamefont {Dalgarno}, \citenamefont {Zgid}, \citenamefont
  {K{\l}os},\ and\ \citenamefont {Cha{\l}asi{\'n}ski}}]{krems2003low}%
  \BibitemOpen
  \bibfield  {author} {\bibinfo {author} {\bibfnamefont {R.}~\bibnamefont
  {Krems}}, \bibinfo {author} {\bibfnamefont {H.}~\bibnamefont {Sadeghpour}},
  \bibinfo {author} {\bibfnamefont {A.}~\bibnamefont {Dalgarno}}, \bibinfo
  {author} {\bibfnamefont {D.}~\bibnamefont {Zgid}}, \bibinfo {author}
  {\bibfnamefont {J.}~\bibnamefont {K{\l}os}}, \ and\ \bibinfo {author}
  {\bibfnamefont {G.}~\bibnamefont {Cha{\l}asi{\'n}ski}},\ }\href@noop {}
  {\bibfield  {journal} {\bibinfo  {journal} {Physical Review A}\ }\textbf
  {\bibinfo {volume} {68}},\ \bibinfo {pages} {051401} (\bibinfo {year}
  {2003}{\natexlab{b}})}\BibitemShut {NoStop}%
\bibitem [{\citenamefont {Cybulski}\ \emph {et~al.}(2005)\citenamefont
  {Cybulski}, \citenamefont {Krems}, \citenamefont {Sadeghpour}, \citenamefont
  {Dalgarno}, \citenamefont {K{\l}os}, \citenamefont {Groenenboom},
  \citenamefont {van~der Avoird}, \citenamefont {Zgid},\ and\ \citenamefont
  {Cha{\l}asi{\'n}ski}}]{cybulski2005interaction}%
  \BibitemOpen
  \bibfield  {author} {\bibinfo {author} {\bibfnamefont {H.}~\bibnamefont
  {Cybulski}}, \bibinfo {author} {\bibfnamefont {R.}~\bibnamefont {Krems}},
  \bibinfo {author} {\bibfnamefont {H.}~\bibnamefont {Sadeghpour}}, \bibinfo
  {author} {\bibfnamefont {A.}~\bibnamefont {Dalgarno}}, \bibinfo {author}
  {\bibfnamefont {J.}~\bibnamefont {K{\l}os}}, \bibinfo {author} {\bibfnamefont
  {G.}~\bibnamefont {Groenenboom}}, \bibinfo {author} {\bibfnamefont
  {A.}~\bibnamefont {van~der Avoird}}, \bibinfo {author} {\bibfnamefont
  {D.}~\bibnamefont {Zgid}}, \ and\ \bibinfo {author} {\bibfnamefont
  {G.}~\bibnamefont {Cha{\l}asi{\'n}ski}},\ }\href@noop {} {\bibfield
  {journal} {\bibinfo  {journal} {The Journal of chemical physics}\ }\textbf
  {\bibinfo {volume} {122}},\ \bibinfo {pages} {094307} (\bibinfo {year}
  {2005})}\BibitemShut {NoStop}%
\bibitem [{\citenamefont {Campbell}\ \emph {et~al.}(2009)\citenamefont
  {Campbell}, \citenamefont {Tscherbul}, \citenamefont {Lu}, \citenamefont
  {Tsikata}, \citenamefont {Krems},\ and\ \citenamefont
  {Doyle}}]{campbell2009mechanism}%
  \BibitemOpen
  \bibfield  {author} {\bibinfo {author} {\bibfnamefont {W.~C.}\ \bibnamefont
  {Campbell}}, \bibinfo {author} {\bibfnamefont {T.~V.}\ \bibnamefont
  {Tscherbul}}, \bibinfo {author} {\bibfnamefont {H.-I.}\ \bibnamefont {Lu}},
  \bibinfo {author} {\bibfnamefont {E.}~\bibnamefont {Tsikata}}, \bibinfo
  {author} {\bibfnamefont {R.~V.}\ \bibnamefont {Krems}}, \ and\ \bibinfo
  {author} {\bibfnamefont {J.~M.}\ \bibnamefont {Doyle}},\ }\href@noop {}
  {\bibfield  {journal} {\bibinfo  {journal} {Physical review letters}\
  }\textbf {\bibinfo {volume} {102}},\ \bibinfo {pages} {013003} (\bibinfo
  {year} {2009})}\BibitemShut {NoStop}%
\bibitem [{\citenamefont {Bethe}(1935)}]{bethe1935theory}%
  \BibitemOpen
  \bibfield  {author} {\bibinfo {author} {\bibfnamefont {H.~A.}\ \bibnamefont
  {Bethe}},\ }\href@noop {} {\bibfield  {journal} {\bibinfo  {journal}
  {Physical Review}\ }\textbf {\bibinfo {volume} {47}},\ \bibinfo {pages} {747}
  (\bibinfo {year} {1935})}\BibitemShut {NoStop}%
\bibitem [{\citenamefont {Wigner}(1948)}]{wigner1948behavior}%
  \BibitemOpen
  \bibfield  {author} {\bibinfo {author} {\bibfnamefont {E.~P.}\ \bibnamefont
  {Wigner}},\ }\href@noop {} {\bibfield  {journal} {\bibinfo  {journal}
  {Physical Review}\ }\textbf {\bibinfo {volume} {73}},\ \bibinfo {pages}
  {1002} (\bibinfo {year} {1948})}\BibitemShut {NoStop}%
\bibitem [{\citenamefont {Langevin}(1905)}]{langevin1905formule}%
  \BibitemOpen
  \bibfield  {author} {\bibinfo {author} {\bibfnamefont {M.}~\bibnamefont
  {Langevin}},\ }in\ \href@noop {} {\emph {\bibinfo {booktitle} {Annales de
  chimie et de physique, Series}}},\ Vol.~\bibinfo {volume} {5}\ (\bibinfo
  {year} {1905})\ pp.\ \bibinfo {pages} {245--288}\BibitemShut {NoStop}%
\bibitem [{\citenamefont {Sold{\'a}n}\ \emph {et~al.}(2002)\citenamefont
  {Sold{\'a}n}, \citenamefont {Cvita{\v{s}}}, \citenamefont {Hutson},
  \citenamefont {Honvault},\ and\ \citenamefont {Launay}}]{soldan2002quantum}%
  \BibitemOpen
  \bibfield  {author} {\bibinfo {author} {\bibfnamefont {P.}~\bibnamefont
  {Sold{\'a}n}}, \bibinfo {author} {\bibfnamefont {M.~T.}\ \bibnamefont
  {Cvita{\v{s}}}}, \bibinfo {author} {\bibfnamefont {J.~M.}\ \bibnamefont
  {Hutson}}, \bibinfo {author} {\bibfnamefont {P.}~\bibnamefont {Honvault}}, \
  and\ \bibinfo {author} {\bibfnamefont {J.-M.}\ \bibnamefont {Launay}},\
  }\href@noop {} {\bibfield  {journal} {\bibinfo  {journal} {Physical review
  letters}\ }\textbf {\bibinfo {volume} {89}},\ \bibinfo {pages} {153201}
  (\bibinfo {year} {2002})}\BibitemShut {NoStop}%
\bibitem [{\citenamefont {Qu{\'e}m{\'e}ner}\ \emph {et~al.}(2004)\citenamefont
  {Qu{\'e}m{\'e}ner}, \citenamefont {Honvault},\ and\ \citenamefont
  {Launay}}]{quemener2004sensitivity}%
  \BibitemOpen
  \bibfield  {author} {\bibinfo {author} {\bibfnamefont {G.}~\bibnamefont
  {Qu{\'e}m{\'e}ner}}, \bibinfo {author} {\bibfnamefont {P.}~\bibnamefont
  {Honvault}}, \ and\ \bibinfo {author} {\bibfnamefont {J.-M.}\ \bibnamefont
  {Launay}},\ }\href@noop {} {\bibfield  {journal} {\bibinfo  {journal} {The
  European Physical Journal D-Atomic, Molecular, Optical and Plasma Physics}\
  }\textbf {\bibinfo {volume} {30}},\ \bibinfo {pages} {201} (\bibinfo {year}
  {2004})}\BibitemShut {NoStop}%
\bibitem [{\citenamefont {Cvita{\v{s}}}\ \emph
  {et~al.}(2005{\natexlab{a}})\citenamefont {Cvita{\v{s}}}, \citenamefont
  {Sold{\'a}n}, \citenamefont {Hutson}, \citenamefont {Honvault},\ and\
  \citenamefont {Launay}}]{cvitavs2005ultracold}%
  \BibitemOpen
  \bibfield  {author} {\bibinfo {author} {\bibfnamefont {M.~T.}\ \bibnamefont
  {Cvita{\v{s}}}}, \bibinfo {author} {\bibfnamefont {P.}~\bibnamefont
  {Sold{\'a}n}}, \bibinfo {author} {\bibfnamefont {J.~M.}\ \bibnamefont
  {Hutson}}, \bibinfo {author} {\bibfnamefont {P.}~\bibnamefont {Honvault}}, \
  and\ \bibinfo {author} {\bibfnamefont {J.-M.}\ \bibnamefont {Launay}},\
  }\href@noop {} {\bibfield  {journal} {\bibinfo  {journal} {Physical review
  letters}\ }\textbf {\bibinfo {volume} {94}},\ \bibinfo {pages} {200402}
  (\bibinfo {year} {2005}{\natexlab{a}})}\BibitemShut {NoStop}%
\bibitem [{\citenamefont {Cvita{\v{s}}}\ \emph
  {et~al.}(2005{\natexlab{b}})\citenamefont {Cvita{\v{s}}}, \citenamefont
  {Sold{\'a}n}, \citenamefont {Hutson}, \citenamefont {Honvault},\ and\
  \citenamefont {Launay}}]{cvitavs2005ultracold2}%
  \BibitemOpen
  \bibfield  {author} {\bibinfo {author} {\bibfnamefont {M.~T.}\ \bibnamefont
  {Cvita{\v{s}}}}, \bibinfo {author} {\bibfnamefont {P.}~\bibnamefont
  {Sold{\'a}n}}, \bibinfo {author} {\bibfnamefont {J.~M.}\ \bibnamefont
  {Hutson}}, \bibinfo {author} {\bibfnamefont {P.}~\bibnamefont {Honvault}}, \
  and\ \bibinfo {author} {\bibfnamefont {J.-M.}\ \bibnamefont {Launay}},\
  }\href@noop {} {\bibfield  {journal} {\bibinfo  {journal} {Physical review
  letters}\ }\textbf {\bibinfo {volume} {94}},\ \bibinfo {pages} {033201}
  (\bibinfo {year} {2005}{\natexlab{b}})}\BibitemShut {NoStop}%
\bibitem [{\citenamefont {Cvita{\v{s}}}\ \emph {et~al.}(2007)\citenamefont
  {Cvita{\v{s}}}, \citenamefont {Sold{\'a}n}, \citenamefont {Hutson},
  \citenamefont {Honvault},\ and\ \citenamefont
  {Launay}}]{cvitavs2007interactions}%
  \BibitemOpen
  \bibfield  {author} {\bibinfo {author} {\bibfnamefont {M.~T.}\ \bibnamefont
  {Cvita{\v{s}}}}, \bibinfo {author} {\bibfnamefont {P.}~\bibnamefont
  {Sold{\'a}n}}, \bibinfo {author} {\bibfnamefont {J.~M.}\ \bibnamefont
  {Hutson}}, \bibinfo {author} {\bibfnamefont {P.}~\bibnamefont {Honvault}}, \
  and\ \bibinfo {author} {\bibfnamefont {J.-M.}\ \bibnamefont {Launay}},\
  }\href@noop {} {\bibfield  {journal} {\bibinfo  {journal} {The Journal of
  chemical physics}\ }\textbf {\bibinfo {volume} {127}},\ \bibinfo {pages}
  {074302} (\bibinfo {year} {2007})}\BibitemShut {NoStop}%
\bibitem [{\citenamefont {Qu{\'e}m{\'e}ner}\ \emph {et~al.}(2007)\citenamefont
  {Qu{\'e}m{\'e}ner}, \citenamefont {Launay},\ and\ \citenamefont
  {Honvault}}]{quemener2007ultracold}%
  \BibitemOpen
  \bibfield  {author} {\bibinfo {author} {\bibfnamefont {G.}~\bibnamefont
  {Qu{\'e}m{\'e}ner}}, \bibinfo {author} {\bibfnamefont {J.-M.}\ \bibnamefont
  {Launay}}, \ and\ \bibinfo {author} {\bibfnamefont {P.}~\bibnamefont
  {Honvault}},\ }\href@noop {} {\bibfield  {journal} {\bibinfo  {journal}
  {Physical Review A}\ }\textbf {\bibinfo {volume} {75}},\ \bibinfo {pages}
  {050701} (\bibinfo {year} {2007})}\BibitemShut {NoStop}%
\bibitem [{\citenamefont {Qu{\'e}m{\'e}ner}\ and\ \citenamefont
  {Bohn}(2010)}]{quemener2010strong}%
  \BibitemOpen
  \bibfield  {author} {\bibinfo {author} {\bibfnamefont {G.}~\bibnamefont
  {Qu{\'e}m{\'e}ner}}\ and\ \bibinfo {author} {\bibfnamefont {J.~L.}\
  \bibnamefont {Bohn}},\ }\href@noop {} {\bibfield  {journal} {\bibinfo
  {journal} {Physical Review A}\ }\textbf {\bibinfo {volume} {81}},\ \bibinfo
  {pages} {022702} (\bibinfo {year} {2010})}\BibitemShut {NoStop}%
\bibitem [{\citenamefont {Qu{\'e}m{\'e}ner}\ \emph {et~al.}(2011)\citenamefont
  {Qu{\'e}m{\'e}ner}, \citenamefont {Bohn}, \citenamefont {Petrov},\ and\
  \citenamefont {Kotochigova}}]{quemener2011universalities}%
  \BibitemOpen
  \bibfield  {author} {\bibinfo {author} {\bibfnamefont {G.}~\bibnamefont
  {Qu{\'e}m{\'e}ner}}, \bibinfo {author} {\bibfnamefont {J.~L.}\ \bibnamefont
  {Bohn}}, \bibinfo {author} {\bibfnamefont {A.}~\bibnamefont {Petrov}}, \ and\
  \bibinfo {author} {\bibfnamefont {S.}~\bibnamefont {Kotochigova}},\
  }\href@noop {} {\bibfield  {journal} {\bibinfo  {journal} {Physical Review
  A}\ }\textbf {\bibinfo {volume} {84}},\ \bibinfo {pages} {062703} (\bibinfo
  {year} {2011})}\BibitemShut {NoStop}%
\bibitem [{\citenamefont {Julienne}\ \emph {et~al.}(2011)\citenamefont
  {Julienne}, \citenamefont {Hanna},\ and\ \citenamefont
  {Idziaszek}}]{julienne2011universal}%
  \BibitemOpen
  \bibfield  {author} {\bibinfo {author} {\bibfnamefont {P.~S.}\ \bibnamefont
  {Julienne}}, \bibinfo {author} {\bibfnamefont {T.~M.}\ \bibnamefont {Hanna}},
  \ and\ \bibinfo {author} {\bibfnamefont {Z.}~\bibnamefont {Idziaszek}},\
  }\href@noop {} {\bibfield  {journal} {\bibinfo  {journal} {Physical Chemistry
  Chemical Physics}\ }\textbf {\bibinfo {volume} {13}},\ \bibinfo {pages}
  {19114} (\bibinfo {year} {2011})}\BibitemShut {NoStop}%
\bibitem [{\citenamefont {Julienne}\ and\ \citenamefont
  {Mies}(1989)}]{julienne1989collisions}%
  \BibitemOpen
  \bibfield  {author} {\bibinfo {author} {\bibfnamefont {P.~S.}\ \bibnamefont
  {Julienne}}\ and\ \bibinfo {author} {\bibfnamefont {F.~H.}\ \bibnamefont
  {Mies}},\ }\href@noop {} {\bibfield  {journal} {\bibinfo  {journal} {JOSA B}\
  }\textbf {\bibinfo {volume} {6}},\ \bibinfo {pages} {2257} (\bibinfo {year}
  {1989})}\BibitemShut {NoStop}%
\bibitem [{\citenamefont {Burke~Jr}\ \emph {et~al.}(1998)\citenamefont
  {Burke~Jr}, \citenamefont {Greene},\ and\ \citenamefont
  {Bohn}}]{burke1998multichannel}%
  \BibitemOpen
  \bibfield  {author} {\bibinfo {author} {\bibfnamefont {J.~P.}\ \bibnamefont
  {Burke~Jr}}, \bibinfo {author} {\bibfnamefont {C.~H.}\ \bibnamefont
  {Greene}}, \ and\ \bibinfo {author} {\bibfnamefont {J.~L.}\ \bibnamefont
  {Bohn}},\ }\href@noop {} {\bibfield  {journal} {\bibinfo  {journal} {Physical
  review letters}\ }\textbf {\bibinfo {volume} {81}},\ \bibinfo {pages} {3355}
  (\bibinfo {year} {1998})}\BibitemShut {NoStop}%
\bibitem [{\citenamefont {Gao}\ \emph {et~al.}(2005)\citenamefont {Gao},
  \citenamefont {Tiesinga}, \citenamefont {Williams},\ and\ \citenamefont
  {Julienne}}]{gao2005multichannel}%
  \BibitemOpen
  \bibfield  {author} {\bibinfo {author} {\bibfnamefont {B.}~\bibnamefont
  {Gao}}, \bibinfo {author} {\bibfnamefont {E.}~\bibnamefont {Tiesinga}},
  \bibinfo {author} {\bibfnamefont {C.~J.}\ \bibnamefont {Williams}}, \ and\
  \bibinfo {author} {\bibfnamefont {P.~S.}\ \bibnamefont {Julienne}},\
  }\href@noop {} {\bibfield  {journal} {\bibinfo  {journal} {Physical Review
  A}\ }\textbf {\bibinfo {volume} {72}},\ \bibinfo {pages} {042719} (\bibinfo
  {year} {2005})}\BibitemShut {NoStop}%
\bibitem [{\citenamefont {Idziaszek}\ and\ \citenamefont
  {Julienne}(2010)}]{idziaszek2010universal}%
  \BibitemOpen
  \bibfield  {author} {\bibinfo {author} {\bibfnamefont {Z.}~\bibnamefont
  {Idziaszek}}\ and\ \bibinfo {author} {\bibfnamefont {P.~S.}\ \bibnamefont
  {Julienne}},\ }\href@noop {} {\bibfield  {journal} {\bibinfo  {journal}
  {Physical review letters}\ }\textbf {\bibinfo {volume} {104}},\ \bibinfo
  {pages} {113202} (\bibinfo {year} {2010})}\BibitemShut {NoStop}%
\bibitem [{\citenamefont {Jachymski}\ \emph {et~al.}(2013)\citenamefont
  {Jachymski}, \citenamefont {Krych}, \citenamefont {Julienne},\ and\
  \citenamefont {Idziaszek}}]{jachymski2013quantum}%
  \BibitemOpen
  \bibfield  {author} {\bibinfo {author} {\bibfnamefont {K.}~\bibnamefont
  {Jachymski}}, \bibinfo {author} {\bibfnamefont {M.}~\bibnamefont {Krych}},
  \bibinfo {author} {\bibfnamefont {P.~S.}\ \bibnamefont {Julienne}}, \ and\
  \bibinfo {author} {\bibfnamefont {Z.}~\bibnamefont {Idziaszek}},\ }\href@noop
  {} {\bibfield  {journal} {\bibinfo  {journal} {Physical review letters}\
  }\textbf {\bibinfo {volume} {110}},\ \bibinfo {pages} {213202} (\bibinfo
  {year} {2013})}\BibitemShut {NoStop}%
\bibitem [{\citenamefont {Julienne}(2009)}]{julienne2009ultracold}%
  \BibitemOpen
  \bibfield  {author} {\bibinfo {author} {\bibfnamefont {P.~S.}\ \bibnamefont
  {Julienne}},\ }\href@noop {} {\bibfield  {journal} {\bibinfo  {journal}
  {Faraday discussions}\ }\textbf {\bibinfo {volume} {142}},\ \bibinfo {pages}
  {361} (\bibinfo {year} {2009})}\BibitemShut {NoStop}%
\bibitem [{\citenamefont {Gao}(2010)}]{gao2010universal}%
  \BibitemOpen
  \bibfield  {author} {\bibinfo {author} {\bibfnamefont {B.}~\bibnamefont
  {Gao}},\ }\href@noop {} {\bibfield  {journal} {\bibinfo  {journal} {Physical
  review letters}\ }\textbf {\bibinfo {volume} {105}},\ \bibinfo {pages}
  {263203} (\bibinfo {year} {2010})}\BibitemShut {NoStop}%
\bibitem [{\citenamefont {{\.Z}uchowski}\ and\ \citenamefont
  {Hutson}(2010)}]{zuchowski2010reactions}%
  \BibitemOpen
  \bibfield  {author} {\bibinfo {author} {\bibfnamefont {P.~S.}\ \bibnamefont
  {{\.Z}uchowski}}\ and\ \bibinfo {author} {\bibfnamefont {J.~M.}\ \bibnamefont
  {Hutson}},\ }\href@noop {} {\bibfield  {journal} {\bibinfo  {journal}
  {Physical Review A}\ }\textbf {\bibinfo {volume} {81}},\ \bibinfo {pages}
  {060703} (\bibinfo {year} {2010})}\BibitemShut {NoStop}%
\bibitem [{\citenamefont {De~Marco}\ \emph {et~al.}(2019)\citenamefont
  {De~Marco}, \citenamefont {Valtolina}, \citenamefont {Matsuda}, \citenamefont
  {Tobias}, \citenamefont {Covey},\ and\ \citenamefont
  {Ye}}]{de2019degenerate}%
  \BibitemOpen
  \bibfield  {author} {\bibinfo {author} {\bibfnamefont {L.}~\bibnamefont
  {De~Marco}}, \bibinfo {author} {\bibfnamefont {G.}~\bibnamefont {Valtolina}},
  \bibinfo {author} {\bibfnamefont {K.}~\bibnamefont {Matsuda}}, \bibinfo
  {author} {\bibfnamefont {W.~G.}\ \bibnamefont {Tobias}}, \bibinfo {author}
  {\bibfnamefont {J.~P.}\ \bibnamefont {Covey}}, \ and\ \bibinfo {author}
  {\bibfnamefont {J.}~\bibnamefont {Ye}},\ }\href@noop {} {\bibfield  {journal}
  {\bibinfo  {journal} {Science}\ }\textbf {\bibinfo {volume} {363}},\ \bibinfo
  {pages} {853} (\bibinfo {year} {2019})}\BibitemShut {NoStop}%
\bibitem [{\citenamefont {Rvachov}\ \emph {et~al.}(2017)\citenamefont
  {Rvachov}, \citenamefont {Son}, \citenamefont {Sommer}, \citenamefont
  {Ebadi}, \citenamefont {Park}, \citenamefont {Zwierlein}, \citenamefont
  {Ketterle},\ and\ \citenamefont {Jamison}}]{rvachov2017long}%
  \BibitemOpen
  \bibfield  {author} {\bibinfo {author} {\bibfnamefont {T.~M.}\ \bibnamefont
  {Rvachov}}, \bibinfo {author} {\bibfnamefont {H.}~\bibnamefont {Son}},
  \bibinfo {author} {\bibfnamefont {A.~T.}\ \bibnamefont {Sommer}}, \bibinfo
  {author} {\bibfnamefont {S.}~\bibnamefont {Ebadi}}, \bibinfo {author}
  {\bibfnamefont {J.~J.}\ \bibnamefont {Park}}, \bibinfo {author}
  {\bibfnamefont {M.~W.}\ \bibnamefont {Zwierlein}}, \bibinfo {author}
  {\bibfnamefont {W.}~\bibnamefont {Ketterle}}, \ and\ \bibinfo {author}
  {\bibfnamefont {A.~O.}\ \bibnamefont {Jamison}},\ }\href@noop {} {\bibfield
  {journal} {\bibinfo  {journal} {Physical review letters}\ }\textbf {\bibinfo
  {volume} {119}},\ \bibinfo {pages} {143001} (\bibinfo {year}
  {2017})}\BibitemShut {NoStop}%
\bibitem [{\citenamefont {Drews}\ \emph {et~al.}(2017)\citenamefont {Drews},
  \citenamefont {Dei{\ss}}, \citenamefont {Jachymski}, \citenamefont
  {Idziaszek},\ and\ \citenamefont {Denschlag}}]{drews2017inelastic}%
  \BibitemOpen
  \bibfield  {author} {\bibinfo {author} {\bibfnamefont {B.}~\bibnamefont
  {Drews}}, \bibinfo {author} {\bibfnamefont {M.}~\bibnamefont {Dei{\ss}}},
  \bibinfo {author} {\bibfnamefont {K.}~\bibnamefont {Jachymski}}, \bibinfo
  {author} {\bibfnamefont {Z.}~\bibnamefont {Idziaszek}}, \ and\ \bibinfo
  {author} {\bibfnamefont {J.~H.}\ \bibnamefont {Denschlag}},\ }\href@noop {}
  {\bibfield  {journal} {\bibinfo  {journal} {Nature communications}\ }\textbf
  {\bibinfo {volume} {8}},\ \bibinfo {pages} {14854} (\bibinfo {year}
  {2017})}\BibitemShut {NoStop}%
\bibitem [{\citenamefont {Ye}\ \emph {et~al.}(2018)\citenamefont {Ye},
  \citenamefont {Guo}, \citenamefont {Gonz{\'a}lez-Mart{\'\i}nez},
  \citenamefont {Qu{\'e}m{\'e}ner},\ and\ \citenamefont
  {Wang}}]{ye2018collisions}%
  \BibitemOpen
  \bibfield  {author} {\bibinfo {author} {\bibfnamefont {X.}~\bibnamefont
  {Ye}}, \bibinfo {author} {\bibfnamefont {M.}~\bibnamefont {Guo}}, \bibinfo
  {author} {\bibfnamefont {M.~L.}\ \bibnamefont {Gonz{\'a}lez-Mart{\'\i}nez}},
  \bibinfo {author} {\bibfnamefont {G.}~\bibnamefont {Qu{\'e}m{\'e}ner}}, \
  and\ \bibinfo {author} {\bibfnamefont {D.}~\bibnamefont {Wang}},\ }\href@noop
  {} {\bibfield  {journal} {\bibinfo  {journal} {Science advances}\ }\textbf
  {\bibinfo {volume} {4}},\ \bibinfo {pages} {eaaq0083} (\bibinfo {year}
  {2018})}\BibitemShut {NoStop}%
\bibitem [{\citenamefont {Gregory}\ \emph {et~al.}(2019)\citenamefont
  {Gregory}, \citenamefont {Frye}, \citenamefont {Blackmore}, \citenamefont
  {Bridge}, \citenamefont {Sawant}, \citenamefont {Hutson},\ and\ \citenamefont
  {Cornish}}]{gregory2019sticky}%
  \BibitemOpen
  \bibfield  {author} {\bibinfo {author} {\bibfnamefont {P.~D.}\ \bibnamefont
  {Gregory}}, \bibinfo {author} {\bibfnamefont {M.~D.}\ \bibnamefont {Frye}},
  \bibinfo {author} {\bibfnamefont {J.~A.}\ \bibnamefont {Blackmore}}, \bibinfo
  {author} {\bibfnamefont {E.~M.}\ \bibnamefont {Bridge}}, \bibinfo {author}
  {\bibfnamefont {R.}~\bibnamefont {Sawant}}, \bibinfo {author} {\bibfnamefont
  {J.~M.}\ \bibnamefont {Hutson}}, \ and\ \bibinfo {author} {\bibfnamefont
  {S.~L.}\ \bibnamefont {Cornish}},\ }\href@noop {} {\bibfield  {journal}
  {\bibinfo  {journal} {Nature communications}\ }\textbf {\bibinfo {volume}
  {10}},\ \bibinfo {pages} {1} (\bibinfo {year} {2019})}\BibitemShut {NoStop}%
\bibitem [{\citenamefont {Mayle}\ \emph {et~al.}(2012)\citenamefont {Mayle},
  \citenamefont {Ruzic},\ and\ \citenamefont {Bohn}}]{mayle2012statistical}%
  \BibitemOpen
  \bibfield  {author} {\bibinfo {author} {\bibfnamefont {M.}~\bibnamefont
  {Mayle}}, \bibinfo {author} {\bibfnamefont {B.~P.}\ \bibnamefont {Ruzic}}, \
  and\ \bibinfo {author} {\bibfnamefont {J.~L.}\ \bibnamefont {Bohn}},\
  }\href@noop {} {\bibfield  {journal} {\bibinfo  {journal} {Physical Review
  A}\ }\textbf {\bibinfo {volume} {85}},\ \bibinfo {pages} {062712} (\bibinfo
  {year} {2012})}\BibitemShut {NoStop}%
\bibitem [{\citenamefont {Mayle}\ \emph {et~al.}(2013)\citenamefont {Mayle},
  \citenamefont {Qu{\'e}m{\'e}ner}, \citenamefont {Ruzic},\ and\ \citenamefont
  {Bohn}}]{mayle2013scattering}%
  \BibitemOpen
  \bibfield  {author} {\bibinfo {author} {\bibfnamefont {M.}~\bibnamefont
  {Mayle}}, \bibinfo {author} {\bibfnamefont {G.}~\bibnamefont
  {Qu{\'e}m{\'e}ner}}, \bibinfo {author} {\bibfnamefont {B.~P.}\ \bibnamefont
  {Ruzic}}, \ and\ \bibinfo {author} {\bibfnamefont {J.~L.}\ \bibnamefont
  {Bohn}},\ }\href@noop {} {\bibfield  {journal} {\bibinfo  {journal} {Physical
  Review A}\ }\textbf {\bibinfo {volume} {87}},\ \bibinfo {pages} {012709}
  (\bibinfo {year} {2013})}\BibitemShut {NoStop}%
\bibitem [{\citenamefont {Idziaszek}\ \emph {et~al.}(2010)\citenamefont
  {Idziaszek}, \citenamefont {Qu{\'e}m{\'e}ner}, \citenamefont {Bohn},\ and\
  \citenamefont {Julienne}}]{idziaszek2010simple}%
  \BibitemOpen
  \bibfield  {author} {\bibinfo {author} {\bibfnamefont {Z.}~\bibnamefont
  {Idziaszek}}, \bibinfo {author} {\bibfnamefont {G.}~\bibnamefont
  {Qu{\'e}m{\'e}ner}}, \bibinfo {author} {\bibfnamefont {J.~L.}\ \bibnamefont
  {Bohn}}, \ and\ \bibinfo {author} {\bibfnamefont {P.~S.}\ \bibnamefont
  {Julienne}},\ }\href@noop {} {\bibfield  {journal} {\bibinfo  {journal}
  {Physical Review A}\ }\textbf {\bibinfo {volume} {82}},\ \bibinfo {pages}
  {020703} (\bibinfo {year} {2010})}\BibitemShut {NoStop}%
\bibitem [{\citenamefont {Mies}(1984)}]{mies1984multichannel}%
  \BibitemOpen
  \bibfield  {author} {\bibinfo {author} {\bibfnamefont {F.~H.}\ \bibnamefont
  {Mies}},\ }\href@noop {} {\bibfield  {journal} {\bibinfo  {journal} {The
  Journal of chemical physics}\ }\textbf {\bibinfo {volume} {80}},\ \bibinfo
  {pages} {2514} (\bibinfo {year} {1984})}\BibitemShut {NoStop}%
\bibitem [{\citenamefont {Mies}\ and\ \citenamefont
  {Julienne}(1984)}]{mies1984multichannelb}%
  \BibitemOpen
  \bibfield  {author} {\bibinfo {author} {\bibfnamefont {F.~H.}\ \bibnamefont
  {Mies}}\ and\ \bibinfo {author} {\bibfnamefont {P.~S.}\ \bibnamefont
  {Julienne}},\ }\href@noop {} {\bibfield  {journal} {\bibinfo  {journal} {The
  Journal of chemical physics}\ }\textbf {\bibinfo {volume} {80}},\ \bibinfo
  {pages} {2526} (\bibinfo {year} {1984})}\BibitemShut {NoStop}%
\bibitem [{\citenamefont {Jankunas}\ \emph {et~al.}(2014)\citenamefont
  {Jankunas}, \citenamefont {Bertsche}, \citenamefont {Jachymski},
  \citenamefont {Hapka},\ and\ \citenamefont
  {Osterwalder}}]{jankunas2014dynamics}%
  \BibitemOpen
  \bibfield  {author} {\bibinfo {author} {\bibfnamefont {J.}~\bibnamefont
  {Jankunas}}, \bibinfo {author} {\bibfnamefont {B.}~\bibnamefont {Bertsche}},
  \bibinfo {author} {\bibfnamefont {K.}~\bibnamefont {Jachymski}}, \bibinfo
  {author} {\bibfnamefont {M.}~\bibnamefont {Hapka}}, \ and\ \bibinfo {author}
  {\bibfnamefont {A.}~\bibnamefont {Osterwalder}},\ }\href@noop {} {\bibfield
  {journal} {\bibinfo  {journal} {The Journal of chemical physics}\ }\textbf
  {\bibinfo {volume} {140}},\ \bibinfo {pages} {244302} (\bibinfo {year}
  {2014})}\BibitemShut {NoStop}%
\bibitem [{\citenamefont {Jankunas}\ \emph {et~al.}(2015)\citenamefont
  {Jankunas}, \citenamefont {Jachymski}, \citenamefont {Hapka},\ and\
  \citenamefont {Osterwalder}}]{jankunas2015observation}%
  \BibitemOpen
  \bibfield  {author} {\bibinfo {author} {\bibfnamefont {J.}~\bibnamefont
  {Jankunas}}, \bibinfo {author} {\bibfnamefont {K.}~\bibnamefont {Jachymski}},
  \bibinfo {author} {\bibfnamefont {M.}~\bibnamefont {Hapka}}, \ and\ \bibinfo
  {author} {\bibfnamefont {A.}~\bibnamefont {Osterwalder}},\ }\href@noop {}
  {\bibfield  {journal} {\bibinfo  {journal} {The Journal of chemical physics}\
  }\textbf {\bibinfo {volume} {142}},\ \bibinfo {pages} {164305} (\bibinfo
  {year} {2015})}\BibitemShut {NoStop}%
\bibitem [{\citenamefont {Jankunas}\ \emph {et~al.}(2016)\citenamefont
  {Jankunas}, \citenamefont {Jachymski}, \citenamefont {Hapka},\ and\
  \citenamefont {Osterwalder}}]{jankunas2016communication}%
  \BibitemOpen
  \bibfield  {author} {\bibinfo {author} {\bibfnamefont {J.}~\bibnamefont
  {Jankunas}}, \bibinfo {author} {\bibfnamefont {K.}~\bibnamefont {Jachymski}},
  \bibinfo {author} {\bibfnamefont {M.}~\bibnamefont {Hapka}}, \ and\ \bibinfo
  {author} {\bibfnamefont {A.}~\bibnamefont {Osterwalder}},\ }\href@noop {}
  {\enquote {\bibinfo {title} {Communication: Importance of rotationally
  inelastic processes in low-energy penning ionization of chf3},}\ } (\bibinfo
  {year} {2016})\BibitemShut {NoStop}%
\bibitem [{\citenamefont {Herbst}\ and\ \citenamefont
  {Yates~Jr}(2013)}]{herbst2013introduction}%
  \BibitemOpen
  \bibfield  {author} {\bibinfo {author} {\bibfnamefont {E.}~\bibnamefont
  {Herbst}}\ and\ \bibinfo {author} {\bibfnamefont {J.~T.}\ \bibnamefont
  {Yates~Jr}},\ }\href@noop {} {\enquote {\bibinfo {title} {Introduction:
  astrochemistry},}\ } (\bibinfo {year} {2013})\BibitemShut {NoStop}%
\bibitem [{\citenamefont {Sahai}\ and\ \citenamefont
  {Nyman}(1997)}]{sahai1997boomerang}%
  \BibitemOpen
  \bibfield  {author} {\bibinfo {author} {\bibfnamefont {R.}~\bibnamefont
  {Sahai}}\ and\ \bibinfo {author} {\bibfnamefont {L.-{\AA}.}\ \bibnamefont
  {Nyman}},\ }\href@noop {} {\bibfield  {journal} {\bibinfo  {journal} {The
  Astrophysical Journal Letters}\ }\textbf {\bibinfo {volume} {487}},\ \bibinfo
  {pages} {L155} (\bibinfo {year} {1997})}\BibitemShut {NoStop}%
\bibitem [{\citenamefont {Herschbach}(2009)}]{herschbach2009molecular}%
  \BibitemOpen
  \bibfield  {author} {\bibinfo {author} {\bibfnamefont {D.}~\bibnamefont
  {Herschbach}},\ }\href@noop {} {\bibfield  {journal} {\bibinfo  {journal}
  {Faraday discussions}\ }\textbf {\bibinfo {volume} {142}},\ \bibinfo {pages}
  {9} (\bibinfo {year} {2009})}\BibitemShut {NoStop}%
\bibitem [{\citenamefont {Chefdeville}\ \emph {et~al.}(2012)\citenamefont
  {Chefdeville}, \citenamefont {Stoecklin}, \citenamefont {Bergeat},
  \citenamefont {Hickson}, \citenamefont {Naulin},\ and\ \citenamefont
  {Costes}}]{chefdeville2012appearance}%
  \BibitemOpen
  \bibfield  {author} {\bibinfo {author} {\bibfnamefont {S.}~\bibnamefont
  {Chefdeville}}, \bibinfo {author} {\bibfnamefont {T.}~\bibnamefont
  {Stoecklin}}, \bibinfo {author} {\bibfnamefont {A.}~\bibnamefont {Bergeat}},
  \bibinfo {author} {\bibfnamefont {K.~M.}\ \bibnamefont {Hickson}}, \bibinfo
  {author} {\bibfnamefont {C.}~\bibnamefont {Naulin}}, \ and\ \bibinfo {author}
  {\bibfnamefont {M.}~\bibnamefont {Costes}},\ }\href@noop {} {\bibfield
  {journal} {\bibinfo  {journal} {Physical Review Letters}\ }\textbf {\bibinfo
  {volume} {109}},\ \bibinfo {pages} {023201} (\bibinfo {year}
  {2012})}\BibitemShut {NoStop}%
\bibitem [{\citenamefont {Chefdeville}\ \emph {et~al.}(2013)\citenamefont
  {Chefdeville}, \citenamefont {Kalugina}, \citenamefont {van~de Meerakker},
  \citenamefont {Naulin}, \citenamefont {Lique},\ and\ \citenamefont
  {Costes}}]{chefdeville2013observation}%
  \BibitemOpen
  \bibfield  {author} {\bibinfo {author} {\bibfnamefont {S.}~\bibnamefont
  {Chefdeville}}, \bibinfo {author} {\bibfnamefont {Y.}~\bibnamefont
  {Kalugina}}, \bibinfo {author} {\bibfnamefont {S.~Y.}\ \bibnamefont {van~de
  Meerakker}}, \bibinfo {author} {\bibfnamefont {C.}~\bibnamefont {Naulin}},
  \bibinfo {author} {\bibfnamefont {F.}~\bibnamefont {Lique}}, \ and\ \bibinfo
  {author} {\bibfnamefont {M.}~\bibnamefont {Costes}},\ }\href@noop {}
  {\bibfield  {journal} {\bibinfo  {journal} {Science}\ }\textbf {\bibinfo
  {volume} {341}},\ \bibinfo {pages} {1094} (\bibinfo {year}
  {2013})}\BibitemShut {NoStop}%
\bibitem [{\citenamefont {Xiao}\ \emph {et~al.}(2011)\citenamefont {Xiao},
  \citenamefont {Xu}, \citenamefont {Liu}, \citenamefont {Wang}, \citenamefont
  {Dong}, \citenamefont {Yang}, \citenamefont {Sun}, \citenamefont {Dai},
  \citenamefont {Zhang},\ and\ \citenamefont {Yang}}]{xiao2011experimental}%
  \BibitemOpen
  \bibfield  {author} {\bibinfo {author} {\bibfnamefont {C.}~\bibnamefont
  {Xiao}}, \bibinfo {author} {\bibfnamefont {X.}~\bibnamefont {Xu}}, \bibinfo
  {author} {\bibfnamefont {S.}~\bibnamefont {Liu}}, \bibinfo {author}
  {\bibfnamefont {T.}~\bibnamefont {Wang}}, \bibinfo {author} {\bibfnamefont
  {W.}~\bibnamefont {Dong}}, \bibinfo {author} {\bibfnamefont {T.}~\bibnamefont
  {Yang}}, \bibinfo {author} {\bibfnamefont {Z.}~\bibnamefont {Sun}}, \bibinfo
  {author} {\bibfnamefont {D.}~\bibnamefont {Dai}}, \bibinfo {author}
  {\bibfnamefont {D.~H.}\ \bibnamefont {Zhang}}, \ and\ \bibinfo {author}
  {\bibfnamefont {X.}~\bibnamefont {Yang}},\ }\href@noop {} {\bibfield
  {journal} {\bibinfo  {journal} {Science}\ }\textbf {\bibinfo {volume}
  {333}},\ \bibinfo {pages} {440} (\bibinfo {year} {2011})}\BibitemShut
  {NoStop}%
\bibitem [{\citenamefont {Willey}\ \emph
  {et~al.}(1988{\natexlab{a}})\citenamefont {Willey}, \citenamefont
  {Crownover}, \citenamefont {Bittner},\ and\ \citenamefont
  {De~Lucia}}]{willey1988very}%
  \BibitemOpen
  \bibfield  {author} {\bibinfo {author} {\bibfnamefont {D.~R.}\ \bibnamefont
  {Willey}}, \bibinfo {author} {\bibfnamefont {R.~L.}\ \bibnamefont
  {Crownover}}, \bibinfo {author} {\bibfnamefont {D.}~\bibnamefont {Bittner}},
  \ and\ \bibinfo {author} {\bibfnamefont {F.~C.}\ \bibnamefont {De~Lucia}},\
  }\href@noop {} {\bibfield  {journal} {\bibinfo  {journal} {The Journal of
  chemical physics}\ }\textbf {\bibinfo {volume} {89}},\ \bibinfo {pages}
  {1923} (\bibinfo {year} {1988}{\natexlab{a}})}\BibitemShut {NoStop}%
\bibitem [{\citenamefont {Nandi}\ \emph {et~al.}(2002)\citenamefont {Nandi},
  \citenamefont {Blanksby}, \citenamefont {Zhang}, \citenamefont {Nimlos},
  \citenamefont {Dayton},\ and\ \citenamefont {Ellison}}]{nandi2002polarized}%
  \BibitemOpen
  \bibfield  {author} {\bibinfo {author} {\bibfnamefont {S.}~\bibnamefont
  {Nandi}}, \bibinfo {author} {\bibfnamefont {S.~J.}\ \bibnamefont {Blanksby}},
  \bibinfo {author} {\bibfnamefont {X.}~\bibnamefont {Zhang}}, \bibinfo
  {author} {\bibfnamefont {M.~R.}\ \bibnamefont {Nimlos}}, \bibinfo {author}
  {\bibfnamefont {D.~C.}\ \bibnamefont {Dayton}}, \ and\ \bibinfo {author}
  {\bibfnamefont {G.~B.}\ \bibnamefont {Ellison}},\ }\href@noop {} {\bibfield
  {journal} {\bibinfo  {journal} {The Journal of Physical Chemistry A}\
  }\textbf {\bibinfo {volume} {106}},\ \bibinfo {pages} {7547} (\bibinfo {year}
  {2002})}\BibitemShut {NoStop}%
\bibitem [{\citenamefont {Morrison}\ \emph {et~al.}(2012)\citenamefont
  {Morrison}, \citenamefont {Agarwal}, \citenamefont {Schaefer~III},\ and\
  \citenamefont {Douberly}}]{morrison2012infrared}%
  \BibitemOpen
  \bibfield  {author} {\bibinfo {author} {\bibfnamefont {A.~M.}\ \bibnamefont
  {Morrison}}, \bibinfo {author} {\bibfnamefont {J.}~\bibnamefont {Agarwal}},
  \bibinfo {author} {\bibfnamefont {H.~F.}\ \bibnamefont {Schaefer~III}}, \
  and\ \bibinfo {author} {\bibfnamefont {G.~E.}\ \bibnamefont {Douberly}},\
  }\href@noop {} {\bibfield  {journal} {\bibinfo  {journal} {The Journal of
  Physical Chemistry A}\ }\textbf {\bibinfo {volume} {116}},\ \bibinfo {pages}
  {5299} (\bibinfo {year} {2012})}\BibitemShut {NoStop}%
\bibitem [{\citenamefont {Schnieder}\ \emph {et~al.}(1995)\citenamefont
  {Schnieder}, \citenamefont {Seekamp-Rahn}, \citenamefont {Borkowski},
  \citenamefont {Wrede}, \citenamefont {Welge}, \citenamefont {Aoiz},
  \citenamefont {Ba{\~n}iares}, \citenamefont {D'Mello}, \citenamefont
  {Herrero}, \citenamefont {Rabanos} \emph
  {et~al.}}]{schnieder1995experimental}%
  \BibitemOpen
  \bibfield  {author} {\bibinfo {author} {\bibfnamefont {L.}~\bibnamefont
  {Schnieder}}, \bibinfo {author} {\bibfnamefont {K.}~\bibnamefont
  {Seekamp-Rahn}}, \bibinfo {author} {\bibfnamefont {J.}~\bibnamefont
  {Borkowski}}, \bibinfo {author} {\bibfnamefont {E.}~\bibnamefont {Wrede}},
  \bibinfo {author} {\bibfnamefont {K.}~\bibnamefont {Welge}}, \bibinfo
  {author} {\bibfnamefont {F.~J.}\ \bibnamefont {Aoiz}}, \bibinfo {author}
  {\bibfnamefont {L.}~\bibnamefont {Ba{\~n}iares}}, \bibinfo {author}
  {\bibfnamefont {M.}~\bibnamefont {D'Mello}}, \bibinfo {author} {\bibfnamefont
  {V.~J.}\ \bibnamefont {Herrero}}, \bibinfo {author} {\bibfnamefont {V.~S.}\
  \bibnamefont {Rabanos}},  \emph {et~al.},\ }\href@noop {} {\bibfield
  {journal} {\bibinfo  {journal} {Science}\ }\textbf {\bibinfo {volume}
  {269}},\ \bibinfo {pages} {207} (\bibinfo {year} {1995})}\BibitemShut
  {NoStop}%
\bibitem [{\citenamefont {Skodje}\ \emph {et~al.}(2000)\citenamefont {Skodje},
  \citenamefont {Skouteris}, \citenamefont {Manolopoulos}, \citenamefont {Lee},
  \citenamefont {Dong},\ and\ \citenamefont {Liu}}]{skodje2000resonance}%
  \BibitemOpen
  \bibfield  {author} {\bibinfo {author} {\bibfnamefont {R.~T.}\ \bibnamefont
  {Skodje}}, \bibinfo {author} {\bibfnamefont {D.}~\bibnamefont {Skouteris}},
  \bibinfo {author} {\bibfnamefont {D.~E.}\ \bibnamefont {Manolopoulos}},
  \bibinfo {author} {\bibfnamefont {S.-H.}\ \bibnamefont {Lee}}, \bibinfo
  {author} {\bibfnamefont {F.}~\bibnamefont {Dong}}, \ and\ \bibinfo {author}
  {\bibfnamefont {K.}~\bibnamefont {Liu}},\ }\href@noop {} {\bibfield
  {journal} {\bibinfo  {journal} {Physical Review Letters}\ }\textbf {\bibinfo
  {volume} {85}},\ \bibinfo {pages} {1206} (\bibinfo {year}
  {2000})}\BibitemShut {NoStop}%
\bibitem [{\citenamefont {Dong}\ \emph {et~al.}(2010)\citenamefont {Dong},
  \citenamefont {Xiao}, \citenamefont {Wang}, \citenamefont {Dai},
  \citenamefont {Yang},\ and\ \citenamefont {Zhang}}]{dong2010transition}%
  \BibitemOpen
  \bibfield  {author} {\bibinfo {author} {\bibfnamefont {W.}~\bibnamefont
  {Dong}}, \bibinfo {author} {\bibfnamefont {C.}~\bibnamefont {Xiao}}, \bibinfo
  {author} {\bibfnamefont {T.}~\bibnamefont {Wang}}, \bibinfo {author}
  {\bibfnamefont {D.}~\bibnamefont {Dai}}, \bibinfo {author} {\bibfnamefont
  {X.}~\bibnamefont {Yang}}, \ and\ \bibinfo {author} {\bibfnamefont {D.~H.}\
  \bibnamefont {Zhang}},\ }\href@noop {} {\bibfield  {journal} {\bibinfo
  {journal} {Science}\ }\textbf {\bibinfo {volume} {327}},\ \bibinfo {pages}
  {1501} (\bibinfo {year} {2010})}\BibitemShut {NoStop}%
\bibitem [{\citenamefont {Wang}\ \emph
  {et~al.}(2013{\natexlab{a}})\citenamefont {Wang}, \citenamefont {Chen},
  \citenamefont {Yang}, \citenamefont {Xiao}, \citenamefont {Sun},
  \citenamefont {Huang}, \citenamefont {Dai}, \citenamefont {Yang},\ and\
  \citenamefont {Zhang}}]{wang2013dynamical}%
  \BibitemOpen
  \bibfield  {author} {\bibinfo {author} {\bibfnamefont {T.}~\bibnamefont
  {Wang}}, \bibinfo {author} {\bibfnamefont {J.}~\bibnamefont {Chen}}, \bibinfo
  {author} {\bibfnamefont {T.}~\bibnamefont {Yang}}, \bibinfo {author}
  {\bibfnamefont {C.}~\bibnamefont {Xiao}}, \bibinfo {author} {\bibfnamefont
  {Z.}~\bibnamefont {Sun}}, \bibinfo {author} {\bibfnamefont {L.}~\bibnamefont
  {Huang}}, \bibinfo {author} {\bibfnamefont {D.}~\bibnamefont {Dai}}, \bibinfo
  {author} {\bibfnamefont {X.}~\bibnamefont {Yang}}, \ and\ \bibinfo {author}
  {\bibfnamefont {D.~H.}\ \bibnamefont {Zhang}},\ }\href@noop {} {\bibfield
  {journal} {\bibinfo  {journal} {Science}\ }\textbf {\bibinfo {volume}
  {342}},\ \bibinfo {pages} {1499} (\bibinfo {year}
  {2013}{\natexlab{a}})}\BibitemShut {NoStop}%
\bibitem [{\citenamefont {Qiu}\ \emph {et~al.}(2006)\citenamefont {Qiu},
  \citenamefont {Ren}, \citenamefont {Che}, \citenamefont {Dai}, \citenamefont
  {Harich}, \citenamefont {Wang}, \citenamefont {Yang}, \citenamefont {Xu},
  \citenamefont {Xie}, \citenamefont {Gustafsson} \emph
  {et~al.}}]{qiu2006observation}%
  \BibitemOpen
  \bibfield  {author} {\bibinfo {author} {\bibfnamefont {M.}~\bibnamefont
  {Qiu}}, \bibinfo {author} {\bibfnamefont {Z.}~\bibnamefont {Ren}}, \bibinfo
  {author} {\bibfnamefont {L.}~\bibnamefont {Che}}, \bibinfo {author}
  {\bibfnamefont {D.}~\bibnamefont {Dai}}, \bibinfo {author} {\bibfnamefont
  {S.~A.}\ \bibnamefont {Harich}}, \bibinfo {author} {\bibfnamefont
  {X.}~\bibnamefont {Wang}}, \bibinfo {author} {\bibfnamefont {X.}~\bibnamefont
  {Yang}}, \bibinfo {author} {\bibfnamefont {C.}~\bibnamefont {Xu}}, \bibinfo
  {author} {\bibfnamefont {D.}~\bibnamefont {Xie}}, \bibinfo {author}
  {\bibfnamefont {M.}~\bibnamefont {Gustafsson}},  \emph {et~al.},\ }\href@noop
  {} {\bibfield  {journal} {\bibinfo  {journal} {Science}\ }\textbf {\bibinfo
  {volume} {311}},\ \bibinfo {pages} {1440} (\bibinfo {year}
  {2006})}\BibitemShut {NoStop}%
\bibitem [{\citenamefont {Kim}\ \emph {et~al.}(2015)\citenamefont {Kim},
  \citenamefont {Weichman}, \citenamefont {Sjolander}, \citenamefont {Neumark},
  \citenamefont {K{\l}os}, \citenamefont {Alexander},\ and\ \citenamefont
  {Manolopoulos}}]{kim2015spectroscopic}%
  \BibitemOpen
  \bibfield  {author} {\bibinfo {author} {\bibfnamefont {J.~B.}\ \bibnamefont
  {Kim}}, \bibinfo {author} {\bibfnamefont {M.~L.}\ \bibnamefont {Weichman}},
  \bibinfo {author} {\bibfnamefont {T.~F.}\ \bibnamefont {Sjolander}}, \bibinfo
  {author} {\bibfnamefont {D.~M.}\ \bibnamefont {Neumark}}, \bibinfo {author}
  {\bibfnamefont {J.}~\bibnamefont {K{\l}os}}, \bibinfo {author} {\bibfnamefont
  {M.~H.}\ \bibnamefont {Alexander}}, \ and\ \bibinfo {author} {\bibfnamefont
  {D.~E.}\ \bibnamefont {Manolopoulos}},\ }\href@noop {} {\bibfield  {journal}
  {\bibinfo  {journal} {Science}\ }\textbf {\bibinfo {volume} {349}},\ \bibinfo
  {pages} {510} (\bibinfo {year} {2015})}\BibitemShut {NoStop}%
\bibitem [{\citenamefont {Wang}\ \emph {et~al.}(2011)\citenamefont {Wang},
  \citenamefont {Lin},\ and\ \citenamefont {Liu}}]{wang2011steric}%
  \BibitemOpen
  \bibfield  {author} {\bibinfo {author} {\bibfnamefont {F.}~\bibnamefont
  {Wang}}, \bibinfo {author} {\bibfnamefont {J.-S.}\ \bibnamefont {Lin}}, \
  and\ \bibinfo {author} {\bibfnamefont {K.}~\bibnamefont {Liu}},\ }\href@noop
  {} {\bibfield  {journal} {\bibinfo  {journal} {Science}\ }\textbf {\bibinfo
  {volume} {331}},\ \bibinfo {pages} {900} (\bibinfo {year}
  {2011})}\BibitemShut {NoStop}%
\bibitem [{\citenamefont {Vogels}\ \emph {et~al.}(2018)\citenamefont {Vogels},
  \citenamefont {Karman}, \citenamefont {K{\l}os}, \citenamefont {Besemer},
  \citenamefont {Onvlee}, \citenamefont {van~der Avoird}, \citenamefont
  {Groenenboom},\ and\ \citenamefont {van~de
  Meerakker}}]{vogels2018scattering}%
  \BibitemOpen
  \bibfield  {author} {\bibinfo {author} {\bibfnamefont {S.~N.}\ \bibnamefont
  {Vogels}}, \bibinfo {author} {\bibfnamefont {T.}~\bibnamefont {Karman}},
  \bibinfo {author} {\bibfnamefont {J.}~\bibnamefont {K{\l}os}}, \bibinfo
  {author} {\bibfnamefont {M.}~\bibnamefont {Besemer}}, \bibinfo {author}
  {\bibfnamefont {J.}~\bibnamefont {Onvlee}}, \bibinfo {author} {\bibfnamefont
  {A.}~\bibnamefont {van~der Avoird}}, \bibinfo {author} {\bibfnamefont
  {G.~C.}\ \bibnamefont {Groenenboom}}, \ and\ \bibinfo {author} {\bibfnamefont
  {S.~Y.}\ \bibnamefont {van~de Meerakker}},\ }\href@noop {} {\bibfield
  {journal} {\bibinfo  {journal} {Nature chemistry}\ }\textbf {\bibinfo
  {volume} {10}},\ \bibinfo {pages} {435} (\bibinfo {year} {2018})}\BibitemShut
  {NoStop}%
\bibitem [{\citenamefont {Von~Zastrow}\ \emph {et~al.}(2014)\citenamefont
  {Von~Zastrow}, \citenamefont {Onvlee}, \citenamefont {Vogels}, \citenamefont
  {Groenenboom}, \citenamefont {Van Der~Avoird},\ and\ \citenamefont {Van
  De~Meerakker}}]{von2014state}%
  \BibitemOpen
  \bibfield  {author} {\bibinfo {author} {\bibfnamefont {A.}~\bibnamefont
  {Von~Zastrow}}, \bibinfo {author} {\bibfnamefont {J.}~\bibnamefont {Onvlee}},
  \bibinfo {author} {\bibfnamefont {S.~N.}\ \bibnamefont {Vogels}}, \bibinfo
  {author} {\bibfnamefont {G.~C.}\ \bibnamefont {Groenenboom}}, \bibinfo
  {author} {\bibfnamefont {A.}~\bibnamefont {Van Der~Avoird}}, \ and\ \bibinfo
  {author} {\bibfnamefont {S.~Y.}\ \bibnamefont {Van De~Meerakker}},\
  }\href@noop {} {\bibfield  {journal} {\bibinfo  {journal} {Nature chemistry}\
  }\textbf {\bibinfo {volume} {6}},\ \bibinfo {pages} {216} (\bibinfo {year}
  {2014})}\BibitemShut {NoStop}%
\bibitem [{\citenamefont {Onvlee}\ \emph {et~al.}(2017)\citenamefont {Onvlee},
  \citenamefont {Gordon}, \citenamefont {Vogels}, \citenamefont {Auth},
  \citenamefont {Karman}, \citenamefont {Nichols}, \citenamefont {van~der
  Avoird}, \citenamefont {Groenenboom}, \citenamefont {Brouard},\ and\
  \citenamefont {van~de Meerakker}}]{onvlee2017imaging}%
  \BibitemOpen
  \bibfield  {author} {\bibinfo {author} {\bibfnamefont {J.}~\bibnamefont
  {Onvlee}}, \bibinfo {author} {\bibfnamefont {S.~D.}\ \bibnamefont {Gordon}},
  \bibinfo {author} {\bibfnamefont {S.~N.}\ \bibnamefont {Vogels}}, \bibinfo
  {author} {\bibfnamefont {T.}~\bibnamefont {Auth}}, \bibinfo {author}
  {\bibfnamefont {T.}~\bibnamefont {Karman}}, \bibinfo {author} {\bibfnamefont
  {B.}~\bibnamefont {Nichols}}, \bibinfo {author} {\bibfnamefont
  {A.}~\bibnamefont {van~der Avoird}}, \bibinfo {author} {\bibfnamefont
  {G.~C.}\ \bibnamefont {Groenenboom}}, \bibinfo {author} {\bibfnamefont
  {M.}~\bibnamefont {Brouard}}, \ and\ \bibinfo {author} {\bibfnamefont
  {S.~Y.}\ \bibnamefont {van~de Meerakker}},\ }\href@noop {} {\bibfield
  {journal} {\bibinfo  {journal} {Nature chemistry}\ }\textbf {\bibinfo
  {volume} {9}},\ \bibinfo {pages} {226} (\bibinfo {year} {2017})}\BibitemShut
  {NoStop}%
\bibitem [{\citenamefont {Lara}\ \emph {et~al.}(2011)\citenamefont {Lara},
  \citenamefont {Dayou}, \citenamefont {Launay}, \citenamefont {Bergeat},
  \citenamefont {Hickson}, \citenamefont {Naulin},\ and\ \citenamefont
  {Costes}}]{lara2011observation}%
  \BibitemOpen
  \bibfield  {author} {\bibinfo {author} {\bibfnamefont {M.}~\bibnamefont
  {Lara}}, \bibinfo {author} {\bibfnamefont {F.}~\bibnamefont {Dayou}},
  \bibinfo {author} {\bibfnamefont {J.-M.}\ \bibnamefont {Launay}}, \bibinfo
  {author} {\bibfnamefont {A.}~\bibnamefont {Bergeat}}, \bibinfo {author}
  {\bibfnamefont {K.~M.}\ \bibnamefont {Hickson}}, \bibinfo {author}
  {\bibfnamefont {C.}~\bibnamefont {Naulin}}, \ and\ \bibinfo {author}
  {\bibfnamefont {M.}~\bibnamefont {Costes}},\ }\href@noop {} {\bibfield
  {journal} {\bibinfo  {journal} {Physical Chemistry Chemical Physics}\
  }\textbf {\bibinfo {volume} {13}},\ \bibinfo {pages} {8127} (\bibinfo {year}
  {2011})}\BibitemShut {NoStop}%
\bibitem [{\citenamefont {Lara}\ \emph {et~al.}(2012)\citenamefont {Lara},
  \citenamefont {Chefdeville}, \citenamefont {Hickson}, \citenamefont
  {Bergeat}, \citenamefont {Naulin}, \citenamefont {Launay},\ and\
  \citenamefont {Costes}}]{lara2012dynamics}%
  \BibitemOpen
  \bibfield  {author} {\bibinfo {author} {\bibfnamefont {M.}~\bibnamefont
  {Lara}}, \bibinfo {author} {\bibfnamefont {S.}~\bibnamefont {Chefdeville}},
  \bibinfo {author} {\bibfnamefont {K.~M.}\ \bibnamefont {Hickson}}, \bibinfo
  {author} {\bibfnamefont {A.}~\bibnamefont {Bergeat}}, \bibinfo {author}
  {\bibfnamefont {C.}~\bibnamefont {Naulin}}, \bibinfo {author} {\bibfnamefont
  {J.-M.}\ \bibnamefont {Launay}}, \ and\ \bibinfo {author} {\bibfnamefont
  {M.}~\bibnamefont {Costes}},\ }\href@noop {} {\bibfield  {journal} {\bibinfo
  {journal} {Physical review letters}\ }\textbf {\bibinfo {volume} {109}},\
  \bibinfo {pages} {133201} (\bibinfo {year} {2012})}\BibitemShut {NoStop}%
\bibitem [{\citenamefont {van~de Meerakker}\ \emph {et~al.}(2012)\citenamefont
  {van~de Meerakker}, \citenamefont {Bethlem}, \citenamefont {Vanhaecke},\ and\
  \citenamefont {Meijer}}]{van2012manipulation}%
  \BibitemOpen
  \bibfield  {author} {\bibinfo {author} {\bibfnamefont {S.~Y.}\ \bibnamefont
  {van~de Meerakker}}, \bibinfo {author} {\bibfnamefont {H.~L.}\ \bibnamefont
  {Bethlem}}, \bibinfo {author} {\bibfnamefont {N.}~\bibnamefont {Vanhaecke}},
  \ and\ \bibinfo {author} {\bibfnamefont {G.}~\bibnamefont {Meijer}},\
  }\href@noop {} {\bibfield  {journal} {\bibinfo  {journal} {Chemical Reviews}\
  }\textbf {\bibinfo {volume} {112}},\ \bibinfo {pages} {4828} (\bibinfo {year}
  {2012})}\BibitemShut {NoStop}%
\bibitem [{\citenamefont {Scharfenberg}\ \emph {et~al.}(2010)\citenamefont
  {Scharfenberg}, \citenamefont {K{\l}os}, \citenamefont {Dagdigian},
  \citenamefont {Alexander}, \citenamefont {Meijer},\ and\ \citenamefont
  {van~de Meerakker}}]{scharfenberg2010state}%
  \BibitemOpen
  \bibfield  {author} {\bibinfo {author} {\bibfnamefont {L.}~\bibnamefont
  {Scharfenberg}}, \bibinfo {author} {\bibfnamefont {J.}~\bibnamefont
  {K{\l}os}}, \bibinfo {author} {\bibfnamefont {P.~J.}\ \bibnamefont
  {Dagdigian}}, \bibinfo {author} {\bibfnamefont {M.~H.}\ \bibnamefont
  {Alexander}}, \bibinfo {author} {\bibfnamefont {G.}~\bibnamefont {Meijer}}, \
  and\ \bibinfo {author} {\bibfnamefont {S.~Y.}\ \bibnamefont {van~de
  Meerakker}},\ }\href@noop {} {\bibfield  {journal} {\bibinfo  {journal}
  {Physical Chemistry Chemical Physics}\ }\textbf {\bibinfo {volume} {12}},\
  \bibinfo {pages} {10660} (\bibinfo {year} {2010})}\BibitemShut {NoStop}%
\bibitem [{\citenamefont {Scharfenberg}\ \emph {et~al.}(2011)\citenamefont
  {Scharfenberg}, \citenamefont {Gubbels}, \citenamefont {Kirste},
  \citenamefont {Groenenboom}, \citenamefont {van~der Avoird}, \citenamefont
  {Meijer},\ and\ \citenamefont {van~de
  Meerakker}}]{scharfenberg2011scattering}%
  \BibitemOpen
  \bibfield  {author} {\bibinfo {author} {\bibfnamefont {L.}~\bibnamefont
  {Scharfenberg}}, \bibinfo {author} {\bibfnamefont {K.~B.}\ \bibnamefont
  {Gubbels}}, \bibinfo {author} {\bibfnamefont {M.}~\bibnamefont {Kirste}},
  \bibinfo {author} {\bibfnamefont {G.~C.}\ \bibnamefont {Groenenboom}},
  \bibinfo {author} {\bibfnamefont {A.}~\bibnamefont {van~der Avoird}},
  \bibinfo {author} {\bibfnamefont {G.}~\bibnamefont {Meijer}}, \ and\ \bibinfo
  {author} {\bibfnamefont {S.~Y.}\ \bibnamefont {van~de Meerakker}},\
  }\href@noop {} {\bibfield  {journal} {\bibinfo  {journal} {The European
  Physical Journal D}\ }\textbf {\bibinfo {volume} {65}},\ \bibinfo {pages}
  {189} (\bibinfo {year} {2011})}\BibitemShut {NoStop}%
\bibitem [{\citenamefont {Kirste}\ \emph {et~al.}(2010)\citenamefont {Kirste},
  \citenamefont {Scharfenberg}, \citenamefont {K{\l}os}, \citenamefont {Lique},
  \citenamefont {Alexander}, \citenamefont {Meijer},\ and\ \citenamefont
  {van~de Meerakker}}]{kirste2010low}%
  \BibitemOpen
  \bibfield  {author} {\bibinfo {author} {\bibfnamefont {M.}~\bibnamefont
  {Kirste}}, \bibinfo {author} {\bibfnamefont {L.}~\bibnamefont
  {Scharfenberg}}, \bibinfo {author} {\bibfnamefont {J.}~\bibnamefont
  {K{\l}os}}, \bibinfo {author} {\bibfnamefont {F.}~\bibnamefont {Lique}},
  \bibinfo {author} {\bibfnamefont {M.~H.}\ \bibnamefont {Alexander}}, \bibinfo
  {author} {\bibfnamefont {G.}~\bibnamefont {Meijer}}, \ and\ \bibinfo {author}
  {\bibfnamefont {S.~Y.}\ \bibnamefont {van~de Meerakker}},\ }\href@noop {}
  {\bibfield  {journal} {\bibinfo  {journal} {Physical Review A}\ }\textbf
  {\bibinfo {volume} {82}},\ \bibinfo {pages} {042717} (\bibinfo {year}
  {2010})}\BibitemShut {NoStop}%
\bibitem [{\citenamefont {Rowe}\ \emph {et~al.}(1984)\citenamefont {Rowe},
  \citenamefont {Dupeyrat}, \citenamefont {Marquette},\ and\ \citenamefont
  {Gaucherel}}]{rowe1984study}%
  \BibitemOpen
  \bibfield  {author} {\bibinfo {author} {\bibfnamefont {B.}~\bibnamefont
  {Rowe}}, \bibinfo {author} {\bibfnamefont {G.}~\bibnamefont {Dupeyrat}},
  \bibinfo {author} {\bibfnamefont {J.}~\bibnamefont {Marquette}}, \ and\
  \bibinfo {author} {\bibfnamefont {P.}~\bibnamefont {Gaucherel}},\ }\href@noop
  {} {\bibfield  {journal} {\bibinfo  {journal} {The Journal of chemical
  physics}\ }\textbf {\bibinfo {volume} {80}},\ \bibinfo {pages} {4915}
  (\bibinfo {year} {1984})}\BibitemShut {NoStop}%
\bibitem [{\citenamefont {Dupeyrat}\ \emph {et~al.}(1985)\citenamefont
  {Dupeyrat}, \citenamefont {Marquette},\ and\ \citenamefont
  {Rowe}}]{dupeyrat1985design}%
  \BibitemOpen
  \bibfield  {author} {\bibinfo {author} {\bibfnamefont {G.}~\bibnamefont
  {Dupeyrat}}, \bibinfo {author} {\bibfnamefont {J.}~\bibnamefont {Marquette}},
  \ and\ \bibinfo {author} {\bibfnamefont {B.}~\bibnamefont {Rowe}},\
  }\href@noop {} {\bibfield  {journal} {\bibinfo  {journal} {The Physics of
  fluids}\ }\textbf {\bibinfo {volume} {28}},\ \bibinfo {pages} {1273}
  (\bibinfo {year} {1985})}\BibitemShut {NoStop}%
\bibitem [{\citenamefont {Sims}\ and\ \citenamefont
  {Smith}(1995)}]{sims1995gas}%
  \BibitemOpen
  \bibfield  {author} {\bibinfo {author} {\bibfnamefont {I.~R.}\ \bibnamefont
  {Sims}}\ and\ \bibinfo {author} {\bibfnamefont {I.~W.}\ \bibnamefont
  {Smith}},\ }\href@noop {} {\bibfield  {journal} {\bibinfo  {journal} {Annual
  Review of Physical Chemistry}\ }\textbf {\bibinfo {volume} {46}},\ \bibinfo
  {pages} {109} (\bibinfo {year} {1995})}\BibitemShut {NoStop}%
\bibitem [{\citenamefont {Smith}\ and\ \citenamefont
  {Rowe}(2000)}]{smith2000reaction}%
  \BibitemOpen
  \bibfield  {author} {\bibinfo {author} {\bibfnamefont {I.~W.}\ \bibnamefont
  {Smith}}\ and\ \bibinfo {author} {\bibfnamefont {B.~R.}\ \bibnamefont
  {Rowe}},\ }\href@noop {} {\bibfield  {journal} {\bibinfo  {journal} {Accounts
  of Chemical Research}\ }\textbf {\bibinfo {volume} {33}},\ \bibinfo {pages}
  {261} (\bibinfo {year} {2000})}\BibitemShut {NoStop}%
\bibitem [{\citenamefont {Smith}(2006)}]{smith2006reactions}%
  \BibitemOpen
  \bibfield  {author} {\bibinfo {author} {\bibfnamefont {I.~W.}\ \bibnamefont
  {Smith}},\ }\href@noop {} {\bibfield  {journal} {\bibinfo  {journal}
  {Angewandte Chemie International Edition}\ }\textbf {\bibinfo {volume}
  {45}},\ \bibinfo {pages} {2842} (\bibinfo {year} {2006})}\BibitemShut
  {NoStop}%
\bibitem [{\citenamefont {James}\ \emph {et~al.}(1998)\citenamefont {James},
  \citenamefont {Sims}, \citenamefont {Smith}, \citenamefont {Alexander},\ and\
  \citenamefont {Yang}}]{james1998combined}%
  \BibitemOpen
  \bibfield  {author} {\bibinfo {author} {\bibfnamefont {P.~L.}\ \bibnamefont
  {James}}, \bibinfo {author} {\bibfnamefont {I.~R.}\ \bibnamefont {Sims}},
  \bibinfo {author} {\bibfnamefont {I.~W.}\ \bibnamefont {Smith}}, \bibinfo
  {author} {\bibfnamefont {M.~H.}\ \bibnamefont {Alexander}}, \ and\ \bibinfo
  {author} {\bibfnamefont {M.}~\bibnamefont {Yang}},\ }\href@noop {} {\bibfield
   {journal} {\bibinfo  {journal} {The Journal of chemical physics}\ }\textbf
  {\bibinfo {volume} {109}},\ \bibinfo {pages} {3882} (\bibinfo {year}
  {1998})}\BibitemShut {NoStop}%
\bibitem [{\citenamefont {Chastaing}\ \emph {et~al.}(1999)\citenamefont
  {Chastaing}, \citenamefont {James}, \citenamefont {Sims},\ and\ \citenamefont
  {Smith}}]{chastaing1999neutral}%
  \BibitemOpen
  \bibfield  {author} {\bibinfo {author} {\bibfnamefont {D.}~\bibnamefont
  {Chastaing}}, \bibinfo {author} {\bibfnamefont {P.~L.}\ \bibnamefont
  {James}}, \bibinfo {author} {\bibfnamefont {I.~R.}\ \bibnamefont {Sims}}, \
  and\ \bibinfo {author} {\bibfnamefont {I.~W.}\ \bibnamefont {Smith}},\
  }\href@noop {} {\bibfield  {journal} {\bibinfo  {journal} {Physical Chemistry
  Chemical Physics}\ }\textbf {\bibinfo {volume} {1}},\ \bibinfo {pages} {2247}
  (\bibinfo {year} {1999})}\BibitemShut {NoStop}%
\bibitem [{\citenamefont {Perreault}\ \emph {et~al.}(2017)\citenamefont
  {Perreault}, \citenamefont {Mukherjee},\ and\ \citenamefont
  {Zare}}]{perreault2017quantum}%
  \BibitemOpen
  \bibfield  {author} {\bibinfo {author} {\bibfnamefont {W.~E.}\ \bibnamefont
  {Perreault}}, \bibinfo {author} {\bibfnamefont {N.}~\bibnamefont
  {Mukherjee}}, \ and\ \bibinfo {author} {\bibfnamefont {R.~N.}\ \bibnamefont
  {Zare}},\ }\href@noop {} {\bibfield  {journal} {\bibinfo  {journal}
  {Science}\ }\textbf {\bibinfo {volume} {358}},\ \bibinfo {pages} {356}
  (\bibinfo {year} {2017})}\BibitemShut {NoStop}%
\bibitem [{\citenamefont {Perreault}\ \emph {et~al.}(2018)\citenamefont
  {Perreault}, \citenamefont {Mukherjee},\ and\ \citenamefont
  {Zare}}]{perreault2018cold}%
  \BibitemOpen
  \bibfield  {author} {\bibinfo {author} {\bibfnamefont {W.~E.}\ \bibnamefont
  {Perreault}}, \bibinfo {author} {\bibfnamefont {N.}~\bibnamefont
  {Mukherjee}}, \ and\ \bibinfo {author} {\bibfnamefont {R.~N.}\ \bibnamefont
  {Zare}},\ }\href@noop {} {\bibfield  {journal} {\bibinfo  {journal} {Nature
  chemistry}\ }\textbf {\bibinfo {volume} {10}},\ \bibinfo {pages} {561}
  (\bibinfo {year} {2018})}\BibitemShut {NoStop}%
\bibitem [{\citenamefont {Perreault}\ \emph {et~al.}(2019)\citenamefont
  {Perreault}, \citenamefont {Mukherjee},\ and\ \citenamefont
  {Zare}}]{perreault2019hd}%
  \BibitemOpen
  \bibfield  {author} {\bibinfo {author} {\bibfnamefont {W.~E.}\ \bibnamefont
  {Perreault}}, \bibinfo {author} {\bibfnamefont {N.}~\bibnamefont
  {Mukherjee}}, \ and\ \bibinfo {author} {\bibfnamefont {R.~N.}\ \bibnamefont
  {Zare}},\ }\href@noop {} {\bibfield  {journal} {\bibinfo  {journal} {The
  Journal of chemical physics}\ }\textbf {\bibinfo {volume} {150}},\ \bibinfo
  {pages} {174301} (\bibinfo {year} {2019})}\BibitemShut {NoStop}%
\bibitem [{\citenamefont {Barnwell}\ \emph {et~al.}(1983)\citenamefont
  {Barnwell}, \citenamefont {Loeser},\ and\ \citenamefont
  {Herschbach}}]{barnwell83}%
  \BibitemOpen
  \bibfield  {author} {\bibinfo {author} {\bibfnamefont {J.~D.}\ \bibnamefont
  {Barnwell}}, \bibinfo {author} {\bibfnamefont {J.~G.}\ \bibnamefont
  {Loeser}}, \ and\ \bibinfo {author} {\bibfnamefont {D.~R.}\ \bibnamefont
  {Herschbach}},\ }\href {\doibase 10.1021/j100238a017} {\bibfield  {journal}
  {\bibinfo  {journal} {The Journal of Physical Chemistry}\ }\textbf {\bibinfo
  {volume} {87}},\ \bibinfo {pages} {2781} (\bibinfo {year}
  {1983})}\BibitemShut {NoStop}%
\bibitem [{\citenamefont {Wu}\ \emph {et~al.}(2017)\citenamefont {Wu},
  \citenamefont {Gantner}, \citenamefont {Koller}, \citenamefont {Zeppenfeld},
  \citenamefont {Chervenkov},\ and\ \citenamefont {Rempe}}]{wu2017cryofuge}%
  \BibitemOpen
  \bibfield  {author} {\bibinfo {author} {\bibfnamefont {X.}~\bibnamefont
  {Wu}}, \bibinfo {author} {\bibfnamefont {T.}~\bibnamefont {Gantner}},
  \bibinfo {author} {\bibfnamefont {M.}~\bibnamefont {Koller}}, \bibinfo
  {author} {\bibfnamefont {M.}~\bibnamefont {Zeppenfeld}}, \bibinfo {author}
  {\bibfnamefont {S.}~\bibnamefont {Chervenkov}}, \ and\ \bibinfo {author}
  {\bibfnamefont {G.}~\bibnamefont {Rempe}},\ }\href@noop {} {\bibfield
  {journal} {\bibinfo  {journal} {Science}\ }\textbf {\bibinfo {volume}
  {358}},\ \bibinfo {pages} {645} (\bibinfo {year} {2017})}\BibitemShut
  {NoStop}%
\bibitem [{\citenamefont {{Cavagnero}}\ and\ \citenamefont
  {{Newell}}(2009)}]{cavagnero2009}%
  \BibitemOpen
  \bibfield  {author} {\bibinfo {author} {\bibfnamefont {M.}~\bibnamefont
  {{Cavagnero}}}\ and\ \bibinfo {author} {\bibfnamefont {C.}~\bibnamefont
  {{Newell}}},\ }\href {\doibase 10.1088/1367-2630/11/5/055040} {\bibfield
  {journal} {\bibinfo  {journal} {New Journal of Physics}\ }\textbf {\bibinfo
  {volume} {11}},\ \bibinfo {eid} {055040} (\bibinfo {year}
  {2009})}\BibitemShut {NoStop}%
\bibitem [{\citenamefont {Willey}\ \emph
  {et~al.}(1988{\natexlab{b}})\citenamefont {Willey}, \citenamefont
  {Crownover}, \citenamefont {Bittner},\ and\ \citenamefont
  {De~Lucia}}]{willey1988very2}%
  \BibitemOpen
  \bibfield  {author} {\bibinfo {author} {\bibfnamefont {D.~R.}\ \bibnamefont
  {Willey}}, \bibinfo {author} {\bibfnamefont {R.~L.}\ \bibnamefont
  {Crownover}}, \bibinfo {author} {\bibfnamefont {D.}~\bibnamefont {Bittner}},
  \ and\ \bibinfo {author} {\bibfnamefont {F.~C.}\ \bibnamefont {De~Lucia}},\
  }\href@noop {} {\bibfield  {journal} {\bibinfo  {journal} {The Journal of
  chemical physics}\ }\textbf {\bibinfo {volume} {89}},\ \bibinfo {pages}
  {6147} (\bibinfo {year} {1988}{\natexlab{b}})}\BibitemShut {NoStop}%
\bibitem [{\citenamefont {Ball}\ and\ \citenamefont
  {De~Lucia}(1998)}]{ball1998direct}%
  \BibitemOpen
  \bibfield  {author} {\bibinfo {author} {\bibfnamefont {C.~D.}\ \bibnamefont
  {Ball}}\ and\ \bibinfo {author} {\bibfnamefont {F.~C.}\ \bibnamefont
  {De~Lucia}},\ }\href@noop {} {\bibfield  {journal} {\bibinfo  {journal}
  {Physical review letters}\ }\textbf {\bibinfo {volume} {81}},\ \bibinfo
  {pages} {305} (\bibinfo {year} {1998})}\BibitemShut {NoStop}%
\bibitem [{\citenamefont {Ball}\ and\ \citenamefont
  {De~Lucia}(1999)}]{ball1999direct}%
  \BibitemOpen
  \bibfield  {author} {\bibinfo {author} {\bibfnamefont {C.~D.}\ \bibnamefont
  {Ball}}\ and\ \bibinfo {author} {\bibfnamefont {F.~C.}\ \bibnamefont
  {De~Lucia}},\ }\href@noop {} {\bibfield  {journal} {\bibinfo  {journal}
  {Chemical physics letters}\ }\textbf {\bibinfo {volume} {300}},\ \bibinfo
  {pages} {227} (\bibinfo {year} {1999})}\BibitemShut {NoStop}%
\bibitem [{\citenamefont {Drayna}\ \emph {et~al.}(2016)\citenamefont {Drayna},
  \citenamefont {Hallas}, \citenamefont {Wang}, \citenamefont {Domingos},
  \citenamefont {Eibenberger}, \citenamefont {Doyle},\ and\ \citenamefont
  {Patterson}}]{drayna2016direct}%
  \BibitemOpen
  \bibfield  {author} {\bibinfo {author} {\bibfnamefont {G.~K.}\ \bibnamefont
  {Drayna}}, \bibinfo {author} {\bibfnamefont {C.}~\bibnamefont {Hallas}},
  \bibinfo {author} {\bibfnamefont {K.}~\bibnamefont {Wang}}, \bibinfo {author}
  {\bibfnamefont {S.~R.}\ \bibnamefont {Domingos}}, \bibinfo {author}
  {\bibfnamefont {S.}~\bibnamefont {Eibenberger}}, \bibinfo {author}
  {\bibfnamefont {J.~M.}\ \bibnamefont {Doyle}}, \ and\ \bibinfo {author}
  {\bibfnamefont {D.}~\bibnamefont {Patterson}},\ }\href@noop {} {\bibfield
  {journal} {\bibinfo  {journal} {Angewandte Chemie International Edition}\
  }\textbf {\bibinfo {volume} {55}},\ \bibinfo {pages} {4957} (\bibinfo {year}
  {2016})}\BibitemShut {NoStop}%
\bibitem [{\citenamefont {Sawyer}\ \emph {et~al.}(2007)\citenamefont {Sawyer},
  \citenamefont {Lev}, \citenamefont {Hudson}, \citenamefont {Stuhl},
  \citenamefont {Lara}, \citenamefont {Bohn},\ and\ \citenamefont
  {Ye}}]{sawyer2007magnetoelectrostatic}%
  \BibitemOpen
  \bibfield  {author} {\bibinfo {author} {\bibfnamefont {B.~C.}\ \bibnamefont
  {Sawyer}}, \bibinfo {author} {\bibfnamefont {B.~L.}\ \bibnamefont {Lev}},
  \bibinfo {author} {\bibfnamefont {E.~R.}\ \bibnamefont {Hudson}}, \bibinfo
  {author} {\bibfnamefont {B.~K.}\ \bibnamefont {Stuhl}}, \bibinfo {author}
  {\bibfnamefont {M.}~\bibnamefont {Lara}}, \bibinfo {author} {\bibfnamefont
  {J.~L.}\ \bibnamefont {Bohn}}, \ and\ \bibinfo {author} {\bibfnamefont
  {J.}~\bibnamefont {Ye}},\ }\href@noop {} {\bibfield  {journal} {\bibinfo
  {journal} {Physical review letters}\ }\textbf {\bibinfo {volume} {98}},\
  \bibinfo {pages} {253002} (\bibinfo {year} {2007})}\BibitemShut {NoStop}%
\bibitem [{\citenamefont {Liu}\ \emph {et~al.}(2017)\citenamefont {Liu},
  \citenamefont {Vashishta}, \citenamefont {Djuricanin}, \citenamefont {Zhou},
  \citenamefont {Zhong}, \citenamefont {Mittertreiner}, \citenamefont {Carty},\
  and\ \citenamefont {Momose}}]{liu2017magnetic}%
  \BibitemOpen
  \bibfield  {author} {\bibinfo {author} {\bibfnamefont {Y.}~\bibnamefont
  {Liu}}, \bibinfo {author} {\bibfnamefont {M.}~\bibnamefont {Vashishta}},
  \bibinfo {author} {\bibfnamefont {P.}~\bibnamefont {Djuricanin}}, \bibinfo
  {author} {\bibfnamefont {S.}~\bibnamefont {Zhou}}, \bibinfo {author}
  {\bibfnamefont {W.}~\bibnamefont {Zhong}}, \bibinfo {author} {\bibfnamefont
  {T.}~\bibnamefont {Mittertreiner}}, \bibinfo {author} {\bibfnamefont
  {D.}~\bibnamefont {Carty}}, \ and\ \bibinfo {author} {\bibfnamefont
  {T.}~\bibnamefont {Momose}},\ }\href@noop {} {\bibfield  {journal} {\bibinfo
  {journal} {Physical review letters}\ }\textbf {\bibinfo {volume} {118}},\
  \bibinfo {pages} {093201} (\bibinfo {year} {2017})}\BibitemShut {NoStop}%
\bibitem [{\citenamefont {Sawyer}\ \emph {et~al.}(2011)\citenamefont {Sawyer},
  \citenamefont {Stuhl}, \citenamefont {Yeo}, \citenamefont {Tscherbul},
  \citenamefont {Hummon}, \citenamefont {Xia}, \citenamefont {K{\l}os},
  \citenamefont {Patterson}, \citenamefont {Doyle},\ and\ \citenamefont
  {Ye}}]{sawyer2011cold}%
  \BibitemOpen
  \bibfield  {author} {\bibinfo {author} {\bibfnamefont {B.~C.}\ \bibnamefont
  {Sawyer}}, \bibinfo {author} {\bibfnamefont {B.~K.}\ \bibnamefont {Stuhl}},
  \bibinfo {author} {\bibfnamefont {M.}~\bibnamefont {Yeo}}, \bibinfo {author}
  {\bibfnamefont {T.~V.}\ \bibnamefont {Tscherbul}}, \bibinfo {author}
  {\bibfnamefont {M.~T.}\ \bibnamefont {Hummon}}, \bibinfo {author}
  {\bibfnamefont {Y.}~\bibnamefont {Xia}}, \bibinfo {author} {\bibfnamefont
  {J.}~\bibnamefont {K{\l}os}}, \bibinfo {author} {\bibfnamefont
  {D.}~\bibnamefont {Patterson}}, \bibinfo {author} {\bibfnamefont {J.~M.}\
  \bibnamefont {Doyle}}, \ and\ \bibinfo {author} {\bibfnamefont
  {J.}~\bibnamefont {Ye}},\ }\href@noop {} {\bibfield  {journal} {\bibinfo
  {journal} {Physical chemistry chemical physics}\ }\textbf {\bibinfo {volume}
  {13}},\ \bibinfo {pages} {19059} (\bibinfo {year} {2011})}\BibitemShut
  {NoStop}%
\bibitem [{\citenamefont {Sawyer}\ \emph {et~al.}(2008)\citenamefont {Sawyer},
  \citenamefont {Stuhl}, \citenamefont {Wang}, \citenamefont {Yeo},\ and\
  \citenamefont {Ye}}]{sawyer2008molecular}%
  \BibitemOpen
  \bibfield  {author} {\bibinfo {author} {\bibfnamefont {B.~C.}\ \bibnamefont
  {Sawyer}}, \bibinfo {author} {\bibfnamefont {B.~K.}\ \bibnamefont {Stuhl}},
  \bibinfo {author} {\bibfnamefont {D.}~\bibnamefont {Wang}}, \bibinfo {author}
  {\bibfnamefont {M.}~\bibnamefont {Yeo}}, \ and\ \bibinfo {author}
  {\bibfnamefont {J.}~\bibnamefont {Ye}},\ }\href@noop {} {\bibfield  {journal}
  {\bibinfo  {journal} {Physical review letters}\ }\textbf {\bibinfo {volume}
  {101}},\ \bibinfo {pages} {203203} (\bibinfo {year} {2008})}\BibitemShut
  {NoStop}%
\bibitem [{\citenamefont {Strebel}\ \emph {et~al.}(2012)\citenamefont
  {Strebel}, \citenamefont {M{\"u}ller}, \citenamefont {Ruff}, \citenamefont
  {Stienkemeier},\ and\ \citenamefont {Mudrich}}]{strebel2012quantum}%
  \BibitemOpen
  \bibfield  {author} {\bibinfo {author} {\bibfnamefont {M.}~\bibnamefont
  {Strebel}}, \bibinfo {author} {\bibfnamefont {T.-O.}\ \bibnamefont
  {M{\"u}ller}}, \bibinfo {author} {\bibfnamefont {B.}~\bibnamefont {Ruff}},
  \bibinfo {author} {\bibfnamefont {F.}~\bibnamefont {Stienkemeier}}, \ and\
  \bibinfo {author} {\bibfnamefont {M.}~\bibnamefont {Mudrich}},\ }\href@noop
  {} {\bibfield  {journal} {\bibinfo  {journal} {Physical Review A}\ }\textbf
  {\bibinfo {volume} {86}},\ \bibinfo {pages} {062711} (\bibinfo {year}
  {2012})}\BibitemShut {NoStop}%
\bibitem [{\citenamefont {Gupta}\ and\ \citenamefont
  {Herschbach}(2001)}]{gupta2001slowing}%
  \BibitemOpen
  \bibfield  {author} {\bibinfo {author} {\bibfnamefont {M.}~\bibnamefont
  {Gupta}}\ and\ \bibinfo {author} {\bibfnamefont {D.}~\bibnamefont
  {Herschbach}},\ }\href@noop {} {\bibfield  {journal} {\bibinfo  {journal}
  {The Journal of Physical Chemistry A}\ }\textbf {\bibinfo {volume} {105}},\
  \bibinfo {pages} {1626} (\bibinfo {year} {2001})}\BibitemShut {NoStop}%
\bibitem [{\citenamefont {Strebel}\ \emph {et~al.}(2010)\citenamefont
  {Strebel}, \citenamefont {Stienkemeier},\ and\ \citenamefont
  {Mudrich}}]{strebel2010improved}%
  \BibitemOpen
  \bibfield  {author} {\bibinfo {author} {\bibfnamefont {M.}~\bibnamefont
  {Strebel}}, \bibinfo {author} {\bibfnamefont {F.}~\bibnamefont
  {Stienkemeier}}, \ and\ \bibinfo {author} {\bibfnamefont {M.}~\bibnamefont
  {Mudrich}},\ }\href@noop {} {\bibfield  {journal} {\bibinfo  {journal}
  {Physical Review A}\ }\textbf {\bibinfo {volume} {81}},\ \bibinfo {pages}
  {033409} (\bibinfo {year} {2010})}\BibitemShut {NoStop}%
\bibitem [{\citenamefont {Thomas}\ \emph {et~al.}(2004)\citenamefont {Thomas},
  \citenamefont {Kj{\ae}rgaard}, \citenamefont {Julienne},\ and\ \citenamefont
  {Wilson}}]{thomas2004imaging}%
  \BibitemOpen
  \bibfield  {author} {\bibinfo {author} {\bibfnamefont {N.~R.}\ \bibnamefont
  {Thomas}}, \bibinfo {author} {\bibfnamefont {N.}~\bibnamefont
  {Kj{\ae}rgaard}}, \bibinfo {author} {\bibfnamefont {P.~S.}\ \bibnamefont
  {Julienne}}, \ and\ \bibinfo {author} {\bibfnamefont {A.~C.}\ \bibnamefont
  {Wilson}},\ }\href@noop {} {\bibfield  {journal} {\bibinfo  {journal}
  {Physical review letters}\ }\textbf {\bibinfo {volume} {93}},\ \bibinfo
  {pages} {173201} (\bibinfo {year} {2004})}\BibitemShut {NoStop}%
\bibitem [{\citenamefont {Weinstein}\ \emph {et~al.}(1998)\citenamefont
  {Weinstein}, \citenamefont {DeCarvalho}, \citenamefont {Guillet},
  \citenamefont {Friedrich},\ and\ \citenamefont
  {Doyle}}]{weinstein1998magnetic}%
  \BibitemOpen
  \bibfield  {author} {\bibinfo {author} {\bibfnamefont {J.~D.}\ \bibnamefont
  {Weinstein}}, \bibinfo {author} {\bibfnamefont {R.}~\bibnamefont
  {DeCarvalho}}, \bibinfo {author} {\bibfnamefont {T.}~\bibnamefont {Guillet}},
  \bibinfo {author} {\bibfnamefont {B.}~\bibnamefont {Friedrich}}, \ and\
  \bibinfo {author} {\bibfnamefont {J.~M.}\ \bibnamefont {Doyle}},\ }\href@noop
  {} {\bibfield  {journal} {\bibinfo  {journal} {Nature}\ }\textbf {\bibinfo
  {volume} {395}},\ \bibinfo {pages} {148} (\bibinfo {year}
  {1998})}\BibitemShut {NoStop}%
\bibitem [{\citenamefont {Maussang}\ \emph {et~al.}(2005)\citenamefont
  {Maussang}, \citenamefont {Egorov}, \citenamefont {Helton}, \citenamefont
  {Nguyen},\ and\ \citenamefont {Doyle}}]{maussang2005zeeman}%
  \BibitemOpen
  \bibfield  {author} {\bibinfo {author} {\bibfnamefont {K.}~\bibnamefont
  {Maussang}}, \bibinfo {author} {\bibfnamefont {D.}~\bibnamefont {Egorov}},
  \bibinfo {author} {\bibfnamefont {J.~S.}\ \bibnamefont {Helton}}, \bibinfo
  {author} {\bibfnamefont {S.~V.}\ \bibnamefont {Nguyen}}, \ and\ \bibinfo
  {author} {\bibfnamefont {J.~M.}\ \bibnamefont {Doyle}},\ }\href@noop {}
  {\bibfield  {journal} {\bibinfo  {journal} {Physical review letters}\
  }\textbf {\bibinfo {volume} {94}},\ \bibinfo {pages} {123002} (\bibinfo
  {year} {2005})}\BibitemShut {NoStop}%
\bibitem [{\citenamefont {Campbell}\ \emph {et~al.}(2007)\citenamefont
  {Campbell}, \citenamefont {Tsikata}, \citenamefont {Lu}, \citenamefont {van
  Buuren},\ and\ \citenamefont {Doyle}}]{campbell2007magnetic}%
  \BibitemOpen
  \bibfield  {author} {\bibinfo {author} {\bibfnamefont {W.~C.}\ \bibnamefont
  {Campbell}}, \bibinfo {author} {\bibfnamefont {E.}~\bibnamefont {Tsikata}},
  \bibinfo {author} {\bibfnamefont {H.-I.}\ \bibnamefont {Lu}}, \bibinfo
  {author} {\bibfnamefont {L.~D.}\ \bibnamefont {van Buuren}}, \ and\ \bibinfo
  {author} {\bibfnamefont {J.~M.}\ \bibnamefont {Doyle}},\ }\href@noop {}
  {\bibfield  {journal} {\bibinfo  {journal} {Physical review letters}\
  }\textbf {\bibinfo {volume} {98}},\ \bibinfo {pages} {213001} (\bibinfo
  {year} {2007})}\BibitemShut {NoStop}%
\bibitem [{\citenamefont {Tsikata}\ \emph {et~al.}(2010)\citenamefont
  {Tsikata}, \citenamefont {Campbell}, \citenamefont {Hummon}, \citenamefont
  {Lu},\ and\ \citenamefont {Doyle}}]{tsikata2010magnetic}%
  \BibitemOpen
  \bibfield  {author} {\bibinfo {author} {\bibfnamefont {E.}~\bibnamefont
  {Tsikata}}, \bibinfo {author} {\bibfnamefont {W.}~\bibnamefont {Campbell}},
  \bibinfo {author} {\bibfnamefont {M.}~\bibnamefont {Hummon}}, \bibinfo
  {author} {\bibfnamefont {H.-I.}\ \bibnamefont {Lu}}, \ and\ \bibinfo {author}
  {\bibfnamefont {J.~M.}\ \bibnamefont {Doyle}},\ }\href@noop {} {\bibfield
  {journal} {\bibinfo  {journal} {New Journal of Physics}\ }\textbf {\bibinfo
  {volume} {12}},\ \bibinfo {pages} {065028} (\bibinfo {year}
  {2010})}\BibitemShut {NoStop}%
\bibitem [{\citenamefont {Egorov}\ \emph {et~al.}(2004)\citenamefont {Egorov},
  \citenamefont {Campbell}, \citenamefont {Friedrich}, \citenamefont {Maxwell},
  \citenamefont {Tsikata}, \citenamefont {Van~Buuren},\ and\ \citenamefont
  {Doyle}}]{egorov2004buffer}%
  \BibitemOpen
  \bibfield  {author} {\bibinfo {author} {\bibfnamefont {D.}~\bibnamefont
  {Egorov}}, \bibinfo {author} {\bibfnamefont {W.}~\bibnamefont {Campbell}},
  \bibinfo {author} {\bibfnamefont {B.}~\bibnamefont {Friedrich}}, \bibinfo
  {author} {\bibfnamefont {S.}~\bibnamefont {Maxwell}}, \bibinfo {author}
  {\bibfnamefont {E.}~\bibnamefont {Tsikata}}, \bibinfo {author} {\bibfnamefont
  {L.}~\bibnamefont {Van~Buuren}}, \ and\ \bibinfo {author} {\bibfnamefont
  {J.}~\bibnamefont {Doyle}},\ }\href@noop {} {\bibfield  {journal} {\bibinfo
  {journal} {The European Physical Journal D-Atomic, Molecular, Optical and
  Plasma Physics}\ }\textbf {\bibinfo {volume} {31}},\ \bibinfo {pages} {307}
  (\bibinfo {year} {2004})}\BibitemShut {NoStop}%
\bibitem [{\citenamefont {Hummon}\ \emph {et~al.}(2011)\citenamefont {Hummon},
  \citenamefont {Tscherbul}, \citenamefont {K{\l}os}, \citenamefont {Lu},
  \citenamefont {Tsikata}, \citenamefont {Campbell}, \citenamefont {Dalgarno},\
  and\ \citenamefont {Doyle}}]{hummon2011cold}%
  \BibitemOpen
  \bibfield  {author} {\bibinfo {author} {\bibfnamefont {M.~T.}\ \bibnamefont
  {Hummon}}, \bibinfo {author} {\bibfnamefont {T.~V.}\ \bibnamefont
  {Tscherbul}}, \bibinfo {author} {\bibfnamefont {J.}~\bibnamefont {K{\l}os}},
  \bibinfo {author} {\bibfnamefont {H.-I.}\ \bibnamefont {Lu}}, \bibinfo
  {author} {\bibfnamefont {E.}~\bibnamefont {Tsikata}}, \bibinfo {author}
  {\bibfnamefont {W.~C.}\ \bibnamefont {Campbell}}, \bibinfo {author}
  {\bibfnamefont {A.}~\bibnamefont {Dalgarno}}, \ and\ \bibinfo {author}
  {\bibfnamefont {J.~M.}\ \bibnamefont {Doyle}},\ }\href@noop {} {\bibfield
  {journal} {\bibinfo  {journal} {Physical review letters}\ }\textbf {\bibinfo
  {volume} {106}},\ \bibinfo {pages} {053201} (\bibinfo {year}
  {2011})}\BibitemShut {NoStop}%
\bibitem [{\citenamefont {Brahms}\ \emph {et~al.}(2011)\citenamefont {Brahms},
  \citenamefont {Tscherbul}, \citenamefont {Zhang}, \citenamefont {K{\l}os},
  \citenamefont {Forrey}, \citenamefont {Au}, \citenamefont {Sadeghpour},
  \citenamefont {Dalgarno}, \citenamefont {Doyle},\ and\ \citenamefont
  {Walker}}]{brahms2011formation}%
  \BibitemOpen
  \bibfield  {author} {\bibinfo {author} {\bibfnamefont {N.}~\bibnamefont
  {Brahms}}, \bibinfo {author} {\bibfnamefont {T.~V.}\ \bibnamefont
  {Tscherbul}}, \bibinfo {author} {\bibfnamefont {P.}~\bibnamefont {Zhang}},
  \bibinfo {author} {\bibfnamefont {J.}~\bibnamefont {K{\l}os}}, \bibinfo
  {author} {\bibfnamefont {R.~C.}\ \bibnamefont {Forrey}}, \bibinfo {author}
  {\bibfnamefont {Y.~S.}\ \bibnamefont {Au}}, \bibinfo {author} {\bibfnamefont
  {H.~R.}\ \bibnamefont {Sadeghpour}}, \bibinfo {author} {\bibfnamefont
  {A.}~\bibnamefont {Dalgarno}}, \bibinfo {author} {\bibfnamefont {J.~M.}\
  \bibnamefont {Doyle}}, \ and\ \bibinfo {author} {\bibfnamefont {T.~G.}\
  \bibnamefont {Walker}},\ }\href@noop {} {\bibfield  {journal} {\bibinfo
  {journal} {Physical Chemistry Chemical Physics}\ }\textbf {\bibinfo {volume}
  {13}},\ \bibinfo {pages} {19125} (\bibinfo {year} {2011})}\BibitemShut
  {NoStop}%
\bibitem [{\citenamefont {Brahms}\ \emph {et~al.}(2010)\citenamefont {Brahms},
  \citenamefont {Tscherbul}, \citenamefont {Zhang}, \citenamefont {K{\l}os},
  \citenamefont {Sadeghpour}, \citenamefont {Dalgarno}, \citenamefont {Doyle},\
  and\ \citenamefont {Walker}}]{brahms2010formation}%
  \BibitemOpen
  \bibfield  {author} {\bibinfo {author} {\bibfnamefont {N.}~\bibnamefont
  {Brahms}}, \bibinfo {author} {\bibfnamefont {T.~V.}\ \bibnamefont
  {Tscherbul}}, \bibinfo {author} {\bibfnamefont {P.}~\bibnamefont {Zhang}},
  \bibinfo {author} {\bibfnamefont {J.}~\bibnamefont {K{\l}os}}, \bibinfo
  {author} {\bibfnamefont {H.~R.}\ \bibnamefont {Sadeghpour}}, \bibinfo
  {author} {\bibfnamefont {A.}~\bibnamefont {Dalgarno}}, \bibinfo {author}
  {\bibfnamefont {J.~M.}\ \bibnamefont {Doyle}}, \ and\ \bibinfo {author}
  {\bibfnamefont {T.~G.}\ \bibnamefont {Walker}},\ }\href@noop {} {\bibfield
  {journal} {\bibinfo  {journal} {Physical review letters}\ }\textbf {\bibinfo
  {volume} {105}},\ \bibinfo {pages} {033001} (\bibinfo {year}
  {2010})}\BibitemShut {NoStop}%
\bibitem [{\citenamefont {Tariq}\ \emph {et~al.}(2013)\citenamefont {Tariq},
  \citenamefont {Al~Taisan}, \citenamefont {Singh},\ and\ \citenamefont
  {Weinstein}}]{tariq2013spectroscopic}%
  \BibitemOpen
  \bibfield  {author} {\bibinfo {author} {\bibfnamefont {N.}~\bibnamefont
  {Tariq}}, \bibinfo {author} {\bibfnamefont {N.}~\bibnamefont {Al~Taisan}},
  \bibinfo {author} {\bibfnamefont {V.}~\bibnamefont {Singh}}, \ and\ \bibinfo
  {author} {\bibfnamefont {J.~D.}\ \bibnamefont {Weinstein}},\ }\href@noop {}
  {\bibfield  {journal} {\bibinfo  {journal} {Physical review letters}\
  }\textbf {\bibinfo {volume} {110}},\ \bibinfo {pages} {153201} (\bibinfo
  {year} {2013})}\BibitemShut {NoStop}%
\bibitem [{\citenamefont {Quiros}\ \emph {et~al.}(2017)\citenamefont {Quiros},
  \citenamefont {Tariq}, \citenamefont {Tscherbul}, \citenamefont {K{\l}os},\
  and\ \citenamefont {Weinstein}}]{quiros2017cold}%
  \BibitemOpen
  \bibfield  {author} {\bibinfo {author} {\bibfnamefont {N.}~\bibnamefont
  {Quiros}}, \bibinfo {author} {\bibfnamefont {N.}~\bibnamefont {Tariq}},
  \bibinfo {author} {\bibfnamefont {T.~V.}\ \bibnamefont {Tscherbul}}, \bibinfo
  {author} {\bibfnamefont {J.}~\bibnamefont {K{\l}os}}, \ and\ \bibinfo
  {author} {\bibfnamefont {J.~D.}\ \bibnamefont {Weinstein}},\ }\href@noop {}
  {\bibfield  {journal} {\bibinfo  {journal} {Physical review letters}\
  }\textbf {\bibinfo {volume} {118}},\ \bibinfo {pages} {213401} (\bibinfo
  {year} {2017})}\BibitemShut {NoStop}%
\bibitem [{\citenamefont {{Fabrikant}}\ \emph {et~al.}(2014)\citenamefont
  {{Fabrikant}}, \citenamefont {{Li}}, \citenamefont {{Fitch}}, \citenamefont
  {{Farrow}}, \citenamefont {{Weinstein}},\ and\ \citenamefont
  {{Lewandowski}}}]{fabrikant2014}%
  \BibitemOpen
  \bibfield  {author} {\bibinfo {author} {\bibfnamefont {M.~I.}\ \bibnamefont
  {{Fabrikant}}}, \bibinfo {author} {\bibfnamefont {T.}~\bibnamefont {{Li}}},
  \bibinfo {author} {\bibfnamefont {N.~J.}\ \bibnamefont {{Fitch}}}, \bibinfo
  {author} {\bibfnamefont {N.}~\bibnamefont {{Farrow}}}, \bibinfo {author}
  {\bibfnamefont {J.~D.}\ \bibnamefont {{Weinstein}}}, \ and\ \bibinfo {author}
  {\bibfnamefont {H.~J.}\ \bibnamefont {{Lewandowski}}},\ }\href {\doibase
  10.1103/PhysRevA.90.033418} {\bibfield  {journal} {\bibinfo  {journal}
  {\pra}\ }\textbf {\bibinfo {volume} {90}},\ \bibinfo {eid} {033418} (\bibinfo
  {year} {2014})}\BibitemShut {NoStop}%
\bibitem [{\citenamefont {{Petzold}}\ \emph {et~al.}(2018)\citenamefont
  {{Petzold}}, \citenamefont {{Kaebert}}, \citenamefont {{Gersema}},
  \citenamefont {{Siercke}},\ and\ \citenamefont {{Ospelkaus}}}]{petzold2018}%
  \BibitemOpen
  \bibfield  {author} {\bibinfo {author} {\bibfnamefont {M.}~\bibnamefont
  {{Petzold}}}, \bibinfo {author} {\bibfnamefont {P.}~\bibnamefont
  {{Kaebert}}}, \bibinfo {author} {\bibfnamefont {P.}~\bibnamefont
  {{Gersema}}}, \bibinfo {author} {\bibfnamefont {M.}~\bibnamefont
  {{Siercke}}}, \ and\ \bibinfo {author} {\bibfnamefont {S.}~\bibnamefont
  {{Ospelkaus}}},\ }\href {\doibase 10.1088/1367-2630/aab9f5} {\bibfield
  {journal} {\bibinfo  {journal} {New Journal of Physics}\ }\textbf {\bibinfo
  {volume} {20}},\ \bibinfo {eid} {042001} (\bibinfo {year}
  {2018})}\BibitemShut {NoStop}%
\bibitem [{\citenamefont {Shuman}\ \emph {et~al.}(2009)\citenamefont {Shuman},
  \citenamefont {Barry}, \citenamefont {Glenn},\ and\ \citenamefont
  {DeMille}}]{shuman2009radiative}%
  \BibitemOpen
  \bibfield  {author} {\bibinfo {author} {\bibfnamefont {E.}~\bibnamefont
  {Shuman}}, \bibinfo {author} {\bibfnamefont {J.}~\bibnamefont {Barry}},
  \bibinfo {author} {\bibfnamefont {D.}~\bibnamefont {Glenn}}, \ and\ \bibinfo
  {author} {\bibfnamefont {D.}~\bibnamefont {DeMille}},\ }\href@noop {}
  {\bibfield  {journal} {\bibinfo  {journal} {Physical review letters}\
  }\textbf {\bibinfo {volume} {103}},\ \bibinfo {pages} {223001} (\bibinfo
  {year} {2009})}\BibitemShut {NoStop}%
\bibitem [{\citenamefont {Shuman}\ \emph {et~al.}(2010)\citenamefont {Shuman},
  \citenamefont {Barry},\ and\ \citenamefont {DeMille}}]{shuman2010laser}%
  \BibitemOpen
  \bibfield  {author} {\bibinfo {author} {\bibfnamefont {E.~S.}\ \bibnamefont
  {Shuman}}, \bibinfo {author} {\bibfnamefont {J.~F.}\ \bibnamefont {Barry}}, \
  and\ \bibinfo {author} {\bibfnamefont {D.}~\bibnamefont {DeMille}},\
  }\href@noop {} {\bibfield  {journal} {\bibinfo  {journal} {Nature}\ }\textbf
  {\bibinfo {volume} {467}},\ \bibinfo {pages} {820} (\bibinfo {year}
  {2010})}\BibitemShut {NoStop}%
\bibitem [{\citenamefont {Zeppenfeld}\ \emph {et~al.}(2012)\citenamefont
  {Zeppenfeld}, \citenamefont {Englert}, \citenamefont {Gl{\"o}ckner},
  \citenamefont {Prehn}, \citenamefont {Mielenz}, \citenamefont {Sommer},
  \citenamefont {van Buuren}, \citenamefont {Motsch},\ and\ \citenamefont
  {Rempe}}]{zeppenfeld2012sisyphus}%
  \BibitemOpen
  \bibfield  {author} {\bibinfo {author} {\bibfnamefont {M.}~\bibnamefont
  {Zeppenfeld}}, \bibinfo {author} {\bibfnamefont {B.~G.}\ \bibnamefont
  {Englert}}, \bibinfo {author} {\bibfnamefont {R.}~\bibnamefont
  {Gl{\"o}ckner}}, \bibinfo {author} {\bibfnamefont {A.}~\bibnamefont {Prehn}},
  \bibinfo {author} {\bibfnamefont {M.}~\bibnamefont {Mielenz}}, \bibinfo
  {author} {\bibfnamefont {C.}~\bibnamefont {Sommer}}, \bibinfo {author}
  {\bibfnamefont {L.~D.}\ \bibnamefont {van Buuren}}, \bibinfo {author}
  {\bibfnamefont {M.}~\bibnamefont {Motsch}}, \ and\ \bibinfo {author}
  {\bibfnamefont {G.}~\bibnamefont {Rempe}},\ }\href@noop {} {\bibfield
  {journal} {\bibinfo  {journal} {Nature}\ }\textbf {\bibinfo {volume} {491}},\
  \bibinfo {pages} {570} (\bibinfo {year} {2012})}\BibitemShut {NoStop}%
\bibitem [{\citenamefont {{Prehn}}\ \emph {et~al.}(2016)\citenamefont
  {{Prehn}}, \citenamefont {{Ibr{\"u}gger}}, \citenamefont {{Gl{\"o}ckner}},
  \citenamefont {{Rempe}},\ and\ \citenamefont {{Zeppenfeld}}}]{prehn2016}%
  \BibitemOpen
  \bibfield  {author} {\bibinfo {author} {\bibfnamefont {A.}~\bibnamefont
  {{Prehn}}}, \bibinfo {author} {\bibfnamefont {M.}~\bibnamefont
  {{Ibr{\"u}gger}}}, \bibinfo {author} {\bibfnamefont {R.}~\bibnamefont
  {{Gl{\"o}ckner}}}, \bibinfo {author} {\bibfnamefont {G.}~\bibnamefont
  {{Rempe}}}, \ and\ \bibinfo {author} {\bibfnamefont {M.}~\bibnamefont
  {{Zeppenfeld}}},\ }\href {\doibase 10.1103/PhysRevLett.116.063005} {\bibfield
   {journal} {\bibinfo  {journal} {\prl}\ }\textbf {\bibinfo {volume} {116}},\
  \bibinfo {eid} {063005} (\bibinfo {year} {2016})}\BibitemShut {NoStop}%
\bibitem [{\citenamefont {Patterson}\ \emph {et~al.}(2013)\citenamefont
  {Patterson}, \citenamefont {Schnell},\ and\ \citenamefont
  {Doyle}}]{patterson2013enantiomer}%
  \BibitemOpen
  \bibfield  {author} {\bibinfo {author} {\bibfnamefont {D.}~\bibnamefont
  {Patterson}}, \bibinfo {author} {\bibfnamefont {M.}~\bibnamefont {Schnell}},
  \ and\ \bibinfo {author} {\bibfnamefont {J.~M.}\ \bibnamefont {Doyle}},\
  }\href@noop {} {\bibfield  {journal} {\bibinfo  {journal} {Nature}\ }\textbf
  {\bibinfo {volume} {497}},\ \bibinfo {pages} {475} (\bibinfo {year}
  {2013})}\BibitemShut {NoStop}%
\bibitem [{\citenamefont {Eibenberger}\ \emph {et~al.}(2017)\citenamefont
  {Eibenberger}, \citenamefont {Doyle},\ and\ \citenamefont
  {Patterson}}]{eibenberger2017enantiomer}%
  \BibitemOpen
  \bibfield  {author} {\bibinfo {author} {\bibfnamefont {S.}~\bibnamefont
  {Eibenberger}}, \bibinfo {author} {\bibfnamefont {J.}~\bibnamefont {Doyle}},
  \ and\ \bibinfo {author} {\bibfnamefont {D.}~\bibnamefont {Patterson}},\
  }\href@noop {} {\bibfield  {journal} {\bibinfo  {journal} {Physical review
  letters}\ }\textbf {\bibinfo {volume} {118}},\ \bibinfo {pages} {123002}
  (\bibinfo {year} {2017})}\BibitemShut {NoStop}%
\bibitem [{\citenamefont {Spaun}\ \emph {et~al.}(2016)\citenamefont {Spaun},
  \citenamefont {Changala}, \citenamefont {Patterson}, \citenamefont {Bjork},
  \citenamefont {Heckl}, \citenamefont {Doyle},\ and\ \citenamefont
  {Ye}}]{spaun2016continuous}%
  \BibitemOpen
  \bibfield  {author} {\bibinfo {author} {\bibfnamefont {B.}~\bibnamefont
  {Spaun}}, \bibinfo {author} {\bibfnamefont {P.~B.}\ \bibnamefont {Changala}},
  \bibinfo {author} {\bibfnamefont {D.}~\bibnamefont {Patterson}}, \bibinfo
  {author} {\bibfnamefont {B.~J.}\ \bibnamefont {Bjork}}, \bibinfo {author}
  {\bibfnamefont {O.~H.}\ \bibnamefont {Heckl}}, \bibinfo {author}
  {\bibfnamefont {J.~M.}\ \bibnamefont {Doyle}}, \ and\ \bibinfo {author}
  {\bibfnamefont {J.}~\bibnamefont {Ye}},\ }\href@noop {} {\bibfield  {journal}
  {\bibinfo  {journal} {Nature}\ }\textbf {\bibinfo {volume} {533}},\ \bibinfo
  {pages} {517} (\bibinfo {year} {2016})}\BibitemShut {NoStop}%
\bibitem [{\citenamefont {Changala}\ \emph {et~al.}(2019)\citenamefont
  {Changala}, \citenamefont {Weichman}, \citenamefont {Lee}, \citenamefont
  {Fermann},\ and\ \citenamefont {Ye}}]{changala2019rovibrational}%
  \BibitemOpen
  \bibfield  {author} {\bibinfo {author} {\bibfnamefont {P.~B.}\ \bibnamefont
  {Changala}}, \bibinfo {author} {\bibfnamefont {M.~L.}\ \bibnamefont
  {Weichman}}, \bibinfo {author} {\bibfnamefont {K.~F.}\ \bibnamefont {Lee}},
  \bibinfo {author} {\bibfnamefont {M.~E.}\ \bibnamefont {Fermann}}, \ and\
  \bibinfo {author} {\bibfnamefont {J.}~\bibnamefont {Ye}},\ }\href@noop {}
  {\bibfield  {journal} {\bibinfo  {journal} {Science}\ }\textbf {\bibinfo
  {volume} {363}},\ \bibinfo {pages} {49} (\bibinfo {year} {2019})}\BibitemShut
  {NoStop}%
\bibitem [{\citenamefont {{Baron}}\ \emph {et~al.}(2014)\citenamefont
  {{Baron}}, \citenamefont {{Campbell}}, \citenamefont {{DeMille}},
  \citenamefont {{Doyle}}, \citenamefont {{Gabrielse}}, \citenamefont
  {{Gurevich}}, \citenamefont {{Hess}}, \citenamefont {{Hutzler}},
  \citenamefont {{Kirilov}}, \citenamefont {{Kozyryev}}, \citenamefont
  {{O'Leary}}, \citenamefont {{Panda}}, \citenamefont {{Parsons}},
  \citenamefont {{Petrik}}, \citenamefont {{Spaun}}, \citenamefont {{Vutha}},\
  and\ \citenamefont {{West}}}]{baron2014}%
  \BibitemOpen
  \bibfield  {author} {\bibinfo {author} {\bibfnamefont {J.}~\bibnamefont
  {{Baron}}}, \bibinfo {author} {\bibfnamefont {W.~C.}\ \bibnamefont
  {{Campbell}}}, \bibinfo {author} {\bibfnamefont {D.}~\bibnamefont
  {{DeMille}}}, \bibinfo {author} {\bibfnamefont {J.~M.}\ \bibnamefont
  {{Doyle}}}, \bibinfo {author} {\bibfnamefont {G.}~\bibnamefont
  {{Gabrielse}}}, \bibinfo {author} {\bibfnamefont {Y.~V.}\ \bibnamefont
  {{Gurevich}}}, \bibinfo {author} {\bibfnamefont {P.~W.}\ \bibnamefont
  {{Hess}}}, \bibinfo {author} {\bibfnamefont {N.~R.}\ \bibnamefont
  {{Hutzler}}}, \bibinfo {author} {\bibfnamefont {E.}~\bibnamefont
  {{Kirilov}}}, \bibinfo {author} {\bibfnamefont {I.}~\bibnamefont
  {{Kozyryev}}}, \bibinfo {author} {\bibfnamefont {B.~R.}\ \bibnamefont
  {{O'Leary}}}, \bibinfo {author} {\bibfnamefont {C.~D.}\ \bibnamefont
  {{Panda}}}, \bibinfo {author} {\bibfnamefont {M.~F.}\ \bibnamefont
  {{Parsons}}}, \bibinfo {author} {\bibfnamefont {E.~S.}\ \bibnamefont
  {{Petrik}}}, \bibinfo {author} {\bibfnamefont {B.}~\bibnamefont {{Spaun}}},
  \bibinfo {author} {\bibfnamefont {A.~C.}\ \bibnamefont {{Vutha}}}, \ and\
  \bibinfo {author} {\bibfnamefont {A.~D.}\ \bibnamefont {{West}}},\ }\href
  {\doibase 10.1126/science.1248213} {\bibfield  {journal} {\bibinfo  {journal}
  {Science}\ }\textbf {\bibinfo {volume} {343}},\ \bibinfo {pages} {269}
  (\bibinfo {year} {2014})}\BibitemShut {NoStop}%
\bibitem [{\citenamefont {{ACME Collaboration}}\ \emph
  {et~al.}(2018)\citenamefont {{ACME Collaboration}}, \citenamefont
  {{Andreev}}, \citenamefont {{Ang}}, \citenamefont {{DeMille}}, \citenamefont
  {{Doyle}}, \citenamefont {{Gabrielse}}, \citenamefont {{Haefner}},
  \citenamefont {{Hutzler}}, \citenamefont {{Lasner}}, \citenamefont
  {{Meisenhelder}}, \citenamefont {{O'Leary}}, \citenamefont {{Panda}},
  \citenamefont {{West}}, \citenamefont {{West}},\ and\ \citenamefont
  {{Wu}}}]{ACME2018}%
  \BibitemOpen
  \bibfield  {author} {\bibinfo {author} {\bibnamefont {{ACME Collaboration}}},
  \bibinfo {author} {\bibfnamefont {V.}~\bibnamefont {{Andreev}}}, \bibinfo
  {author} {\bibfnamefont {D.~G.}\ \bibnamefont {{Ang}}}, \bibinfo {author}
  {\bibfnamefont {D.}~\bibnamefont {{DeMille}}}, \bibinfo {author}
  {\bibfnamefont {J.~M.}\ \bibnamefont {{Doyle}}}, \bibinfo {author}
  {\bibfnamefont {G.}~\bibnamefont {{Gabrielse}}}, \bibinfo {author}
  {\bibfnamefont {J.}~\bibnamefont {{Haefner}}}, \bibinfo {author}
  {\bibfnamefont {N.~R.}\ \bibnamefont {{Hutzler}}}, \bibinfo {author}
  {\bibfnamefont {Z.}~\bibnamefont {{Lasner}}}, \bibinfo {author}
  {\bibfnamefont {C.}~\bibnamefont {{Meisenhelder}}}, \bibinfo {author}
  {\bibfnamefont {B.~R.}\ \bibnamefont {{O'Leary}}}, \bibinfo {author}
  {\bibfnamefont {C.~D.}\ \bibnamefont {{Panda}}}, \bibinfo {author}
  {\bibfnamefont {A.~D.}\ \bibnamefont {{West}}}, \bibinfo {author}
  {\bibfnamefont {E.~P.}\ \bibnamefont {{West}}}, \ and\ \bibinfo {author}
  {\bibfnamefont {X.}~\bibnamefont {{Wu}}},\ }\href {\doibase
  10.1038/s41586-018-0599-8} {\bibfield  {journal} {\bibinfo  {journal} {\nat}\
  }\textbf {\bibinfo {volume} {562}},\ \bibinfo {pages} {355} (\bibinfo {year}
  {2018})}\BibitemShut {NoStop}%
\bibitem [{\citenamefont {Chen}(2019)}]{chen2019new}%
  \BibitemOpen
  \bibfield  {author} {\bibinfo {author} {\bibfnamefont {G.}~\bibnamefont
  {Chen}},\ }\emph {\bibinfo {title} {A New Tool For Col Ion-Molecule
  Chemistry}},\ \href@noop {} {Ph.D. thesis},\ \bibinfo  {school} {UCLA}
  (\bibinfo {year} {2019})\BibitemShut {NoStop}%
\bibitem [{\citenamefont {Greenberg}(2020)}]{greenberg2020cold}%
  \BibitemOpen
  \bibfield  {author} {\bibinfo {author} {\bibfnamefont {J.}~\bibnamefont
  {Greenberg}},\ }\emph {\bibinfo {title} {Cold, Controlled, Ion-molecule
  Reactions}},\ \href@noop {} {Ph.D. thesis},\ \bibinfo  {school} {University
  of Colorado at Boulder} (\bibinfo {year} {2020})\BibitemShut {NoStop}%
\bibitem [{\citenamefont {Wei}\ \emph {et~al.}(2012)\citenamefont {Wei},
  \citenamefont {Lyuksyutov},\ and\ \citenamefont
  {Herschbach}}]{wei2012merged}%
  \BibitemOpen
  \bibfield  {author} {\bibinfo {author} {\bibfnamefont {Q.}~\bibnamefont
  {Wei}}, \bibinfo {author} {\bibfnamefont {I.}~\bibnamefont {Lyuksyutov}}, \
  and\ \bibinfo {author} {\bibfnamefont {D.}~\bibnamefont {Herschbach}},\
  }\href@noop {} {\bibfield  {journal} {\bibinfo  {journal} {The Journal of
  chemical physics}\ }\textbf {\bibinfo {volume} {137}},\ \bibinfo {pages}
  {054202} (\bibinfo {year} {2012})}\BibitemShut {NoStop}%
\bibitem [{\citenamefont {Lavert-Ofir}\ \emph {et~al.}(2014)\citenamefont
  {Lavert-Ofir}, \citenamefont {Shagam}, \citenamefont {Henson}, \citenamefont
  {Gersten}, \citenamefont {K{\l}os}, \citenamefont {{\.Z}uchowski},
  \citenamefont {Narevicius},\ and\ \citenamefont
  {Narevicius}}]{lavert2014observation}%
  \BibitemOpen
  \bibfield  {author} {\bibinfo {author} {\bibfnamefont {E.}~\bibnamefont
  {Lavert-Ofir}}, \bibinfo {author} {\bibfnamefont {Y.}~\bibnamefont {Shagam}},
  \bibinfo {author} {\bibfnamefont {A.~B.}\ \bibnamefont {Henson}}, \bibinfo
  {author} {\bibfnamefont {S.}~\bibnamefont {Gersten}}, \bibinfo {author}
  {\bibfnamefont {J.}~\bibnamefont {K{\l}os}}, \bibinfo {author} {\bibfnamefont
  {P.~S.}\ \bibnamefont {{\.Z}uchowski}}, \bibinfo {author} {\bibfnamefont
  {J.}~\bibnamefont {Narevicius}}, \ and\ \bibinfo {author} {\bibfnamefont
  {E.}~\bibnamefont {Narevicius}},\ }\href@noop {} {\bibfield  {journal}
  {\bibinfo  {journal} {Nature chemistry}\ }\textbf {\bibinfo {volume} {6}},\
  \bibinfo {pages} {332} (\bibinfo {year} {2014})}\BibitemShut {NoStop}%
\bibitem [{\citenamefont {Shagam}\ \emph
  {et~al.}(2015{\natexlab{b}})\citenamefont {Shagam}, \citenamefont {Klein},
  \citenamefont {Skomorowski}, \citenamefont {Yun}, \citenamefont {Averbukh},
  \citenamefont {Koch},\ and\ \citenamefont {Narevicius}}]{Shagam2015}%
  \BibitemOpen
  \bibfield  {author} {\bibinfo {author} {\bibfnamefont {Y.}~\bibnamefont
  {Shagam}}, \bibinfo {author} {\bibfnamefont {A.}~\bibnamefont {Klein}},
  \bibinfo {author} {\bibfnamefont {W.}~\bibnamefont {Skomorowski}}, \bibinfo
  {author} {\bibfnamefont {R.}~\bibnamefont {Yun}}, \bibinfo {author}
  {\bibfnamefont {V.}~\bibnamefont {Averbukh}}, \bibinfo {author}
  {\bibfnamefont {C.~P.}\ \bibnamefont {Koch}}, \ and\ \bibinfo {author}
  {\bibfnamefont {E.}~\bibnamefont {Narevicius}},\ }\href {\doibase
  10.1038/nchem.2359} {\bibfield  {journal} {\bibinfo  {journal} {Nature
  Chemistry}\ }\textbf {\bibinfo {volume} {7}},\ \bibinfo {pages} {921}
  (\bibinfo {year} {2015}{\natexlab{b}})}\BibitemShut {NoStop}%
\bibitem [{\citenamefont {Bibelnik}\ \emph {et~al.}(2019)\citenamefont
  {Bibelnik}, \citenamefont {Gersten}, \citenamefont {Henson}, \citenamefont
  {Lavert-Ofir}, \citenamefont {Shagam}, \citenamefont {Skomorowski},
  \citenamefont {Koch},\ and\ \citenamefont {Narevicius}}]{bibelnik2019cold}%
  \BibitemOpen
  \bibfield  {author} {\bibinfo {author} {\bibfnamefont {N.}~\bibnamefont
  {Bibelnik}}, \bibinfo {author} {\bibfnamefont {S.}~\bibnamefont {Gersten}},
  \bibinfo {author} {\bibfnamefont {A.~B.}\ \bibnamefont {Henson}}, \bibinfo
  {author} {\bibfnamefont {E.}~\bibnamefont {Lavert-Ofir}}, \bibinfo {author}
  {\bibfnamefont {Y.}~\bibnamefont {Shagam}}, \bibinfo {author} {\bibfnamefont
  {W.}~\bibnamefont {Skomorowski}}, \bibinfo {author} {\bibfnamefont {C.~P.}\
  \bibnamefont {Koch}}, \ and\ \bibinfo {author} {\bibfnamefont
  {E.}~\bibnamefont {Narevicius}},\ }\href@noop {} {\bibfield  {journal}
  {\bibinfo  {journal} {Molecular Physics}\ }\textbf {\bibinfo {volume}
  {117}},\ \bibinfo {pages} {2128} (\bibinfo {year} {2019})}\BibitemShut
  {NoStop}%
\bibitem [{\citenamefont {Bertsche}\ \emph {et~al.}(2014)\citenamefont
  {Bertsche}, \citenamefont {Jankunas},\ and\ \citenamefont
  {Osterwalder}}]{bertsche2014low}%
  \BibitemOpen
  \bibfield  {author} {\bibinfo {author} {\bibfnamefont {B.}~\bibnamefont
  {Bertsche}}, \bibinfo {author} {\bibfnamefont {J.}~\bibnamefont {Jankunas}},
  \ and\ \bibinfo {author} {\bibfnamefont {A.}~\bibnamefont {Osterwalder}},\
  }\href@noop {} {\bibfield  {journal} {\bibinfo  {journal} {CHIMIA
  International Journal for Chemistry}\ }\textbf {\bibinfo {volume} {68}},\
  \bibinfo {pages} {256} (\bibinfo {year} {2014})}\BibitemShut {NoStop}%
\bibitem [{\citenamefont {van~de Meerakker}\ and\ \citenamefont
  {Meijer}(2009)}]{van2009collision}%
  \BibitemOpen
  \bibfield  {author} {\bibinfo {author} {\bibfnamefont {S.~Y.}\ \bibnamefont
  {van~de Meerakker}}\ and\ \bibinfo {author} {\bibfnamefont {G.}~\bibnamefont
  {Meijer}},\ }\href@noop {} {\bibfield  {journal} {\bibinfo  {journal}
  {Faraday Discussions}\ }\textbf {\bibinfo {volume} {142}},\ \bibinfo {pages}
  {113} (\bibinfo {year} {2009})}\BibitemShut {NoStop}%
\bibitem [{\citenamefont {van~der Poel}\ and\ \citenamefont
  {Bethlem}(2018)}]{van2018detailed}%
  \BibitemOpen
  \bibfield  {author} {\bibinfo {author} {\bibfnamefont {A.~P.}\ \bibnamefont
  {van~der Poel}}\ and\ \bibinfo {author} {\bibfnamefont {H.~L.}\ \bibnamefont
  {Bethlem}},\ }\href@noop {} {\bibfield  {journal} {\bibinfo  {journal} {EPJ
  techniques and instrumentation}\ }\textbf {\bibinfo {volume} {5}},\ \bibinfo
  {pages} {6} (\bibinfo {year} {2018})}\BibitemShut {NoStop}%
\bibitem [{\citenamefont {Van Der~Poel}\ \emph {et~al.}(2018)\citenamefont {Van
  Der~Poel}, \citenamefont {Zieger}, \citenamefont {Van De~Meerakker},
  \citenamefont {Loreau}, \citenamefont {Van Der~Avoird},\ and\ \citenamefont
  {Bethlem}}]{van2018cold}%
  \BibitemOpen
  \bibfield  {author} {\bibinfo {author} {\bibfnamefont {A.~P.}\ \bibnamefont
  {Van Der~Poel}}, \bibinfo {author} {\bibfnamefont {P.~C.}\ \bibnamefont
  {Zieger}}, \bibinfo {author} {\bibfnamefont {S.~Y.}\ \bibnamefont {Van
  De~Meerakker}}, \bibinfo {author} {\bibfnamefont {J.}~\bibnamefont {Loreau}},
  \bibinfo {author} {\bibfnamefont {A.}~\bibnamefont {Van Der~Avoird}}, \ and\
  \bibinfo {author} {\bibfnamefont {H.~L.}\ \bibnamefont {Bethlem}},\
  }\href@noop {} {\bibfield  {journal} {\bibinfo  {journal} {Physical review
  letters}\ }\textbf {\bibinfo {volume} {120}},\ \bibinfo {pages} {033402}
  (\bibinfo {year} {2018})}\BibitemShut {NoStop}%
\bibitem [{\citenamefont {Heiner}\ \emph {et~al.}(2007)\citenamefont {Heiner},
  \citenamefont {Carty}, \citenamefont {Meijer},\ and\ \citenamefont
  {Bethlem}}]{heiner2007molecular}%
  \BibitemOpen
  \bibfield  {author} {\bibinfo {author} {\bibfnamefont {C.~E.}\ \bibnamefont
  {Heiner}}, \bibinfo {author} {\bibfnamefont {D.}~\bibnamefont {Carty}},
  \bibinfo {author} {\bibfnamefont {G.}~\bibnamefont {Meijer}}, \ and\ \bibinfo
  {author} {\bibfnamefont {H.~L.}\ \bibnamefont {Bethlem}},\ }\href@noop {}
  {\bibfield  {journal} {\bibinfo  {journal} {Nature Physics}\ }\textbf
  {\bibinfo {volume} {3}},\ \bibinfo {pages} {115} (\bibinfo {year}
  {2007})}\BibitemShut {NoStop}%
\bibitem [{\citenamefont {Heiner}\ \emph {et~al.}(2009)\citenamefont {Heiner},
  \citenamefont {Bethlem},\ and\ \citenamefont
  {Meijer}}]{heiner2009synchrotron}%
  \BibitemOpen
  \bibfield  {author} {\bibinfo {author} {\bibfnamefont {C.~E.}\ \bibnamefont
  {Heiner}}, \bibinfo {author} {\bibfnamefont {H.~L.}\ \bibnamefont {Bethlem}},
  \ and\ \bibinfo {author} {\bibfnamefont {G.}~\bibnamefont {Meijer}},\
  }\href@noop {} {\bibfield  {journal} {\bibinfo  {journal} {Chemical Physics
  Letters}\ }\textbf {\bibinfo {volume} {473}},\ \bibinfo {pages} {1} (\bibinfo
  {year} {2009})}\BibitemShut {NoStop}%
\bibitem [{\citenamefont {Crompvoets}\ \emph {et~al.}(2001)\citenamefont
  {Crompvoets}, \citenamefont {Bethlem}, \citenamefont {Jongma},\ and\
  \citenamefont {Meijer}}]{crompvoets2001prototype}%
  \BibitemOpen
  \bibfield  {author} {\bibinfo {author} {\bibfnamefont {F.~M.}\ \bibnamefont
  {Crompvoets}}, \bibinfo {author} {\bibfnamefont {H.~L.}\ \bibnamefont
  {Bethlem}}, \bibinfo {author} {\bibfnamefont {R.~T.}\ \bibnamefont {Jongma}},
  \ and\ \bibinfo {author} {\bibfnamefont {G.}~\bibnamefont {Meijer}},\
  }\href@noop {} {\bibfield  {journal} {\bibinfo  {journal} {Nature}\ }\textbf
  {\bibinfo {volume} {411}},\ \bibinfo {pages} {174} (\bibinfo {year}
  {2001})}\BibitemShut {NoStop}%
\bibitem [{\citenamefont {Crompvoets}\ \emph {et~al.}(2004)\citenamefont
  {Crompvoets}, \citenamefont {Bethlem}, \citenamefont {K{\"u}pper},
  \citenamefont {van Roij},\ and\ \citenamefont
  {Meijer}}]{crompvoets2004dynamics}%
  \BibitemOpen
  \bibfield  {author} {\bibinfo {author} {\bibfnamefont {F.~M.}\ \bibnamefont
  {Crompvoets}}, \bibinfo {author} {\bibfnamefont {H.~L.}\ \bibnamefont
  {Bethlem}}, \bibinfo {author} {\bibfnamefont {J.}~\bibnamefont {K{\"u}pper}},
  \bibinfo {author} {\bibfnamefont {A.~J.}\ \bibnamefont {van Roij}}, \ and\
  \bibinfo {author} {\bibfnamefont {G.}~\bibnamefont {Meijer}},\ }\href@noop {}
  {\bibfield  {journal} {\bibinfo  {journal} {Physical Review A}\ }\textbf
  {\bibinfo {volume} {69}},\ \bibinfo {pages} {063406} (\bibinfo {year}
  {2004})}\BibitemShut {NoStop}%
\bibitem [{\citenamefont {Zieger}\ \emph {et~al.}(2010)\citenamefont {Zieger},
  \citenamefont {van~de Meerakker}, \citenamefont {Heiner}, \citenamefont
  {Bethlem}, \citenamefont {van Roij},\ and\ \citenamefont
  {Meijer}}]{zieger2010multiple}%
  \BibitemOpen
  \bibfield  {author} {\bibinfo {author} {\bibfnamefont {P.~C.}\ \bibnamefont
  {Zieger}}, \bibinfo {author} {\bibfnamefont {S.~Y.}\ \bibnamefont {van~de
  Meerakker}}, \bibinfo {author} {\bibfnamefont {C.~E.}\ \bibnamefont
  {Heiner}}, \bibinfo {author} {\bibfnamefont {H.~L.}\ \bibnamefont {Bethlem}},
  \bibinfo {author} {\bibfnamefont {A.~J.}\ \bibnamefont {van Roij}}, \ and\
  \bibinfo {author} {\bibfnamefont {G.}~\bibnamefont {Meijer}},\ }\href@noop {}
  {\bibfield  {journal} {\bibinfo  {journal} {Physical Review Letters}\
  }\textbf {\bibinfo {volume} {105}},\ \bibinfo {pages} {173001} (\bibinfo
  {year} {2010})}\BibitemShut {NoStop}%
\bibitem [{\citenamefont {Loreau}\ and\ \citenamefont {Van~der
  Avoird}(2015)}]{loreau2015scattering}%
  \BibitemOpen
  \bibfield  {author} {\bibinfo {author} {\bibfnamefont {J.}~\bibnamefont
  {Loreau}}\ and\ \bibinfo {author} {\bibfnamefont {A.}~\bibnamefont {Van~der
  Avoird}},\ }\href@noop {} {\bibfield  {journal} {\bibinfo  {journal} {The
  Journal of Chemical Physics}\ }\textbf {\bibinfo {volume} {143}},\ \bibinfo
  {pages} {184303} (\bibinfo {year} {2015})}\BibitemShut {NoStop}%
\bibitem [{\citenamefont {Thorsheim}\ \emph {et~al.}(1987)\citenamefont
  {Thorsheim}, \citenamefont {Weiner},\ and\ \citenamefont
  {Julienne}}]{thorsheim1987laser}%
  \BibitemOpen
  \bibfield  {author} {\bibinfo {author} {\bibfnamefont {H.}~\bibnamefont
  {Thorsheim}}, \bibinfo {author} {\bibfnamefont {J.}~\bibnamefont {Weiner}}, \
  and\ \bibinfo {author} {\bibfnamefont {P.~S.}\ \bibnamefont {Julienne}},\
  }\href@noop {} {\bibfield  {journal} {\bibinfo  {journal} {Physical review
  letters}\ }\textbf {\bibinfo {volume} {58}},\ \bibinfo {pages} {2420}
  (\bibinfo {year} {1987})}\BibitemShut {NoStop}%
\bibitem [{\citenamefont {Deiglmayr}\ \emph {et~al.}(2008)\citenamefont
  {Deiglmayr}, \citenamefont {Grochola}, \citenamefont {Repp}, \citenamefont
  {M{\"o}rtlbauer}, \citenamefont {Gl{\"u}ck}, \citenamefont {Lange},
  \citenamefont {Dulieu}, \citenamefont {Wester},\ and\ \citenamefont
  {Weidem{\"u}ller}}]{deiglmayr2008formation}%
  \BibitemOpen
  \bibfield  {author} {\bibinfo {author} {\bibfnamefont {J.}~\bibnamefont
  {Deiglmayr}}, \bibinfo {author} {\bibfnamefont {A.}~\bibnamefont {Grochola}},
  \bibinfo {author} {\bibfnamefont {M.}~\bibnamefont {Repp}}, \bibinfo {author}
  {\bibfnamefont {K.}~\bibnamefont {M{\"o}rtlbauer}}, \bibinfo {author}
  {\bibfnamefont {C.}~\bibnamefont {Gl{\"u}ck}}, \bibinfo {author}
  {\bibfnamefont {J.}~\bibnamefont {Lange}}, \bibinfo {author} {\bibfnamefont
  {O.}~\bibnamefont {Dulieu}}, \bibinfo {author} {\bibfnamefont
  {R.}~\bibnamefont {Wester}}, \ and\ \bibinfo {author} {\bibfnamefont
  {M.}~\bibnamefont {Weidem{\"u}ller}},\ }\href@noop {} {\bibfield  {journal}
  {\bibinfo  {journal} {Physical review letters}\ }\textbf {\bibinfo {volume}
  {101}},\ \bibinfo {pages} {133004} (\bibinfo {year} {2008})}\BibitemShut
  {NoStop}%
\bibitem [{\citenamefont {Lang}\ \emph {et~al.}(2008)\citenamefont {Lang},
  \citenamefont {Winkler}, \citenamefont {Strauss}, \citenamefont {Grimm},\
  and\ \citenamefont {Denschlag}}]{lang2008ultracold}%
  \BibitemOpen
  \bibfield  {author} {\bibinfo {author} {\bibfnamefont {F.}~\bibnamefont
  {Lang}}, \bibinfo {author} {\bibfnamefont {K.}~\bibnamefont {Winkler}},
  \bibinfo {author} {\bibfnamefont {C.}~\bibnamefont {Strauss}}, \bibinfo
  {author} {\bibfnamefont {R.}~\bibnamefont {Grimm}}, \ and\ \bibinfo {author}
  {\bibfnamefont {J.~H.}\ \bibnamefont {Denschlag}},\ }\href@noop {} {\bibfield
   {journal} {\bibinfo  {journal} {Physical Review Letters}\ }\textbf {\bibinfo
  {volume} {101}},\ \bibinfo {pages} {133005} (\bibinfo {year}
  {2008})}\BibitemShut {NoStop}%
\bibitem [{\citenamefont {Ni}\ \emph {et~al.}(2008)\citenamefont {Ni},
  \citenamefont {Ospelkaus}, \citenamefont {De~Miranda}, \citenamefont {Pe'Er},
  \citenamefont {Neyenhuis}, \citenamefont {Zirbel}, \citenamefont
  {Kotochigova}, \citenamefont {Julienne}, \citenamefont {Jin},\ and\
  \citenamefont {Ye}}]{ni2008high}%
  \BibitemOpen
  \bibfield  {author} {\bibinfo {author} {\bibfnamefont {K.-K.}\ \bibnamefont
  {Ni}}, \bibinfo {author} {\bibfnamefont {S.}~\bibnamefont {Ospelkaus}},
  \bibinfo {author} {\bibfnamefont {M.}~\bibnamefont {De~Miranda}}, \bibinfo
  {author} {\bibfnamefont {A.}~\bibnamefont {Pe'Er}}, \bibinfo {author}
  {\bibfnamefont {B.}~\bibnamefont {Neyenhuis}}, \bibinfo {author}
  {\bibfnamefont {J.}~\bibnamefont {Zirbel}}, \bibinfo {author} {\bibfnamefont
  {S.}~\bibnamefont {Kotochigova}}, \bibinfo {author} {\bibfnamefont
  {P.}~\bibnamefont {Julienne}}, \bibinfo {author} {\bibfnamefont
  {D.}~\bibnamefont {Jin}}, \ and\ \bibinfo {author} {\bibfnamefont
  {J.}~\bibnamefont {Ye}},\ }\href@noop {} {\bibfield  {journal} {\bibinfo
  {journal} {science}\ }\textbf {\bibinfo {volume} {322}},\ \bibinfo {pages}
  {231} (\bibinfo {year} {2008})}\BibitemShut {NoStop}%
\bibitem [{\citenamefont {Hutson}\ and\ \citenamefont
  {Soldan}(2006)}]{hutson2006molecule}%
  \BibitemOpen
  \bibfield  {author} {\bibinfo {author} {\bibfnamefont {J.~M.}\ \bibnamefont
  {Hutson}}\ and\ \bibinfo {author} {\bibfnamefont {P.}~\bibnamefont
  {Soldan}},\ }\href@noop {} {\bibfield  {journal} {\bibinfo  {journal}
  {International Reviews in Physical Chemistry}\ }\textbf {\bibinfo {volume}
  {25}},\ \bibinfo {pages} {497} (\bibinfo {year} {2006})}\BibitemShut
  {NoStop}%
\bibitem [{\citenamefont {K{\"o}hler}\ \emph {et~al.}(2006)\citenamefont
  {K{\"o}hler}, \citenamefont {G{\'o}ral},\ and\ \citenamefont
  {Julienne}}]{kohler2006production}%
  \BibitemOpen
  \bibfield  {author} {\bibinfo {author} {\bibfnamefont {T.}~\bibnamefont
  {K{\"o}hler}}, \bibinfo {author} {\bibfnamefont {K.}~\bibnamefont
  {G{\'o}ral}}, \ and\ \bibinfo {author} {\bibfnamefont {P.~S.}\ \bibnamefont
  {Julienne}},\ }\href@noop {} {\bibfield  {journal} {\bibinfo  {journal}
  {Reviews of modern physics}\ }\textbf {\bibinfo {volume} {78}},\ \bibinfo
  {pages} {1311} (\bibinfo {year} {2006})}\BibitemShut {NoStop}%
\bibitem [{\citenamefont {Donley}\ \emph {et~al.}(2002)\citenamefont {Donley},
  \citenamefont {Claussen}, \citenamefont {Thompson},\ and\ \citenamefont
  {Wieman}}]{donley2002atom}%
  \BibitemOpen
  \bibfield  {author} {\bibinfo {author} {\bibfnamefont {E.~A.}\ \bibnamefont
  {Donley}}, \bibinfo {author} {\bibfnamefont {N.~R.}\ \bibnamefont
  {Claussen}}, \bibinfo {author} {\bibfnamefont {S.~T.}\ \bibnamefont
  {Thompson}}, \ and\ \bibinfo {author} {\bibfnamefont {C.~E.}\ \bibnamefont
  {Wieman}},\ }\href@noop {} {\bibfield  {journal} {\bibinfo  {journal}
  {Nature}\ }\textbf {\bibinfo {volume} {417}},\ \bibinfo {pages} {529}
  (\bibinfo {year} {2002})}\BibitemShut {NoStop}%
\bibitem [{\citenamefont {Chin}\ \emph {et~al.}(2003)\citenamefont {Chin},
  \citenamefont {Kerman}, \citenamefont {Vuleti{\'c}},\ and\ \citenamefont
  {Chu}}]{chin2003sensitive}%
  \BibitemOpen
  \bibfield  {author} {\bibinfo {author} {\bibfnamefont {C.}~\bibnamefont
  {Chin}}, \bibinfo {author} {\bibfnamefont {A.~J.}\ \bibnamefont {Kerman}},
  \bibinfo {author} {\bibfnamefont {V.}~\bibnamefont {Vuleti{\'c}}}, \ and\
  \bibinfo {author} {\bibfnamefont {S.}~\bibnamefont {Chu}},\ }\href@noop {}
  {\bibfield  {journal} {\bibinfo  {journal} {Physical review letters}\
  }\textbf {\bibinfo {volume} {90}},\ \bibinfo {pages} {033201} (\bibinfo
  {year} {2003})}\BibitemShut {NoStop}%
\bibitem [{\citenamefont {Herbig}\ \emph {et~al.}(2003)\citenamefont {Herbig},
  \citenamefont {Kraemer}, \citenamefont {Mark}, \citenamefont {Weber},
  \citenamefont {Chin}, \citenamefont {N{\"a}gerl},\ and\ \citenamefont
  {Grimm}}]{herbig2003preparation}%
  \BibitemOpen
  \bibfield  {author} {\bibinfo {author} {\bibfnamefont {J.}~\bibnamefont
  {Herbig}}, \bibinfo {author} {\bibfnamefont {T.}~\bibnamefont {Kraemer}},
  \bibinfo {author} {\bibfnamefont {M.}~\bibnamefont {Mark}}, \bibinfo {author}
  {\bibfnamefont {T.}~\bibnamefont {Weber}}, \bibinfo {author} {\bibfnamefont
  {C.}~\bibnamefont {Chin}}, \bibinfo {author} {\bibfnamefont {H.-C.}\
  \bibnamefont {N{\"a}gerl}}, \ and\ \bibinfo {author} {\bibfnamefont
  {R.}~\bibnamefont {Grimm}},\ }\href@noop {} {\bibfield  {journal} {\bibinfo
  {journal} {Science}\ }\textbf {\bibinfo {volume} {301}},\ \bibinfo {pages}
  {1510} (\bibinfo {year} {2003})}\BibitemShut {NoStop}%
\bibitem [{\citenamefont {D{\"u}rr}\ \emph {et~al.}(2004)\citenamefont
  {D{\"u}rr}, \citenamefont {Volz}, \citenamefont {Marte},\ and\ \citenamefont
  {Rempe}}]{durr2004observation}%
  \BibitemOpen
  \bibfield  {author} {\bibinfo {author} {\bibfnamefont {S.}~\bibnamefont
  {D{\"u}rr}}, \bibinfo {author} {\bibfnamefont {T.}~\bibnamefont {Volz}},
  \bibinfo {author} {\bibfnamefont {A.}~\bibnamefont {Marte}}, \ and\ \bibinfo
  {author} {\bibfnamefont {G.}~\bibnamefont {Rempe}},\ }\href@noop {}
  {\bibfield  {journal} {\bibinfo  {journal} {Physical review letters}\
  }\textbf {\bibinfo {volume} {92}},\ \bibinfo {pages} {020406} (\bibinfo
  {year} {2004})}\BibitemShut {NoStop}%
\bibitem [{\citenamefont {Regal}\ \emph {et~al.}(2003)\citenamefont {Regal},
  \citenamefont {Ticknor}, \citenamefont {Bohn},\ and\ \citenamefont
  {Jin}}]{regal2003creation}%
  \BibitemOpen
  \bibfield  {author} {\bibinfo {author} {\bibfnamefont {C.~A.}\ \bibnamefont
  {Regal}}, \bibinfo {author} {\bibfnamefont {C.}~\bibnamefont {Ticknor}},
  \bibinfo {author} {\bibfnamefont {J.~L.}\ \bibnamefont {Bohn}}, \ and\
  \bibinfo {author} {\bibfnamefont {D.~S.}\ \bibnamefont {Jin}},\ }\href@noop
  {} {\bibfield  {journal} {\bibinfo  {journal} {Nature}\ }\textbf {\bibinfo
  {volume} {424}},\ \bibinfo {pages} {47} (\bibinfo {year} {2003})}\BibitemShut
  {NoStop}%
\bibitem [{\citenamefont {Strecker}\ \emph {et~al.}(2003)\citenamefont
  {Strecker}, \citenamefont {Partridge},\ and\ \citenamefont
  {Hulet}}]{strecker2003conversion}%
  \BibitemOpen
  \bibfield  {author} {\bibinfo {author} {\bibfnamefont {K.~E.}\ \bibnamefont
  {Strecker}}, \bibinfo {author} {\bibfnamefont {G.~B.}\ \bibnamefont
  {Partridge}}, \ and\ \bibinfo {author} {\bibfnamefont {R.~G.}\ \bibnamefont
  {Hulet}},\ }\href@noop {} {\bibfield  {journal} {\bibinfo  {journal}
  {Physical review letters}\ }\textbf {\bibinfo {volume} {91}},\ \bibinfo
  {pages} {080406} (\bibinfo {year} {2003})}\BibitemShut {NoStop}%
\bibitem [{\citenamefont {Cubizolles}\ \emph {et~al.}(2003)\citenamefont
  {Cubizolles}, \citenamefont {Bourdel}, \citenamefont {Kokkelmans},
  \citenamefont {Shlyapnikov},\ and\ \citenamefont
  {Salomon}}]{cubizolles2003production}%
  \BibitemOpen
  \bibfield  {author} {\bibinfo {author} {\bibfnamefont {J.}~\bibnamefont
  {Cubizolles}}, \bibinfo {author} {\bibfnamefont {T.}~\bibnamefont {Bourdel}},
  \bibinfo {author} {\bibfnamefont {S.}~\bibnamefont {Kokkelmans}}, \bibinfo
  {author} {\bibfnamefont {G.}~\bibnamefont {Shlyapnikov}}, \ and\ \bibinfo
  {author} {\bibfnamefont {C.}~\bibnamefont {Salomon}},\ }\href@noop {}
  {\bibfield  {journal} {\bibinfo  {journal} {Physical review letters}\
  }\textbf {\bibinfo {volume} {91}},\ \bibinfo {pages} {240401} (\bibinfo
  {year} {2003})}\BibitemShut {NoStop}%
\bibitem [{\citenamefont {Regal}\ \emph {et~al.}(2004)\citenamefont {Regal},
  \citenamefont {Greiner},\ and\ \citenamefont {Jin}}]{regal2004lifetime}%
  \BibitemOpen
  \bibfield  {author} {\bibinfo {author} {\bibfnamefont {C.}~\bibnamefont
  {Regal}}, \bibinfo {author} {\bibfnamefont {M.}~\bibnamefont {Greiner}}, \
  and\ \bibinfo {author} {\bibfnamefont {D.}~\bibnamefont {Jin}},\ }\href@noop
  {} {\bibfield  {journal} {\bibinfo  {journal} {Physical review letters}\
  }\textbf {\bibinfo {volume} {92}},\ \bibinfo {pages} {083201} (\bibinfo
  {year} {2004})}\BibitemShut {NoStop}%
\bibitem [{\citenamefont {Jochim}\ \emph
  {et~al.}(2003{\natexlab{a}})\citenamefont {Jochim}, \citenamefont
  {Bartenstein}, \citenamefont {Altmeyer}, \citenamefont {Hendl}, \citenamefont
  {Chin}, \citenamefont {Denschlag},\ and\ \citenamefont
  {Grimm}}]{jochim2003pure}%
  \BibitemOpen
  \bibfield  {author} {\bibinfo {author} {\bibfnamefont {S.}~\bibnamefont
  {Jochim}}, \bibinfo {author} {\bibfnamefont {M.}~\bibnamefont {Bartenstein}},
  \bibinfo {author} {\bibfnamefont {A.}~\bibnamefont {Altmeyer}}, \bibinfo
  {author} {\bibfnamefont {G.}~\bibnamefont {Hendl}}, \bibinfo {author}
  {\bibfnamefont {C.}~\bibnamefont {Chin}}, \bibinfo {author} {\bibfnamefont
  {J.~H.}\ \bibnamefont {Denschlag}}, \ and\ \bibinfo {author} {\bibfnamefont
  {R.}~\bibnamefont {Grimm}},\ }\href@noop {} {\bibfield  {journal} {\bibinfo
  {journal} {Physical review letters}\ }\textbf {\bibinfo {volume} {91}},\
  \bibinfo {pages} {240402} (\bibinfo {year} {2003}{\natexlab{a}})}\BibitemShut
  {NoStop}%
\bibitem [{\citenamefont {Petrov}\ \emph {et~al.}(2004)\citenamefont {Petrov},
  \citenamefont {Salomon},\ and\ \citenamefont
  {Shlyapnikov}}]{petrov2004weakly}%
  \BibitemOpen
  \bibfield  {author} {\bibinfo {author} {\bibfnamefont {D.}~\bibnamefont
  {Petrov}}, \bibinfo {author} {\bibfnamefont {C.}~\bibnamefont {Salomon}}, \
  and\ \bibinfo {author} {\bibfnamefont {G.~V.}\ \bibnamefont {Shlyapnikov}},\
  }\href@noop {} {\bibfield  {journal} {\bibinfo  {journal} {Physical Review
  Letters}\ }\textbf {\bibinfo {volume} {93}},\ \bibinfo {pages} {090404}
  (\bibinfo {year} {2004})}\BibitemShut {NoStop}%
\bibitem [{\citenamefont {Greiner}\ \emph {et~al.}(2003)\citenamefont
  {Greiner}, \citenamefont {Regal},\ and\ \citenamefont
  {Jin}}]{greiner2003emergence}%
  \BibitemOpen
  \bibfield  {author} {\bibinfo {author} {\bibfnamefont {M.}~\bibnamefont
  {Greiner}}, \bibinfo {author} {\bibfnamefont {C.~A.}\ \bibnamefont {Regal}},
  \ and\ \bibinfo {author} {\bibfnamefont {D.~S.}\ \bibnamefont {Jin}},\
  }\href@noop {} {\bibfield  {journal} {\bibinfo  {journal} {Nature}\ }\textbf
  {\bibinfo {volume} {426}},\ \bibinfo {pages} {537} (\bibinfo {year}
  {2003})}\BibitemShut {NoStop}%
\bibitem [{\citenamefont {Jochim}\ \emph
  {et~al.}(2003{\natexlab{b}})\citenamefont {Jochim}, \citenamefont
  {Bartenstein}, \citenamefont {Altmeyer}, \citenamefont {Hendl}, \citenamefont
  {Riedl}, \citenamefont {Chin}, \citenamefont {Denschlag},\ and\ \citenamefont
  {Grimm}}]{jochim2003bose}%
  \BibitemOpen
  \bibfield  {author} {\bibinfo {author} {\bibfnamefont {S.}~\bibnamefont
  {Jochim}}, \bibinfo {author} {\bibfnamefont {M.}~\bibnamefont {Bartenstein}},
  \bibinfo {author} {\bibfnamefont {A.}~\bibnamefont {Altmeyer}}, \bibinfo
  {author} {\bibfnamefont {G.}~\bibnamefont {Hendl}}, \bibinfo {author}
  {\bibfnamefont {S.}~\bibnamefont {Riedl}}, \bibinfo {author} {\bibfnamefont
  {C.}~\bibnamefont {Chin}}, \bibinfo {author} {\bibfnamefont {J.~H.}\
  \bibnamefont {Denschlag}}, \ and\ \bibinfo {author} {\bibfnamefont
  {R.}~\bibnamefont {Grimm}},\ }\href@noop {} {\bibfield  {journal} {\bibinfo
  {journal} {Science}\ }\textbf {\bibinfo {volume} {302}},\ \bibinfo {pages}
  {2101} (\bibinfo {year} {2003}{\natexlab{b}})}\BibitemShut {NoStop}%
\bibitem [{\citenamefont {Zwierlein}\ \emph {et~al.}(2003)\citenamefont
  {Zwierlein}, \citenamefont {Stan}, \citenamefont {Schunck}, \citenamefont
  {Raupach}, \citenamefont {Gupta}, \citenamefont {Hadzibabic},\ and\
  \citenamefont {Ketterle}}]{zwierlein2003observation}%
  \BibitemOpen
  \bibfield  {author} {\bibinfo {author} {\bibfnamefont {M.~W.}\ \bibnamefont
  {Zwierlein}}, \bibinfo {author} {\bibfnamefont {C.~A.}\ \bibnamefont {Stan}},
  \bibinfo {author} {\bibfnamefont {C.~H.}\ \bibnamefont {Schunck}}, \bibinfo
  {author} {\bibfnamefont {S.~M.}\ \bibnamefont {Raupach}}, \bibinfo {author}
  {\bibfnamefont {S.}~\bibnamefont {Gupta}}, \bibinfo {author} {\bibfnamefont
  {Z.}~\bibnamefont {Hadzibabic}}, \ and\ \bibinfo {author} {\bibfnamefont
  {W.}~\bibnamefont {Ketterle}},\ }\href@noop {} {\bibfield  {journal}
  {\bibinfo  {journal} {Physical review letters}\ }\textbf {\bibinfo {volume}
  {91}},\ \bibinfo {pages} {250401} (\bibinfo {year} {2003})}\BibitemShut
  {NoStop}%
\bibitem [{\citenamefont {Wang}\ \emph
  {et~al.}(2013{\natexlab{b}})\citenamefont {Wang}, \citenamefont {Heo},
  \citenamefont {Rvachov}, \citenamefont {Cotta},\ and\ \citenamefont
  {Ketterle}}]{wang2013deviation}%
  \BibitemOpen
  \bibfield  {author} {\bibinfo {author} {\bibfnamefont {T.~T.}\ \bibnamefont
  {Wang}}, \bibinfo {author} {\bibfnamefont {M.-S.}\ \bibnamefont {Heo}},
  \bibinfo {author} {\bibfnamefont {T.~M.}\ \bibnamefont {Rvachov}}, \bibinfo
  {author} {\bibfnamefont {D.~A.}\ \bibnamefont {Cotta}}, \ and\ \bibinfo
  {author} {\bibfnamefont {W.}~\bibnamefont {Ketterle}},\ }\href@noop {}
  {\bibfield  {journal} {\bibinfo  {journal} {Physical review letters}\
  }\textbf {\bibinfo {volume} {110}},\ \bibinfo {pages} {173203} (\bibinfo
  {year} {2013}{\natexlab{b}})}\BibitemShut {NoStop}%
\bibitem [{\citenamefont {Xu}\ \emph {et~al.}(2003)\citenamefont {Xu},
  \citenamefont {Mukaiyama}, \citenamefont {Abo-Shaeer}, \citenamefont {Chin},
  \citenamefont {Miller},\ and\ \citenamefont {Ketterle}}]{xu2003formation}%
  \BibitemOpen
  \bibfield  {author} {\bibinfo {author} {\bibfnamefont {K.}~\bibnamefont
  {Xu}}, \bibinfo {author} {\bibfnamefont {T.}~\bibnamefont {Mukaiyama}},
  \bibinfo {author} {\bibfnamefont {J.}~\bibnamefont {Abo-Shaeer}}, \bibinfo
  {author} {\bibfnamefont {J.~K.}\ \bibnamefont {Chin}}, \bibinfo {author}
  {\bibfnamefont {D.}~\bibnamefont {Miller}}, \ and\ \bibinfo {author}
  {\bibfnamefont {W.}~\bibnamefont {Ketterle}},\ }\href@noop {} {\bibfield
  {journal} {\bibinfo  {journal} {Physical review letters}\ }\textbf {\bibinfo
  {volume} {91}},\ \bibinfo {pages} {210402} (\bibinfo {year}
  {2003})}\BibitemShut {NoStop}%
\bibitem [{\citenamefont {Wynar}\ \emph {et~al.}(2000)\citenamefont {Wynar},
  \citenamefont {Freeland}, \citenamefont {Han}, \citenamefont {Ryu},\ and\
  \citenamefont {Heinzen}}]{wynar2000molecules}%
  \BibitemOpen
  \bibfield  {author} {\bibinfo {author} {\bibfnamefont {R.}~\bibnamefont
  {Wynar}}, \bibinfo {author} {\bibfnamefont {R.}~\bibnamefont {Freeland}},
  \bibinfo {author} {\bibfnamefont {D.}~\bibnamefont {Han}}, \bibinfo {author}
  {\bibfnamefont {C.}~\bibnamefont {Ryu}}, \ and\ \bibinfo {author}
  {\bibfnamefont {D.}~\bibnamefont {Heinzen}},\ }\href@noop {} {\bibfield
  {journal} {\bibinfo  {journal} {Science}\ }\textbf {\bibinfo {volume}
  {287}},\ \bibinfo {pages} {1016} (\bibinfo {year} {2000})}\BibitemShut
  {NoStop}%
\bibitem [{\citenamefont {Knoop}\ \emph {et~al.}(2009)\citenamefont {Knoop},
  \citenamefont {Ferlaino}, \citenamefont {Mark}, \citenamefont {Berninger},
  \citenamefont {Sch{\"o}bel}, \citenamefont {N{\"a}gerl},\ and\ \citenamefont
  {Grimm}}]{knoop2009observation}%
  \BibitemOpen
  \bibfield  {author} {\bibinfo {author} {\bibfnamefont {S.}~\bibnamefont
  {Knoop}}, \bibinfo {author} {\bibfnamefont {F.}~\bibnamefont {Ferlaino}},
  \bibinfo {author} {\bibfnamefont {M.}~\bibnamefont {Mark}}, \bibinfo {author}
  {\bibfnamefont {M.}~\bibnamefont {Berninger}}, \bibinfo {author}
  {\bibfnamefont {H.}~\bibnamefont {Sch{\"o}bel}}, \bibinfo {author}
  {\bibfnamefont {H.-C.}\ \bibnamefont {N{\"a}gerl}}, \ and\ \bibinfo {author}
  {\bibfnamefont {R.}~\bibnamefont {Grimm}},\ }\href@noop {} {\bibfield
  {journal} {\bibinfo  {journal} {Nature Physics}\ }\textbf {\bibinfo {volume}
  {5}},\ \bibinfo {pages} {227} (\bibinfo {year} {2009})}\BibitemShut {NoStop}%
\bibitem [{\citenamefont {Mukaiyama}\ \emph {et~al.}(2004)\citenamefont
  {Mukaiyama}, \citenamefont {Abo-Shaeer}, \citenamefont {Xu}, \citenamefont
  {Chin},\ and\ \citenamefont {Ketterle}}]{mukaiyama2004dissociation}%
  \BibitemOpen
  \bibfield  {author} {\bibinfo {author} {\bibfnamefont {T.}~\bibnamefont
  {Mukaiyama}}, \bibinfo {author} {\bibfnamefont {J.}~\bibnamefont
  {Abo-Shaeer}}, \bibinfo {author} {\bibfnamefont {K.}~\bibnamefont {Xu}},
  \bibinfo {author} {\bibfnamefont {J.~K.}\ \bibnamefont {Chin}}, \ and\
  \bibinfo {author} {\bibfnamefont {W.}~\bibnamefont {Ketterle}},\ }\href@noop
  {} {\bibfield  {journal} {\bibinfo  {journal} {Physical review letters}\
  }\textbf {\bibinfo {volume} {92}},\ \bibinfo {pages} {180402} (\bibinfo
  {year} {2004})}\BibitemShut {NoStop}%
\bibitem [{\citenamefont {Zahzam}\ \emph {et~al.}(2006)\citenamefont {Zahzam},
  \citenamefont {Vogt}, \citenamefont {Mudrich}, \citenamefont {Comparat},\
  and\ \citenamefont {Pillet}}]{zahzam2006atom}%
  \BibitemOpen
  \bibfield  {author} {\bibinfo {author} {\bibfnamefont {N.}~\bibnamefont
  {Zahzam}}, \bibinfo {author} {\bibfnamefont {T.}~\bibnamefont {Vogt}},
  \bibinfo {author} {\bibfnamefont {M.}~\bibnamefont {Mudrich}}, \bibinfo
  {author} {\bibfnamefont {D.}~\bibnamefont {Comparat}}, \ and\ \bibinfo
  {author} {\bibfnamefont {P.}~\bibnamefont {Pillet}},\ }\href@noop {}
  {\bibfield  {journal} {\bibinfo  {journal} {Physical review letters}\
  }\textbf {\bibinfo {volume} {96}},\ \bibinfo {pages} {023202} (\bibinfo
  {year} {2006})}\BibitemShut {NoStop}%
\bibitem [{\citenamefont {Staanum}\ \emph {et~al.}(2006)\citenamefont
  {Staanum}, \citenamefont {Kraft}, \citenamefont {Lange}, \citenamefont
  {Wester},\ and\ \citenamefont {Weidem{\"u}ller}}]{staanum2006experimental}%
  \BibitemOpen
  \bibfield  {author} {\bibinfo {author} {\bibfnamefont {P.}~\bibnamefont
  {Staanum}}, \bibinfo {author} {\bibfnamefont {S.~D.}\ \bibnamefont {Kraft}},
  \bibinfo {author} {\bibfnamefont {J.}~\bibnamefont {Lange}}, \bibinfo
  {author} {\bibfnamefont {R.}~\bibnamefont {Wester}}, \ and\ \bibinfo {author}
  {\bibfnamefont {M.}~\bibnamefont {Weidem{\"u}ller}},\ }\href@noop {}
  {\bibfield  {journal} {\bibinfo  {journal} {Physical review letters}\
  }\textbf {\bibinfo {volume} {96}},\ \bibinfo {pages} {023201} (\bibinfo
  {year} {2006})}\BibitemShut {NoStop}%
\bibitem [{\citenamefont {Knoop}\ \emph {et~al.}(2010)\citenamefont {Knoop},
  \citenamefont {Ferlaino}, \citenamefont {Berninger}, \citenamefont {Mark},
  \citenamefont {N{\"a}gerl}, \citenamefont {Grimm}, \citenamefont {D'incao},\
  and\ \citenamefont {Esry}}]{knoop2010magnetically}%
  \BibitemOpen
  \bibfield  {author} {\bibinfo {author} {\bibfnamefont {S.}~\bibnamefont
  {Knoop}}, \bibinfo {author} {\bibfnamefont {F.}~\bibnamefont {Ferlaino}},
  \bibinfo {author} {\bibfnamefont {M.}~\bibnamefont {Berninger}}, \bibinfo
  {author} {\bibfnamefont {M.}~\bibnamefont {Mark}}, \bibinfo {author}
  {\bibfnamefont {H.-C.}\ \bibnamefont {N{\"a}gerl}}, \bibinfo {author}
  {\bibfnamefont {R.}~\bibnamefont {Grimm}}, \bibinfo {author} {\bibfnamefont
  {J.}~\bibnamefont {D'incao}}, \ and\ \bibinfo {author} {\bibfnamefont
  {B.}~\bibnamefont {Esry}},\ }\href@noop {} {\bibfield  {journal} {\bibinfo
  {journal} {Physical review letters}\ }\textbf {\bibinfo {volume} {104}},\
  \bibinfo {pages} {053201} (\bibinfo {year} {2010})}\BibitemShut {NoStop}%
\bibitem [{\citenamefont {Zenesini}\ \emph {et~al.}(2014)\citenamefont
  {Zenesini}, \citenamefont {Huang}, \citenamefont {Berninger}, \citenamefont
  {N{\"a}gerl}, \citenamefont {Ferlaino},\ and\ \citenamefont
  {Grimm}}]{zenesini2014resonant}%
  \BibitemOpen
  \bibfield  {author} {\bibinfo {author} {\bibfnamefont {A.}~\bibnamefont
  {Zenesini}}, \bibinfo {author} {\bibfnamefont {B.}~\bibnamefont {Huang}},
  \bibinfo {author} {\bibfnamefont {M.}~\bibnamefont {Berninger}}, \bibinfo
  {author} {\bibfnamefont {H.-C.}\ \bibnamefont {N{\"a}gerl}}, \bibinfo
  {author} {\bibfnamefont {F.}~\bibnamefont {Ferlaino}}, \ and\ \bibinfo
  {author} {\bibfnamefont {R.}~\bibnamefont {Grimm}},\ }\href@noop {}
  {\bibfield  {journal} {\bibinfo  {journal} {Physical Review A}\ }\textbf
  {\bibinfo {volume} {90}},\ \bibinfo {pages} {022704} (\bibinfo {year}
  {2014})}\BibitemShut {NoStop}%
\bibitem [{\citenamefont {Hudson}\ \emph {et~al.}(2008)\citenamefont {Hudson},
  \citenamefont {Gilfoy}, \citenamefont {Kotochigova}, \citenamefont {Sage},\
  and\ \citenamefont {DeMille}}]{hudson2008inelastic}%
  \BibitemOpen
  \bibfield  {author} {\bibinfo {author} {\bibfnamefont {E.~R.}\ \bibnamefont
  {Hudson}}, \bibinfo {author} {\bibfnamefont {N.~B.}\ \bibnamefont {Gilfoy}},
  \bibinfo {author} {\bibfnamefont {S.}~\bibnamefont {Kotochigova}}, \bibinfo
  {author} {\bibfnamefont {J.~M.}\ \bibnamefont {Sage}}, \ and\ \bibinfo
  {author} {\bibfnamefont {D.}~\bibnamefont {DeMille}},\ }\href@noop {}
  {\bibfield  {journal} {\bibinfo  {journal} {Physical review letters}\
  }\textbf {\bibinfo {volume} {100}},\ \bibinfo {pages} {203201} (\bibinfo
  {year} {2008})}\BibitemShut {NoStop}%
\bibitem [{\citenamefont {Zirbel}\ \emph {et~al.}(2008)\citenamefont {Zirbel},
  \citenamefont {Ni}, \citenamefont {Ospelkaus}, \citenamefont {D'Incao},
  \citenamefont {Wieman}, \citenamefont {Ye},\ and\ \citenamefont
  {Jin}}]{zirbel2008collisional}%
  \BibitemOpen
  \bibfield  {author} {\bibinfo {author} {\bibfnamefont {J.}~\bibnamefont
  {Zirbel}}, \bibinfo {author} {\bibfnamefont {K.-K.}\ \bibnamefont {Ni}},
  \bibinfo {author} {\bibfnamefont {S.}~\bibnamefont {Ospelkaus}}, \bibinfo
  {author} {\bibfnamefont {J.}~\bibnamefont {D'Incao}}, \bibinfo {author}
  {\bibfnamefont {C.}~\bibnamefont {Wieman}}, \bibinfo {author} {\bibfnamefont
  {J.}~\bibnamefont {Ye}}, \ and\ \bibinfo {author} {\bibfnamefont
  {D.}~\bibnamefont {Jin}},\ }\href@noop {} {\bibfield  {journal} {\bibinfo
  {journal} {Physical review letters}\ }\textbf {\bibinfo {volume} {100}},\
  \bibinfo {pages} {143201} (\bibinfo {year} {2008})}\BibitemShut {NoStop}%
\bibitem [{\citenamefont {Deiglmayr}\ \emph {et~al.}(2011)\citenamefont
  {Deiglmayr}, \citenamefont {Repp}, \citenamefont {Wester}, \citenamefont
  {Dulieu},\ and\ \citenamefont {Weidem{\"u}ller}}]{deiglmayr2011inelastic}%
  \BibitemOpen
  \bibfield  {author} {\bibinfo {author} {\bibfnamefont {J.}~\bibnamefont
  {Deiglmayr}}, \bibinfo {author} {\bibfnamefont {M.}~\bibnamefont {Repp}},
  \bibinfo {author} {\bibfnamefont {R.}~\bibnamefont {Wester}}, \bibinfo
  {author} {\bibfnamefont {O.}~\bibnamefont {Dulieu}}, \ and\ \bibinfo {author}
  {\bibfnamefont {M.}~\bibnamefont {Weidem{\"u}ller}},\ }\href@noop {}
  {\bibfield  {journal} {\bibinfo  {journal} {Physical Chemistry Chemical
  Physics}\ }\textbf {\bibinfo {volume} {13}},\ \bibinfo {pages} {19101}
  (\bibinfo {year} {2011})}\BibitemShut {NoStop}%
\bibitem [{\citenamefont {Rui}\ \emph {et~al.}(2017)\citenamefont {Rui},
  \citenamefont {Yang}, \citenamefont {Liu}, \citenamefont {Zhang},
  \citenamefont {Liu}, \citenamefont {Nan}, \citenamefont {Chen}, \citenamefont
  {Zhao},\ and\ \citenamefont {Pan}}]{rui2017controlled}%
  \BibitemOpen
  \bibfield  {author} {\bibinfo {author} {\bibfnamefont {J.}~\bibnamefont
  {Rui}}, \bibinfo {author} {\bibfnamefont {H.}~\bibnamefont {Yang}}, \bibinfo
  {author} {\bibfnamefont {L.}~\bibnamefont {Liu}}, \bibinfo {author}
  {\bibfnamefont {D.-C.}\ \bibnamefont {Zhang}}, \bibinfo {author}
  {\bibfnamefont {Y.-X.}\ \bibnamefont {Liu}}, \bibinfo {author} {\bibfnamefont
  {J.}~\bibnamefont {Nan}}, \bibinfo {author} {\bibfnamefont {Y.-A.}\
  \bibnamefont {Chen}}, \bibinfo {author} {\bibfnamefont {B.}~\bibnamefont
  {Zhao}}, \ and\ \bibinfo {author} {\bibfnamefont {J.-W.}\ \bibnamefont
  {Pan}},\ }\href@noop {} {\bibfield  {journal} {\bibinfo  {journal} {Nature
  Physics}\ }\textbf {\bibinfo {volume} {13}},\ \bibinfo {pages} {699}
  (\bibinfo {year} {2017})}\BibitemShut {NoStop}%
\bibitem [{\citenamefont {Chin}\ \emph {et~al.}(2005)\citenamefont {Chin},
  \citenamefont {Kraemer}, \citenamefont {Mark}, \citenamefont {Herbig},
  \citenamefont {Waldburger}, \citenamefont {N{\"a}gerl},\ and\ \citenamefont
  {Grimm}}]{chin2005observation}%
  \BibitemOpen
  \bibfield  {author} {\bibinfo {author} {\bibfnamefont {C.}~\bibnamefont
  {Chin}}, \bibinfo {author} {\bibfnamefont {T.}~\bibnamefont {Kraemer}},
  \bibinfo {author} {\bibfnamefont {M.}~\bibnamefont {Mark}}, \bibinfo {author}
  {\bibfnamefont {J.}~\bibnamefont {Herbig}}, \bibinfo {author} {\bibfnamefont
  {P.}~\bibnamefont {Waldburger}}, \bibinfo {author} {\bibfnamefont {H.-C.}\
  \bibnamefont {N{\"a}gerl}}, \ and\ \bibinfo {author} {\bibfnamefont
  {R.}~\bibnamefont {Grimm}},\ }\href@noop {} {\bibfield  {journal} {\bibinfo
  {journal} {Physical review letters}\ }\textbf {\bibinfo {volume} {94}},\
  \bibinfo {pages} {123201} (\bibinfo {year} {2005})}\BibitemShut {NoStop}%
\bibitem [{\citenamefont {Ferlaino}\ \emph {et~al.}(2008)\citenamefont
  {Ferlaino}, \citenamefont {Knoop}, \citenamefont {Mark}, \citenamefont
  {Berninger}, \citenamefont {Sch{\"o}bel}, \citenamefont {N{\"a}gerl},\ and\
  \citenamefont {Grimm}}]{ferlaino2008collisions}%
  \BibitemOpen
  \bibfield  {author} {\bibinfo {author} {\bibfnamefont {F.}~\bibnamefont
  {Ferlaino}}, \bibinfo {author} {\bibfnamefont {S.}~\bibnamefont {Knoop}},
  \bibinfo {author} {\bibfnamefont {M.}~\bibnamefont {Mark}}, \bibinfo {author}
  {\bibfnamefont {M.}~\bibnamefont {Berninger}}, \bibinfo {author}
  {\bibfnamefont {H.}~\bibnamefont {Sch{\"o}bel}}, \bibinfo {author}
  {\bibfnamefont {H.-C.}\ \bibnamefont {N{\"a}gerl}}, \ and\ \bibinfo {author}
  {\bibfnamefont {R.}~\bibnamefont {Grimm}},\ }\href@noop {} {\bibfield
  {journal} {\bibinfo  {journal} {Physical review letters}\ }\textbf {\bibinfo
  {volume} {101}},\ \bibinfo {pages} {023201} (\bibinfo {year}
  {2008})}\BibitemShut {NoStop}%
\bibitem [{\citenamefont {Wang}\ \emph {et~al.}(2019)\citenamefont {Wang},
  \citenamefont {Ye}, \citenamefont {Guo}, \citenamefont {Blume},\ and\
  \citenamefont {Wang}}]{wang2019observation}%
  \BibitemOpen
  \bibfield  {author} {\bibinfo {author} {\bibfnamefont {F.}~\bibnamefont
  {Wang}}, \bibinfo {author} {\bibfnamefont {X.}~\bibnamefont {Ye}}, \bibinfo
  {author} {\bibfnamefont {M.}~\bibnamefont {Guo}}, \bibinfo {author}
  {\bibfnamefont {D.}~\bibnamefont {Blume}}, \ and\ \bibinfo {author}
  {\bibfnamefont {D.}~\bibnamefont {Wang}},\ }\href@noop {} {\bibfield
  {journal} {\bibinfo  {journal} {Physical Review A}\ }\textbf {\bibinfo
  {volume} {100}},\ \bibinfo {pages} {042706} (\bibinfo {year}
  {2019})}\BibitemShut {NoStop}%
\bibitem [{\citenamefont {Hoffmann}\ \emph {et~al.}(2018)\citenamefont
  {Hoffmann}, \citenamefont {Paintner}, \citenamefont {Limmer}, \citenamefont
  {Petrov},\ and\ \citenamefont {Denschlag}}]{hoffmann2018reaction}%
  \BibitemOpen
  \bibfield  {author} {\bibinfo {author} {\bibfnamefont {D.~K.}\ \bibnamefont
  {Hoffmann}}, \bibinfo {author} {\bibfnamefont {T.}~\bibnamefont {Paintner}},
  \bibinfo {author} {\bibfnamefont {W.}~\bibnamefont {Limmer}}, \bibinfo
  {author} {\bibfnamefont {D.~S.}\ \bibnamefont {Petrov}}, \ and\ \bibinfo
  {author} {\bibfnamefont {J.~H.}\ \bibnamefont {Denschlag}},\ }\href@noop {}
  {\bibfield  {journal} {\bibinfo  {journal} {Nature communications}\ }\textbf
  {\bibinfo {volume} {9}},\ \bibinfo {pages} {5244} (\bibinfo {year}
  {2018})}\BibitemShut {NoStop}%
\bibitem [{\citenamefont {Takekoshi}\ \emph {et~al.}(2014)\citenamefont
  {Takekoshi}, \citenamefont {Reichs{\"o}llner}, \citenamefont {Schindewolf},
  \citenamefont {Hutson}, \citenamefont {Le~Sueur}, \citenamefont {Dulieu},
  \citenamefont {Ferlaino}, \citenamefont {Grimm},\ and\ \citenamefont
  {N{\"a}gerl}}]{takekoshi2014ultracold}%
  \BibitemOpen
  \bibfield  {author} {\bibinfo {author} {\bibfnamefont {T.}~\bibnamefont
  {Takekoshi}}, \bibinfo {author} {\bibfnamefont {L.}~\bibnamefont
  {Reichs{\"o}llner}}, \bibinfo {author} {\bibfnamefont {A.}~\bibnamefont
  {Schindewolf}}, \bibinfo {author} {\bibfnamefont {J.~M.}\ \bibnamefont
  {Hutson}}, \bibinfo {author} {\bibfnamefont {C.~R.}\ \bibnamefont
  {Le~Sueur}}, \bibinfo {author} {\bibfnamefont {O.}~\bibnamefont {Dulieu}},
  \bibinfo {author} {\bibfnamefont {F.}~\bibnamefont {Ferlaino}}, \bibinfo
  {author} {\bibfnamefont {R.}~\bibnamefont {Grimm}}, \ and\ \bibinfo {author}
  {\bibfnamefont {H.-C.}\ \bibnamefont {N{\"a}gerl}},\ }\href@noop {}
  {\bibfield  {journal} {\bibinfo  {journal} {Physical review letters}\
  }\textbf {\bibinfo {volume} {113}},\ \bibinfo {pages} {205301} (\bibinfo
  {year} {2014})}\BibitemShut {NoStop}%
\bibitem [{\citenamefont {Guo}\ \emph {et~al.}(2018)\citenamefont {Guo},
  \citenamefont {Ye}, \citenamefont {He}, \citenamefont
  {Gonz{\'a}lez-Mart{\'\i}nez}, \citenamefont {Vexiau}, \citenamefont
  {Qu{\'e}m{\'e}ner},\ and\ \citenamefont {Wang}}]{guo2018dipolar}%
  \BibitemOpen
  \bibfield  {author} {\bibinfo {author} {\bibfnamefont {M.}~\bibnamefont
  {Guo}}, \bibinfo {author} {\bibfnamefont {X.}~\bibnamefont {Ye}}, \bibinfo
  {author} {\bibfnamefont {J.}~\bibnamefont {He}}, \bibinfo {author}
  {\bibfnamefont {M.~L.}\ \bibnamefont {Gonz{\'a}lez-Mart{\'\i}nez}}, \bibinfo
  {author} {\bibfnamefont {R.}~\bibnamefont {Vexiau}}, \bibinfo {author}
  {\bibfnamefont {G.}~\bibnamefont {Qu{\'e}m{\'e}ner}}, \ and\ \bibinfo
  {author} {\bibfnamefont {D.}~\bibnamefont {Wang}},\ }\href@noop {} {\bibfield
   {journal} {\bibinfo  {journal} {Physical Review X}\ }\textbf {\bibinfo
  {volume} {8}},\ \bibinfo {pages} {041044} (\bibinfo {year}
  {2018})}\BibitemShut {NoStop}%
\bibitem [{\citenamefont {Yang}\ \emph {et~al.}(2019)\citenamefont {Yang},
  \citenamefont {Zhang}, \citenamefont {Liu}, \citenamefont {Liu},
  \citenamefont {Nan}, \citenamefont {Zhao},\ and\ \citenamefont
  {Pan}}]{yang2019observation}%
  \BibitemOpen
  \bibfield  {author} {\bibinfo {author} {\bibfnamefont {H.}~\bibnamefont
  {Yang}}, \bibinfo {author} {\bibfnamefont {D.-C.}\ \bibnamefont {Zhang}},
  \bibinfo {author} {\bibfnamefont {L.}~\bibnamefont {Liu}}, \bibinfo {author}
  {\bibfnamefont {Y.-X.}\ \bibnamefont {Liu}}, \bibinfo {author} {\bibfnamefont
  {J.}~\bibnamefont {Nan}}, \bibinfo {author} {\bibfnamefont {B.}~\bibnamefont
  {Zhao}}, \ and\ \bibinfo {author} {\bibfnamefont {J.-W.}\ \bibnamefont
  {Pan}},\ }\href@noop {} {\bibfield  {journal} {\bibinfo  {journal} {Science}\
  }\textbf {\bibinfo {volume} {363}},\ \bibinfo {pages} {261} (\bibinfo {year}
  {2019})}\BibitemShut {NoStop}%
\bibitem [{\citenamefont {De~Miranda}\ \emph {et~al.}(2011)\citenamefont
  {De~Miranda}, \citenamefont {Chotia}, \citenamefont {Neyenhuis},
  \citenamefont {Wang}, \citenamefont {Qu{\'e}m{\'e}ner}, \citenamefont
  {Ospelkaus}, \citenamefont {Bohn}, \citenamefont {Ye},\ and\ \citenamefont
  {Jin}}]{de2011controlling}%
  \BibitemOpen
  \bibfield  {author} {\bibinfo {author} {\bibfnamefont {M.}~\bibnamefont
  {De~Miranda}}, \bibinfo {author} {\bibfnamefont {A.}~\bibnamefont {Chotia}},
  \bibinfo {author} {\bibfnamefont {B.}~\bibnamefont {Neyenhuis}}, \bibinfo
  {author} {\bibfnamefont {D.}~\bibnamefont {Wang}}, \bibinfo {author}
  {\bibfnamefont {G.}~\bibnamefont {Qu{\'e}m{\'e}ner}}, \bibinfo {author}
  {\bibfnamefont {S.}~\bibnamefont {Ospelkaus}}, \bibinfo {author}
  {\bibfnamefont {J.}~\bibnamefont {Bohn}}, \bibinfo {author} {\bibfnamefont
  {J.}~\bibnamefont {Ye}}, \ and\ \bibinfo {author} {\bibfnamefont
  {D.}~\bibnamefont {Jin}},\ }\href@noop {} {\bibfield  {journal} {\bibinfo
  {journal} {Nature Physics}\ }\textbf {\bibinfo {volume} {7}},\ \bibinfo
  {pages} {502} (\bibinfo {year} {2011})}\BibitemShut {NoStop}%
\bibitem [{\citenamefont {Danzl}\ \emph {et~al.}(2010)\citenamefont {Danzl},
  \citenamefont {Mark}, \citenamefont {Haller}, \citenamefont {Gustavsson},
  \citenamefont {Hart}, \citenamefont {Aldegunde}, \citenamefont {Hutson},\
  and\ \citenamefont {N{\"a}gerl}}]{danzl2010ultracold}%
  \BibitemOpen
  \bibfield  {author} {\bibinfo {author} {\bibfnamefont {J.~G.}\ \bibnamefont
  {Danzl}}, \bibinfo {author} {\bibfnamefont {M.~J.}\ \bibnamefont {Mark}},
  \bibinfo {author} {\bibfnamefont {E.}~\bibnamefont {Haller}}, \bibinfo
  {author} {\bibfnamefont {M.}~\bibnamefont {Gustavsson}}, \bibinfo {author}
  {\bibfnamefont {R.}~\bibnamefont {Hart}}, \bibinfo {author} {\bibfnamefont
  {J.}~\bibnamefont {Aldegunde}}, \bibinfo {author} {\bibfnamefont {J.~M.}\
  \bibnamefont {Hutson}}, \ and\ \bibinfo {author} {\bibfnamefont {H.-C.}\
  \bibnamefont {N{\"a}gerl}},\ }\href@noop {} {\bibfield  {journal} {\bibinfo
  {journal} {Nature Physics}\ }\textbf {\bibinfo {volume} {6}},\ \bibinfo
  {pages} {265} (\bibinfo {year} {2010})}\BibitemShut {NoStop}%
\bibitem [{\citenamefont {Chotia}\ \emph {et~al.}(2012)\citenamefont {Chotia},
  \citenamefont {Neyenhuis}, \citenamefont {Moses}, \citenamefont {Yan},
  \citenamefont {Covey}, \citenamefont {Foss-Feig}, \citenamefont {Rey},
  \citenamefont {Jin},\ and\ \citenamefont {Ye}}]{chotia2012long}%
  \BibitemOpen
  \bibfield  {author} {\bibinfo {author} {\bibfnamefont {A.}~\bibnamefont
  {Chotia}}, \bibinfo {author} {\bibfnamefont {B.}~\bibnamefont {Neyenhuis}},
  \bibinfo {author} {\bibfnamefont {S.~A.}\ \bibnamefont {Moses}}, \bibinfo
  {author} {\bibfnamefont {B.}~\bibnamefont {Yan}}, \bibinfo {author}
  {\bibfnamefont {J.~P.}\ \bibnamefont {Covey}}, \bibinfo {author}
  {\bibfnamefont {M.}~\bibnamefont {Foss-Feig}}, \bibinfo {author}
  {\bibfnamefont {A.~M.}\ \bibnamefont {Rey}}, \bibinfo {author} {\bibfnamefont
  {D.~S.}\ \bibnamefont {Jin}}, \ and\ \bibinfo {author} {\bibfnamefont
  {J.}~\bibnamefont {Ye}},\ }\href@noop {} {\bibfield  {journal} {\bibinfo
  {journal} {Physical review letters}\ }\textbf {\bibinfo {volume} {108}},\
  \bibinfo {pages} {080405} (\bibinfo {year} {2012})}\BibitemShut {NoStop}%
\bibitem [{\citenamefont {Zhu}\ \emph {et~al.}(2014)\citenamefont {Zhu},
  \citenamefont {Gadway}, \citenamefont {Foss-Feig}, \citenamefont
  {Schachenmayer}, \citenamefont {Wall}, \citenamefont {Hazzard}, \citenamefont
  {Yan}, \citenamefont {Moses}, \citenamefont {Covey}, \citenamefont {Jin}
  \emph {et~al.}}]{zhu2014suppressing}%
  \BibitemOpen
  \bibfield  {author} {\bibinfo {author} {\bibfnamefont {B.}~\bibnamefont
  {Zhu}}, \bibinfo {author} {\bibfnamefont {B.}~\bibnamefont {Gadway}},
  \bibinfo {author} {\bibfnamefont {M.}~\bibnamefont {Foss-Feig}}, \bibinfo
  {author} {\bibfnamefont {J.}~\bibnamefont {Schachenmayer}}, \bibinfo {author}
  {\bibfnamefont {M.}~\bibnamefont {Wall}}, \bibinfo {author} {\bibfnamefont
  {K.~R.}\ \bibnamefont {Hazzard}}, \bibinfo {author} {\bibfnamefont
  {B.}~\bibnamefont {Yan}}, \bibinfo {author} {\bibfnamefont {S.~A.}\
  \bibnamefont {Moses}}, \bibinfo {author} {\bibfnamefont {J.~P.}\ \bibnamefont
  {Covey}}, \bibinfo {author} {\bibfnamefont {D.~S.}\ \bibnamefont {Jin}},
  \emph {et~al.},\ }\href@noop {} {\bibfield  {journal} {\bibinfo  {journal}
  {Physical review letters}\ }\textbf {\bibinfo {volume} {112}},\ \bibinfo
  {pages} {070404} (\bibinfo {year} {2014})}\BibitemShut {NoStop}%
\bibitem [{\citenamefont {Dei{\ss}}\ \emph {et~al.}(2015)\citenamefont
  {Dei{\ss}}, \citenamefont {Drews}, \citenamefont {Denschlag}, \citenamefont
  {Bouloufa-Maafa}, \citenamefont {Vexiau},\ and\ \citenamefont
  {Dulieu}}]{deiss2015polarizability}%
  \BibitemOpen
  \bibfield  {author} {\bibinfo {author} {\bibfnamefont {M.}~\bibnamefont
  {Dei{\ss}}}, \bibinfo {author} {\bibfnamefont {B.}~\bibnamefont {Drews}},
  \bibinfo {author} {\bibfnamefont {J.~H.}\ \bibnamefont {Denschlag}}, \bibinfo
  {author} {\bibfnamefont {N.}~\bibnamefont {Bouloufa-Maafa}}, \bibinfo
  {author} {\bibfnamefont {R.}~\bibnamefont {Vexiau}}, \ and\ \bibinfo {author}
  {\bibfnamefont {O.}~\bibnamefont {Dulieu}},\ }\href@noop {} {\bibfield
  {journal} {\bibinfo  {journal} {New Journal of Physics}\ }\textbf {\bibinfo
  {volume} {17}},\ \bibinfo {pages} {065019} (\bibinfo {year}
  {2015})}\BibitemShut {NoStop}%
\bibitem [{\citenamefont {Barry}\ \emph {et~al.}(2014)\citenamefont {Barry},
  \citenamefont {McCarron}, \citenamefont {Norrgard}, \citenamefont
  {Steinecker},\ and\ \citenamefont {DeMille}}]{barry2014magneto}%
  \BibitemOpen
  \bibfield  {author} {\bibinfo {author} {\bibfnamefont {J.}~\bibnamefont
  {Barry}}, \bibinfo {author} {\bibfnamefont {D.}~\bibnamefont {McCarron}},
  \bibinfo {author} {\bibfnamefont {E.}~\bibnamefont {Norrgard}}, \bibinfo
  {author} {\bibfnamefont {M.}~\bibnamefont {Steinecker}}, \ and\ \bibinfo
  {author} {\bibfnamefont {D.}~\bibnamefont {DeMille}},\ }\href@noop {}
  {\bibfield  {journal} {\bibinfo  {journal} {Nature}\ }\textbf {\bibinfo
  {volume} {512}},\ \bibinfo {pages} {286} (\bibinfo {year}
  {2014})}\BibitemShut {NoStop}%
\bibitem [{\citenamefont {Norrgard}\ \emph {et~al.}(2016)\citenamefont
  {Norrgard}, \citenamefont {McCarron}, \citenamefont {Steinecker},
  \citenamefont {Tarbutt},\ and\ \citenamefont
  {DeMille}}]{norrgard2016submillikelvin}%
  \BibitemOpen
  \bibfield  {author} {\bibinfo {author} {\bibfnamefont {E.}~\bibnamefont
  {Norrgard}}, \bibinfo {author} {\bibfnamefont {D.}~\bibnamefont {McCarron}},
  \bibinfo {author} {\bibfnamefont {M.}~\bibnamefont {Steinecker}}, \bibinfo
  {author} {\bibfnamefont {M.}~\bibnamefont {Tarbutt}}, \ and\ \bibinfo
  {author} {\bibfnamefont {D.}~\bibnamefont {DeMille}},\ }\href@noop {}
  {\bibfield  {journal} {\bibinfo  {journal} {Physical review letters}\
  }\textbf {\bibinfo {volume} {116}},\ \bibinfo {pages} {063004} (\bibinfo
  {year} {2016})}\BibitemShut {NoStop}%
\bibitem [{\citenamefont {McCarron}\ \emph {et~al.}(2018)\citenamefont
  {McCarron}, \citenamefont {Steinecker}, \citenamefont {Zhu},\ and\
  \citenamefont {DeMille}}]{mccarron2018magnetic}%
  \BibitemOpen
  \bibfield  {author} {\bibinfo {author} {\bibfnamefont {D.}~\bibnamefont
  {McCarron}}, \bibinfo {author} {\bibfnamefont {M.}~\bibnamefont
  {Steinecker}}, \bibinfo {author} {\bibfnamefont {Y.}~\bibnamefont {Zhu}}, \
  and\ \bibinfo {author} {\bibfnamefont {D.}~\bibnamefont {DeMille}},\
  }\href@noop {} {\bibfield  {journal} {\bibinfo  {journal} {Physical review
  letters}\ }\textbf {\bibinfo {volume} {121}},\ \bibinfo {pages} {013202}
  (\bibinfo {year} {2018})}\BibitemShut {NoStop}%
\bibitem [{\citenamefont {Hummon}\ \emph {et~al.}(2013)\citenamefont {Hummon},
  \citenamefont {Yeo}, \citenamefont {Stuhl}, \citenamefont {Collopy},
  \citenamefont {Xia},\ and\ \citenamefont {Ye}}]{hummon20132d}%
  \BibitemOpen
  \bibfield  {author} {\bibinfo {author} {\bibfnamefont {M.~T.}\ \bibnamefont
  {Hummon}}, \bibinfo {author} {\bibfnamefont {M.}~\bibnamefont {Yeo}},
  \bibinfo {author} {\bibfnamefont {B.~K.}\ \bibnamefont {Stuhl}}, \bibinfo
  {author} {\bibfnamefont {A.~L.}\ \bibnamefont {Collopy}}, \bibinfo {author}
  {\bibfnamefont {Y.}~\bibnamefont {Xia}}, \ and\ \bibinfo {author}
  {\bibfnamefont {J.}~\bibnamefont {Ye}},\ }\href@noop {} {\bibfield  {journal}
  {\bibinfo  {journal} {Physical Review Letters}\ }\textbf {\bibinfo {volume}
  {110}},\ \bibinfo {pages} {143001} (\bibinfo {year} {2013})}\BibitemShut
  {NoStop}%
\bibitem [{\citenamefont {Yeo}\ \emph {et~al.}(2015)\citenamefont {Yeo},
  \citenamefont {Hummon}, \citenamefont {Collopy}, \citenamefont {Yan},
  \citenamefont {Hemmerling}, \citenamefont {Chae}, \citenamefont {Doyle},\
  and\ \citenamefont {Ye}}]{yeo2015rotational}%
  \BibitemOpen
  \bibfield  {author} {\bibinfo {author} {\bibfnamefont {M.}~\bibnamefont
  {Yeo}}, \bibinfo {author} {\bibfnamefont {M.~T.}\ \bibnamefont {Hummon}},
  \bibinfo {author} {\bibfnamefont {A.~L.}\ \bibnamefont {Collopy}}, \bibinfo
  {author} {\bibfnamefont {B.}~\bibnamefont {Yan}}, \bibinfo {author}
  {\bibfnamefont {B.}~\bibnamefont {Hemmerling}}, \bibinfo {author}
  {\bibfnamefont {E.}~\bibnamefont {Chae}}, \bibinfo {author} {\bibfnamefont
  {J.~M.}\ \bibnamefont {Doyle}}, \ and\ \bibinfo {author} {\bibfnamefont
  {J.}~\bibnamefont {Ye}},\ }\href@noop {} {\bibfield  {journal} {\bibinfo
  {journal} {Physical review letters}\ }\textbf {\bibinfo {volume} {114}},\
  \bibinfo {pages} {223003} (\bibinfo {year} {2015})}\BibitemShut {NoStop}%
\bibitem [{\citenamefont {Collopy}\ \emph {et~al.}(2018)\citenamefont
  {Collopy}, \citenamefont {Ding}, \citenamefont {Wu}, \citenamefont
  {Finneran}, \citenamefont {Anderegg}, \citenamefont {Augenbraun},
  \citenamefont {Doyle},\ and\ \citenamefont {Ye}}]{collopy20183d}%
  \BibitemOpen
  \bibfield  {author} {\bibinfo {author} {\bibfnamefont {A.~L.}\ \bibnamefont
  {Collopy}}, \bibinfo {author} {\bibfnamefont {S.}~\bibnamefont {Ding}},
  \bibinfo {author} {\bibfnamefont {Y.}~\bibnamefont {Wu}}, \bibinfo {author}
  {\bibfnamefont {I.~A.}\ \bibnamefont {Finneran}}, \bibinfo {author}
  {\bibfnamefont {L.}~\bibnamefont {Anderegg}}, \bibinfo {author}
  {\bibfnamefont {B.~L.}\ \bibnamefont {Augenbraun}}, \bibinfo {author}
  {\bibfnamefont {J.~M.}\ \bibnamefont {Doyle}}, \ and\ \bibinfo {author}
  {\bibfnamefont {J.}~\bibnamefont {Ye}},\ }\href@noop {} {\bibfield  {journal}
  {\bibinfo  {journal} {Physical review letters}\ }\textbf {\bibinfo {volume}
  {121}},\ \bibinfo {pages} {213201} (\bibinfo {year} {2018})}\BibitemShut
  {NoStop}%
\bibitem [{\citenamefont {Zhelyazkova}\ \emph {et~al.}(2014)\citenamefont
  {Zhelyazkova}, \citenamefont {Cournol}, \citenamefont {Wall}, \citenamefont
  {Matsushima}, \citenamefont {Hudson}, \citenamefont {Hinds}, \citenamefont
  {Tarbutt},\ and\ \citenamefont {Sauer}}]{zhelyazkova2014laser}%
  \BibitemOpen
  \bibfield  {author} {\bibinfo {author} {\bibfnamefont {V.}~\bibnamefont
  {Zhelyazkova}}, \bibinfo {author} {\bibfnamefont {A.}~\bibnamefont
  {Cournol}}, \bibinfo {author} {\bibfnamefont {T.~E.}\ \bibnamefont {Wall}},
  \bibinfo {author} {\bibfnamefont {A.}~\bibnamefont {Matsushima}}, \bibinfo
  {author} {\bibfnamefont {J.~J.}\ \bibnamefont {Hudson}}, \bibinfo {author}
  {\bibfnamefont {E.}~\bibnamefont {Hinds}}, \bibinfo {author} {\bibfnamefont
  {M.}~\bibnamefont {Tarbutt}}, \ and\ \bibinfo {author} {\bibfnamefont
  {B.}~\bibnamefont {Sauer}},\ }\href@noop {} {\bibfield  {journal} {\bibinfo
  {journal} {Physical Review A}\ }\textbf {\bibinfo {volume} {89}},\ \bibinfo
  {pages} {053416} (\bibinfo {year} {2014})}\BibitemShut {NoStop}%
\bibitem [{\citenamefont {Williams}\ \emph {et~al.}(2018)\citenamefont
  {Williams}, \citenamefont {Caldwell}, \citenamefont {Fitch}, \citenamefont
  {Truppe}, \citenamefont {Rodewald}, \citenamefont {Hinds}, \citenamefont
  {Sauer},\ and\ \citenamefont {Tarbutt}}]{williams2018magnetic}%
  \BibitemOpen
  \bibfield  {author} {\bibinfo {author} {\bibfnamefont {H.}~\bibnamefont
  {Williams}}, \bibinfo {author} {\bibfnamefont {L.}~\bibnamefont {Caldwell}},
  \bibinfo {author} {\bibfnamefont {N.}~\bibnamefont {Fitch}}, \bibinfo
  {author} {\bibfnamefont {S.}~\bibnamefont {Truppe}}, \bibinfo {author}
  {\bibfnamefont {J.}~\bibnamefont {Rodewald}}, \bibinfo {author}
  {\bibfnamefont {E.}~\bibnamefont {Hinds}}, \bibinfo {author} {\bibfnamefont
  {B.}~\bibnamefont {Sauer}}, \ and\ \bibinfo {author} {\bibfnamefont
  {M.}~\bibnamefont {Tarbutt}},\ }\href@noop {} {\bibfield  {journal} {\bibinfo
   {journal} {Physical review letters}\ }\textbf {\bibinfo {volume} {120}},\
  \bibinfo {pages} {163201} (\bibinfo {year} {2018})}\BibitemShut {NoStop}%
\bibitem [{\citenamefont {Anderegg}\ \emph {et~al.}(2018)\citenamefont
  {Anderegg}, \citenamefont {Augenbraun}, \citenamefont {Bao}, \citenamefont
  {Burchesky}, \citenamefont {Cheuk}, \citenamefont {Ketterle},\ and\
  \citenamefont {Doyle}}]{anderegg2018laser}%
  \BibitemOpen
  \bibfield  {author} {\bibinfo {author} {\bibfnamefont {L.}~\bibnamefont
  {Anderegg}}, \bibinfo {author} {\bibfnamefont {B.~L.}\ \bibnamefont
  {Augenbraun}}, \bibinfo {author} {\bibfnamefont {Y.}~\bibnamefont {Bao}},
  \bibinfo {author} {\bibfnamefont {S.}~\bibnamefont {Burchesky}}, \bibinfo
  {author} {\bibfnamefont {L.~W.}\ \bibnamefont {Cheuk}}, \bibinfo {author}
  {\bibfnamefont {W.}~\bibnamefont {Ketterle}}, \ and\ \bibinfo {author}
  {\bibfnamefont {J.~M.}\ \bibnamefont {Doyle}},\ }\href@noop {} {\bibfield
  {journal} {\bibinfo  {journal} {Nature Physics}\ ,\ \bibinfo {pages} {1}}
  (\bibinfo {year} {2018})}\BibitemShut {NoStop}%
\bibitem [{\citenamefont {Caldwell}\ \emph {et~al.}(2019)\citenamefont
  {Caldwell}, \citenamefont {Devlin}, \citenamefont {Williams}, \citenamefont
  {Fitch}, \citenamefont {Hinds}, \citenamefont {Sauer},\ and\ \citenamefont
  {Tarbutt}}]{caldwell2019deep}%
  \BibitemOpen
  \bibfield  {author} {\bibinfo {author} {\bibfnamefont {L.}~\bibnamefont
  {Caldwell}}, \bibinfo {author} {\bibfnamefont {J.}~\bibnamefont {Devlin}},
  \bibinfo {author} {\bibfnamefont {H.}~\bibnamefont {Williams}}, \bibinfo
  {author} {\bibfnamefont {N.}~\bibnamefont {Fitch}}, \bibinfo {author}
  {\bibfnamefont {E.}~\bibnamefont {Hinds}}, \bibinfo {author} {\bibfnamefont
  {B.}~\bibnamefont {Sauer}}, \ and\ \bibinfo {author} {\bibfnamefont
  {M.}~\bibnamefont {Tarbutt}},\ }\href@noop {} {\bibfield  {journal} {\bibinfo
   {journal} {Physical review letters}\ }\textbf {\bibinfo {volume} {123}},\
  \bibinfo {pages} {033202} (\bibinfo {year} {2019})}\BibitemShut {NoStop}%
\bibitem [{\citenamefont {Lim}\ \emph {et~al.}(2018)\citenamefont {Lim},
  \citenamefont {Almond}, \citenamefont {Trigatzis}, \citenamefont {Devlin},
  \citenamefont {Fitch}, \citenamefont {Sauer}, \citenamefont {Tarbutt},\ and\
  \citenamefont {Hinds}}]{lim2018laser}%
  \BibitemOpen
  \bibfield  {author} {\bibinfo {author} {\bibfnamefont {J.}~\bibnamefont
  {Lim}}, \bibinfo {author} {\bibfnamefont {J.}~\bibnamefont {Almond}},
  \bibinfo {author} {\bibfnamefont {M.}~\bibnamefont {Trigatzis}}, \bibinfo
  {author} {\bibfnamefont {J.}~\bibnamefont {Devlin}}, \bibinfo {author}
  {\bibfnamefont {N.}~\bibnamefont {Fitch}}, \bibinfo {author} {\bibfnamefont
  {B.}~\bibnamefont {Sauer}}, \bibinfo {author} {\bibfnamefont
  {M.}~\bibnamefont {Tarbutt}}, \ and\ \bibinfo {author} {\bibfnamefont
  {E.}~\bibnamefont {Hinds}},\ }\href@noop {} {\bibfield  {journal} {\bibinfo
  {journal} {Physical review letters}\ }\textbf {\bibinfo {volume} {120}},\
  \bibinfo {pages} {123201} (\bibinfo {year} {2018})}\BibitemShut {NoStop}%
\bibitem [{\citenamefont {Chen}\ \emph {et~al.}(2017)\citenamefont {Chen},
  \citenamefont {Bu},\ and\ \citenamefont {Yan}}]{chen2017radiative}%
  \BibitemOpen
  \bibfield  {author} {\bibinfo {author} {\bibfnamefont {T.}~\bibnamefont
  {Chen}}, \bibinfo {author} {\bibfnamefont {W.}~\bibnamefont {Bu}}, \ and\
  \bibinfo {author} {\bibfnamefont {B.}~\bibnamefont {Yan}},\ }\href@noop {}
  {\bibfield  {journal} {\bibinfo  {journal} {Physical Review A}\ }\textbf
  {\bibinfo {volume} {96}},\ \bibinfo {pages} {053401} (\bibinfo {year}
  {2017})}\BibitemShut {NoStop}%
\bibitem [{\citenamefont {Xu}\ \emph {et~al.}(2019)\citenamefont {Xu},
  \citenamefont {Xia}, \citenamefont {Yin}, \citenamefont {Gu}, \citenamefont
  {Xia},\ and\ \citenamefont {Yin}}]{xu2019determination}%
  \BibitemOpen
  \bibfield  {author} {\bibinfo {author} {\bibfnamefont {S.}~\bibnamefont
  {Xu}}, \bibinfo {author} {\bibfnamefont {M.}~\bibnamefont {Xia}}, \bibinfo
  {author} {\bibfnamefont {Y.}~\bibnamefont {Yin}}, \bibinfo {author}
  {\bibfnamefont {R.}~\bibnamefont {Gu}}, \bibinfo {author} {\bibfnamefont
  {Y.}~\bibnamefont {Xia}}, \ and\ \bibinfo {author} {\bibfnamefont
  {J.}~\bibnamefont {Yin}},\ }\href@noop {} {\bibfield  {journal} {\bibinfo
  {journal} {The Journal of chemical physics}\ }\textbf {\bibinfo {volume}
  {150}},\ \bibinfo {pages} {084302} (\bibinfo {year} {2019})}\BibitemShut
  {NoStop}%
\bibitem [{\citenamefont {Ketterle}\ and\ \citenamefont
  {Van~Druten}(1996)}]{ketterle1996evaporative}%
  \BibitemOpen
  \bibfield  {author} {\bibinfo {author} {\bibfnamefont {W.}~\bibnamefont
  {Ketterle}}\ and\ \bibinfo {author} {\bibfnamefont {N.}~\bibnamefont
  {Van~Druten}},\ }in\ \href@noop {} {\emph {\bibinfo {booktitle} {Advances in
  atomic, molecular, and optical physics}}},\ Vol.~\bibinfo {volume} {37}\
  (\bibinfo  {publisher} {Elsevier},\ \bibinfo {year} {1996})\ pp.\ \bibinfo
  {pages} {181--236}\BibitemShut {NoStop}%
\bibitem [{\citenamefont {Janssen}\ \emph
  {et~al.}(2011{\natexlab{b}})\citenamefont {Janssen}, \citenamefont
  {{\.Z}uchowski}, \citenamefont {van~der Avoird}, \citenamefont {Hutson},\
  and\ \citenamefont {Groenenboom}}]{janssen2011cold}%
  \BibitemOpen
  \bibfield  {author} {\bibinfo {author} {\bibfnamefont {L.~M.}\ \bibnamefont
  {Janssen}}, \bibinfo {author} {\bibfnamefont {P.~S.}\ \bibnamefont
  {{\.Z}uchowski}}, \bibinfo {author} {\bibfnamefont {A.}~\bibnamefont {van~der
  Avoird}}, \bibinfo {author} {\bibfnamefont {J.~M.}\ \bibnamefont {Hutson}}, \
  and\ \bibinfo {author} {\bibfnamefont {G.~C.}\ \bibnamefont {Groenenboom}},\
  }\href@noop {} {\bibfield  {journal} {\bibinfo  {journal} {The Journal of
  chemical physics}\ }\textbf {\bibinfo {volume} {134}},\ \bibinfo {pages}
  {124309} (\bibinfo {year} {2011}{\natexlab{b}})}\BibitemShut {NoStop}%
\bibitem [{\citenamefont {Suleimanov}\ \emph {et~al.}(2012)\citenamefont
  {Suleimanov}, \citenamefont {Tscherbul},\ and\ \citenamefont
  {Krems}}]{suleimanov2012efficient}%
  \BibitemOpen
  \bibfield  {author} {\bibinfo {author} {\bibfnamefont {Y.~V.}\ \bibnamefont
  {Suleimanov}}, \bibinfo {author} {\bibfnamefont {T.}~\bibnamefont
  {Tscherbul}}, \ and\ \bibinfo {author} {\bibfnamefont {R.}~\bibnamefont
  {Krems}},\ }\href@noop {} {\bibfield  {journal} {\bibinfo  {journal} {The
  Journal of Chemical Physics}\ }\textbf {\bibinfo {volume} {137}},\ \bibinfo
  {pages} {024103} (\bibinfo {year} {2012})}\BibitemShut {NoStop}%
\bibitem [{\citenamefont {Janssen}\ \emph {et~al.}(2013)\citenamefont
  {Janssen}, \citenamefont {van~der Avoird},\ and\ \citenamefont
  {Groenenboom}}]{janssen2013quantum}%
  \BibitemOpen
  \bibfield  {author} {\bibinfo {author} {\bibfnamefont {L.~M.}\ \bibnamefont
  {Janssen}}, \bibinfo {author} {\bibfnamefont {A.}~\bibnamefont {van~der
  Avoird}}, \ and\ \bibinfo {author} {\bibfnamefont {G.~C.}\ \bibnamefont
  {Groenenboom}},\ }\href@noop {} {\bibfield  {journal} {\bibinfo  {journal}
  {Physical review letters}\ }\textbf {\bibinfo {volume} {110}},\ \bibinfo
  {pages} {063201} (\bibinfo {year} {2013})}\BibitemShut {NoStop}%
\bibitem [{\citenamefont {Parazzoli}\ \emph {et~al.}(2011)\citenamefont
  {Parazzoli}, \citenamefont {Fitch}, \citenamefont {\.{Z}uchowski},
  \citenamefont {Hutson},\ and\ \citenamefont
  {Lewandowski}}]{parazzoli2011large}%
  \BibitemOpen
  \bibfield  {author} {\bibinfo {author} {\bibfnamefont {L.~P.}\ \bibnamefont
  {Parazzoli}}, \bibinfo {author} {\bibfnamefont {N.~J.}\ \bibnamefont
  {Fitch}}, \bibinfo {author} {\bibfnamefont {P.~S.}\ \bibnamefont
  {\.{Z}uchowski}}, \bibinfo {author} {\bibfnamefont {J.~M.}\ \bibnamefont
  {Hutson}}, \ and\ \bibinfo {author} {\bibfnamefont {H.~J.}\ \bibnamefont
  {Lewandowski}},\ }\href@noop {} {\bibfield  {journal} {\bibinfo  {journal}
  {Physical review letters}\ }\textbf {\bibinfo {volume} {106}},\ \bibinfo
  {pages} {193201} (\bibinfo {year} {2011})}\BibitemShut {NoStop}%
\bibitem [{\citenamefont {Stuhl}\ \emph {et~al.}(2012)\citenamefont {Stuhl},
  \citenamefont {Hummon}, \citenamefont {Yeo}, \citenamefont
  {Qu{\'e}m{\'e}ner}, \citenamefont {Bohn},\ and\ \citenamefont
  {Ye}}]{stuhl2012evaporative}%
  \BibitemOpen
  \bibfield  {author} {\bibinfo {author} {\bibfnamefont {B.~K.}\ \bibnamefont
  {Stuhl}}, \bibinfo {author} {\bibfnamefont {M.~T.}\ \bibnamefont {Hummon}},
  \bibinfo {author} {\bibfnamefont {M.}~\bibnamefont {Yeo}}, \bibinfo {author}
  {\bibfnamefont {G.}~\bibnamefont {Qu{\'e}m{\'e}ner}}, \bibinfo {author}
  {\bibfnamefont {J.~L.}\ \bibnamefont {Bohn}}, \ and\ \bibinfo {author}
  {\bibfnamefont {J.}~\bibnamefont {Ye}},\ }\href@noop {} {\bibfield  {journal}
  {\bibinfo  {journal} {Nature}\ }\textbf {\bibinfo {volume} {492}},\ \bibinfo
  {pages} {396} (\bibinfo {year} {2012})}\BibitemShut {NoStop}%
\bibitem [{\citenamefont {Larson}\ \emph {et~al.}(1986)\citenamefont {Larson},
  \citenamefont {Bergquist}, \citenamefont {Bollinger}, \citenamefont {Itano},\
  and\ \citenamefont {Wineland}}]{larson1986sympathetic}%
  \BibitemOpen
  \bibfield  {author} {\bibinfo {author} {\bibfnamefont {D.~J.}\ \bibnamefont
  {Larson}}, \bibinfo {author} {\bibfnamefont {J.~C.}\ \bibnamefont
  {Bergquist}}, \bibinfo {author} {\bibfnamefont {J.~J.}\ \bibnamefont
  {Bollinger}}, \bibinfo {author} {\bibfnamefont {W.~M.}\ \bibnamefont
  {Itano}}, \ and\ \bibinfo {author} {\bibfnamefont {D.~J.}\ \bibnamefont
  {Wineland}},\ }\href@noop {} {\bibfield  {journal} {\bibinfo  {journal}
  {Physical review letters}\ }\textbf {\bibinfo {volume} {57}},\ \bibinfo
  {pages} {70} (\bibinfo {year} {1986})}\BibitemShut {NoStop}%
\bibitem [{\citenamefont {Myatt}\ \emph {et~al.}(1997)\citenamefont {Myatt},
  \citenamefont {Burt}, \citenamefont {Ghrist}, \citenamefont {Cornell},\ and\
  \citenamefont {Wieman}}]{myatt1997production}%
  \BibitemOpen
  \bibfield  {author} {\bibinfo {author} {\bibfnamefont {C.}~\bibnamefont
  {Myatt}}, \bibinfo {author} {\bibfnamefont {E.}~\bibnamefont {Burt}},
  \bibinfo {author} {\bibfnamefont {R.}~\bibnamefont {Ghrist}}, \bibinfo
  {author} {\bibfnamefont {E.~A.}\ \bibnamefont {Cornell}}, \ and\ \bibinfo
  {author} {\bibfnamefont {C.}~\bibnamefont {Wieman}},\ }\href@noop {}
  {\bibfield  {journal} {\bibinfo  {journal} {Physical Review Letters}\
  }\textbf {\bibinfo {volume} {78}},\ \bibinfo {pages} {586} (\bibinfo {year}
  {1997})}\BibitemShut {NoStop}%
\bibitem [{\citenamefont {Schreck}\ \emph {et~al.}(2001)\citenamefont
  {Schreck}, \citenamefont {Ferrari}, \citenamefont {Corwin}, \citenamefont
  {Cubizolles}, \citenamefont {Khaykovich}, \citenamefont {Mewes},\ and\
  \citenamefont {Salomon}}]{schreck2001sympathetic}%
  \BibitemOpen
  \bibfield  {author} {\bibinfo {author} {\bibfnamefont {F.}~\bibnamefont
  {Schreck}}, \bibinfo {author} {\bibfnamefont {G.}~\bibnamefont {Ferrari}},
  \bibinfo {author} {\bibfnamefont {K.}~\bibnamefont {Corwin}}, \bibinfo
  {author} {\bibfnamefont {J.}~\bibnamefont {Cubizolles}}, \bibinfo {author}
  {\bibfnamefont {L.}~\bibnamefont {Khaykovich}}, \bibinfo {author}
  {\bibfnamefont {M.-O.}\ \bibnamefont {Mewes}}, \ and\ \bibinfo {author}
  {\bibfnamefont {C.}~\bibnamefont {Salomon}},\ }\href@noop {} {\bibfield
  {journal} {\bibinfo  {journal} {Physical Review A}\ }\textbf {\bibinfo
  {volume} {64}},\ \bibinfo {pages} {011402} (\bibinfo {year}
  {2001})}\BibitemShut {NoStop}%
\bibitem [{\citenamefont {{\.Z}uchowski}\ and\ \citenamefont
  {Hutson}(2008)}]{zuchowski2008prospects}%
  \BibitemOpen
  \bibfield  {author} {\bibinfo {author} {\bibfnamefont {P.~S.}\ \bibnamefont
  {{\.Z}uchowski}}\ and\ \bibinfo {author} {\bibfnamefont {J.~M.}\ \bibnamefont
  {Hutson}},\ }\href@noop {} {\bibfield  {journal} {\bibinfo  {journal}
  {Physical Review A}\ }\textbf {\bibinfo {volume} {78}},\ \bibinfo {pages}
  {022701} (\bibinfo {year} {2008})}\BibitemShut {NoStop}%
\bibitem [{\citenamefont {{\.Z}uchowski}\ and\ \citenamefont
  {Hutson}(2009)}]{zuchowski2009low}%
  \BibitemOpen
  \bibfield  {author} {\bibinfo {author} {\bibfnamefont {P.~S.}\ \bibnamefont
  {{\.Z}uchowski}}\ and\ \bibinfo {author} {\bibfnamefont {J.~M.}\ \bibnamefont
  {Hutson}},\ }\href@noop {} {\bibfield  {journal} {\bibinfo  {journal}
  {Physical Review A}\ }\textbf {\bibinfo {volume} {79}},\ \bibinfo {pages}
  {062708} (\bibinfo {year} {2009})}\BibitemShut {NoStop}%
\bibitem [{\citenamefont {Barletta}\ \emph {et~al.}(2008)\citenamefont
  {Barletta}, \citenamefont {Tennyson},\ and\ \citenamefont
  {Barker}}]{barletta2008creating}%
  \BibitemOpen
  \bibfield  {author} {\bibinfo {author} {\bibfnamefont {P.}~\bibnamefont
  {Barletta}}, \bibinfo {author} {\bibfnamefont {J.}~\bibnamefont {Tennyson}},
  \ and\ \bibinfo {author} {\bibfnamefont {P.}~\bibnamefont {Barker}},\
  }\href@noop {} {\bibfield  {journal} {\bibinfo  {journal} {Physical Review
  A}\ }\textbf {\bibinfo {volume} {78}},\ \bibinfo {pages} {052707} (\bibinfo
  {year} {2008})}\BibitemShut {NoStop}%
\bibitem [{\citenamefont {Barker}\ \emph {et~al.}(2009)\citenamefont {Barker},
  \citenamefont {Purcell}, \citenamefont {Douglas}, \citenamefont {Barletta},
  \citenamefont {Coppendale}, \citenamefont {Maher-McWilliams},\ and\
  \citenamefont {Tennyson}}]{barker2009sympathetic}%
  \BibitemOpen
  \bibfield  {author} {\bibinfo {author} {\bibfnamefont {P.}~\bibnamefont
  {Barker}}, \bibinfo {author} {\bibfnamefont {S.}~\bibnamefont {Purcell}},
  \bibinfo {author} {\bibfnamefont {P.}~\bibnamefont {Douglas}}, \bibinfo
  {author} {\bibfnamefont {P.}~\bibnamefont {Barletta}}, \bibinfo {author}
  {\bibfnamefont {N.}~\bibnamefont {Coppendale}}, \bibinfo {author}
  {\bibfnamefont {C.}~\bibnamefont {Maher-McWilliams}}, \ and\ \bibinfo
  {author} {\bibfnamefont {J.}~\bibnamefont {Tennyson}},\ }\href@noop {}
  {\bibfield  {journal} {\bibinfo  {journal} {Faraday discussions}\ }\textbf
  {\bibinfo {volume} {142}},\ \bibinfo {pages} {175} (\bibinfo {year}
  {2009})}\BibitemShut {NoStop}%
\bibitem [{\citenamefont {Barletta}\ \emph {et~al.}(2010)\citenamefont
  {Barletta}, \citenamefont {Tennyson},\ and\ \citenamefont
  {Barker}}]{barletta2010direct}%
  \BibitemOpen
  \bibfield  {author} {\bibinfo {author} {\bibfnamefont {P.}~\bibnamefont
  {Barletta}}, \bibinfo {author} {\bibfnamefont {J.}~\bibnamefont {Tennyson}},
  \ and\ \bibinfo {author} {\bibfnamefont {P.}~\bibnamefont {Barker}},\
  }\href@noop {} {\bibfield  {journal} {\bibinfo  {journal} {New Journal of
  Physics}\ }\textbf {\bibinfo {volume} {12}},\ \bibinfo {pages} {113002}
  (\bibinfo {year} {2010})}\BibitemShut {NoStop}%
\bibitem [{\citenamefont {Lara}\ \emph {et~al.}(2006)\citenamefont {Lara},
  \citenamefont {Bohn}, \citenamefont {Potter}, \citenamefont {Sold{\'a}n},\
  and\ \citenamefont {Hutson}}]{lara2006ultracold}%
  \BibitemOpen
  \bibfield  {author} {\bibinfo {author} {\bibfnamefont {M.}~\bibnamefont
  {Lara}}, \bibinfo {author} {\bibfnamefont {J.~L.}\ \bibnamefont {Bohn}},
  \bibinfo {author} {\bibfnamefont {D.}~\bibnamefont {Potter}}, \bibinfo
  {author} {\bibfnamefont {P.}~\bibnamefont {Sold{\'a}n}}, \ and\ \bibinfo
  {author} {\bibfnamefont {J.~M.}\ \bibnamefont {Hutson}},\ }\href@noop {}
  {\bibfield  {journal} {\bibinfo  {journal} {Physical review letters}\
  }\textbf {\bibinfo {volume} {97}},\ \bibinfo {pages} {183201} (\bibinfo
  {year} {2006})}\BibitemShut {NoStop}%
\bibitem [{\citenamefont {Lara}\ \emph {et~al.}(2007)\citenamefont {Lara},
  \citenamefont {Bohn}, \citenamefont {Potter}, \citenamefont {Sold{\'a}n},\
  and\ \citenamefont {Hutson}}]{lara2007cold}%
  \BibitemOpen
  \bibfield  {author} {\bibinfo {author} {\bibfnamefont {M.}~\bibnamefont
  {Lara}}, \bibinfo {author} {\bibfnamefont {J.~L.}\ \bibnamefont {Bohn}},
  \bibinfo {author} {\bibfnamefont {D.~E.}\ \bibnamefont {Potter}}, \bibinfo
  {author} {\bibfnamefont {P.}~\bibnamefont {Sold{\'a}n}}, \ and\ \bibinfo
  {author} {\bibfnamefont {J.~M.}\ \bibnamefont {Hutson}},\ }\href@noop {}
  {\bibfield  {journal} {\bibinfo  {journal} {Physical Review A}\ }\textbf
  {\bibinfo {volume} {75}},\ \bibinfo {pages} {012704} (\bibinfo {year}
  {2007})}\BibitemShut {NoStop}%
\bibitem [{\citenamefont {Tacconi}\ \emph {et~al.}(2007)\citenamefont
  {Tacconi}, \citenamefont {Gonzalez-Sanchez}, \citenamefont {Bodo},\ and\
  \citenamefont {Gianturco}}]{tacconi2007collisions}%
  \BibitemOpen
  \bibfield  {author} {\bibinfo {author} {\bibfnamefont {M.}~\bibnamefont
  {Tacconi}}, \bibinfo {author} {\bibfnamefont {L.}~\bibnamefont
  {Gonzalez-Sanchez}}, \bibinfo {author} {\bibfnamefont {E.}~\bibnamefont
  {Bodo}}, \ and\ \bibinfo {author} {\bibfnamefont {F.}~\bibnamefont
  {Gianturco}},\ }\href@noop {} {\bibfield  {journal} {\bibinfo  {journal}
  {Physical Review A}\ }\textbf {\bibinfo {volume} {76}},\ \bibinfo {pages}
  {032702} (\bibinfo {year} {2007})}\BibitemShut {NoStop}%
\bibitem [{\citenamefont {Sold{\'a}n}\ \emph {et~al.}(2009)\citenamefont
  {Sold{\'a}n}, \citenamefont {{\.Z}uchowski},\ and\ \citenamefont
  {Hutson}}]{soldan2009prospects}%
  \BibitemOpen
  \bibfield  {author} {\bibinfo {author} {\bibfnamefont {P.}~\bibnamefont
  {Sold{\'a}n}}, \bibinfo {author} {\bibfnamefont {P.~S.}\ \bibnamefont
  {{\.Z}uchowski}}, \ and\ \bibinfo {author} {\bibfnamefont {J.~M.}\
  \bibnamefont {Hutson}},\ }\href@noop {} {\bibfield  {journal} {\bibinfo
  {journal} {Faraday Discussions}\ }\textbf {\bibinfo {volume} {142}},\
  \bibinfo {pages} {191} (\bibinfo {year} {2009})}\BibitemShut {NoStop}%
\bibitem [{\citenamefont {Wallis}\ and\ \citenamefont
  {Hutson}(2009)}]{wallis2009production}%
  \BibitemOpen
  \bibfield  {author} {\bibinfo {author} {\bibfnamefont {A.~O.}\ \bibnamefont
  {Wallis}}\ and\ \bibinfo {author} {\bibfnamefont {J.~M.}\ \bibnamefont
  {Hutson}},\ }\href@noop {} {\bibfield  {journal} {\bibinfo  {journal}
  {Physical review letters}\ }\textbf {\bibinfo {volume} {103}},\ \bibinfo
  {pages} {183201} (\bibinfo {year} {2009})}\BibitemShut {NoStop}%
\bibitem [{\citenamefont {Gonz{\'a}lez-Mart{\'\i}nez}\ and\ \citenamefont
  {Hutson}(2011)}]{gonzalez2011effect}%
  \BibitemOpen
  \bibfield  {author} {\bibinfo {author} {\bibfnamefont {M.~L.}\ \bibnamefont
  {Gonz{\'a}lez-Mart{\'\i}nez}}\ and\ \bibinfo {author} {\bibfnamefont {J.~M.}\
  \bibnamefont {Hutson}},\ }\href@noop {} {\bibfield  {journal} {\bibinfo
  {journal} {Physical Review A}\ }\textbf {\bibinfo {volume} {84}},\ \bibinfo
  {pages} {052706} (\bibinfo {year} {2011})}\BibitemShut {NoStop}%
\bibitem [{\citenamefont {Wallis}\ \emph {et~al.}(2011)\citenamefont {Wallis},
  \citenamefont {Longdon}, \citenamefont {{\.Z}uchowski},\ and\ \citenamefont
  {Hutson}}]{wallis2011prospects}%
  \BibitemOpen
  \bibfield  {author} {\bibinfo {author} {\bibfnamefont {A.~O.}\ \bibnamefont
  {Wallis}}, \bibinfo {author} {\bibfnamefont {E.~J.}\ \bibnamefont {Longdon}},
  \bibinfo {author} {\bibfnamefont {P.~S.}\ \bibnamefont {{\.Z}uchowski}}, \
  and\ \bibinfo {author} {\bibfnamefont {J.~M.}\ \bibnamefont {Hutson}},\
  }\href@noop {} {\bibfield  {journal} {\bibinfo  {journal} {The European
  Physical Journal D}\ }\textbf {\bibinfo {volume} {65}},\ \bibinfo {pages}
  {151} (\bibinfo {year} {2011})}\BibitemShut {NoStop}%
\bibitem [{\citenamefont {{\.Z}uchowski}\ and\ \citenamefont
  {Hutson}(2011)}]{zuchowski2011cold}%
  \BibitemOpen
  \bibfield  {author} {\bibinfo {author} {\bibfnamefont {P.~S.}\ \bibnamefont
  {{\.Z}uchowski}}\ and\ \bibinfo {author} {\bibfnamefont {J.~M.}\ \bibnamefont
  {Hutson}},\ }\href@noop {} {\bibfield  {journal} {\bibinfo  {journal}
  {Physical Chemistry Chemical Physics}\ }\textbf {\bibinfo {volume} {13}},\
  \bibinfo {pages} {3669} (\bibinfo {year} {2011})}\BibitemShut {NoStop}%
\bibitem [{\citenamefont {Gonz{\'a}lez-Mart{\'\i}nez}\ and\ \citenamefont
  {Hutson}(2013)}]{gonzalez2013ultracold}%
  \BibitemOpen
  \bibfield  {author} {\bibinfo {author} {\bibfnamefont {M.~L.}\ \bibnamefont
  {Gonz{\'a}lez-Mart{\'\i}nez}}\ and\ \bibinfo {author} {\bibfnamefont {J.~M.}\
  \bibnamefont {Hutson}},\ }\href@noop {} {\bibfield  {journal} {\bibinfo
  {journal} {Physical review letters}\ }\textbf {\bibinfo {volume} {111}},\
  \bibinfo {pages} {203004} (\bibinfo {year} {2013})}\BibitemShut {NoStop}%
\bibitem [{\citenamefont {Tokunaga}\ \emph {et~al.}(2011)\citenamefont
  {Tokunaga}, \citenamefont {Skomorowski}, \citenamefont {{\.Z}uchowski},
  \citenamefont {Moszynski}, \citenamefont {Hutson}, \citenamefont {Hinds},\
  and\ \citenamefont {Tarbutt}}]{tokunaga2011prospects}%
  \BibitemOpen
  \bibfield  {author} {\bibinfo {author} {\bibfnamefont {S.~K.}\ \bibnamefont
  {Tokunaga}}, \bibinfo {author} {\bibfnamefont {W.}~\bibnamefont
  {Skomorowski}}, \bibinfo {author} {\bibfnamefont {P.~S.}\ \bibnamefont
  {{\.Z}uchowski}}, \bibinfo {author} {\bibfnamefont {R.}~\bibnamefont
  {Moszynski}}, \bibinfo {author} {\bibfnamefont {J.~M.}\ \bibnamefont
  {Hutson}}, \bibinfo {author} {\bibfnamefont {E.}~\bibnamefont {Hinds}}, \
  and\ \bibinfo {author} {\bibfnamefont {M.}~\bibnamefont {Tarbutt}},\
  }\href@noop {} {\bibfield  {journal} {\bibinfo  {journal} {The European
  Physical Journal D}\ }\textbf {\bibinfo {volume} {65}},\ \bibinfo {pages}
  {141} (\bibinfo {year} {2011})}\BibitemShut {NoStop}%
\bibitem [{\citenamefont {Tscherbul}\ \emph {et~al.}(2011)\citenamefont
  {Tscherbul}, \citenamefont {K{\l}os},\ and\ \citenamefont
  {Buchachenko}}]{tscherbul2011ultracold}%
  \BibitemOpen
  \bibfield  {author} {\bibinfo {author} {\bibfnamefont {T.}~\bibnamefont
  {Tscherbul}}, \bibinfo {author} {\bibfnamefont {J.}~\bibnamefont {K{\l}os}},
  \ and\ \bibinfo {author} {\bibfnamefont {A.}~\bibnamefont {Buchachenko}},\
  }\href@noop {} {\bibfield  {journal} {\bibinfo  {journal} {Physical Review
  A}\ }\textbf {\bibinfo {volume} {84}},\ \bibinfo {pages} {040701} (\bibinfo
  {year} {2011})}\BibitemShut {NoStop}%
\bibitem [{\citenamefont {Tscherbul}\ \emph {et~al.}(2007)\citenamefont
  {Tscherbul}, \citenamefont {K{\l}os}, \citenamefont {Rajchel},\ and\
  \citenamefont {Krems}}]{tscherbul2007fine}%
  \BibitemOpen
  \bibfield  {author} {\bibinfo {author} {\bibfnamefont {T.}~\bibnamefont
  {Tscherbul}}, \bibinfo {author} {\bibfnamefont {J.}~\bibnamefont {K{\l}os}},
  \bibinfo {author} {\bibfnamefont {L.}~\bibnamefont {Rajchel}}, \ and\
  \bibinfo {author} {\bibfnamefont {R.}~\bibnamefont {Krems}},\ }\href@noop {}
  {\bibfield  {journal} {\bibinfo  {journal} {Physical Review A}\ }\textbf
  {\bibinfo {volume} {75}},\ \bibinfo {pages} {033416} (\bibinfo {year}
  {2007})}\BibitemShut {NoStop}%
\bibitem [{\citenamefont {Lim}\ \emph {et~al.}(2015)\citenamefont {Lim},
  \citenamefont {Frye}, \citenamefont {Hutson},\ and\ \citenamefont
  {Tarbutt}}]{lim2015modeling}%
  \BibitemOpen
  \bibfield  {author} {\bibinfo {author} {\bibfnamefont {J.}~\bibnamefont
  {Lim}}, \bibinfo {author} {\bibfnamefont {M.~D.}\ \bibnamefont {Frye}},
  \bibinfo {author} {\bibfnamefont {J.~M.}\ \bibnamefont {Hutson}}, \ and\
  \bibinfo {author} {\bibfnamefont {M.}~\bibnamefont {Tarbutt}},\ }\href@noop
  {} {\bibfield  {journal} {\bibinfo  {journal} {Physical Review A}\ }\textbf
  {\bibinfo {volume} {92}},\ \bibinfo {pages} {053419} (\bibinfo {year}
  {2015})}\BibitemShut {NoStop}%
\bibitem [{\citenamefont {Morita}\ \emph {et~al.}(2018)\citenamefont {Morita},
  \citenamefont {Kosicki}, \citenamefont {{\.Z}uchowski},\ and\ \citenamefont
  {Tscherbul}}]{morita2018atom}%
  \BibitemOpen
  \bibfield  {author} {\bibinfo {author} {\bibfnamefont {M.}~\bibnamefont
  {Morita}}, \bibinfo {author} {\bibfnamefont {M.~B.}\ \bibnamefont {Kosicki}},
  \bibinfo {author} {\bibfnamefont {P.~S.}\ \bibnamefont {{\.Z}uchowski}}, \
  and\ \bibinfo {author} {\bibfnamefont {T.~V.}\ \bibnamefont {Tscherbul}},\
  }\href@noop {} {\bibfield  {journal} {\bibinfo  {journal} {Physical Review
  A}\ }\textbf {\bibinfo {volume} {98}},\ \bibinfo {pages} {042702} (\bibinfo
  {year} {2018})}\BibitemShut {NoStop}%
\bibitem [{\citenamefont {Morita}\ \emph {et~al.}(2019)\citenamefont {Morita},
  \citenamefont {Krems},\ and\ \citenamefont
  {Tscherbul}}]{morita2019universal}%
  \BibitemOpen
  \bibfield  {author} {\bibinfo {author} {\bibfnamefont {M.}~\bibnamefont
  {Morita}}, \bibinfo {author} {\bibfnamefont {R.~V.}\ \bibnamefont {Krems}}, \
  and\ \bibinfo {author} {\bibfnamefont {T.~V.}\ \bibnamefont {Tscherbul}},\
  }\href@noop {} {\bibfield  {journal} {\bibinfo  {journal} {Physical review
  letters}\ }\textbf {\bibinfo {volume} {123}},\ \bibinfo {pages} {013401}
  (\bibinfo {year} {2019})}\BibitemShut {NoStop}%
\bibitem [{\citenamefont {Morita}\ \emph {et~al.}(2017)\citenamefont {Morita},
  \citenamefont {K{\l}os}, \citenamefont {Buchachenko},\ and\ \citenamefont
  {Tscherbul}}]{morita2017cold}%
  \BibitemOpen
  \bibfield  {author} {\bibinfo {author} {\bibfnamefont {M.}~\bibnamefont
  {Morita}}, \bibinfo {author} {\bibfnamefont {J.}~\bibnamefont {K{\l}os}},
  \bibinfo {author} {\bibfnamefont {A.~A.}\ \bibnamefont {Buchachenko}}, \ and\
  \bibinfo {author} {\bibfnamefont {T.~V.}\ \bibnamefont {Tscherbul}},\
  }\href@noop {} {\bibfield  {journal} {\bibinfo  {journal} {Physical Review
  A}\ }\textbf {\bibinfo {volume} {95}},\ \bibinfo {pages} {063421} (\bibinfo
  {year} {2017})}\BibitemShut {NoStop}%
\bibitem [{\citenamefont {Lavert-Ofir}\ \emph
  {et~al.}(2011{\natexlab{a}})\citenamefont {Lavert-Ofir}, \citenamefont
  {Gersten}, \citenamefont {Henson}, \citenamefont {Shani}, \citenamefont
  {David}, \citenamefont {Narevicius},\ and\ \citenamefont
  {Narevicius}}]{lavert2011moving}%
  \BibitemOpen
  \bibfield  {author} {\bibinfo {author} {\bibfnamefont {E.}~\bibnamefont
  {Lavert-Ofir}}, \bibinfo {author} {\bibfnamefont {S.}~\bibnamefont
  {Gersten}}, \bibinfo {author} {\bibfnamefont {A.~B.}\ \bibnamefont {Henson}},
  \bibinfo {author} {\bibfnamefont {I.}~\bibnamefont {Shani}}, \bibinfo
  {author} {\bibfnamefont {L.}~\bibnamefont {David}}, \bibinfo {author}
  {\bibfnamefont {J.}~\bibnamefont {Narevicius}}, \ and\ \bibinfo {author}
  {\bibfnamefont {E.}~\bibnamefont {Narevicius}},\ }\href@noop {} {\bibfield
  {journal} {\bibinfo  {journal} {New Journal of Physics}\ }\textbf {\bibinfo
  {volume} {13}},\ \bibinfo {pages} {103030} (\bibinfo {year}
  {2011}{\natexlab{a}})}\BibitemShut {NoStop}%
\bibitem [{\citenamefont {Lavert-Ofir}\ \emph
  {et~al.}(2011{\natexlab{b}})\citenamefont {Lavert-Ofir}, \citenamefont
  {David}, \citenamefont {Henson}, \citenamefont {Gersten}, \citenamefont
  {Narevicius},\ and\ \citenamefont {Narevicius}}]{lavert2011stopping}%
  \BibitemOpen
  \bibfield  {author} {\bibinfo {author} {\bibfnamefont {E.}~\bibnamefont
  {Lavert-Ofir}}, \bibinfo {author} {\bibfnamefont {L.}~\bibnamefont {David}},
  \bibinfo {author} {\bibfnamefont {A.~B.}\ \bibnamefont {Henson}}, \bibinfo
  {author} {\bibfnamefont {S.}~\bibnamefont {Gersten}}, \bibinfo {author}
  {\bibfnamefont {J.}~\bibnamefont {Narevicius}}, \ and\ \bibinfo {author}
  {\bibfnamefont {E.}~\bibnamefont {Narevicius}},\ }\href@noop {} {\bibfield
  {journal} {\bibinfo  {journal} {Physical Chemistry Chemical Physics}\
  }\textbf {\bibinfo {volume} {13}},\ \bibinfo {pages} {18948} (\bibinfo {year}
  {2011}{\natexlab{b}})}\BibitemShut {NoStop}%
\bibitem [{\citenamefont {Trimeche}\ \emph {et~al.}(2011)\citenamefont
  {Trimeche}, \citenamefont {Bera}, \citenamefont {Cromi{\`e}res},
  \citenamefont {Robert},\ and\ \citenamefont
  {Vanhaecke}}]{trimeche2011trapping}%
  \BibitemOpen
  \bibfield  {author} {\bibinfo {author} {\bibfnamefont {A.}~\bibnamefont
  {Trimeche}}, \bibinfo {author} {\bibfnamefont {M.~N.}\ \bibnamefont {Bera}},
  \bibinfo {author} {\bibfnamefont {J.-P.}\ \bibnamefont {Cromi{\`e}res}},
  \bibinfo {author} {\bibfnamefont {J.}~\bibnamefont {Robert}}, \ and\ \bibinfo
  {author} {\bibfnamefont {N.}~\bibnamefont {Vanhaecke}},\ }\href@noop {}
  {\bibfield  {journal} {\bibinfo  {journal} {The European Physical Journal D}\
  }\textbf {\bibinfo {volume} {65}},\ \bibinfo {pages} {263} (\bibinfo {year}
  {2011})}\BibitemShut {NoStop}%
\bibitem [{\citenamefont {Meek}\ \emph {et~al.}(2008)\citenamefont {Meek},
  \citenamefont {Bethlem}, \citenamefont {Conrad},\ and\ \citenamefont
  {Meijer}}]{meek2008trapping}%
  \BibitemOpen
  \bibfield  {author} {\bibinfo {author} {\bibfnamefont {S.~A.}\ \bibnamefont
  {Meek}}, \bibinfo {author} {\bibfnamefont {H.~L.}\ \bibnamefont {Bethlem}},
  \bibinfo {author} {\bibfnamefont {H.}~\bibnamefont {Conrad}}, \ and\ \bibinfo
  {author} {\bibfnamefont {G.}~\bibnamefont {Meijer}},\ }\href@noop {}
  {\bibfield  {journal} {\bibinfo  {journal} {Physical review letters}\
  }\textbf {\bibinfo {volume} {100}},\ \bibinfo {pages} {153003} (\bibinfo
  {year} {2008})}\BibitemShut {NoStop}%
\bibitem [{\citenamefont {Meek}\ \emph
  {et~al.}(2009{\natexlab{a}})\citenamefont {Meek}, \citenamefont {Conrad},\
  and\ \citenamefont {Meijer}}]{meek2009trapping}%
  \BibitemOpen
  \bibfield  {author} {\bibinfo {author} {\bibfnamefont {S.~A.}\ \bibnamefont
  {Meek}}, \bibinfo {author} {\bibfnamefont {H.}~\bibnamefont {Conrad}}, \ and\
  \bibinfo {author} {\bibfnamefont {G.}~\bibnamefont {Meijer}},\ }\href@noop {}
  {\bibfield  {journal} {\bibinfo  {journal} {Science}\ }\textbf {\bibinfo
  {volume} {324}},\ \bibinfo {pages} {1699} (\bibinfo {year}
  {2009}{\natexlab{a}})}\BibitemShut {NoStop}%
\bibitem [{\citenamefont {Meek}\ \emph
  {et~al.}(2009{\natexlab{b}})\citenamefont {Meek}, \citenamefont {Conrad},\
  and\ \citenamefont {Meijer}}]{meek2009stark}%
  \BibitemOpen
  \bibfield  {author} {\bibinfo {author} {\bibfnamefont {S.~A.}\ \bibnamefont
  {Meek}}, \bibinfo {author} {\bibfnamefont {H.}~\bibnamefont {Conrad}}, \ and\
  \bibinfo {author} {\bibfnamefont {G.}~\bibnamefont {Meijer}},\ }\href@noop {}
  {\bibfield  {journal} {\bibinfo  {journal} {New Journal of Physics}\ }\textbf
  {\bibinfo {volume} {11}},\ \bibinfo {pages} {055024} (\bibinfo {year}
  {2009}{\natexlab{b}})}\BibitemShut {NoStop}%
\bibitem [{\citenamefont {Osterwalder}\ \emph {et~al.}(2010)\citenamefont
  {Osterwalder}, \citenamefont {Meek}, \citenamefont {Hammer}, \citenamefont
  {Haak},\ and\ \citenamefont {Meijer}}]{osterwalder2010deceleration}%
  \BibitemOpen
  \bibfield  {author} {\bibinfo {author} {\bibfnamefont {A.}~\bibnamefont
  {Osterwalder}}, \bibinfo {author} {\bibfnamefont {S.~A.}\ \bibnamefont
  {Meek}}, \bibinfo {author} {\bibfnamefont {G.}~\bibnamefont {Hammer}},
  \bibinfo {author} {\bibfnamefont {H.}~\bibnamefont {Haak}}, \ and\ \bibinfo
  {author} {\bibfnamefont {G.}~\bibnamefont {Meijer}},\ }\href@noop {}
  {\bibfield  {journal} {\bibinfo  {journal} {Physical Review A}\ }\textbf
  {\bibinfo {volume} {81}},\ \bibinfo {pages} {051401} (\bibinfo {year}
  {2010})}\BibitemShut {NoStop}%
\bibitem [{\citenamefont {Meek}\ \emph {et~al.}(2011)\citenamefont {Meek},
  \citenamefont {Parsons}, \citenamefont {Heyne}, \citenamefont
  {Platschkowski}, \citenamefont {Haak}, \citenamefont {Meijer},\ and\
  \citenamefont {Osterwalder}}]{meek2011traveling}%
  \BibitemOpen
  \bibfield  {author} {\bibinfo {author} {\bibfnamefont {S.~A.}\ \bibnamefont
  {Meek}}, \bibinfo {author} {\bibfnamefont {M.~F.}\ \bibnamefont {Parsons}},
  \bibinfo {author} {\bibfnamefont {G.}~\bibnamefont {Heyne}}, \bibinfo
  {author} {\bibfnamefont {V.}~\bibnamefont {Platschkowski}}, \bibinfo {author}
  {\bibfnamefont {H.}~\bibnamefont {Haak}}, \bibinfo {author} {\bibfnamefont
  {G.}~\bibnamefont {Meijer}}, \ and\ \bibinfo {author} {\bibfnamefont
  {A.}~\bibnamefont {Osterwalder}},\ }\href@noop {} {\bibfield  {journal}
  {\bibinfo  {journal} {Review of scientific instruments}\ }\textbf {\bibinfo
  {volume} {82}},\ \bibinfo {pages} {093108} (\bibinfo {year}
  {2011})}\BibitemShut {NoStop}%
\bibitem [{\citenamefont {Akerman}\ \emph {et~al.}(2017)\citenamefont
  {Akerman}, \citenamefont {Karpov}, \citenamefont {Segev}, \citenamefont
  {Bibelnik}, \citenamefont {Narevicius},\ and\ \citenamefont
  {Narevicius}}]{akerman2017trapping}%
  \BibitemOpen
  \bibfield  {author} {\bibinfo {author} {\bibfnamefont {N.}~\bibnamefont
  {Akerman}}, \bibinfo {author} {\bibfnamefont {M.}~\bibnamefont {Karpov}},
  \bibinfo {author} {\bibfnamefont {Y.}~\bibnamefont {Segev}}, \bibinfo
  {author} {\bibfnamefont {N.}~\bibnamefont {Bibelnik}}, \bibinfo {author}
  {\bibfnamefont {J.}~\bibnamefont {Narevicius}}, \ and\ \bibinfo {author}
  {\bibfnamefont {E.}~\bibnamefont {Narevicius}},\ }\href@noop {} {\bibfield
  {journal} {\bibinfo  {journal} {Physical review letters}\ }\textbf {\bibinfo
  {volume} {119}},\ \bibinfo {pages} {073204} (\bibinfo {year}
  {2017})}\BibitemShut {NoStop}%
\bibitem [{\citenamefont {Segev}\ \emph {et~al.}(2019)\citenamefont {Segev},
  \citenamefont {Pitzer}, \citenamefont {Karpov}, \citenamefont {Akerman},
  \citenamefont {Narevicius},\ and\ \citenamefont
  {Narevicius}}]{segev2019collisions}%
  \BibitemOpen
  \bibfield  {author} {\bibinfo {author} {\bibfnamefont {Y.}~\bibnamefont
  {Segev}}, \bibinfo {author} {\bibfnamefont {M.}~\bibnamefont {Pitzer}},
  \bibinfo {author} {\bibfnamefont {M.}~\bibnamefont {Karpov}}, \bibinfo
  {author} {\bibfnamefont {N.}~\bibnamefont {Akerman}}, \bibinfo {author}
  {\bibfnamefont {J.}~\bibnamefont {Narevicius}}, \ and\ \bibinfo {author}
  {\bibfnamefont {E.}~\bibnamefont {Narevicius}},\ }\href@noop {} {\bibfield
  {journal} {\bibinfo  {journal} {Nature}\ }\textbf {\bibinfo {volume} {572}},\
  \bibinfo {pages} {189} (\bibinfo {year} {2019})}\BibitemShut {NoStop}%
\end{thebibliography}%

\end{document}